\documentclass[11pt,a4paper,titlepage,tablecaptionbelow,twoside]{scrreprt}

\usepackage{graphicx, epsfig, amssymb} 
\usepackage{amsmath, amsfonts}
\usepackage{bm} 
\usepackage{cite} 
\usepackage[format=plain,labelfont=bf]{caption}

\usepackage[linktocpage]{hyperref}
\usepackage[caption=false]{subfig}
\usepackage[usenames]{color}

\def\Lie{\mathcal{L}}
\def\A{\mathcal{A}}
\def\E{\mathcal{E}}
\def\H{\mathcal{H}}
\def\M{\mathcal{M}}

\def\al{\alpha}
\def\be{\beta}
\def\de{\delta}
\def\eps{\epsilon}
\def\ga{\gamma}
\def\ka{\kappa}
\def\la{\lambda}
\def\si{\sigma}
\def\ze{\zeta}
\def\De{\Delta}
\def\Ga{\Gamma}
\def\Si{\Sigma}
\def\p{\partial}
\def\na{\nabla}
\def\non{\nonumber}

\def\ba{\bar{a}}

\def\bA{\bar{A}}
\def\bD{\bar{D}}
\def\bK{\bar{K}}
\def\bR{\bar{R}}
\def\bga{\bar{\gamma}}
\def\bmu{\bar{\mu}}

\def\hA{\hat{A}}
\def\hga{\hat{\gamma}}
\def\ti{\tilde}
\def\tg{\tilde{\gamma}}
\def\tA{\tilde{A}}
\def\tK{\tilde{K}}
\def\tG{\tilde{\Gamma}}
\def\tD{\tilde{D}}
\def\nk{n_K}
\def\nq{n_Q}

\renewcommand{\vec}[1]{\boldsymbol{#1}}

\begin{document}
\begin{titlepage}
\begin{tabular}{p{3cm}p{14cm}}
\includegraphics[width=0.15\textwidth]{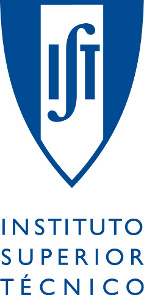} &
\vspace{-4cm}
\centering
{\Huge{Universidade T\'ecnica de Lisboa \newline
\newline
Instituto Superior T\'ecnico 
}}
\end{tabular}

\begin{center}
\vspace{0.5cm}

{\Huge{Black hole dynamics in generic spacetimes
}}\\

\vspace{1.5cm}
{\Large{Helvi Witek}}
\vspace{1.0cm}

Supervisor: Doctor V\'itor Manuel dos Santos Cardoso \\
\vspace{1.0cm}
Thesis approved in public session to obtain the PhD Degree in \\
Physics\\
\vspace{0.5cm}
Jury final classification:\\
Pass with distinction\\
\vspace{1.0cm}
Jury\\
\begin{table}[htpb!]
\begin{tabular}{ll}
Chairperson: & Chairman of the IST Scientific Board \\ 
Members of the Committee: 
 & Doctor Frans Pretorius \\
 & Doctor Jos\'e Pizarro de Sande e Lemos \\
 & Doctor Carlos Alberto Ruivo Herdeiro \\
 & Doctor V\'itor Manuel dos Santos Cardoso \\
 & Doctor Leonardo Gualtieri \\
 & Doctor Ulrich Sperhake 
\end{tabular}
\end{table}
\vspace{1.0cm}

{\Large{2012}}\\

\end{center}
\end{titlepage}
\cleardoublepage
\begin{titlepage}
\begin{tabular}{p{3cm}p{14cm}}
\includegraphics[width=0.12\textwidth]{Logo_IST_color.jpg} &
\vspace{-3cm}
\centering
{\huge{Universidade T\'ecnica de Lisboa \newline
\newline
Instituto Superior T\'ecnico 
}}
\end{tabular}

\begin{center}
\vspace{0.1cm}

{\huge{Black hole dynamics in generic spacetimes
}}\\

\vspace{0.5cm}
{\Large{Helvi Witek}}
\vspace{0.3cm}

Supervisor: Doctor V\'itor Manuel dos Santos Cardoso \\
\vspace{0.5cm}
Thesis approved in public session to obtain the PhD Degree in \\
Physics\\
\vspace{0.3cm}
Jury final classification:\\
Pass with distinction\\
\vspace{0.3cm}
{\small{Jury}}\\
\begin{table}[htpb!]
\begin{tabular}{lll}
{\small{Chairperson:}} & {\small{Chairman of the IST Scientific Board}} & \\
{\small{Members of}} & & \\
{\small{the Committee: }}
 & {\small{Doctor Frans Pretorius,}} &{\small{Professor Catedr\'atico da Princeton}} \\
                                     & &{\small{University, Estados Unidos da Am\'erica }}\\
 & {\small{Doctor Jos\'e Pizarro de Sande e Lemos,}} &{\small{Professor Catedr\'atico do Instituto Superior}} \\
        & & {\small{T\'ecnico, da Universidade T\'ecnica de Lisboa }}\\
 & {\small{Doctor Carlos Alberto Ruivo Herdeiro,}} & {\small{Professor Auxiliar (com Agrega\c{c}\~{a}o),}}\\
        & &{\small{da Universidade de Aveiro}}\\
 & {\small{Doctor  V\'itor Manuel dos Santos Cardoso,}} & {\small{Professor Auxiliar do Instituto Superior}}\\
        & &{\small{T\'{e}cnico, da Universidade T\'{e}cnica de Lisboa}} \\
 & {\small{Doctor Leonardo Gualtieri,}} & {\small{Investigador da Universit\`a degli Studi}}\\
        & &{\small{di Roma ``La Sapienza'', It\'alia}} \\
 & {\small{Doctor Ulrich Sperhake,}} & {\small{Investigador do Instituto Superior T\'ecnico}}\\
        & &{\small{(CENTRA), da Universidade T\'ecnica de Lisboa}} 
\end{tabular}
\end{table}
\begin{table}[htpb!]
\begin{tabular}{ll}
{\small{Funding Institution: }}& {\small{Funda\c{c}\~ao Para a Ci\^{e}ncia e a T\'ecnologia (FCT) -- Portugal,}}\\
        & {\small{grant SFRH/BD/46061/2008}}
\end{tabular}
\end{table}

{\large{2012}}\\

\end{center}
\end{titlepage}
\cleardoublepage
\newpage
\pagenumbering{Roman}
\chapter*{Resumo}
\subsection*{T\'itulo: Din\^amica de buracos negros em espa\c{c}os-tempo gen\'ericos}
\subsection*{Nome: Helvi Witek}
\subsection*{Doutoramento em F\'isica}
\subsection*{Orientador: Doutor V\'itor Manuel dos Santos Cardoso}
\subsection*{Resumo:}

A din\^amica de buracos negros desempenha um papel fulcral em astrof\'isica, f\'isica de altas energias
e f\'isica fundamental. Esta tese foca-se na din\^amica de buracos negros em espa\c{c}os-tempo gen\'ericos,
em particular na extens\~{a}o de m\'etodos de relatividade num\'erica para espa\c{c}os-tempo de 
diferente dimensionalidade ou com outro tipo de comportamento assimpt\'otico.

No \^ambito da f\'isica de altas energias, apresentamos um c\'odigo num\'erico, o \textsc{HD-Lean}, desenvolvido para
simular a interac\c{c}\~{a}o de buracos negros em espa\c{c}os-tempo gen\'ericos. Este c\'odigo foi usado
para investigar a colis\~{a}o frontal de buracos negros em espa\c{c}os-tempo de dimensionalidade
D=4,5,6, bem como calcular a emiss\~{a}o de radia\c{c}\~{a}o gravitacional para este processo.
No contexto da dualidade gauge/gravity, estudamos a din\^amica de buracos negros 
em ``caixas'', com o objectivo de simular o comportamento de espa\c{c}os-tempo anti-de Sitter.

No \^{a}mbito da astrof\'isica, bin\'arias de buracos negros s\~{a}o uma das principais fontes 
de radia\c{c}\~{a}o gravitacional. O nosso estudo mostrou pela primeira vez que a colis\~{a}o de 
dois buracos negros de massas $M_1,M_2,\,M2\ll M$ pode tamb\'em ser feita em Relatividade num\'erica,
apesar das diferentes escalas envolvidas no processo. Al\'em do mais mostramos acordo excelente
com c\'alculos perturbativos, fechando de alguma forma o fosso entre um e outro m\'etodo.

Finalmente, na \'ultima parte da tese revisitamos o problema do espalhamento de ondas
em buracos negros com rota\c{c}\~{a}o. Campos escalares com massa d\~{a}o origem a uma instabilidade superradiante,
com potenciais consequ\^{e}ncias observacionais se a sua massa estiver no intervalo $10^{-22}-10^{-10}eV$.

\paragraph{Palavras-chave:} 
Buracos negros, 
relatividade num\'erica,
relatividade geral,
f\'isica de altas energias,
espa\c{c}os-tempo de altas dimensionalidade,
ondas gravitacionais,
dualidade gauge/ gravity,
estabilidade de buracos negros,
instabilidade de superradi\^{a}ncia,
axi\~{o}es

\newpage\mbox{}
\chapter*{Abstract}
\subsection*{Title: Black hole dynamics in generic spacetimes}
\subsection*{Name: Helvi Witek}
\subsection*{PhD in Physics}
\subsection*{Supervisor: Doutor V\'itor Manuel dos Santos Cardoso}
\subsection*{Abstract:}

The dynamics of black hole spacetimes play a crucial role in astrophysics, high energy physics
and fundamental physics. In this thesis I have investigated the dynamics of black holes 
in generic spacetimes by extending established numerical relativity methods to 
higher dimensional or non-asymptotically flat spacetimes.

In the high-energy context, I have developed \textsc{HD-Lean}, a $3+1$ numerical code to evolve BHs 
in higher dimensional spacetimes. \textsc{HD-Lean} was used to simulate (low energy) 
head-on collision of black holes in $D=5$ and $D=6$ spacetime dimensions 
and to understand the gravitational wave emission throughout the process.
In the specific context of the gauge/ gravity duality, I have performed important
steps towards numerical evolutions in asymptotically AdS spacetimes by investigating BH configurations confined by a box, 
thus mimicking the AdS asymptotics.

In an astrophysical setting, most realistic BH binary configurations consist of unequal-mass binaries. 
I have shown that numerical relativity is able to perform the study
of very low-mass ratio binaries, with mass-ratios down to $1/100$. Our results are in excellent agreement
with perturbative calculations, thus finally bridging the gap between the two different techniques.

The last part of this thesis revisits the behaviour of 
massive scalar fields in the background of BHs. Massive scalar fields around Kerr BHs 
give rise to the superradiant instability and might therefore have important 
astrophysical and observational implications, if their mass is in the range $10^{-22}-10^{-10}eV$. 
These ultra-light fields might be realized in the ``axiverse'' scenario, consisting of 
ultra-light bosonic states emerging from string-theory compactifications.

\paragraph{Keywords:} 
Black holes, 
numerical relativity,
general relativity,
high energy physics,
higher dimensional gravity,
gravitational waves,
gauge/ gravity duality,
stability of black holes,
superradiant instability,
axions


\newpage\mbox{}
\chapter*{Acknowlegdements}
I am indepted to a number of people without whose support this work would not have been possible.

First of all, I would like to warmly thank my PhD supervisor Prof. V\'itor Cardoso 
for many illuminating discussions and all the physics that I've learned in the last four years. 
I am grateful for your constant support and encouragement and the many opportunities that you have made possible.
It has been a great pleasure to work with you.
You are an exceptional supervisor and I could not have wished for a better one. Thank you, V\'itor.

I am grateful to my closest collaborators 
Leonardo Gualtieri, Carlos Herdeiro, Andrea Nerozzi, Ulrich Sperhake, and Miguel Zilh\~{a}o,
for our many fruitful discussions on this exciting project. It has been a great pleasure to 
collaborate with you.

I would like to thank my long-time collaborator David Hilditch for our many enlightening discussions 
and interesting projects.

I am indebted to Andrea Nerozzi and Paolo Pani for carefully proof-reading this thesis.

It is my pleasure to thank all members of CENTRA and especially our gravity group who provided an exceptional
and inspiring working environement.
Special thanks to Dulce Concei\c{c}\~{a}o and Rita Sousa who have been a constant support and  
never got tired to help me through the burocracy jungle.
I would like to thank Sergio Almeida for building up and maintaining our cluster ``Baltasar'' 
and for being a great aid in all computer related problems.

I am thankful to my office-mates Andrea Nerozzi, Jorge Rocha and Antonino Flachi for 
our many inspiring discussions and for creating such a great working environment.

I would like to thank 
Emanuele Berti, Terence Delsate, Akihiro Ishibashi, Hirotada Okawa, Paolo Pani and Jan Steinhoff 
for many fruitful and interesting dicussions.

I acknowlegde the kind hospitality of the 
TAPIR group at Caltech,
the gravitational physics group at OleMiss, the University of Mississippi,
the Numerical Relativity group at the Friedrich-Schiller University Jena, 
and the Yukawa Institute for Theoretical Physics at Kyoto University,
where parts of this work have been accomplished.

I warmly thank my family, Kerstin and Johannes and Manfred and Helga Witek, 
for your constant support, inspiration and unconditional love. I would not stand where I am today without you.

I am happy to thank my long-standing friends Tine, Anke and Stephan for your unconditional friendship
and support.


\newpage\mbox{}
\tableofcontents 

\chapter*{Preamble}\label{chapter:Preamble}
The research presented in this thesis has been carried out at the 
Centro Multidisciplinar de Astrof\'isica (CENTRA) at the Instituto Superior T\'ecnico / 
Universidade T\'ecnica de Lisboa. 

I declare that this thesis is not substantially the same as any that I have submitted for a 
degree, diploma or other qualification at any other university and that no part of it has 
already been or is concurrently submitted for any such degree, diploma or other qualification.

Chapter~\ref{chapter:UnequalMass4D}, presenting head-on collisions of unequal mass black holes 
with small mass ratios in four dimensional spacetimes, is the result of the collaboration with 
Ulrich~Sperhake, Vitor~Cardoso, Christian~D.~Ott and Erik~Schnetter \cite{Sperhake:2011ik}.

Chapters~\ref{chapter:HigherDimF} and~\ref{chapter:HigherDimR}, summarizing 
a numerical framework for the evolution of BHs in higher dimensional spacetimes
and presenting corresponding results of BH collisions,
are the outcome of the collaboration with 
Vitor~Cardoso, Leonardo~Gualtieri, Carlos~Herdeiro, Andrea~Nerozzi, 
Ulrich~Sperhake, Miguel~Zilhao and Marcus~Ansorg published in 
\cite{Zilhao:2010sr,Witek:2010xi,Witek:2010az,Zilhao:2011yc}
as well as ongoing work in collaboration with 
David~Hilditch, Hirotada~Okawa and Masaru~Shibata \cite{Hilditch2012,HDFcoll2012}.

Chapter~\ref{chapter:BBbox}, devoted to the investigation of BH binaries enclosed by 
a perfectly reflecting mirror mimicking AdS spacetimes,
is the result of the collaboration with 
Vitor~Cardoso, Carlos~Herdeiro, Andrea~Nerozzi, Ulrich~Sperhake and Miguel Zilhao
\cite{Witek:2010qc}.

Chapter~\ref{chapter:Stability}, exploring stability properties of BHs 
in the presence of a massive scalar field,
which has been part of a larger project in collaboration with
Vitor~Cardoso, Akihiro~Ishibashi and Ulrich~Sperhake \cite{Witek:2012tr}.

Most of these chapters have been published. Some chapters present results of ongoing work in progress as indicated.
The publications included in this thesis are 
\begin{enumerate}
\item
U.~Sperhake, V.~Cardoso, C.~D.~Ott, E.~Schnetter and \underline{H.~Witek},
``Extreme black hole simulations: collisions of unequal mass black holes and the point particle limit,''
Phys.\ Rev.\ D {\bf 84} (2011) 084038, [arXiv:1105.5391 [gr-qc]]. (Chapter~\ref{chapter:UnequalMass4D})
\item
M.~Zilhao, \underline{H.~Witek}, U.~Sperhake, V.~Cardoso, L.~Gualtieri, C.~Herdeiro and A.~Nerozzi,
``Numerical relativity for D dimensional axially symmetric space-times: formalism and code tests,''
Phys.\ Rev.\ D {\bf 81} (2010) 084052, [arXiv:1001.2302 [gr-qc]]. 
(part of Chapters~\ref{chapter:HigherDimF} and~\ref{chapter:HigherDimR})
\item
\underline{H.~Witek}, M.~Zilhao, L.~Gualtieri, V.~Cardoso, C.~Herdeiro, A.~Nerozzi and U.~Sperhake,
``Numerical relativity for D dimensional space-times: head-on collisions of black holes and gravitational wave extraction,''
Phys.\ Rev.\ D {\bf 82} (2010) 104014, [arXiv:1006.3081 [gr-qc]].
(part of Chapters~\ref{chapter:HigherDimF} and~\ref{chapter:HigherDimR})
\item
\underline{H.~Witek}, V.~Cardoso, L.~Gualtieri, C.~Herdeiro, U.~Sperhake and M.~Zilhao,
``Head-on collisions of unequal mass black holes in D=5 dimensions,''
Phys.\ Rev.\ D {\bf 83} (2011) 044017, [arXiv:1011.0742 [gr-qc]].
(part of Chapters~\ref{chapter:HigherDimF} and~\ref{chapter:HigherDimR})
\item 
M.~Zilhao, M.~Ansorg, V.~Cardoso, L.~Gualtieri, C.~Herdeiro, U.~Sperhake and \underline{H.~Witek},
``Higher-dimensional puncture initial data,''
Phys.\ Rev.\ D {\bf 84} (2011) 084039, [arXiv:1109.2149 [gr-qc]].
(part of Chapters~\ref{chapter:HigherDimF} and~\ref{chapter:HigherDimR})
\item
\underline{H.~Witek}, V.~Cardoso, C.~Herdeiro, A.~Nerozzi, U.~Sperhake and M.~Zilhao,
``Black holes in a box: towards the numerical evolution of black holes in AdS,''
Phys.\ Rev.\ D {\bf 82} (2010) 104037, [arXiv:1004.4633 [hep-th]].
(Chapter~\ref{chapter:BBbox})
\item
\underline{H.~Witek}, D.~Hilditch and U.~Sperhake,
``Stability of the puncture method with a generalized BSSN formulation,''
Phys.\ Rev.\ D {\bf 83}, (2011) 104041, [arXiv:1011.4407 [gr-qc]].
(part of Chapter~\ref{chapter:NRframework}
\item
\underline{H.~Witek}, V.~Cardoso, A.~Ishibashi and U.~Sperhake,
``Superradiant instabilities in astrophysical systems,''
Phys.\ Rev.\ D {\bf 87}, 043513 (2013), [arXiv:1212.0551 [gr-qc]].
(part of Chapter~\ref{chapter:Stability})
\end{enumerate}

\clearpage
\pagenumbering{arabic}

\chapter{Introduction}\label{chapter:Intro}
Black holes (BHs) are among the most intriguing predictions of General Relativity (GR) 
and alternative theories of gravity.
They play a key role in many areas of physics reaching from astrophysics, cosmology and gravitational wave astronomy
over high energy physics and the gauge/gravity duality to fundamental mathematical questions concerning their
existence and stability properties.
Of utmost interest are the dynamics of BHs in the fully non-linear, strong-field regime of GR.
Because of the complex structure of Einstein's equations, a set of $D(D+1)/2$ coupled, non-linear partial differential equations (PDEs),
this regime can only be explored by employing Numerical Relativity (NR) techniques.
Since the outstanding breakthrough in 2005 and 2006 \cite{Pretorius:2005gq,Campanelli:2005dd,Baker:2005vv}
NR has grown into a mature research field
capable to model a plethora of exciting phenomena 
(see e.g.
\cite{Pretorius:2007nq,Centrella:2010mx,Hinder:2010vn,Sperhake:2011xk,
Yoshino:2011zz,Yoshino:2011zza,Cardoso:2012qm,Sperhake:2013qa}
for recent reviews).

In this thesis I have explored BH dynamics in generic spacetimes,
including BH collisions in four and higher dimensional asymptotically flat spacetimes,
BH inspirals in a box, thus  mimicking asymptotically Anti-de Sitter (AdS) spacetimes,
and the stability of BHs against massive scalar field perturbations.
In the following I give a brief overview and motivation for each of these topics.

\paragraph{Astrophysics and gravitational wave astronomy}

Since the first discovery of BHs as solutions to Einstein's Equations by Schwarzschild in 1916
they have outgrown their status of merely exotic theoretical predictions of GR 
and are nowadays known to make up an important class of astronomical objects.
Supermassive BHs (SMBHs) with $10^6-10^9M_{\odot}$ are expected to be hosted at the center of most 
galaxies~\cite{Rees:1984si,Begelman:1980vb,Ferrarese:2004qr};
in fact there is strong observational evidence for our very own SMBH in Sagittarius ${\rm{A}}^{\ast}$
with $M\sim 4.2\cdot10^{6}M_{\odot}$ in the centre of the Milky Way.
Their solar-mass counterparts with $3-30M_{\odot}$
are part of the galaxies population~\cite{McClintock:2009as,Antoniadis:2013pzd,Seoane:2013qna}.
Their motion is dominated by gravitational interactions and the collision of two BHs is among the 
most violent scenarios in Nature, releasing a burst of energy in form of gravitational radiation. 
Moreover, gravitational waves emitted throughout such a collision provide the unique opportunity
to explore and test the strong-field regime of GR or extensions thereof -- an as yet still open 
issue in gravitational physics.
Once detected, gravitational waves 
will open up a completely new window for astronomical observations,
thus offering invaluable astrophysical insight into strong-field phenomena of gravity
that are accessable in no other way.
First detections of gravitational waves are expected within the next decade. 
Gravitational wave observatories, such as pulsar timing arrays~\cite{epta, ipta, nanograv}
or ground-based interferometric GW detectors such as 
LIGO \cite{Abbott:2007kv}, Virgo \cite{Acernese:2008zzf} and GEO600 \cite{Luck:2006ug} 
are in place.
The second generation, advanced LIGO-VIRGO detector network is expected 
to be up and running in 2015 \cite{Abadie:2011kd,aLIGO}
while KAGRA~\cite{Aso:2013eba,Somiya:2011np} is under construction.
Third generation ground-based detectors such as the Einstein Telescope~\cite{Punturo:2010zz}
are under development 
and space-based detectors in the spirit of the eLISA/NGO \cite{AmaroSeoane:2012km}
operation might become available in the future.
Although BH binaries are suspected to be among the strongest sources for GWs, their detection poses an
extremely challenging task.
The expected GW signal arriving at Earth is extremely weak;
the dimensionless strain, or amplitude, is expected to be of the order $h=10^{-20}$.
For the LIGO detectors with an arm length of $L\sim3km$ this strain translates into 
a variation in length of $\Delta L \sim 10^{-17}m$ -- two orders of magnitude smaller than the size of a proton.
In order to facilitate the measurement of these very weak signals buried under all kinds of noise,
such as seismic noise from the Earth, thermic noise of the test masses and radiation pressure from the laser light,
matched-filtering techniques are mandatory.
This technique aims at filtering real signals out of the noise background via the comparison 
with theoretical templates.
Therefore a detailed theoretical understanding of the GW signal from the inspiral, merger and ringdown phase 
is of utmost importance.
While the early stages of an inspiraling BH binary, when $v/c\ll1$, 
can be modelled well by analytic post-Newtonian (PN) and effective one-body (EOB) approaches
and the final, after-merger ringdown phase is approximated well by perturbative approaches,
both (semi-analytic) methods break-down in the highly dynamical merger phase.
In order to accurately model the actual collision between the two BHs, the fully non-linear 
system of Einstein's equations has to be solved by employing numerical relativity methods.

Since the remarkable breakthrough in 2005 \cite{Pretorius:2005gq} 
followed by \cite{Campanelli:2005dd,Baker:2005vv} shortly afterwards,
NR has grown into a mature research field now able to evolve numerous configurations
of BH binaries.

In Chapter~\ref{chapter:NRframework}, 
I will give a brief overview on the most commonly employed techniques in NR
in $4$-dimensional, asymptotically flat spacetimes -- the setup of choice  to simulate
astrophysical BH configurations.
In particular, I will briefly summarize the $3+1$-splitting of spacetime
into $3$-dimensional spatial hypersurfaces that will be evolved in time.
Subsequently I review the $3+1$-decomposition of Einstein's equations and their particular 
formulation yielding (numerically) long-term stable simulations, the construction of initial data
and the extraction of physically meaningful quantities.

In Chapter~\ref{chapter:UnequalMass4D} I present numerical simulations
of a particular BH binary setup. Specifically, I have evolved head-on collisions of unequal mass
BHs with a mass ratio varying from $q=1,\dots,1/100$.
Thus, I have been able to fill the gap between point particle calculations valid in the small
mass ratio regime $q\ll1$ and a fully numerical modelling of the intermediate to comparable mass ratio regime.
%

\paragraph{High energy physics}
In recent years a further exciting application of BH physics in the highly dynamical non-linear regime has come into
focus of the NR community: the intriguing possibility of BH production in high-energy particle colliders or
interactions of ultra-high energetic cosmic rays with the atmosphere.
If particles collide at very high energies such that the center of mass energy is well beyond the Planck scale,
gravity becomes the dominant interaction. 
Then, due to Thorne's hoop conjecture \cite{Thorne:1972ji} and ``no-hair'' theorem like arguments
\cite{Chrusciel:2012jk},
the particular nature of these particles should not be important to describe their collision or, 
figuratively speaking, {\textit{matter does not matter}}. 
A fully non-linear, numerical study by Choptuik \& Pretorius \cite{Choptuik:2009ww}
and East \& Pretorius~\cite{East:2012mb}
has provided compelling evidence for this statement.
The authors have simulated ultra-relativistic particle collisions by 
the collision of, respectively, bosonic solitons and fluid balls with generic initial boost
and observed BH formation above a critical boost parameter of $\gamma_c=2.9$ in the former
and $\gamma_{c}=8.5$ in the latter case.
Therefore, the trans-Planckian scattering of particles is well described by scattering processes of 
highly boosted black holes.
This fact gains a lot of importance
in the context of so-called TeV gravity scenarios. 
In order to solve the hierarchy problem, higher dimensional theories of gravity have been proposed,
in which the fundamental Planck scale is of the order of TeV
\cite{Antoniadis:1990ew,ArkaniHamed:1998rs,Antoniadis:1998ig,Randall:1999ee,Randall:1999vf}.
This offers the exciting possibility that BHs could be produced, e.g., at the Large Hadron Collider (LHC) 
or in ultra high energy cosmic ray interactions with the atmosphere 
\cite{Banks:1999gd,Giddings:2001bu,Dimopoulos:2001hw,Ahn:2002mj,
Ahn:2003qn,Feng:2001ib, Cardoso:2004zi, Cavaglia:2002si,Kanti:2004nr,Solodukhin:2002ui}. 

Recent searches for micro-BHs at Atlas~\cite{Aad:2012ic} and CMS~\cite{Chatrchyan:2012taa}
put new constraints on BH production and excluded semi-classical or quantum BHs with 
$M\sim4.3\ldots6.3TeV$~\cite{Chatrchyan:2013xva}.

Of particular importance is the computation of the energy released in form of
gravitational radiation during the collision, which in turn
determines the mass and angular momentum of the newly born BH.
A second, crucial parameter is the cross-section of BH collision
to provide estimates on the BH production rate.
Lower bounds on the BH mass have been obtained from area theorem arguments 
\cite{Yoshino:2002br,Yoshino:2002tx}.
Estimates on the radiated energy have been provided by 
computations of (highly) relativistic point particles falling into a higher dimensional BH
\cite{Berti:2003si,Berti:2010gx},
as well as shock wave collisions describing the ultra-relativistic regime 
\cite{Herdeiro:2011ck,Coelho:2012sy,Coelho:2012AA,Sampaio:2013faa}.
Instead, I have aimed at  complementary calculations of the 
energy emitted throughout the collision of BHs in higher dimensional spacetimes.
In particular, I have attempted the challenging goal to evolve numerically 
collisions of comparable mass BHs in the fully dynamical, non-linear regime,
because proton-proton collisions are the most likely type of collisions yielding BH formation.
In Chapter~\ref{chapter:HigherDimF} I present the framework based on the dimensional reduction and 
the numerical implementation 
enabling to explore higher dimensional BH spacetimes numerically.
In Chapter~\ref{chapter:HigherDimR} I summarize our results of these simulations,
in particular benchmark tests of the code as well as 
head-on collisions of equal-mass BHs in $D=5$ and $D=6$ spacetime dimensions and 
of unequal-mass BHs with comparable mass ratios in $D=5$.

\paragraph{Gauge/gravity duality}
Another fascinating application of BH physics is provided by the gauge/gravity duality
\cite{Maldacena:1997re,Witten:1998qj}.
Loosely speaking, this correspondence relates a theory of gravity in an Anti-de Sitter (AdS) spacetime to a gauge theory 
``living'' on the boundary of this AdS spacetime. 
For example, it has been found that the Hawking-Page phase 
transition of black holes in AdS, i.e., the transistion between thermally stable and unstable BHs
\cite{Hawking:1982dh},
is dual to the deconfinement/ confinement transition in a strongly coupled, thermal gauge theory
\cite{Witten:1998qj}.
Therefore, the analysis of BHs in AdS provides an appealing tool to study particles
in a gauge theory.
Eventually, these calculations may find applications in the description
and understanding of high energy collisions of heavy ions in the Relativistic Heavy Ion Collider (RHIC).
In the past couple of years first (fully) numerical works on the gravity side, i.e., of asymptotically AdS spacetimes
have been presented.
These studies include:
\noindent{(i)} simulations of shock wave collisions in AdS \cite{Chesler:2010bi,Chesler:2013cqa},
\noindent{(ii)} evolutions of scalar fields in AdS which has
lead to turbulent-like instabilities and BH formation \cite{Bizon:2011gg}
or time-periodic solutions~\cite{Maliborski:2013jca}
and \noindent{(iii)}
numerical studies of prompt BH formation through the ringdown phase which results in a static BH \cite{Bantilan:2012vu}. 
Because of the active role played by the AdS boundary, perturbations of 
(small) Kerr-AdS BHs are expected to give rise to superradiant instabilities
\cite{Hawking:1999dp,Cardoso:2004hs,Cardoso:2006wa}.
It has been shown in \cite{Cardoso:2004hs} that this superradiant instability 
can be understood as ``BH bomb'' like instability first suggested by
Press \& Teukolsky \cite{Press:1972zz} and, in a different setup, by Zel'dovich \cite{zeldovich1,zeldovich2}.

The key idea is that a wave packet with frequency $\omega$ impinging on a Kerr BH 
gets amplified due to superradiant scattering if its frequency satisfies
the superradiant condition $\omega<m\Omega_H$, 
where $m$ is the azimuthal wave quantum number and $\Omega_H$ the angular momentum
of the BH horizon.
Now, in the original gedanken experiment \cite{zeldovich1,zeldovich2,Press:1972zz},
the system has been surrounded by a perfectly reflecting cavity, at which the wave is reflected.
Upon subsequent amplification due to superradiant scattering and reflection at the mirror,
the field grows without bound and triggers the system to become unstable.

Whereas this setup at first glance appears somewhat artificial, such a ``mirror'' is 
naturally provided by the AdS boundary, 
thus yielding the superradiant instability of small Kerr-AdS BHs 
\cite{Hawking:1999dp,Cardoso:2004hs,Cardoso:2006wa}.

Furthermore it has been stated \cite{Cardoso:2004hs} that the global BH-AdS spacetime can be
{\textit{mimicked}} by an asymptotically flat BH spacetime {\textit{in a finite-size box}},
which plays the role of the active boundary.

Because asymptotically AdS spacetime pose a challenging problem 
I have started with a toy model in order to study the effects of the boundary.
Therefore, I have exploited the aforementioned relation and performed numerical simulations 
in $3+1$-dimensions
of BH binaries surrounded by a mirror-like box. 
The framework and results are presented in Chapter~\ref{chapter:BBbox}.
The goal of this project has been twofold:
\noindent{(i)} it has represented first fundamental steps towards 
actual simulations of fully dynamical BH-AdS spacetime;
and \noindent{(ii)} it has offered the unique potential to study,
fully non-linearly,
a ``BH bomb'' like setup \cite{Press:1972zz}
and the interaction of a rotating BH with gravitational perturbations.

\paragraph{Stability of BHs in astrophysical systems}
The ``BH bomb'' scenario in real physical setups cannot only be achieved by asymptotically AdS spacetimes but also
by massive fields surrounding Kerr BHs.
In this case, the mass term $\mu_S$ effectively acts as a mirror if 
$\omega\le\mu_S$, where $\omega$ is the frequency of the field 
\cite{Damour:1976kh,Zouros:1979iw,Detweiler:1980uk,Furuhashi:2004jk,Dolan:2007mj,Cardoso:2005vk,Hod:2009cp,Barranco:2012qs}.

Whereas this effect is very weak for known standard model particles in astrophysical environments,
it might become astrophysically significant 
if there exist fields with small, but non-vanishing mass.
Fascinating candidates are so-called axions.
Axions are ultra-light bosonic states emerging in string theory compactifications,
which have not been ruled out by experiments.
Recently the ``axiverse'' scenario has been suggested, implying the existence of a plethora of axionic particles 
covering a mass range from $10^{-33}eV\le\mu_S\le10^{-8}eV$ 
\cite{Arvanitaki:2009fg,Arvanitaki:2010sy,Kodama:2011zc}.
The existence of such ultra-light bosonic states might have a vast number of 
astrophysical implications and observational signatures.
These implications include modifications of the cosmic microwave background (CMB),
or modification of a gravitational wave signal due to the creation 
of ``gravitational atoms'' -- bosonic clouds in the vicinity of BHs. 
If the axion mass is in the range $10^{-22}eV\le\mu_S\le10^{-10}eV$, 
the time scales for the superradiant instability becomes astrophysically significant
and its interaction with astrophysical BHs is expected to result in observable signatures
\cite{Arvanitaki:2009fg,Arvanitaki:2010sy,Cardoso:2011xi,Yunes:2011aa,Alsing:2011er,Kodama:2011zc,Yoshino:2012kn,Mocanu:2012fd}.
Thus, axions have the fascinating potential to facilitate high precision BH physics 
\cite{Arvanitaki:2009fg,Arvanitaki:2010sy,Kodama:2011zc}.

With this plethora of applications in mind, I have investigated massive
scalar fields in $4$-dimensional, asymptotically flat BH backgrounds, 
which I present in Chapter~\ref{chapter:Stability}.
The main motivation to revisit the superradiant instabilities of massive scalar fields,
that has been studied extensively in the frequency domain
\cite{Press:1972zz,Damour:1976kh,Cardoso:2004nk,Cardoso:2005vk,Dolan:2007mj,Rosa:2009ei,Cardoso:2011xi,Yoshino:2012kn},
in the {\textit{time domain}} 
has been twofold:
\noindent{(i)}
By developing the code presented in Chapter~\ref{chapter:Stability} 
to evolve massive scalar fields in BH backgrounds, I have
accomplished first important steps towards more generic setups.
These scenarios include
non-linear self-interaction terms of the scalar field as well as a fully non-linear evolution
modelling backreaction effects and are subject to future investigations.

\noindent{(ii)}
Before the most recent studies by Yoshino \& Kodama \cite{Yoshino:2012kn}
and Dolan~\cite{Dolan:2012yt}
there has been only one attempt to investigate
the massive scalar field instability in the time domain
by Strafuss \& Khanna \cite{Strafuss:2004qc}. 
Some of their conclusions, however, are in stark contradiction to previous
computations in the frequency domain~\cite{Cardoso:2005vk,Dolan:2007mj}.
I have found a mechanism that has the potential to explain these contracdictory results.
I will discuss the underlying effect in detail in Chapter~\ref{chapter:Stability}
based on Ref.~\cite{Witek:2012tr}.


\chapter{Review of Numerical Relativity in $D=4$}\label{chapter:NRframework}
The basis of most NR schemes is the reformulation of Einstein's equations as time 
evolution problem. At the core of this formulation lies the splitting of spacetime into
spacelike hypersurfaces labelled by a time parameter. 
In this chapter I give a brief review of the key concepts of the $3+1$-approach 
NR.
For the sake of illustration I focus on $4$-dimensional spacetimes
and will discuss modifications of the approach to
For a detailed overview and summary of the subject I refer the reader to
\cite{York1979,Baumgarte:2002jm,Pretorius:2007nq,Gourgoulhon:2007ue,Alcubierre:2008,Centrella:2010mx,
Hinder:2010vn,Sperhake:2011xk,Sarbach:2012pr} and references therein.

Whereas spacetime is usually considered as an entity, where space and time are on equal footing,
it is mandatory to split them in order to investigate the time evolution of a physical
system with some given initial configuration.
There are various alternative ways to perform this splitting, such as the 
characteristic approach, in which the spacetime is foliated by null hypersurfaces 
(see e.g. \cite{Winicour:2005ge} for a review)
or the generalized harmonic formulation, in which the Einstein's equations are formulated 
as a set of wave equations together with harmonic coordinates \cite{Pretorius:2004jg,Pretorius:2006tp}.
The most commonly employed approach in NR is the $3+1$-approach
first introduced by Arnowitt, Deser and Misner \cite{Arnowitt:1962hi} and later reformulated by York \cite{York1979}.
In this approach the equations of motion (EoM) are formulated as a Cauchy problem, i.e., 
as an initial (boundary) value problem.

In this chapter, I will summarize the $3+1$-decomposition of the spacetime 
and Einstein's equations in Sec.~\ref{sec:NRdecomp}.
In Sec.~\ref{sec:NRInitData} I give an overview on techniques to construct initial data for BHs spacetimes
within this framework. 
In Sec.~\ref{sec:NRgauge} I review the most common gauge choices describing the employed coordinate system.
In Sec.~\ref{sec:NRgBSSN} I will summarize the generalized Baumgarte-Shapiro-Shibata-Nakamura scheme
of the $3+1$-form of Einstein's equations, which represents a well-posed initial value problem of
the system.
Finally,  in Sec.~\ref{sec:NRphys}, I summarize techniques to extract physically meaningful quantities.

\section{3+1-decomposition}\label{sec:NRdecomp}
\subsection{Foliation}\label{ssec:NRfoliation}
\begin{figure}[htpb!]
\begin{center}
\includegraphics[width=0.5\textwidth]{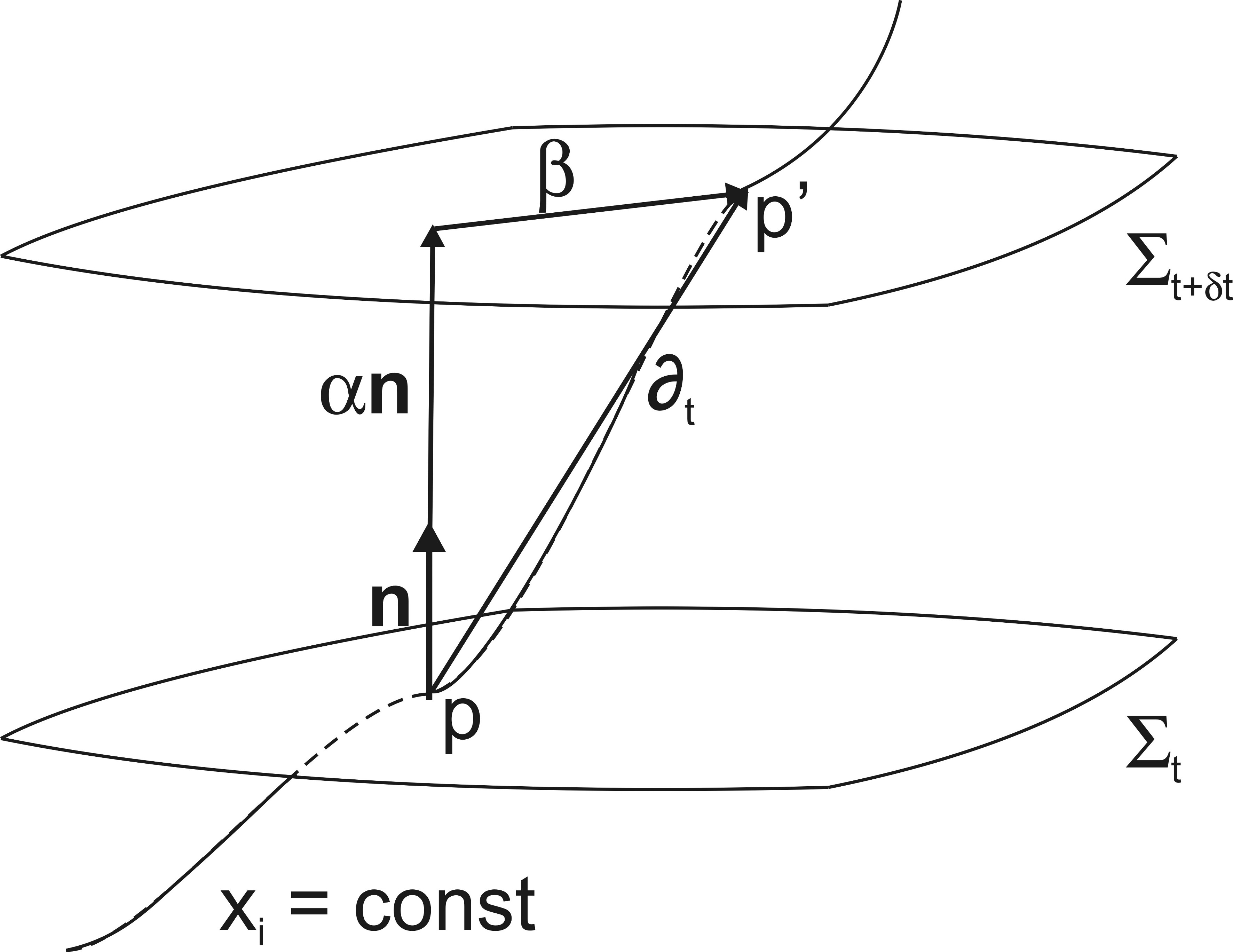}
\end{center}
\caption{\label{fig:NRfoliation}
Illustration of the foliation of a spacetime $\M$ into spatial hypersurfaces $\Si_t$ and $\Si_{t+\de t}$.
Depicted is also the vector $t^{\mu} = \al n^{\mu} + \be^{\mu}$ pointing from a point $p\in \Si_t$ to 
a point $p'\in \Si_{t+\de t}$ on the next hypersurface.
}
\end{figure}

Any globally hyperbolic (here $4$-dimensional) spacetime $\M$ can be foliated into 
purely spatial, $3$-dimensional hypersurfaces $\Si_t$ labeled by the timelike parameter $t$,
as depicted in Fig.~\ref{fig:NRfoliation}.
Then, the geometry of this region of spacetime is determined by
the induced $3$-metric $\ga_{ij}$, which measures the proper distance $dl^2 = \ga_{ij} dx^i dx^j$ within the
spatial hypersurface $\Si_t$,
the lapse function $\al$, which measures the elapsed proper time as measured by an observer
moving along the normal vector $n^{\mu}$ between two hypersurfaces (Eulerian observer),
and the shift vector $\be^{\mu} = (0,\be^i)$ which determines the relative velocity between 
the Eulerian observer and lines of constant spatial coordinates.
The lapse function $\al$ and the shift vector $\be^i$ encode the information about the coordinates 
of the spacetime and are a manifestion of the coordinate degree of freedom in GR, thus freely specifiable.
The vector $t^{\mu}$ pointing from a point $p$ in $\Si_t$ to a point $p'$ on the next slice $\Si_{t+\de t}$ 
is constructed from
\begin{align}
\label{eq:NRtime}
t^{\mu} = \al n^{\mu} + \beta^{\mu}
\,,
\end{align}
where $n^{\mu}$ is the vector normal to the spatial hypersurface $\Si_t$ with
\begin{align}
n_{\mu} = &  (-\al,0,0,0)
\,,\quad
n^{\mu} = \frac{1}{\al}(1,-\be^i)
\,,\quad
n_{\mu} n^{\mu} = -1
\,.
\end{align}
Note, that we have choses the last relation such that $n^{\mu}$ is timelike.
In other words, the lapse function $\al$ relates the coordinate time $t$ that labels the slices of the foliation 
to the time measured by an Eulerian observer. Then, the expression of the $4$-acceleration $a_{\mu}$ of this
observer in terms of the lapse writes
\begin{align}
\label{eq:NRacc}
a_{\mu} = & n^{\nu}\na_{\nu} n_{\mu} 
= \frac{1}{\al} D_{\mu}\al = D_{\mu} \ln\al
\,.
\end{align}
The line element of the $4$-dimensional spacetime is obtained from
\begin{align}
\label{eq:NRlineeleg}
ds^2 = & g_{\mu\nu}dx^{\mu}dx^{\nu} 
= (-\al^2 + \be^k\be_k) dt^2 + 2\beta_i dt dx^i + \ga_{ij} dx^i dx^j
\,.
\end{align}
From the line element~\eqref{eq:NRlineeleg} I can read of the components of the spacetime metric $g_{\mu\nu}$
in terms of the $3$-metric $\ga_{ij}$, the lapse $\al$ and shift $\be^i$ 
\begin{align}
g_{\mu\nu}  = &
\begin{pmatrix} -\al^2 + \be_k\be^k & \be_j \\ \be_i & \ga_{ij}  \end{pmatrix}
\,,\quad
g^{\mu\nu}  =
\frac{1}{\al^2} \begin{pmatrix} -1 & \be^j \\ \be^i & \al^2\ga^{ij} -\be^i\be^j  \end{pmatrix}
\,.
\end{align}
The $3$-metric $\ga_{ij}$ is defined as the metric induced on the spatial hypersurface $\Si_t$ as
\begin{align}
\label{eq:NRmetrics}
\ga_{\mu\nu} = & g_{\mu\nu} + n_{\mu}n_{\nu}\,,\quad
\ga^{\mu\nu} =   g^{\mu\nu} + n^{\mu}n^{\nu}
\,.
\end{align}
Relation~\eqref{eq:NRmetrics} defines the projection operator $\perp$
\begin{align}
\label{eq:NRproj}
\perp^{\mu}{}_{\nu} = &
\ga^{\mu}{}_{\nu} = \de^{\mu}{}_{\nu} + n^{\mu}n_{\nu}
\,,
\end{align}
where $\ga^{\mu}{}_{\nu} n^{\nu} = 0$ holds.
Any $4$-dimensional tensor $T^{\mu_1 ... \mu_p}{}_{\nu_1 ...\nu_q}$ of the spacetime $\M$ can be decomposed
into a purely spatial part, a part that is contracted only with the normal vector $n^{\mu}$ and components
resulting from mixed projections.
Let's exemplarily consider a rank-$2$ tensor $T_{\mu\nu}$ with spatial component 
$\mathcal{S}_{\mu\nu} = \ga^{\al}{}_{\mu}\ga^{\be}{}_{\nu} T_{\al\be}$, 
normal component $\mathcal{N} = T_{\mu\nu} n^{\mu}n^{\nu}$ 
and mixed projections
$\mathcal{T}_{\mu} = \ga^{\al}{}_{\mu}T_{\al\nu}n^{\nu}$.
The full spacetime tensor $T_{\mu\nu}$ is then reconstructed from
\begin{align}
T_{\mu\nu} = & \mathcal{S}_{\mu\nu} + \mathcal{T}_{\mu} n_{\nu}
             + \mathcal{T}_{\nu} n_{\mu} + \mathcal{N} n_{\mu} n_{\nu}
\,.
\end{align}
This operation can be straightforwardly applied to any rank $(p,q)$-tensor of the spacetime $\M$.
In particular, its spatial components 
are obtained by applying the projection operator~\eqref{eq:NRproj}
\begin{align}
\perp T^{\mu_1 ... \mu_p}{}_{\nu_1 ...\nu_q} = &
\ga^{\mu_1}{}_{\ka_1} ...\ga^{\mu_p}{}_{\ka_p}
\ga^{\la_1}{}_{\nu_1} ...\ga^{\la_q}{}_{\nu_q} 
T^{\ka_1...\ka_p}{}_{\la_1...\la_q}
\,.
\end{align}
The covariant derivative $D_i$ associated with the spatial metric $\ga_{ij}$ of a tensor field
$T^{\mu_1 ... \mu_p}{}_{\nu_1 ...\nu_q}$ on the hypersurface $\Si_t$ is given by
\begin{align}
D_{\rho} T^{\mu_1 ... \mu_p}{}_{\nu_1 ...\nu_q} = &
\ga^{\mu_1}{}_{\ka_1} ...\ga^{\mu_p}{}_{\ka_p}
\ga^{\la_1}{}_{\nu_1} ...\ga^{\la_q}{}_{\nu_q}
\ga^{\si}{}_{\rho} \na_{\si} T^{\ka_1...\ka_p}{}_{\la_1...\la_q}
\end{align}
in terms of the covariant derivative $\na_{\mu}$ associated with the spacetime metric $g_{\mu\nu}$.
The metric compatible, torsion-free connection coefficients (Christoffel symbols) associated 
with the $3$-metric $\ga_{ij}$ are given by
\begin{align}
\Ga^{k}{}_{ij} = & \frac{1}{2}\ga^{kl}\left( \p_i \ga_{lj} + \p_j \ga_{il} - \p_l \ga_{ij} \right)
\,.
\end{align}
The directional derivative of a vector $v^{\mu}$ along a vector $u^{\nu}$ is given by
\begin{align}
u^{\nu}D_{\nu} v^{\mu} = & u^{\nu}\na_{\nu}v^{\mu} - n^{\mu}u^{\nu}v^{\la}\na_{\nu} n_{\la}
\,.
\end{align}
%

\subsection{Intrinsic and Extrinsic curvature}\label{ssec:NRcurvature}
\begin{figure}[htpb!]
\begin{center}
\includegraphics[width=0.5\textwidth]{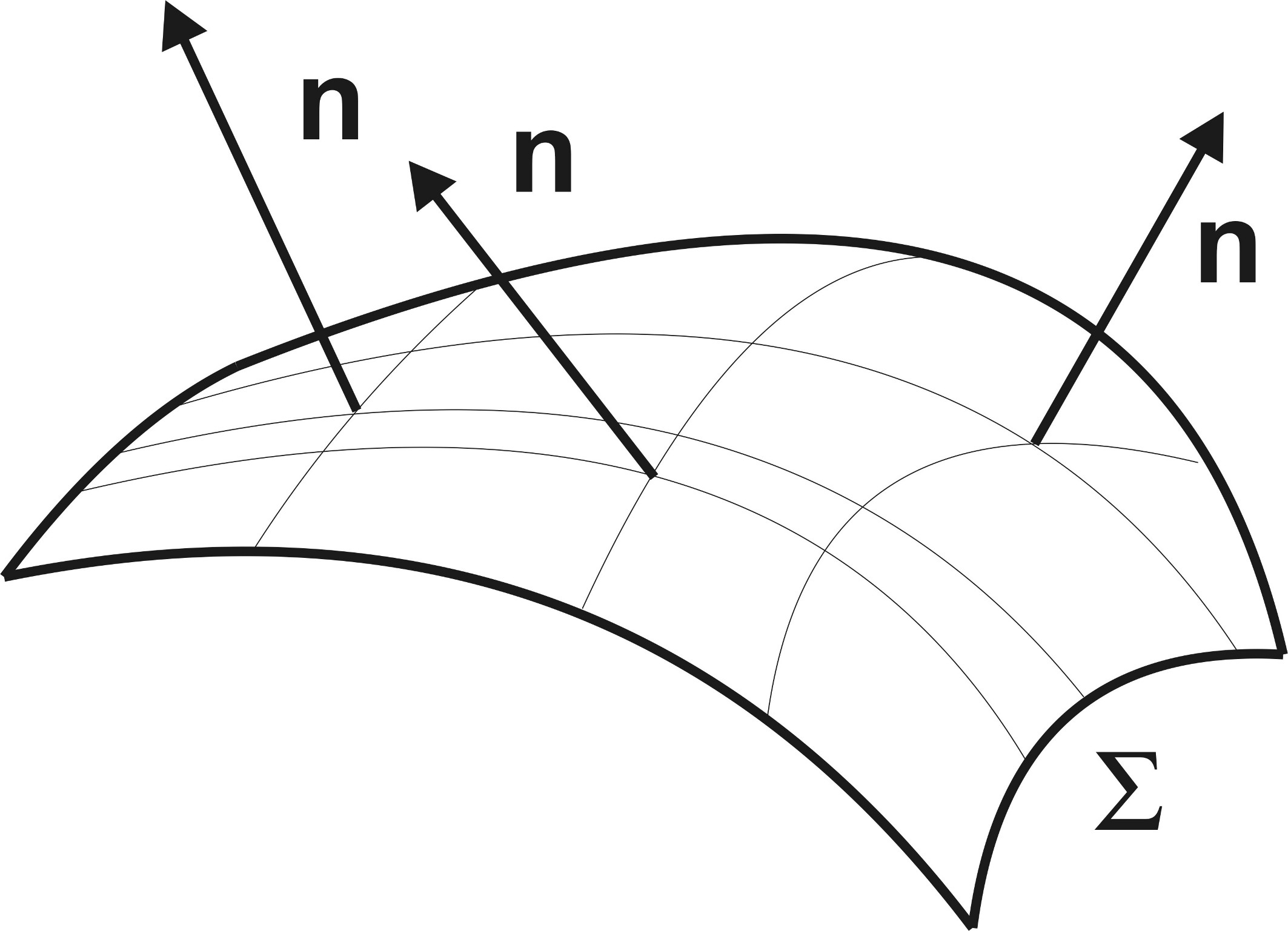}
\end{center}
\caption{\label{fig:NRextrcurv} 
Illustration of a spatial hypersurface $\Si_t$. The normal vector $n^{\mu}$ changes its
direction when it is displaced on the the hypersurface thus yielding a non-vanishing extrinsic curvature.
}
\end{figure}

The intrinsic curvature of the spatial hypersurface $\Si_t$ is determined by the $3$-dimensional
Riemann tensor $\,^{(3)}\!R^k{}_{lij}$.
It measures the non-commutativity of two successive (spatial) covariant
derivatives $D_i$ associated with the $3$-metric $\ga_{ij}$ 
and is expressed by the Ricci identity
\begin{align}
\label{eq:NRricciidentity}
(D_i D_j - D_j D_i) v^k = &\,^{(3)}\!R^{k}{}_{lij}v^l
\,,
\end{align}
for any spatial vector $v^k\in\Sigma_{t}$.
In contrast, the extrinsic curvature $K_{ij}$ describes how a spatial slice $\Si_t$ is embedded into
the spacetime manifold $\M$, i.e.,
how the direction of the normal vector $n^{\mu}$ changes as it is transported along the hypersurface $\Si_t$,
as illustrated in Fig.~\ref{fig:NRextrcurv}.
Then, the extrinsic curvature is defined as 
\begin{align}
\label{eq:NRdefK}
K_{\mu\nu} = & -\ga^{\la}{}_{\mu} \na_{\la}n_{\nu}
= - \na_{\mu}n_{\nu} - n_{\mu} n^{\la}\na_{\la} n_{\nu}
= - \na_{\mu}n_{\nu} - n_{\mu} a_{\nu}
\,.
\end{align}
One can show that the extrinsic curvature is symmetric $K_{\mu\nu}=K_{\nu\mu}$ and 
purely spatial, i.e., $K_{\mu\nu}n^{\mu} = K_{\mu\nu}n^{\nu} = 0$.
Hence, from now on I will only consider its spatial components $K_{ij}$.
Furthermore, the extrinsic curvature $K_{ij}$ is related to the Lie derivative 
of the $3$-metric $\ga_{ij}$ 
along the normal vector $n^{\mu}$.
If I apply the definition of the Lie derivative of a tensor field $T^{\mu_1 ... \mu_p}{}_{\nu_1 ...\nu_q}$ along
a vector $u^{\la}$ given by
\begin{align}
\label{eq:NRdefLie}
\Lie_{u} T^{\mu_1 ... \mu_p}{}_{\nu_1 ...\nu_q} 
= &   u^{\la}\na_{\la} T^{\mu_1 ... \mu_p}{}_{\nu_1 ...\nu_q}
    - T^{\la ... \mu_p}{}_{\nu_1 ...\nu_q} \na_{\la} u^{\mu_1} - ...
    + T^{\mu_1 ... \mu_p}{}_{\la ...\nu_q} \na_{\nu_1} u^{\la} + ...
\,,
\end{align}
to the spatial metric I obtain
\begin{align}
\label{eq:NRevolga0}
\Lie_{\al n} \ga_{\mu\nu} = & \al n^{\la}\na_{\la}\ga_{\mu\nu}
+ \ga_{\la\nu} \na_{\mu}(\al n^{\la}) + \ga_{\mu\la} \na_{\nu}(\al n^{\la})
= -2\al K_{\mu\nu} 
\,.
\end{align}
This expression provides yet another interpretation of the extrinsic curvature
as the ``velocity'' or ``time derivative'' of the spatial metric $\ga_{ij}$ as seen by an Eulerian observer.
Because the relation~\eqref{eq:NRevolga0} has been derived only from 
geometrical concepts it is purely kinematic.
The dynamics of the system will come into the game by considering the $3+1$-decomposition of 
the EoM provided by Einstein's equations, which yields an evolution equation for the extrinsic curvature $K_{ij}$.
%
\subsection{Projections of the Riemann tensor and the Gauss-Codazzi equations}\label{ssec:NRprojRiem}
Before applying the $3+1$-split to Einstein's equations I first consider the different 
projections of the $4$-dimensional Riemann tensor $\,^{(4)}\!R^{\mu}{}_{\nu\rho\si}$.
The possible, non-zero projections of the $4$-dimensional Riemann tensor $\,^{(4)}\!R^{\mu}{}_{\nu\rho\si}$
include the full (spatial) projection onto the hypersurface,
the contraction of one index with the vector normal to the hypersurface and 
the spatial projection of the three remaining indices 
as well as the projection twice along the normal and onto the hypersurface.
All remaining combinations of projections vanish due to the symmetries of the Riemann tensor.
First, I consider the second (spatial) derivative of a purely spatial vector $v_{\mu}$, which
can be rewritten as
\begin{align}
\label{eq:NRaux1}
D_{\mu}D_{\nu} v_{\la} = &
 \ga^{\al}{}_{\la}\ga^{\rho}{}_{\mu}\ga^{\be}{}_{\nu} \na_{\rho}\na_{\be}v_{\al}
+\ga^{\al}{}_{\la}\ga^{\rho}{}_{\mu}\ga^{\de}{}_{\nu} \na_{\be}v_{\al} \na_{\rho} \ga^{\be}{}_{\de}
+\ga^{\rho}{}_{\la}\ga^{\de}{}_{\mu}\ga^{\be}{}_{\nu} \na_{\be}v_{\al} \na_{\de} \ga^{\al}{}_{\rho}
\non\\
= & 
 \ga^{\al}{}_{\la}\ga^{\rho}{}_{\mu}\ga^{\be}{}_{\nu} \na_{\rho}\na_{\be}v_{\al}
-\ga^{\al}{}_{\la} K_{\mu\nu}n^{\be}\na_{\be}v_{\al}
-K_{\la\mu}K^{\al}{}_{\nu} v_{\al}
\,,
\end{align}
where I have used the relation
$\na_{\rho}\ga^{\al}{}_{\be} = -n^{\al} K_{\be \rho} - n_{\be} K^{\al}{}_{\rho} - \frac{1}{\al}n_{\be} n_{\rho} D^{\al}\al
-\frac{1}{\al}n^{\al}n_{\rho} D_{\be}\al$.
If I insert Eq.~\eqref{eq:NRaux1} into the Ricci identity, Eq.~\eqref{eq:NRricciidentity}, I obtain
\begin{equation}
\label{eq:NRGauss}
\boxed{
\ga^{\si}{}_{\al} \ga^{\be}{}_{\mu} \ga^{\ka}{}_{\nu} \ga^{\la}{}_{\rho}
\,^{(4)}\!R^{\al}{}_{\be\ka\la} = 
\,^{(3)}\!R^{\si}{}_{\mu\nu\rho} + K_{\mu\rho}K^{\si}{}_{\nu} - K_{\nu\rho} K^{\si}{}_{\mu}
}
\,,
\end{equation}
which determines the Riemann tensor fully projected onto the spatial hypersurface.

Next, I contract the Riemann tensor once with the vector normal to the spatial hypersurface $\Si_t$ and 
project the remaining three free indices onto the hypersurface.
By employing the Ricci identity, Eq.~\eqref{eq:NRricciidentity}, and the definition of the
extrinsic curvature, Eq.~\eqref{eq:NRdefK}, I obtain
\begin{align}
\ga^{\al}{}_{\mu}\ga^{\be}{}_{\nu}\ga^{\si}{}_{\rho} \,^{(4)}\!R_{\la\al\be\si}n^{\la} = &
\ga^{\al}{}_{\mu}\ga^{\be}{}_{\nu}\ga^{\si}{}_{\rho}\left(
- \na_{\al} K_{\be\si} - \na_{\al}(n_{\be} a_{\si}) + \na_{\be} K_{\al\si} + \na_{\be}(n_{\al}a_{\si})\right)
\,,
\end{align}
which yields 
\begin{equation}
\label{eq:NRCodazzi}
\boxed{
\ga^{\al}{}_{\mu}\ga^{\be}{}_{\nu}\ga^{\si}{}_{\rho} \,^{(4)}\!R_{\la\al\be\si}n^{\la} = 
D_{\nu}K_{\mu\rho} - D_{\mu} K_{\nu\rho}
}
\,.
\end{equation}
Finally, I contract the Riemann tensor twice with the normal vector $n^{\mu}$
and project it twice onto the hypersurface $\Si_t$.
This operation leads to
\begin{equation}
\label{eq:NRRicciRel}
\boxed{
\ga^{\al}{}_{\mu}\ga^{\be}{}_{\nu} \,^{(4)}\!R_{\la\al\ka\be}n^{\la} n^{\ka} =
\Lie_{n}K_{\mu\nu} + K_{\mu\la}K^{\la}{}_{\nu} + \frac{1}{\al}D_{\mu}D_{\nu}\al
}
\,.
\end{equation}
Eqs.~\eqref{eq:NRGauss}, \eqref{eq:NRCodazzi} and~\eqref{eq:NRRicciRel}
are known as {\textit{Gauss-Codazzi equations}}.

\newpage
\subsection{$3+1$-decomposition of Einstein's equations}\label{ssec:NREE3p1}
In GR, the dynamics of a ($4$-dimensional) spacetime $\M$ are prescribed by Einstein's equations
\begin{align}
\label{eq:NREE0}
\,^{(4)}\!G_{\mu\nu} = &\,^{(4)}\!R_{\mu\nu} - \frac{1}{2}g_{\mu\nu} \,^{(4)}\!R = 8\pi T_{\mu\nu}
\,,
\end{align}
where $\,^{(4)}\!R_{\mu\nu}$ and $\,^{(4)}\!R$ are the $4$-dimensional Ricci tensor and scalar, respectively,
and $T_{\mu\nu}$ is the stress-energy tensor.
In order to perform the $3+1$-decomposition of Einstein's equations I write the EoM~\eqref{eq:NREE0} in the form
\begin{subequations}
\begin{align}
\label{eq:NREE1}
E_{1,\mu\nu} = & \,^{(4)}\!R_{\mu\nu} - \frac{1}{2}g_{\mu\nu} \,^{(4)}\!R - 8\pi T_{\mu\nu} = 0
\,,\quad\text{or}
\,\\
\label{eq:NREE2}
E_{2,\mu\nu} = & \,^{(4)}\!R_{\mu\nu} - 8\pi \left( T_{\mu\nu} - \frac{1}{2} g_{\mu\nu} T\right) = 0
\,.
\end{align}
\end{subequations}

By contracting Eq.~\eqref{eq:NREE1} twice with the vector $n^{\mu}$ normal to the spatial hypersurface,
i.e., $E_{1,\mu\nu} n^{\mu} n^{\nu} = 0$,
and inserting the Gauss relation~\eqref{eq:NRGauss} I get the {\textit{Hamiltonian}} constraint
\begin{equation}
\label{eq:NRHamiltonian}
\boxed{
\H = \,^{(3)}\!R - K_{ij} K^{ij} + K^2 - 16\pi \rho = 0
}
\,,
\end{equation}
where $\rho = T_{\mu\nu} n^{\mu} n^{\nu}$ is the energy density.

If I project Eq.~\eqref{eq:NREE1} once onto the hypersurface $\Si_t$ and once perpendicular to it, i.e.,
$\ga^{\la}{}_{\mu} E_{1,\la\nu} n^{\nu} = 0$,
and substitute the $4$-dimensional, projected Ricci tensor by the Gauss-Codazzi relation~\eqref{eq:NRCodazzi}
I obtain the {\textit{momentum}} constraint
\begin{equation}
\label{eq:NRMomentum}
\boxed{
\M_i = D_j K^{j}{}_i - D_i K - 8\pi j_i = 0
}
\,,
\end{equation}
where $j_{\mu} = - \ga^{\la}{}_{\mu} T_{\la\nu} n^{\nu}$ is the energy momentum flux.
One can show straight-forwardly that the momentum density is purely spatial
$j_{\mu} n^{\mu} = - \ga^{\la}{}_{\mu} T_{\la\nu} n^{\nu} n^{\mu} = 0$.

Finally, I consider the fully spatial projection of Eq.~\eqref{eq:NREE2}, 
$\ga^{\ka}{}_{\mu}\ga^{\la}{}_{\nu}E_{2,\ka\la} = 0$, 
and insert Eqs.~\eqref{eq:NRGauss} and~\eqref{eq:NRRicciRel}
to eliminate the Ricci tensor.
Performing this operation yields
\begin{align}
\E_{ij} = & 
- \Lie_{n} K_{ij} - \frac{1}{\al}D_iD_j\al 
-2K^{k}{}_{i} K_{kj} + K K_{ij} + \,^{(3)}\!R_{ij} 
+ 4\pi\left( \ga_{ij}(S-\rho) - 2 S_{ij} \right)
= 0
\,,
\end{align}
where $S_{ij} = \ga^{\ka}{}_i \ga^{\la}{}_j T_{\ka\la}$ is the purely spatial matter stress tensor.
By re-organizing the terms and applying Eq.~\eqref{eq:NRtime} 
I obtain the time evolution equation for the extrinsic curvature
\begin{equation}
\label{eq:NRevolK}
\boxed{
(\p_t - \Lie_{\be}) K_{ij} = - D_iD_j\al 
+ \al\left(  \,^{(3)}\!R_{ij} - 2 K^{k}{}_{i} K_{kj} + K K_{ij} \right) 
+ 4\pi \al \left( \ga_{ij}(S-\rho) - 2 S_{ij} \right)
}
\,.
\end{equation}
The time evolution equation for the metric results from Eqs.~\eqref{eq:NRtime} and~\eqref{eq:NRevolga0} and writes
\begin{equation}
\label{eq:NRevolga}
\boxed{
(\p_t - \Lie_{\be}) \ga_{ij} =  -2\al K_{ij} 
}
\end{equation}

In summary, the full set of Einstein's equations~\eqref{eq:NREE0} in $3+1$-form are provided by the
time evolution equations for the $3$-metric, Eq.~\eqref{eq:NRevolga}, and extrinsic curvature, Eq.~\eqref{eq:NRevolK},
as well as the Hamiltonian and momentum constraints, Eqs.~\eqref{eq:NRHamiltonian} and~\eqref{eq:NRMomentum},
respectively.
In the NR community, this particular set of equations is usually referred to as {\textit{ADM}} equations, 
although they are not in the Hamiltonian form as originally introduced by Arnowitt, Deser \& Misner
\cite{Arnowitt:1962hi}
but rather in the form derived by York \cite{York1979}.
In the following sections I will only consider $3$-dimensional quantities and will therefore drop the superscript
$\,^{(3)}\![...]$.

\section{Initial Data}\label{sec:NRInitData}
The $3+1$-decomposition of Einstein's equations results in a set of evolution equations for the spatial metric $\ga_{ij}$ 
and the extrinsic curvature $K_{ij}$, Eqs.~\eqref{eq:NRevolga} and~\eqref{eq:NRevolK}, respectively,
and a set of constraints to the system
which have to be satisfied for all times.
In a free evolution scheme, the Hamiltonian and momentum constraints, 
Eqs.~\eqref{eq:NRHamiltonian} and~\eqref{eq:NRMomentum}, 
are solved only to provide initial data for the $3$-metric and curvature, 
thus representing the initial configuration for the physical system of interest. 
Although conceptually straightforward, the initial data construction is a non-trivial task. The constraint equations
are a set of four coupled, elliptic PDEs which, in general, are difficult to solve.
In the following I briefly summarize
one of the most common procedures to solve the constraint equations, namely the conformal decomposition
first introduced by Lichnerowicz and York \cite{Lichnerowicz1944,York:1971hw,York:1972sj}.
For further methods and more detailed discussions I refer the reader to 
\cite{Cook:2000vr,Alcubierre:2008,Gourgoulhon:2007ue} and references therein.

\subsection{York-Lichnerowicz decomposition and CTT approach}\label{ssec:NRCTTM}
In the conformal transverse traceless approach for the construction of initial data 
I first conformally decompose the (physical) $3$-metric $\ga_{ij}$
\begin{align}
\label{eq:NRconfmetric0}
\ga_{ij} = & \psi^4 \tilde{\ga}_{ij}
\,,
\end{align}
into the conformal metric $\tilde{\ga}_{ij}$ which I assume to be given and the conformal factor $\psi$.
Note, that with this rescaling $\tilde{\ga}=\det\tilde{\ga}_{ij}=1$.
The extrinsic curvature is split into its trace $K$ and tracefree part $A_{ij}$
\begin{align}
\label{eq:NRKsplit}
K_{ij} = & A_{ij} + \frac{1}{3}\ga_{ij} K
\,.
\end{align}
Additionally, the tracefree part $A_{ij}$ of the extrinsic curvature is conformally rescaled according to
\begin{align}
\label{eq:NRArescale}
A^{ij} = \psi^{-10} \hat{A}^{ij}
\,,\qquad
A_{ij} = \psi^{-2}\hat{A}_{ij}
\,.
\end{align}
By employing the conformal decomposition, Eqs.~\eqref{eq:NRconfmetric0},~\eqref{eq:NRKsplit},~\eqref{eq:NRArescale},
the Hamiltonian constraint~\eqref{eq:NRHamiltonian} becomes an elliptic equation for the 
conformal factor $\psi$
\begin{align}
\label{eq:NRconfHam}
\H = &   \tilde{D}^i\tilde{D}_i\psi -\frac{1}{8}\psi\tilde{R} +\frac{1}{8}\psi^{-7}\hat{A}_{ij}\hat{A}^{ij} 
       - \frac{1}{12}\psi^5 K^2 + 2\pi \psi^5 \rho = 0
\,,
\end{align}
where $\tilde{D}_i$ and $\tilde{R}$ denote the covariant derivative and Ricci scalar with respect to the
conformal metric $\tilde{\ga}_{ij}$.
The momentum constraint~\eqref{eq:NRMomentum} becomes
\begin{align}
\label{eq:NRconfMom}
\M^i = &
\tilde{D}_j\hat{A}^{ij} - \frac{2}{3}\psi^{6}\tilde{D}^i K - 8\pi\psi^{10} j^i = 0
\,.
\end{align}
In order to solve the momentum constraint~\eqref{eq:NRconfMom}, I further split the traceless part of the extrinsic
curvature $\hat{A}^{ij}$ into a transverse and longitudinal part \cite{York1979,York:1973ia}
\begin{align}
\label{eq:NRAtt}
\hat{A}^{ij} = & (L X)^{ij} + \hat{A}^{ij}_{TT}
\,,
\end{align}
where $\hat{A}^{ij}_{TT}$ is transverse and traceless, i.e.,
\begin{align}
\tilde{D}_j\hat{A}^{ij}_{TT} = &0
\quad\text{and}\quad
\tilde{\gamma}_{ij}\hat{A}^{ij}_{TT} = 0
\,.
\end{align}
The longitudinal part $(L X)^{ij}$ is a conformal Killing form associated with the conformal metric acting on a vector field $X^i$ according to
\begin{align}
(L X)^{ij} =& \tilde{D}^iX^j + \tilde{D}^jX^i -\frac{2}{3}\tilde{\gamma}^{ij}\tilde{D}_k X^k
\,.
\end{align}
Then, the constraints reduce to
\begin{align}
\label{eq:NRHamfinal}
\tilde{D}^i\tilde{D}_i\psi -\frac{1}{8}\psi\tilde{R} 
+\frac{1}{8}\psi^{-7}\left((LX)_{ij} + \hat{A}^{TT}_{ij}\right) \left((LX)^{ij} + \hat{A}^{ij}_{TT}\right) 
-\frac{1}{12}\psi^5K^2 +2\pi \psi^5 \rho = & 0
\,,\\
\label{eq:NRmomfinal}
\tilde{D}_j(LX)^{ij} - \frac{2}{3}\psi^{6}\tilde{D}^i K - 8\pi\psi^{10} j^i = & 0
\,.
\end{align}
Assuming that the conformal metric $\tilde{\ga}_{ij}$, the trace of the extrinsic curvature $K$ and the
transverse traceless part of the extrinsic curvature $\hat{A}^{ij}_{TT}$ are given,
the constraint equations~\eqref{eq:NRHamfinal} and~\eqref{eq:NRmomfinal} can be solved for
the conformal factor $\psi$ and the vector $X^i$ and the 
physical quantities $(\ga_{ij},K_{ij})$ can be reconstructed on the initial time slice.
%
\subsection{Initial data for multiple BH systems -- the puncture approach}\label{ssec:NRBHinit}
\begin{figure}[htpb!]
\begin{center}
\includegraphics[width=0.62\textwidth]{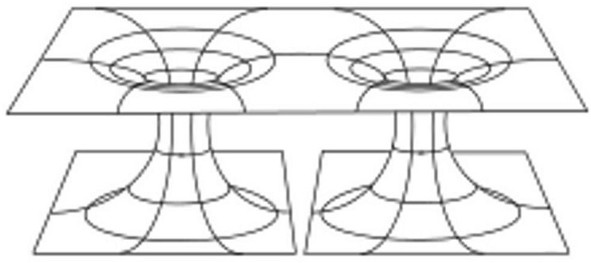}
\end{center}
\caption{\label{fig:NRBLID}
Brill-Lindquist type wormhole topology of a spacetime containing
two BHs. In this case our universe is joined to two distinct univeres via
wormholes.
}
\end{figure}

In this section I apply the previously discussed approach to solve the constraint equations to
the case of BH spacetimes in which I are most interested in the present work.
Therefore I consider the $3+1$-Einstein equations in vacuum, implying that the matter 
variables $(\rho, j_i, S_{ij})$ vanish.
The currently most popular method to implement BH configurations is the 
{\textit{moving puncture}} approach~\cite{Brandt:1997tf,Baker:2006yw,Campanelli:2005dd}. 
The initial data shows a Brill-Lindquist wormhole topology \cite{Brill:1963yv},
depicted in Fig.~\ref{fig:NRBLID}.
The spatial slice consists of two or more copies of $\mathbb{R}^3$
with points $r_i$ removed. 
Thus, one obtains several asymptotic flat regions connected by Einstein-Rosen bridges
and each asymptotically flat end is compactified and identified with a single point $r_i$ on $\mathbb{R}^3$. 
These coordinate singularities at the points $r_i$ are referred to as ``puncture''. 
In more practical terms, one uses an isotropic radial coordinate
$r_{\rm{iso}} = \tfrac{m^2}{4r}$, such that the sphere $r_{\rm{iso}} = \tfrac{m}{2}$
remains invariant.
Puncture data enables us to represent BH spacetimes without excision. 
The singularities are not cut off the numerical grid during the evolution but are avoided by the gauge choice.
In the case of spacetimes containing $N$ BHs
each singular point represents infinity in a different asymptotically flat region,
so that our universe is connected with $N$ different universes through Einstein-Rosen brigdes (wormholes).

Notice, that Eqs.~\eqref{eq:NRHamfinal} and~\eqref{eq:NRmomfinal} still constitute a system
of four coupled PDEs.
Therefore, I apply further simplifications to the initial data problem. 
In particular, I consider Brill-Lindquist type initial data, which describes a system of 
$N$ BHs momentarily at rest on the initial time slice,
and Bowen-York initial data, which sets up a system of $N$ BHs with linear momentum and spin.

\paragraph{Brill-Lindquist type initial data}
I first consider time symmetric initial data, implying $K_{ij}=0$, in a
conformally flat spacetime, i.e., $\tilde{\ga}_{ij} = \eta_{ij}$.
Then, the momentum constraint is trivially satisfied and the Hamiltonian constraint reduces to
a Laplace equation in flat space for the conformal factor $\psi_0$ at $t=0$
\begin{align}
\label{eq:NRLaplacePsi}
\hat{D}^i \hat{D}_i\psi_0 = & 0
\,,
\end{align}
where $\hat{D}_i$ is the flat-space covariant derivative.

The boundary condition for an asymptotically flat spacetime requires $\psi=1$ at infinity.
Then, the ansatz for a single Schwarzschild BH is 
\begin{align}
\label{eq:NRBLansatz}
\psi_0 = 1 + \frac{m}{2r}
\,,
\end{align}
where $m$ is the bare mass parameter of the BH.
Because the Laplace equation~\eqref{eq:NRLaplacePsi} is linear, I obtain the solution
representing a spacetime of $N$ BHs momentarily at rest by superposition 
\begin{align}
\label{eq:NRBLsol}
\psi_0 = & 1 + \sum_{i=1}^N \frac{m_i}{2|r - r_i|}
\,,
\end{align}
where $m_i$ is the bare mass parameter of the $i^{th}$ BH located at $r_i$.
This solution is known as {\textit{Brill-Lindquist}} initial data \cite{Brill:1963yv,Lindquist1963}.

\paragraph{Bowen-York initial data}
Whereas Brill-Lindquist type data sets up time symmetric data representing 
a spacetime of $N$ BHs momentarily at rest, this assumption is relaxed in the case of {\textit{Bowen-York}} initial data.
Thus, it will set up more realistic configurations of BHs with linear momentum $P_i$ and spin $S_i$.
In this approach I assume $K_{ij} \ne 0$, employ the flat metric as the conformal metric and 
impose the maximal slicing condition, i.e., 
\begin{align}
\label{eq:NRinitdataaux}
\gamma_{ij} = & \psi^4_0\tilde{\ga}_{ij} = \psi^4_0 \eta_{ij}
\,,\quad
K = 0
\,,\quad 
K_{ij} = \psi^{-2}_0\hat{A}_{ij} 
\,,
\end{align}
where $\psi_0$ denotes the conformal factor on the initial time slice.
Then, the Hamiltonian and momentum constraints decouple.
Specifically, the latter becomes 
\begin{align}
\partial_j \hat{A}^{ij} = & 0
\,.
\end{align}
and is solved by setting \cite{Brandt:1997tf}
\begin{align}
\label{eq:NRBYsol}
\hat{A}^{ij} = & \sum_{a=1}^N \hat{A}_{PS(a)}^{ij}
\,,
\end{align}
with the {\textit{Bowen-York}} solution \cite{Bowen:1980yu}
\begin{align}
\label{eq:BYEC}
\hat{A}_{PS(a)}^{ij} = & 
  \frac{3}{2r^2}\left(P_a^{i} n^{j} + P_a^{j} n^{i} - (g^{ij} - n^{i} n^{j})P_a^{k} n_{k} \right) 
+ \frac{3}{r^3} \left( \epsilon^{ikl} S_{k_a} n_{l} n^{j} + \epsilon^{jkl}S_{k_a} n_{l} n^{i}\right)
\,.
\end{align}
$P^i_a$ and $S_{i_a}$ are identified as the momentum and the spin of the $a^{th}$ BH, respectively. 
This can be seen by computing the ADM linear and angular momentum at infinity 
(see, e.g., \cite{Alcubierre:2008} and references therin)
\begin{subequations}
\begin{align}
P^i = & \frac{1}{8\pi}\lim_{r\rightarrow\infty} \oint dS (K^{i}{}_{k} - K \delta^{i}{}_{k}) u^k
\,,\\
J^i = & \frac{1}{16\pi}\lim_{r\rightarrow\infty} \oint dS \eps^{ijk}x_j K_{kl} u^l
\,.
\end{align}
\end{subequations} 
These integrals are performed over spheres of constant radius $r$, where $u^i$ denotes the normal 
vector to the sphere.
If I now consider the conformal factor at infinity $\psi_{\infty}=1$ and insert Eq.~\eqref{eq:BYEC},
I find that $P^i_a$ and $S_{i_a}$ are indeed the linear and angular momenta of the spacetime.

The Hamiltonian constraint is given by Eq.~\eqref{eq:NRconfHam}
with $K=0$ and $\hat{A}^{ij}$ computed according to Eq.~\eqref{eq:NRBYsol}.
The solution is given by the ansatz~\cite{Brandt:1997tf}
\begin{align}
\label{eq:NRBYansatz}
\psi_0 = & u + \psi_{BL}
\,,\quad\text{with}\quad
\psi_{BL} = \sum_{i=1}^N \frac{m_i}{2|r - r_i|}
\,,
\end{align}
where $u$ is a regular function which is $C^2$ at the puncture and $C^{\infty}$ elsewhere.
The singular behaviour is absorbed in the term $\psi_{BL}$.
Inserting ansatz~\eqref{eq:NRBYansatz} into the Hamiltonian constraint~\eqref{eq:NRconfHam} 
yields an elliptic equation for the function $u$.
After finding a solution for $u$, the conformal factor and, thus, the physical metric $\ga_{ij}$ can be 
reconstructed whereas the extrinsic curvature is given by Eqs.~\eqref{eq:NRinitdataaux} and~\eqref{eq:NRBYsol}, 
describing the initial configuration of a system of $N$ BHs each with linear momentum $P_i$ and spin
$S_i$.

\section{Gauge conditions}\label{sec:NRgauge}
As we have seen in Sec.~\ref{sec:NRdecomp}, the coordinates, which are encoded by the lapse function $\al$ and
the shift vector $\be^i$ can be choosen freely.
Nevertheless, for the sake of applications such as numerical simulations they need to be specified 
in order to close the system of equations.
In the following we give a brief review of the most common choices of the slicing condition fixing the lapse
function $\al$ and the shift condition prescribing $\be^i$.

\subsection{Choice of foliation}\label{ssec:NRslicing}
\paragraph{Geodesic Slicing}
The simplest option for the choice of foliation is given by the {\textit{geodesic slicing}} 
condition $\al = 1$.
This choice of the lapse function $\al$ yields a vanishing 4-acceleration $a_{\mu}$ 
of the Eulerian observer as can been seen from Eq.~\eqref{eq:NRacc}
implying that the worldlines of this observer are geodesics. 
However, from a practical viewpoint this choice of the foliation is a poor one 
because in a Schwarzschild spacetime every Eulerian observer starting from rest will freely fall 
into the singularity at a finite time, thus resulting in a break-down of any numerical simulation.

\paragraph{Maximal Slicing}
In order to avoid the pathologies of the geodesic slicing condition
we impose a condition on the trace of the extrinsic curvature rather than on the lapse function $\al$. 
Specifically, we choose $K=0$, which is known as {\it maximal slicing} condition.
In strong field regions this condition will tend to hold back the evolution of the slice 
such that the proper time increases ``more slowly''.
Whereas the entire spacetime outside the event horizon is covered by this foliation 
the slices pile up near the BH region such that they never reach the singularity. 
This behaviour is referred to as {\it singularity avoidance}. 
Consequently, the proper time of an Eulerian observer between two hypersurfaces 
tends to zero with increasing (coordinate) time $t$. 
This implies that $\al$ goes to zero as $t$ goes to infinity and is called {\it collapse of the lapse}.

\paragraph{Harmonic Slicing}
The harmonic condition for spacetime coordinates $(x^{\mu})$ is given by $\Box_g x^{\mu} = 0$,
where $\Box_g = \na_{\mu}\na^{\mu}$ is the d'Alembertian associated with the spacetime metric $g_{\mu\nu}$.
Imposing the harmonic condition only on the time coordinate $x^0$
while the remaining spatial coordinates are allowed to be choosen freely
yields the {\textit{harmonic slicing}} condition
\begin{align}
(\p_t - \Lie_{\beta})\al = & - \al^2 K
\,.
\end{align}
Like the maximal slicing condition the harmonic slicing also proves to be singularity avoiding.

A further succesful approach is achieved by employing generalized harmonic coordinates, 
in which the wave equation for each coordinate is allowed to contain a source term, i.e.
$\Box_g x^{\mu} = H^{\mu}$.
These gauge-driving source terms $H^{\mu}$ can either be specified algebraically or 
evolved in such a way that hyperbolicity is preserved 
\cite{Friedrich1985, Garfinkle:2001ni, Lindblom:2005qh, Pretorius:2006tp}.
The first successful numerical simulations of orbiting BH binaries considered a source term for 
the time coordinate that effectively kept the lapse close unity, while the spatial coordinates remained harmonic
\cite{Pretorius:2005gq,Pretorius:2006tp}.
This was accomplished by evolving the source term itself, according to
\begin{align}
\Box_g H_0 = & \frac{1}{\al} \left( -\xi_1 (\al - 1) + \xi_2 (\p_t - \be^k\p_k)H_0 \right)
\,,
\end{align}
where $\xi_{1,2}$ are constants. 
\paragraph{$1+log$-Slicing}
The currently most common slicing condition used in many $3+1$ numerical relativity codes is the 
$1+log$-slicing condition defined by
\begin{align}
\label{eq:NR1plog}
(\p_t - \Lie_{\be}) \al = & - 2\al K
\,.
\end{align}
In normal coordinates, i.e., for vanishing shift vector $\beta^i=0$, 
this equation can be rewritten as 
\begin{align}
\frac{\partial}{\partial t}\alpha = \frac{\partial}{\partial t}\ln\gamma
\end{align}
and yields
\begin{align}
\label{eq:logslic}
\alpha = f(x) + \ln\gamma
\,,
\end{align}
hence the naming $1+log$-slicing.
This slicing condition has strong singularity avoiding properties and has been found 
to mimic the maximal slicing condition 
while simultaneously being computationally cheaper
\cite{Anninos:1995am}.

\paragraph{Generalization of Slicing Conditions - Bona-Mass{\'o} Family of Slicing Conditions}
The slicing conditions introduced so far have been generalized by Bona et al. 
\cite{Bona:1994dr} 
and later further modified by \cite{Alcubierre:2002kk} 
to 
\begin{align}
\label{eq:NRMSL}
(\p_t - \Lie_{\be}) \al = & -\al^2 f(\al) (K-K_0)
\,,
\end{align}
where $f(\alpha)$ is an arbitrary, positive funtion of the lapse $\al$ and $K_0$ is the value of the 
trace of the extrinisic curvature on the initial time slice.
In the case of $f(\al) = 0$ and $\al = 1$ the Bona-Mass{\'o} slicing condition~\eqref{eq:NRMSL}
reduces to the geodesic slicing. 
The harmonic slicing corresponds to the choice $f(\alpha)=1$,
whereas the prescription $f(\alpha) = \frac{2}{\alpha}$ leads to the $1+log$-slicing. 

\subsection{Choice of spatial coordinates}
\paragraph{Minimal Distortion}
The intention of the {\it minimal distortion} condition is the construction of a shift condition 
that minimizes the change of the conformal metric in time \cite{Smarr1978,Smarr:1977uf}.
Therefore one introduces the distortion tensor $Q_{ij}$ that is defined as 
\begin{align}
Q_{ij} = & \p_t\ga_{ij} - \frac{1}{3}\ga_{ij}\ga^{kl}\p_t\ga_{kl}
\,.
\end{align}
Because $Q_{ij}$ is tracefree it does not take into account the change of the volume of this domain but only its shear. 
Then, minimizing the integral of $Q_{ij}Q^{ij}$ over the spatial hypersurface with respect to the shift 
gives the {\textit{minimal distortion condition}}
\begin{align}
D_j Q^{ij} = & 0
\,,
\end{align}
which implies $D_j\p_t\tilde{\gamma}^{ij} = 0$.
In the weak field limit this condition includes the standard transverse traceless gauge 
of the linearized theory \cite{Smarr1978}. 
Smarr and York \cite{Smarr1978} introduced the so-called {\textit{radiation gauge}} 
which combines the minimal distortion with the maximal slicing condition $K=0$,
thus describing radiative spacetimes.

If we introduce the conformal connection function 
$\tilde{\Ga}^i = \tilde{\ga}^{jk} \tilde{\Ga}^{i}{}_{jk}=-\p_j\tilde{\ga}^{ij}$ 
that will play a key role in the 
BSSN formulation introduced in Sec.~\ref{sec:NRgBSSN} below,
we can relate the minimal distortion condition to the so-called {\textit{$\Ga$-freezing condition}}
\begin{align}
\label{eq:NRGammaFreeze}
\p_t \tilde{\Ga}^i = & 0
\,,
\end{align}
modulo lower order terms $\sim\p_k\tilde{\ga}_{ij}$.

\paragraph{Gamma-Driver Conditions}
Because the minimal distortion and $\Ga$-freezing conditions are of elliptic type, they are computationally
expensive to solve at each timestep.
Therefore, modifications thereof have been considered, that are of hyperbolic type and can straightforwardly
be implemented as time evolution equations.

Of particular relevance for numerical implementations using the BSSN evolution system is the 
{\textit{$\Ga$-driver}} condition.
The $\Ga$-driver condition is a modification of the $\Ga$-freezing condition 
first introduced in \cite{Alcubierre:2002kk} and is given by
\begin{align}
\label{eq:NRgammadriver0}
\p_t\be^i = B^i
\,,\quad
\p_t B^i = \xi_{\Ga}\p_t \tilde{\Ga}^i - \eta_{\be} B^i
\,,
\end{align}
where 
$\tilde{\Ga}^i = \tilde{\ga}^{jk} \tilde{\Ga}^{i}{}_{jk} = -\p_j\tilde{\ga}^{ij}$ is the conformal connection function, 
$\xi_{\Ga}$ is a positive scalar function, 
$\eta_{\be}$ is a damping parameter and 
$B^i$ is an auxiliary function.
It has been found crucial to add the dissipation term $\eta_{\be} B^i$ in order
to avoid strong oscillations in the shift \cite{Alcubierre:2002kk}. 
This system of equations can be integrated to \cite{vanMeter:2006vi}
\begin{align}
\label{eq:NRgammadriver}
\p_t\be^i = \xi_{\Ga} \tilde{\Ga}^i - \eta_{\be} \be^i + \be^k\p_k \be^i
\,,
\end{align}
where we have added the advection term $\be^k\p_k \be^i$.
The $\Ga$-driver condition in one of the presented forms
together with the $1+log$-slicing, Eq.~\eqref{eq:NR1plog}, are known as
{\textit{moving puncture}} approach or also {\textit{puncture gauge}}. 
The puncture gauge has been widely used in NR codes
employing the so-called BSSN evolution system, described 
in the next section.

\newpage
\section{Generalized BSSN formulation}\label{sec:NRgBSSN}
For a long time, numerical methods based on the ADM 
equations~\eqref{eq:NRHamiltonian},~\eqref{eq:NRMomentum},~\eqref{eq:NRevolK},
and~\eqref{eq:NRevolga},
suffered from a number of problems, most prominently from being numerically unstable (in $3+1$) allowing only 
for short evolutions of BH spacetimes.
In fact, a detailed analysis of the properties of the underlying PDE system reveals that the ADM equations
are in general only weakly hyperbolic and therefore exhibit an ill-posed initial value problem
(see e.g. \cite{Gundlach:2006tw,Alcubierre:2008,Sarbach:2012pr} and references therein).
Therefore, a large variety of alternative or modified versions of the evolution equations,
suitable choices of the coordinate gauge and different treatments of the BH singularities 
have been investigated.
A plethora of different well-posed formulations of the $3+1$-Einstein's equations have been proposed,
including the BSSN system \cite{Shibata:1995we,Baumgarte:1998te}
and generalizations thereof \cite{Witek:2010es,Laguna:2002zc},
as well as reductions to a first order BSSN version \cite{Brown:2012me},
the Nagy-Ortiz-Reula (NOR) \cite{Nagy:2004td} and Z4 scheme \cite{Bona:2003fj,Gundlach:2005eh},
a conformally decomposed version of Z4 \cite{Bernuzzi:2009ex, Weyhausen:2011cg,Cao:2011fu,Alic:2011gg},
the Bona-Mass{\'o} formulation \cite{Bona:1994dr,Bona:1997hp,Bona:1992zz}, 
the Kidder-Scheel-Teukolsky family \cite{Kidder:2001tz,Sarbach:2002gr}
and a Hamiltonian formulation \cite{Hilditch:2010wp},
to name but a few.

The currently most prominent and succesful techniques include the 
Generalized Harmonic (GH) formulation employed in the original 
breakthrough in 2005 \cite{Pretorius:2005gq,Pretorius:2006tp}, 
in which Einstein's equations are cast into a set of wave equations together with 
the generalized harmonic gauge,
and the moving puncture technique which yielded first succesful simulations
of BH binary spacetimes shortly afterwards \cite{Campanelli:2005dd,Baker:2005vv}.
The latter method makes use of the BSSN formulation of the $3+1$-Einstein's 
equations \cite{Shibata:1995we,Baumgarte:1998te} 
and the puncture gauge \cite{Alcubierre:2002kk,Campanelli:2005dd,Baker:2005vv}.

In this section we wish to review the key aspects of a generalized BSSN formalism
introduced and studied in great detail in \cite{Witek:2010es} 
based on the BSSN system \cite{Shibata:1995we,Baumgarte:1998te}
and modifications first considered in \cite{Laguna:2002zc}.
Whereas the BSSN equations traditionally are introduced as a conformal decomposition and
readjustment of the ADM equations (see e.g. \cite{Baumgarte:1998te,Beyer:2004sv,Alcubierre:2002kk}),
we here follow a slightly different approach \cite{Witek:2010es}.
Specifically, we consider the addition of physical and definition-differential constraints which alter the
characteristic structure of the ADM equations, thus guaranteeing well-posedness of the system,
accompanied by a conformal decomposition changing the evolved variables into a 
form convenient for numerical simulations.

\subsection{Constraint addition}
First we define the differential constraint
\begin{align}
\label{eq:NRGi0}
G_i = & f_i - \ga^{jk}\left(\p_k\ga_{ij} - \frac{1}{3}\p_i\ga_{jk} \right)
\,,
\end{align}
which will be related to the relevant BSSN variables below.
The ADM equations, 
providing evolution equations for the physical $3$-metric $\ga_{ij}$, the extrinsic curvature $K_{ij}$ and $f_i$,
are readjusted to \cite{Witek:2010es,Gundlach:2006tw}
\begin{subequations}
\label{eq:NRADMmod}
\begin{align}
\p_t \ga_{ij} = & [\textrm{ADM}]
\,,\\
\p_t K_{ij}   = & [\textrm{ADM}] + \al \p_{(i}G_{j)} - \frac{1}{3}\al \ga_{ij}\left(\H + \ga^{kl}\p_k G_l \right)
\,,\\
\p_t f_i      = & \p_t\left( \ga^{jk}\p_k\ga_{ij} - \frac{1}{3}\ga^{jk} \p_i\ga_{jk} \right)_{[\textrm{ADM}]} 
                 + 2\al\M_i - 2\al G^j A_{ij} 
\non\\ &
                 + \Lie_{\be} G_i + \ga_{ij} G^k \p_k\be^j - \frac{2}{3} G_i \p_j\be^j
\,,
\end{align}
\end{subequations}
where $[\textrm{ADM}]$ denotes the original ADM evolution equations~\eqref{eq:NRevolga} and~\eqref{eq:NRevolK},
$\H$ is the Hamiltonian constraint~\eqref{eq:NRHamiltonian} and $\M_i$ the momentum constraint~\eqref{eq:NRMomentum}.

\subsection{Conformal decomposition and densitization}
The generalized BSSN system \cite{Witek:2010es} employs conformal variables as dynamical variables
\begin{subequations}
\label{eq:NRgBSSNvars}
\begin{align}
\chi = & \ga^{-1/3}
\,,\quad
\tg_{ij} = \ga^{-1/3}\ga_{ij} = \chi \ga_{ij}
\,,\\
\tK = & \chi^{-3 \nk /2} K
\,,\quad
\tA_{ij} = \chi^{1-3 \nk /2} \left( K_{ij} - \frac{1}{3}\ga_{ij} K \right)
\,,\\
\tG^i = & \tg^{jk}\tG^{i}{}_{jk} = - \p_j\tg^{ij}
\,,
\end{align}
\end{subequations}
where 
$\chi$ and $\tg_{ij}$ are the conformal factor and metric,
$\ga$ is the determinant of the physical $3$-metric,
$\tK$ and $\tA_{ij}$ are the densitized trace and tracefree part of the extrinsic curvature
and $\tG^i$ is the conformal connection function.
The densitization parameter $n_K$ parametrizes the change of conformal variables.
Note, that by construction the determinant of the conformal metric $\tg = 1$.
The key difference between the original BSSN evolution scheme and the generalized version
is the densitization of the curvature variables. 
Notice, that we recover the BSSN system if the parameter $\nk$ vanishes.
The definition of the conformal variables implies a set of additional algebraic constraints
\begin{align}
\label{eq:NRgBSSNaddcon}
T = & \tg^{ij} \tA_{ij} = 0
\,,\quad
D = \ln(\tg) = 0
\,,
\end{align}
representing the requirements that $\tA_{ij}$ remains tracefree
and the determinant of the conformal metric $\tg = 1$.
The definition constraint, Eq.~\eqref{eq:NRGi0}, becomes
\begin{align}
\label{eq:NRGi}
G_i = & 
f_i - \ga^{jk}\left(\p_k\ga_{ij} - \frac{1}{3}\p_i\ga_{jk} \right)
= \tg_{ij} \tG^i - \tg^{jk}\p_k\tg_{ij}
= 0
\,.
\end{align}
The evolution equations for the dynamical variables~\eqref{eq:NRgBSSNvars}
are obtained by taking the time derivative of their definitions,
substituting the evolution equations~\eqref{eq:NRADMmod}
and rewriting everything in terms of the conformal quantities.
This procedure  yields the evolution equations
\begin{subequations}
\label{eq:NRgBSSNevol}
\begin{align}
\p_t \chi = &  \frac{2}{3} \al \chi^{1+\frac{3}{2}\nk} \tK
             + \be^i\p_i\chi - \frac{2}{3} \chi \p_k\be^k
\,,\\
\p_t \tg_{ij} = &
        - 2 \al \chi^{\frac{3}{2}\nk} \tA_{ij}
        + \tg_{ik} \p_{j} \be^k + \tg_{kj} \p_i \be^k
        - \frac{2}{3} \tg_{ij} \p_k\be^k
\,,\\
\p_t \tK = & 
        - \chi^{-\frac{3}{2}\nk} D^iD_i\al
        + \al \chi^{\frac{3}{2}\nk} 
          \left[ \tA^{ij}\tA_{ij} + \frac{1-3\nk}{3} \tK^2 \right] 
        + \be^i\p_i\tK + \nk\tK\p_k\be^k 
\,,\\
\p_t \tA_{ij} = & 
         -\chi^{1-\frac{3}{2}\nk}[D_iD_j\al]^{\textrm{tf}}
        + \al \chi^{1-\frac{3}{2}\nk} [R_{ij}]^{\textrm{tf}} 
        + \al \chi^{\frac{3}{2}\nk}
          \left[ (1-\nk) \tK\tA_{ij} - 2\tA_{ik}\tA^k{}_j \right] 
\non\\ &
        + \be^k\p_k\tA_{ij} 
        + \tA_{ik}\p_{j}\be^k + \tA_{kj}\p_{i}\be^k
        + \left(\nk-\frac{2}{3}\right)\tA_{ij}\p_k\be^k 
\,,\\
\p_t \tG^i = &
- 2 \al \chi^{\frac{3}{2}\nk} \tg^{ik} \left( \frac{2}{3}\p_k\tK + \nk \chi^{-1} \tK \p_k\chi \right)
- \chi^{\frac{3}{2}\nk} \tA^{ik}\left( 3\al\chi^{-1}\p_k\chi + 2 \p_k\al \right)
\non \\ &
+ 2 \al \chi^{\frac{3}{2}\nk} \tA^{kl}\tG^{i}{}_{kl}
+ \be^k \p_k\tG^i + \frac{2}{3}\tG^i \p_k\be^k - \tG^k \p_k \be^i
+ \frac{1}{3}\tg^{ik}\p_k\p_l\be^l + \tg^{kl}\p_k\p_l\be^i
\,,
\end{align}
\end{subequations}
where $[\ldots]^{\textrm{tf}}$ denotes the tracefree part
with respect to the physical metric.
The Ricci tensor is given by
\begin{subequations}
\label{eq:NRgBSSNricci}
\begin{align}
R_{ij} = & \ti{R}_{ij} + R^{\chi}_{ij}
\,,\\
\ti{R}_{ij} = &
- \frac{1}{2} \tg^{kl}\p_k\p_l \tg_{ij}
+ \frac{1}{2} \tg_{ik}\p_j\tG^k + \frac{1}{2} \tg_{kj}\p_i \tG^k
+ \frac{1}{2} \tG^k \left( \tG_{ijk} + \tG_{jik} \right)
\non\\ &
+ \tg^{kl} \tG^{m}{}_{ik} \tG_{jlm}
+ \tg^{kl} \tG^{m}{}_{jk} \left( \tG_{ilm} + \tG_{mil} \right)
\,\\
R^{\chi}_{ij} = & 
  \frac{1}{2\chi}   \tD_i \tD_j \chi + \frac{1}{2\chi} \tg_{ij} \tD^k \tD_k\chi
- \frac{1}{4\chi^2} \tD_i \chi \tD_j \chi - \frac{3}{4\chi^2} \tg_{ij} \tD^k\chi \tD_k\chi
\,,
\end{align}
\end{subequations}
where $\ti{D}_i$ is the covariant derivative with respect to the conformal metric $\tg_{ij}$.
The Christoffel symbols $\Ga^{i}{}_{jk}$ and $\tG^{i}{}_{jk}$ associated with the physical and conformal
metric, respectively, are related via 
\begin{align}
\Ga^{i}{}_{jk} = & \tG^{i}{}_{jk} 
- \frac{1}{2\chi}\left( \de^{k}{}_j\p_i\chi + \de^{k}{}_i \p_j\chi - \tg_{ij}\tg^{kl}\p_l\chi \right)
\,.
\end{align}
The second derivative of the lapse in Eqs.~\eqref{eq:NRgBSSNevol} is given by
\begin{align}
D_i D_j \al = & \ti{D}_i \ti{D}_j \al
+ \frac{1}{2\chi}\left( \p_i\al \p_j\chi + \p_j\al \p_i\chi - \tg_{ij}\tg^{kl}\p_k\al \p_l\chi \right)
\,.
\end{align}
The physical constraints, Eqs.~\eqref{eq:NRHamiltonian} and~\eqref{eq:NRMomentum}, become
\begin{subequations}
\label{eq:NRgBSSNconstraints}
\begin{align}
\H = & R - \chi^{3\nk}\left( \tA_{ij} \tA^{ij} - \frac{2}{3}\tK^2 \right) = 0
\,,\\
\M_i = & - \tG^j \tA_{ij} - \frac{2}{3}\p_i\tK - \nk \frac{\tK}{\chi} \p_i\chi
- \frac{3(1-\nk)}{2\chi}\tA^{j}{}_i\p_j\chi
+ \tg^{jk}\p_j\tA_{ik} - \tA^{j}{}_k \tG^{k}{}_{ij}
= 0
\,.
\end{align}
\end{subequations}

\subsection{Gauge conditions}
The system of evolution equations~\eqref{eq:NRgBSSNevol} is closed by the choice of gauge conditions
for the lapse function $\al$ and shift vector $\be^i$.
The BSSN system has been proven to be a well-posed, numerically robust formulation of the initial value problem
when combined with the moving puncture approach.
The moving puncture approach
employs the $1+log$-slicing condition, Eq.~\eqref{eq:NR1plog}, and the $\Gamma$-driver shift condition, 
in the version of Eqs.~\eqref{eq:NRgammadriver0} or Eq.~\eqref{eq:NRgammadriver}.
In the generalized BSSN system we consider the densitized lapse function $Q$, related to the lapse
function $\al$ via
\begin{align}
Q = & \ga^{-\nq/2} \al = \chi^{3\nq/2} \al
\,,
\end{align}
as dynamical variable \cite{Witek:2010es}. $\nq$ is the densitization parameter and as such parametrizes
the particular choice of the dynamical variable.
The $1+log$-slicing condition~\eqref{eq:NR1plog} becomes
\begin{align}
\label{eq:NRQevol}
\p_t Q = &
  \chi^{3/2(\nk - \nq)} Q \tK \left( - 2 \chi^{3/2\nq} + \nq Q \right)
+ \be^k\p_k Q - \nq Q \p_k\be^k
\,,
\end{align}
where we have added the advection term.
It has been proven \cite{Witek:2010es} that the generalized BSSN evolution system, Eqs.~\eqref{eq:NRgBSSNevol},
together with the moving puncture gauge conditions, 
Eqs.~\eqref{eq:NRQevol} and~\eqref{eq:NRgammadriver0}
exhibits a strongly hyperbolic formulation of the $3+1$-Einstein's equations and
therefore satisfies the necessary conditions to provide a well-posed initial value problem.

\section{Extraction of meaningful quantities}\label{sec:NRphys}
\subsection{Gravitational wave extraction}\label{ssec:waveextraction}
One of the key goals in performing binary BH evolutions in astrophysical and other applications
is the extraction of gravitational radiation.
In order to gain information about gravitational waves two distinct methods have been developed:
\noindent{(i)} The approach first introduced by Regge, Wheeler and Zerilli \cite{Regge:1957td,Zerilli:1970se}
considers perturbations of the Schwarzschild metric. From these metric perturbations one can 
compute a gauge invariant master function which encodes the information about gravitational radiation.
\noindent{(ii)} The second approach is based on the tetrad formalism introduced by 
Newman and Penrose \cite{Newman:1961qr}. In particular one computes the (coordinate) gauge
independent Weyl scalars~\footnote{But note, that they do depend on the choice od the tetrad.}, which are contractions of the Weyl tensor with the principal null directions
of the tetrad. In particular, one can show that in a specific choice of this tetrad 
the Weyl scalars $\Psi_4$ and $\Psi_0$ carry information
about out- and ingoing gravitational radiation~\cite{Wald:1984,Chandra:1983,
Alcubierre:2008,Centrella:2010mx, Hinder:2011xx}
.
Note also, that Teukolsky's analysis of the stability of the Kerr BH \cite{Teukolsky:1973ha} employs the 
Newman-Penrose formalism. 

In this section we restrict ourselves to the summary of the Newman-Penrose approach \cite{Newman:1961qr}.
We will study an extension of the Regge-Wheeler-Zerilli formalism to higher dimensional spacetimes 
in detail in Sec.~\ref{sec:HDFWaveExtr} and provide a connection to the four dimensional case in this context.
The relation between both methods has been investigated, e.g., in Refs.~\cite{Lousto:2005xu,
Zenginoglu:2009ey,Reisswig:2010cd,Witek:2010xi}.

Nowadays, the most common approach to extract
information about gravitational radiation 
in numerical simulations of $4$-dimensional BH binaries 
is the computation of the Weyl scalar $\Psi_4$. It has the asymptotic property of being equal to
the outgoing radiation if the complex null-tetrad is chosen properly. 
In Chapter~\ref{chapter:BBbox} we intend to study BH binaries surrounded by a mirror-like box,
thus mimicking BH-AdS spacetimes.
This toy model has the same local geometry as vacuum models and 
for a sufficiently large box size we expect the standard
tetrad to yield a good approximation to the gravitational wave information. 
Because of the imposed reflecting boundary conditions we also deal with
ingoing waves in our simulations, once they are reflected from the box boundary.
Therefore, we will also consider the Weyl scalar $\Psi_0$ 
in order to account for the ingoing contribution.

For self-containedness,
Appendix~\ref{sec:appWaveExtraction} reviews the electromagnetic decomposition
of the Weyl tensor and in particular the construction of the relevant quantities
for our study, $\Psi_0$ and $\Psi_4$. 

To be explicit, we define a spherical coordinate system centered at
the center-of-mass of the binary with orthonormal basis
$(\hat{r},\hat\theta,\hat\phi)$.
The coordinates are chosen such that the azimuthal axis is aligned with
the orbital angular momentum
and the binary orbits are in the direction of increasing azimuthal coordinate. 
Our definitions and notation are the same as in
\cite{Friedrich:1996hq,Sperhake:2006cy}.
To define our complex null-tetrad, we use the timelike unit vector normal
to a given hypersurface $\hat{n}$ and the radial unit vector $\hat{r}$
to define an ingoing ($\vec{k}$) and outgoing null vector ($\vec{\ell}$) by
\begin{align}
\label{eq:Boxnulltetrad1}
\vec{k} \equiv & \frac1{\sqrt{2}}(\hat{n} + \hat{r})
\,,\quad
\vec{\ell} \equiv  \frac1{\sqrt{2}}(\hat{n} - \hat{r})
\,.
\end{align}
We define the complex null vector $\vec{m}$ and its complex conjugate by
\begin{align}
\label{eq:Boxnulltetrad2}
\vec{m} \equiv & \frac1{\sqrt{2}}(\hat\phi + i\hat\theta)
\,, \quad 
\vec{\bar{m}} \equiv \frac1{\sqrt{2}}(\hat\phi - i\hat\theta)
\,.
\end{align}
In terms of this tetrad, we define $\Psi_0$ and $\Psi_4$ as
\begin{subequations}
\begin{align}
\label{eq:BoxPsi0_Weyl}
\Psi_0 \equiv & C_{\alpha\beta\gamma\delta}k^\alpha m^{\beta}k^\gamma m^\delta
\,, \\
\label{eq:BoxPsi4_Weyl}
\Psi_4 \equiv & C_{\alpha\beta\gamma\delta}\ell^\alpha \bar{m}^{\beta}\ell^\gamma \bar{m}^\delta
\,,
\end{align}
\end{subequations}
where $C_{\alpha\beta\gamma\delta}$ is the Weyl tensor.
To relate $\Psi_0$ and $\Psi_4$ to the amplitudes of the gravitational
waves, we note that in the transverse-traceless (TT) gauge, assuming the
functional form $f(t \pm r)$ for the ingoing or outgoing waves, we have
\begin{align}
\frac{1}{4} (\ddot{h}^{TT}_{\hat\theta\hat\theta}-\ddot{h}^{TT}_{\hat\phi\hat\phi}) = &
  -R_{\hat{n}\hat\theta\hat{n}\hat\theta} 
= \mp R_{\hat{n}\hat\phi\hat{r}\hat\phi}
= -R_{\hat{r}\hat\theta\hat{r}\hat\theta} 
= R_{\hat{n}\hat\phi\hat{n}\hat\phi}
= \pm R_{\hat{n}\hat\theta\hat{r}\hat\theta} 
= R_{\hat{r}\hat\phi\hat{r}\hat\phi}
\,, \non\\
\frac{1}{2} \ddot{h}^{TT}_{\hat\theta\hat\phi} = &
  -R_{\hat{n}\hat\theta\hat{n}\hat\phi} 
= -R_{\hat{r}\hat\theta\hat{r}\hat\phi} 
= \pm R_{\hat{n}\hat\theta\hat{r}\hat\phi} 
= \pm R_{\hat{r}\hat\theta\hat{n}\hat\phi}
\nonumber\,.
\end{align}
A `dot' denotes derivative with respect to the argument $(t\pm r)$.  Following
standard conventions~\cite{Chandra:1983,Wald:1984,Alcubierre:2008}, we take the $h_+$ and $h_\times$ polarisations
of the gravitational waves to be given by
\begin{align} 
\ddot{h}_+ = &  \frac12(\ddot{h}^{TT}_{\hat\theta\hat\theta} - \ddot{h}^{TT}_{\hat\phi\hat\phi})
\,,\quad
\ddot{h}_\times = \ddot{h}^{TT}_{\hat\theta\hat\phi}
\,.
\end{align}
Then, we find that in vacuum regions of the spacetime, for outgoing waves
$ \ddot{h}_{+,\times}= \ddot{h}_{+,\times}(t-r)$ 
\begin{align} 
\label{eq:BoxPsi4_ddh_defn} 
\Psi_0 = & 0 
\,,\quad
\Psi_4 = \ddot{h}_+ + i\ddot{h}_\times
\,,
\end{align}
while for ingoing waves $ \ddot{h}_{+,\times}= \ddot{h}_{+,\times}(t+r)$ 
\begin{align}
\label{eq:BoxPsi4_ddh_defn2} 
\Psi_0 = & \ddot{h}_+ - i\ddot{h}_\times
\,,\quad
\Psi_4 =  0
\,.
\end{align}
The fact that $\Psi_4$ ($\Psi_0$) are zero for ingoing (outgoing) waves is
consistent with the leading order in perturbation theory. The correct
expression and
in particular its dependence on the radial coordinate is given by the
solution of the
Teukolsky master equation at large distances from the source, which
states that for
outgoing waves \cite{Teukolsky:1973ha}
\begin{align} 
\label{asymptotics}
\Psi_0 \approx & \frac{e^{i\omega(t-r)}}{r^5}
\,,\quad
\Psi_4 \approx   \frac{e^{i\omega(t-r)}}{r}
\,,
\end{align}
while for ingoing waves
\begin{align}
\Psi_0 \approx & \frac{e^{i\omega(t+r)}}{r}
\,,\quad
\Psi_4 \approx   \frac{e^{i\omega(t+r)}}{r^5}
\,.
\end{align}
We decompose the resulting NP scalar NP scalarss $\Psi_4$ ($\Psi_0$) into modes
by projection onto spherical harmonics of spin-weight $s=-2$ ($s=2$)
according to
\begin{subequations}
\label{psi4dec} 
\begin{align}
Mr\Psi_4 = & Mr\,\sum_{l=2}^\infty \sum_{m=-l}^l
             \,{_{-2}}Y_{lm}(\theta\,,\phi)\, \psi^4_{lm}
\,, \\
Mr\Psi_0 = & Mr\,\sum_{l=2}^\infty \sum_{m=-l}^l \,{_{2}}Y_{lm}(\theta\,,\phi)\, \psi^0_{lm}
\,.
\end{align}
\end{subequations}
$_{-2}Y_{l m}(\theta,\phi)$ and $_{2}Y_{l m}(\theta,\phi)$
are $s=\mp2$ spin-weighted spherical harmonics \cite{Goldberg:1966uu,Berti:2005gp}
defined as
\begin{align}
_sY_{l m}(\theta,\phi) \equiv &
(-1)^s\sqrt{\frac{2l+1}{4\pi}}d^l_{m(-s)}(\theta)e^{im\phi}
\,,
\end{align}
where $d^l_{ms}$ is the Wigner $d$-function 
\begin{align}
\label{eq:Box3}
d^l_{ms}(\theta)  \equiv &
  \sum_{t=C_1}^{C_2}\frac{(-1)^t\sqrt{(l+m)!(l-m)!(l+s)!(l-s)!}}
  {(l+m-t)!(l-s-t)!t!(t+s-m)!} 
  (\cos\theta/2)^{2l+m-s-2t}(\sin\theta/2)^{2t+s-m}\,,
\end{align}
and where $C_1= \max(0,m-s)$ and $C_2=\min(l+m,l-s)$. 
Here $M$ is the ADM mass of the system, computed from the initial data and assuming this is an asymptotically flat spacetime.

In the numerical code, the null-tetrad is constructed from a Cartesian
orthonormal triad $(u,v,w)$ and the timelike vector $\hat{n}$ is orthonormal to
$t={\rm constant\ }$ hypersurfaces.
The spacetime is evolved in time $t$ using Cartesian coordinates $x,y,z$.
In practice, we compute the Newman-Penrose scalars $\Psi_0$ and $\Psi_4$
using the electromagnetic decomposition of the Weyl tensor
according to Eqs.~\eqref{eq:BoxelmagnPsi}
on the entire Cartesian grid.
Then, they are interpolated onto coordinate spheres of different extraction
radii $r_{\rm ex}$ with a uniform distribution of points in 
$(\theta,\phi)$.
All the waveform related data from the simulations presented
in the course of this work are taken from such samplings of
$\Psi_0(t,r=r_{\rm ex},\theta,\phi)$ and $\Psi_4(t,r=r_{\rm ex}, \theta,\phi)$.
A more detailed description is given in Appendix~\ref{sec:appWaveExtraction}.

Given the Newman-Penrose scalar $\Psi_4$, we can compute the radiated energy, linear and angular momentum from the radiation
content~\cite{Alcubierre:2008,Ruiz:2007yx}, and references therein:
\begin{subequations}
\label{eq:Boxenergywave}
\begin{align}
\frac{dE}{dt}  = &  \lim_{r\rightarrow \infty} \frac{r^2}{16\pi}
      \int_{\Omega} \left| \int_{-\infty}^{t} \Psi_4 d\tilde{t}
      \right|^2 d{\Omega} 
\,, \\
\frac{dP_i}{dt} = & -\lim_{r \rightarrow \infty} \frac{r^2}{16\pi}
      \int_{\Omega} \ell_i \left| \int_{-\infty}^{t} \Psi_4 d\tilde{t}
      \right|^2 d{\Omega} 
\,, \\
\frac{dJ_z}{dt} = & -\lim_{r \rightarrow \infty} \frac{r^2}{16\pi}
        \mathrm{Re}\left[ \int_{\Omega} \left( 
          \int_{-\infty}^t \Psi_4 d\tilde{t} \right)
          \partial_{\phi} \left( \int_{-\infty}^t \int_{-\infty}^{\hat{t}} \bar{\Psi}_4
          d\tilde{t} d\hat{t}
        \right)d{\Omega} \right] 
\,,
\end{align}
\end{subequations}
where
\begin{align}
\ell_i = &\left(-\sin \theta \cos \phi,\,-\sin \theta \sin \phi,\,
           -\cos \theta \right)
\,.\non
\end{align}
The definitions above are based on time integrals which start in the infinite
past (at retarded time $t=-\infty$), and thus capture the complete
gravitational wave signal.  Starting the time integrations at $t=-\infty$
corresponds to the limit of infinite extraction radius on the initial time
slice --- the slice would then extend all the way to spatial infinity, no part
of the waveform would be lost, and it would take an infinite time for the
waves to reach the extraction sphere. This situation cannot be handled
with the current numerical
codes; we therefore work with finite extraction radii. 

The mass and angular momentum of the final BH can be estimated from
balance arguments.
Given the parameters $(P_{y_{i}},d)$ in the Bowen-York initial data, we
straightforwardly calculate the total initial angular momentum as
\begin{align}
J_{\rm ini} = & L_{\rm ini} = d P_{y_i}
\,,
\end{align}
if the initial spin of each BH is zero.
Ansorg's \textsc{TwoPunctures} \cite{Ansorg:2004ds} initial data solver
directly provides the total ADM mass $M$ of the system and we obtain
radiated energy and angular momentum $E_{\rm rad}$ and $J_{\rm rad}$
from the gravitational wave signal. In case of a merger, this gives
us the final angular momentum and mass of the BH
\begin{subequations}
\begin{align} 
M_{\rm fin}  =& M - E_{\rm rad}
\,, \\
J_{\rm fin}  =& J_{\rm ini} - J_{\rm rad}
\,.
\end{align}
\end{subequations}
The dimensionless spin parameter of the final BH follows directly from
\begin{align}
j_{\rm fin} = & \frac{J_{\rm fin}}{M_{\rm fin}^2}
\,.
\end{align}
%
\subsection{Apparent horizon properties}\label{sec:AH}
A further method to characterize the evolution of a BH spacetime is provided 
by the properties of the apparent horizon (AH) of the final BH itself.
Since this relies only on local
quantities, it does not depend upon the spacetime being asymptotically
flat. In order to monitor the mass and spin of the final BH, we use
Thornburg's Apparent
Horizon Finder \textsc{AHFinderDirect}
\cite{Thornburg:2003sf,Thornburg:1995cp} in different ways, which also
allow us to obtain uncertainty estimates:
\begin{enumerate}
\item
  The irreducible mass $M_{\rm irr}$, calculated from the AH area, 
  enables us to calculate the final BH mass $M_{\rm BH}$~\cite{Christodoulou:1970wf}
  \begin{align}
  \label{eq:Boxchristodoulou}
  M_{\rm BH}^2 = & M_{{\rm irr}}^2 + \frac{J^2}{4M_{{\rm irr}}^2}
  \,.
  \end{align}

  This relation provides a method to check the internal
  consistency of the result for the final BH spin as calculated from the above
  balance arguments. For this purpose we set $M_{\rm BH}=M_{\rm fin}$ and solve
  Eq.~\eqref{eq:Boxchristodoulou} for the spin
  \begin{align}
  j_{\rm fin}^2 = & \frac{J^2}{M_{\rm fin}^2}
         = 4\frac{M_{\rm irr}^2}{M_{\rm fin}^2}\left(1 - \frac{M_{\rm irr}^2}{M_{\rm fin}^2}\right)
  \,.
  \end{align}

  For comparison we also compute the spin of the final BH from the two
  following estimates:

  \item
    We measure the ratio $C_r(j) = C_p/C_e$ of polar to
  equatorial circumference of the final BH \cite{Anninos:1994pa}.
  If we assume the final
  object to be a Kerr BH, this ratio is
  $C_r=\frac{2}{\pi}\sqrt{1-\beta^2}\,E(\beta^2)$, where
  $\beta^2\equiv\,j^2M/(2r_+)$, $E(\beta^2)$ is a complete elliptic
  integral and $r_+/M=1+\sqrt{1-j^2}$. This expression can be inverted
  to find the dimensionless spin parameter, $j_{C_r}$, of the final BH.

\item
  The equatorial circumference of a Kerr BH is $C_e=4\pi
  M$.  Therefore $2\pi A_{AH}/C_e^2=1+\sqrt{1-j_{AH}^2}$, where
  $A_{AH}$ is the area of the apparent horizon.  Thus, the AH area and
  the equatorial circumference
  can be used to estimate the spin of the final BH from
  \cite{Kiuchi:2009jt}
  \begin{align}
  \label{eq:BoxspinArea}
  j_{AH} = & \sqrt{ 1 - \left(\frac{2\pi A_{AH}}{C_e^2} - 1 \right)^2 } 
  \,.
  \end{align}
\end{enumerate}
%


\chapter{Collisions of unequal mass black holes and the point particle limit}
\label{chapter:UnequalMass4D}
\section{Introduction}\label{sec:UM4Dintro}
Black holes play a key role
in a variety of processes in astrophysics, gravitational wave
physics and high-energy physics.  Following the 2005 breakthroughs
\cite{Pretorius:2005gq,Campanelli:2005dd,Baker:2005vv},
NR has been an essential tool in the modeling
of BH binaries in the strong-field regime. At the same
time it has become clear that detailed studies of BH
systems often involve a close interplay between fully non-linear
numerical simulations and (semi)-analytic
approximation techniques of various
types. For example, the generation of gravitational wave (GW)
template banks for use in the analysis of observational data from
operational laser interferometric GW detectors such as LIGO, VIRGO, GEO600 and KAGRA
or future-planned observatories such as the Einstein Telescope
requires the combination of NR with post-Newtonian
or other techniques; see Refs. \cite{Berti:2007fi,Centrella:2010mx,
Sperhake:2011xk,
Damour:2010zb,Pan:2009wj,Santamaria:2010yb,Abadie:2011kd,MacDonald:2011ne,Bernuzzi:2011aj}
and references therein. Post-Newtonian studies have also played an
important role in the guidance of numerical investigations of the
BH recoil, most notably in the discovery of the so-called
{\em superkicks} and their possible suppression due to spin alignment
\cite{Gonzalez:2007hi,Campanelli:2007cga,Campanelli:2007ew, Kesden:2010yp,
Kesden:2010ji,Berti:2012zp,Gerosa:2013laa}. 
In the context of high-energy collisions of BHs,
linearization tools such as the zero-frequency limit or point particle
calculations provide valuable insight into the scattering threshold and
GW emission of BH collisions in four and higher-dimensional
spacetimes \cite{Berti:2010ce}.  Of particular relevance for 
future spaceborne missions in the spirit of LISA, 
are extreme-mass-ratio inspirals (EMRIs).
They represent a particularly
difficult challenge to NR and their modeling
relies heavily on perturbative methods and self-force calculations; see
Refs. \cite{Hinderer:2008dm,Barack:2009ux,Canizares:2009ay,Yunes:2010zj,
Sundararajan:2010sr,Poisson:2011nh} and references therein.

With the above as motivation, it is vital to obtain a detailed
understanding of the range of validity of the various types of
approximation methods. At the same time, these methods provide valuable
tools to calibrate the accuracy of numerically generated solutions to the
Einstein equations. In this chapter, based on \cite{Sperhake:2011ik}, we provide such a study
for the case of a classical calculation in GR, the head-on
infall of a point-particle (PP) into a BH \cite{Davis:1971gg}.

In recent years, NR has started probing the
intermediate mass-ratio regime 
(i) by evolving the final orbits of (approximately) quasi-circular inspirals of BH binaries with
mass-ratio $q\equiv m_2/m_1=1/10$ \cite{Gonzalez:2008bi, Lousto:2010tb};
(ii) by comparing numerical results with perturbative calculations employing
the fully numerical BH trajectories for mass ratios up to $q=1/20$
\cite{Lousto:2010qx};
(iii) by evolving the last orbits of a quasi-circular inspiral with mass ratios up to $q=1/100$
\cite{Lousto:2010ut,Nakano:2011pb,Lousto:2013oza}.
Recently, calculations of EMRIs with $q=1/10^{6}$ in full GR have been presented~\cite{East:2013iwa}
Nevertheless, in this chapter based on Ref.~\cite{Sperhake:2011ik}
I focus on head-on collision of BHs with $q\geq1/100$, for two reasons: (i) the lower computational cost due to the higher
degree of spacetime symmetry and the absence of the lengthy inspiral
phase and (ii) the availability of high-precision results in the PP limit.

In our study we will make extensive use of the calculation by Davis {\em
et al.} \cite{Davis:1971gg} who model the collision of
a small object of mass $m$ with a BH of mass $M \gg m$ in the PP limit.  In the
original calculation the particle was falling from rest at infinity,
and the total radiated energy was found to be
\begin{align}
\label{eq:UM4DDRPP}
E^{\rm rad}_{\rm PP} = & 0.0104 \frac{m^2}{M}
\,.
\end{align}
This setting has been generalized to arbitrary initial distance and boost,
in which case initial data and consequent spurious radiation play a role
\cite{Lousto:2004pr,Lousto:1996sx,Cardoso:2002ay,Berti:2010ce,Mitsou:2010jv}.

Fully numerical results for BH head-on collisions obtained
in the equal and comparable mass regime have been compared with PP
predictions and results obtained in the close-limit approximation
\cite{Andrade1997} by Anninos et al \cite{Anninos:1994gp,
Anninos:1998wt}. These studies demonstrated agreement for the radiated
energy and linear momentum bearing in mind the accuracies achievable at
the time. The waveforms presented therein, however, exhibit a
significant signal starting at $t-R_{\rm ex}=0$; see for example Fig.~1
in \cite{Anninos:1998wt}. This contribution most likely arises from
spurious radiation inherent in the initial data due to the small initial
separation and its impact on the quantitative conclusions is not
entirely clear.

At the time when the present research was carried out no 
comparisons between PP calculations and fully numerical results for mass ratios in a truly perturbative
regime have been available (but note Ref.~\cite{East:2013iwa}).
By simulating BH binaries up to a mass ratio of $q=1/100$
we fill this gap and identify those aspects of the PP predictions
which describe BH dynamics well in general and which only hold
in the extreme mass-ratio limit. From a different point of view, the
agreement with the PP calculations represents an important
validation of the fully numerical calculations in the regime of
high-mass ratios. In this context we emphasize that we are able to
accurately extract 
radiated GW energies of the order of $10^{-6}~M$
and linear momenta corresponding to recoil velocities of a few dozens of m/s
from binary BH simulations.
We note, however, that even smaller amounts of energy have been extracted from
general relativistic simulations of stellar core collapse;
see e.~g.~\cite{Reisswig:2010cd}.


\vspace{0.2cm}
\section{Numerical Setup and Analysis Tools}\label{sec:UM4Dsetup}
The numerical simulations of unequal-mass BH collisions
starting from rest have been performed with the \textsc{Lean} 
code~\cite{Sperhake:2006cy,Sperhake:2007gu}.
The \textsc{Lean} code is based on the \textsc{Cactus} computational
toolkit \cite{Goodale02a, cactus} and uses the \textsc{Carpet} mesh refinement
package \cite{Schnetter:2003rb, carpet},
 the apparent horizon finder \textsc{AHFinderDirect}
\cite{Thornburg:1995cp,Thornburg:2003sf} and the \textsc{TwoPuncture}
initial data solver~\cite{Ansorg:2004ds}. The $3+1$ Einstein's
equations are evolved using the BSSN \cite{Shibata:1995we,
Baumgarte:1998te} formulation, together with the moving puncture
approach \cite{Baker:2005vv,Campanelli:2005dd}, discussed in Sec.~\ref{sec:NRgBSSN}.
The gauge conditions are determined by the puncture gauge, 
i.e., the 1+log slicing and $\Gamma$ driver shift condition \cite{Alcubierre:2002kk} (cf. Sec.~\ref{sec:NRgauge}).
The systems are set up using Brill-Lindquist initial data, summarized in Sec.~\ref{sec:NRInitData}.
We have evolved BH binaries with mass ratios $q \equiv m_2/m_1
= 1, 1/2, 1/3, 1/4, 1/10$ and $1/100$, where $m_i$ is the bare mass parameter
of the $i$-th BH.

We use the Newman-Penrose scalar $\Psi_4$ to measure gravitational radiation
at extraction radii $R_{\rm ex}$, chosen in a range of $40~M$ to $90~M$
from the center of the collision. We decompose
$\Psi_4$ into multipoles $\psi_{lm}$ using $s=-2$ spin-weighted spherical harmonics $-2$, ${_{-2}}Y_{lm}$, 
according to Eq.~\eqref{psi4dec}.
Due to the symmetry properties of the systems under consideration,
the only non-vanishing multipoles all have $m=0$ in a suitably chosen frame, and are purely real,
corresponding to a single polarization state $h_+$. In the equal-mass limit,
the additional symmetry causes all multipoles with odd $l$ to vanish
identically. The energy spectrum and luminosity of the radiation
are given by
\begin{subequations}
\begin{align}
\label{eq:UM4Dspectrum} 
\frac{dE}{d\omega} = & \sum_l \frac{1}{16\pi^2}\frac{|\hat{\psi}_{l0}(\omega)|^2}{\omega^2}
\equiv \sum_l \frac{dE_l}{d\omega}\,,
\,,\\
\label{eq:UM4Dluminosity}
\frac{dE}{dt} = & \sum_l \frac{1}{16\pi M^2} \left|\int_{-\infty}^{t} \psi_{l0}(\tilde{t}) d\tilde{t} \right|^2
\equiv \sum \frac{dE_l}{dt}
\,,
\end{align}
\end{subequations}
respectively, where a hat denotes the Fourier transform and $\psi_{l0}$
is evaluated on a sphere at
infinity.

\vspace{0.2cm}
\section{Simulations and uncertainties}\label{sec:UM4Dsims}
We have performed a series of simulations of head-on collisions with mass
ratio ranging from $q=1$ to $q=1/100$ with initial coordinate separation
$d$ and proper horizon-to-horizon separation $L$ as given in
Table~\ref{tab:UM4Dmodels}.
\begin{table}
\begin{center}
\begin{tabular}{r|rr|cccc|c}
  \hline
  $q$ & $d/M$ & $L/M$ & $E^{\rm rad}/M$ & \multicolumn{3}{c}{$E^{\rm rad}_{l=2,3,4}(\%)$} & $v/({\rm km}/{\rm s})$ \\
  \hline
    1 & 10.24 & 12.48 & $5.32 \times 10^{-4}$ & 99.6 & 0    & 0.03 & 0 \\
    1 & 12.74 & 16.76 & $5.39 \times 10^{-4}$ & 99.3 & 0    & 0.03 & 0 \\
    1 & 17.51 & 21.82 & $5.56 \times 10^{-4}$ & 99.4 & 0    & 0.03 & 0 \\
  1/2 & 12.74 & 16.69 & $4.33 \times 10^{-4}$ & 98.1 & 1.28 & 0.07 & 3.71 \\
  1/3 & 12.74 & 16.60 & $3.11 \times 10^{-4}$ & 96.7 & 2.83 & 0.16 & 3.97 \\
  1/4 &  7.31 & 10.57 & $2.16 \times 10^{-4}$ & 95.8 & 3.85 & 0.25 & 3.65 \\
  1/4 & 12.74 & 16.53 & $2.28 \times 10^{-4}$ & 95.4 & 4.14 & 0.28 & 3.72 \\
  1/4 & 17.51 & 21.61 & $2.33 \times 10^{-4}$ & 95.6 & 4.13 & 0.27 & 3.83 \\
 1/10 & 12.72 & 16.28 & $6.05 \times 10^{-5}$ & 92.1 & 7.09 & 0.67 & 1.31 \\
 1/10 & 16.72 & 20.55 & $6.16 \times 10^{-5}$ & 92.5 & 7.23 & 0.70 & 1.33 \\
 1/10 & 20.72 & 24.76 & $6.29 \times 10^{-5}$ & 92.0 & 7.15 & 0.67 & 1.34 \\
1/100 &  7.15 &  9.58 & $9.10 \times 10^{-7}$ & 88.1 & 9.01 & 1.15 & 0.0243 \\
1/100 & 11.87 & 15.08 & $9.65 \times 10^{-7}$ & 88.0 & 9.87 & 1.46 & 0.0248 \\
1/100 & 13.85 & 17.21 & $9.94 \times 10^{-7}$ & 87.8 &10.11 & 1.46 & 0.0256 \\
1/100 & 15.08 & 18.53 & $1.012 \times 10^{-6}$ & 87.7 &10.05 & 1.51 & 0.0260 \\
\hline
\end{tabular}
\end{center}
\caption{Mass ratio $q$, coordinate and proper separation $d$ and $L$,
respectively, as well as radiated energy $E_{\rm rad}$ with percentage
distribution in the $l=2$, $l=3$ and $l=4$ multipoles and recoil velocity
$v$ for the set of binary models evolved numerically.
}
\label{tab:UM4Dmodels}
\end{table}
We describe the grid setup used for these simulations in terms of the
number $n_{\rm rl}$ of refinement levels, the
radius $R$ of the computational domain, the resolution $H$ used in the
wave extraction zone, typically the third refinement level counted from the outside, 
the radius $r$ in units of the smaller hole's mass $m_2$
of the innermost refinement level centered on the individual
punctures
and the resolution $h/m_2$ of the innermost
refinement level. The values for these parameters are summarized for all
mass ratios in Table~\ref{tab:UM4Dgrids}.
Because of the vastly different length scales in case of the small mass
ratios $q=1/10$ and  $q=(1/100)$ we include, respectively, $2$ or $5$ additional
refinement levels around the smaller BH. Thus we are able to evolve the system
at feasible computational costs.
\begin{table}
\begin{center}
\begin{tabular}{r|ccccc}
  \hline
  $q$ & $n_{\rm rl}$ & $R/M$ & $H/M$ & $r/m_2$ & $h/m_2$ \\
  \hline
    1 & 9 & 512 & 0.76 & 2 & 1/21 \\
  1/2 & 9 & 341 & 0.51 & 2 & 1/21 \\
  1/3 & 9 & 256 & 0.76 & 2 & 1/21 \\
  1/4 & 9 & 205 & (1.22,~1.07,~0.95) & 1 & (1/21,~1/24,~1/27) \\
 1/10 &12 & 303 & 0.73 & 0.625 & 1/64 \\
1/100 &15 & 223 & (1.01,~0.63,~0.51) & 0.625 & (1/40,~1/64,~1/80) \\
  \hline
\end{tabular}
\end{center}
\caption{Grid setup used for the different mass ratios $q$. The number
         of refinement levels is given by $n_{\rm rl}$,
         $R$ is the
         radius of the computational domain, $H$ the resolution in the
         wave extraction zone, $r$ the radius of the innermost refinement
         box around the individual punctures and $h$ the resolution
         used on that level. The additional low and high resolution for
         $q=1/4$ and $q=1/100$ have been used for the convergence studies.
}
\label{tab:UM4Dgrids}
\end{table}

Our results are affected by three main sources of uncertainties: finite
extraction radius, discretization and, for small initial separations of the
binary, spurious initial radiation. We reduce
the error arising from finite extraction radius by measuring the waveform
components at several radii, and fitting them to an expression of the form
$\psi_{lm}(r,t)=\psi_{lm}^{(0)}(t)+\psi_{lm}^{(1)}(t)/r$.
The waveform ``at infinity'' $\psi_{lm}^{(0)}(t)$ is the quantity reported
throughout this work and used to calculate related quantities, such as the
radiated energy. The uncertainty in this extrapolated value is estimated by
performing a second fit including also a quadratic term $\psi_{lm}^{(2)}/r^2$,
and taking the difference between the first- and second-order fits. The
resulting uncertainty increases as we decrease the mass ratio $q$ and
is $1-4~\%$ for the total radiated energy and the
$l=2$ waveform and energy, and $3-5~\%$ for the subdominant multipoles
and the radiated linear momentum.

In order to estimate the discretization error of our simulations, we
have performed a convergence analysis for models $(q=1/4,~L=16.53~M)$
and $(q=1/100,~L=9.58~M)$ using the three resolutions listed 
in Table~\ref{tab:UM4Dgrids}.
\begin{figure}
\begin{center}
\includegraphics[clip=true,width=0.75\textwidth]{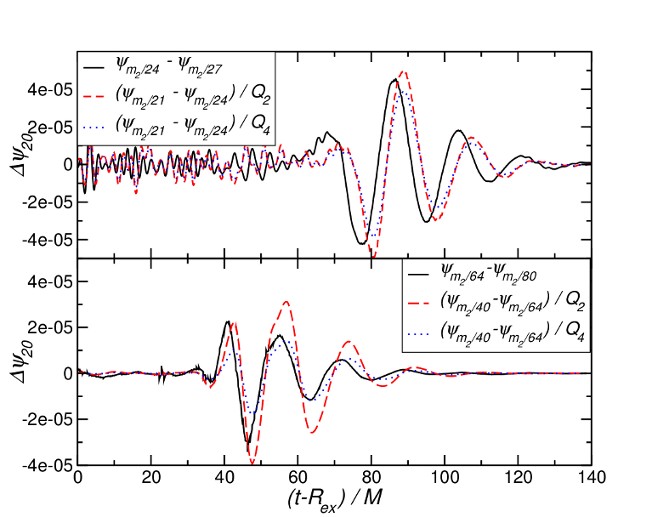}
\end{center}
\caption{Convergence analysis for the $l=2$ multipole of the gravitational
         wave signal for simulation $q=1/4,~D=16.53~M$ (upper panel) and
         simulation $q=1/100,~D=9.58$ (lower panel). In both cases we
         show the higher resolution differences (solid black)
         together with the lower
         resolution result rescaled for second (dashed red lines) and
         fourth-order convergence (dotted blue lines).
}
\label{fig:UM4Dconv_psi20}
\end{figure}
The resulting convergence plots for the $l=2$ multipole of the wave signal
is shown in Fig.~\ref{fig:UM4Dconv_psi20} and demonstrates convergence between
second and fourth order.
With regard to the analysis below, we note in particular that
the $q=1/100$ case exhibits second order convergence in the plunge-merger
signal around $t-R_{\rm ex} \approx 40~M$ but is close to fourth-order
convergence for the remainder of the waveform. Bearing in mind that
the plunge-merger transition represents the most dynamic part of the
evolution and that the second-order ingredients in the code are associated
with the prolongation of grid functions at the refinement
boundaries in time, this observation is
compatible with the numerical discretization. We observe similar
convergence properties for the $l=3$ multipole, but overall convergence
close to fourth-order for the radiated energy and linear momentum,
presumably because the accumulated errors are dominated by the
fourth-order contributions observed for most of the signal.
The resulting numerical uncertainties for $q=1/100$ are about $10~\%$
in the waveform for the plunge-merger transition and $5~\%$ for the
remainder of the signal as well as $6~\%$ for the radiated energy
and $8~\%$ for the linear momentum emitted in gravitational waves.
We note that in both cases the discretization error leads to
an overestimate of the radiated quantities.
For $q=1/4$ we observe significantly smaller uncertainties in the
range of $2~\%$ for all quantities.

Finally, we comment on the unphysical gravitational radiation inherent in
the conformally flat puncture initial data.
In order to extract physically meaningful information, one
has to separate the spurious radiation from the radiation generated by the
collision itself. This is done by ``waiting'' for the spurious radiation to
radiate off the computational domain, and then discarding the early,
contaminated part of the wave signal. For small values of the
initial separation, however, the binary will merge before the
spurious radiation has had enough time to leave the system, and physical
and unphysical contributions to the wave signal partially overlap
and cannot be cleanly distinguished. For our set of simulations, this
problem arises only in the case $q=1/100$, $L=9.58~M$, where it introduces
an additional error of about $2~\%$ to the radiated energy and momentum.

\vspace{0.2cm}
\section{Results}
All collisions summarized in Table \ref{tab:UM4Dmodels} result in the formation
of a single BH plus gravitational radiation, i.~e.~there is no indication
of violation of the cosmic censorship conjecture. The final BH is born
distorted, and eventually rings down to a Schwarzschild solution via
emission of quasinormal modes \cite{Berti:2009kk}.
\begin{figure}
\begin{center}
\begin{tabular}{ccc}
\hspace{-0.5cm}
\includegraphics[width=0.365\textwidth]{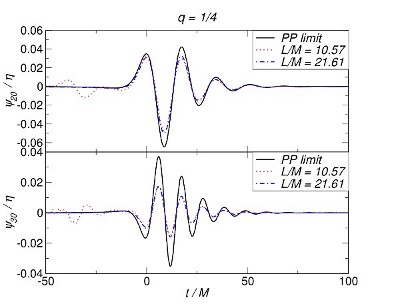} &
\hspace{-0.9cm}
\includegraphics[width=0.365\textwidth]{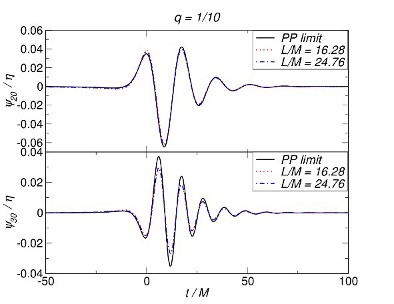} &
\hspace{-0.9cm}
\includegraphics[width=0.365\textwidth]{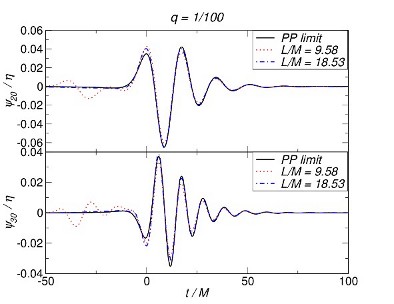}
\end{tabular}
\end{center}
\caption{\label{fig:UM4DwaveformsPP} (Color online)
Rescaled waveforms for mass ratios
$q=1/4$ (left panel), $q=1/10$ (mid panel) and $q=1/100$ (right panel)
for $l=2$ (upper) and $l=3$ (lower half of each panel),
for two different initial separations. Also shown is the waveform in the 
PP limit (black solid lines).
}
\end{figure}

We illustrate the $l=2$ and $l=3$
wave signal in Fig.~\ref{fig:UM4DwaveformsPP} for the
$l=2$ and $l=3$ multipoles obtained for the mass ratios $q=1/4$ (top),
$q=1/10$ (center) and $q=1/100$ (bottom).
In each panel the solid (black) curves represent
the PP prediction for infall from infinity whereas the
dotted (red) and dash-dotted (blue) curves show the numerical results
for different values of the finite initial separation.
To leading order, the gravitational radiation output of BH collisions
scales with the square of the reduced mass
$\mu\equiv M\eta$ of the system, where $\eta=q/(q+1)^2$ is
the dimensionless, symmetric mass ratio \cite{Davis:1971gg}.
For comparison of the numerical
results with PP predictions, we therefore rescale the
former by the corresponding powers of $\eta$, 
quadratic for energy and linear for the waveforms in Fig.~\ref{fig:UM4DwaveformsPP}.

The waveforms show interesting features. For small initial separations,
the early part of the waveform is contaminated by ``spurious''
radiation; cf.~the dotted (red) curve in the top and bottom panels of
Fig.~\ref{fig:UM4DwaveformsPP}. As the initial separation increases,
however, this problem disappears, because the longer infall duration of
the binary provides sufficient time for the unphysical radiation to propagate
off the grid; cf.~the dash-dotted (blue) curves. A closer inspection
of the $q=1/100$ case yields excellent agreement between the numerical
and PP predictions except for the plunge-merger transition
around $t\approx 0$ in the figure. From the discussion in
Sec.~\ref{sec:UM4Dsims}, however, we recall that the discretization
error is particularly large in this regime. In fact, for the $q=1/100$
model studied in Sec.~\ref{sec:UM4Dsims}, a second-order Richardson
extrapolation predicts about a $10~\%$ reduction in the amplitude
around the first strong maximum in the $l=2$ waveform which is very
close in magnitude and sign to the deviation of the numerical
from the PP result.
As demonstrated by the upper central panel in Fig.~\ref{fig:UM4DwaveformsPP},
we find equally good agreement of the numerical $l=2$ multipole with PP
predictions for the less extreme mass ratio $q=1/10$ 
and only a small deviation for the larger mass ratio $q=1/4$ 
(upper top panel in Fig.~\ref{fig:UM4DwaveformsPP}).
Our findings thus confirm over a wide range of mass ratios the 
observation by Ref.~\cite{Anninos:1994gp}, that there is a weak
dependence of the re-scaled waveforms on the mass ratio.
The $l=3$ mode, on the other hand, is a good discriminator between high- and
low-mass ratios.
This behavior was qualitatively expected, as
higher multipoles are suppressed in the equal-mass case;
by symmetry the $l=3$ mode is absent when the masses are equal.
It is interesting,
however, that even for what one might call a small mass ratio, $q=1/10$,
higher multipoles are still visibly suppressed.
%
\begin{table}
\begin{center}
\begin{tabular}{cccccc}
\hline
$q$                                  & 1/1    & 1/4    & 1/10   & 1/100  & PP\\ \hline
$E^{\rm rad}_{\infty}/(M\eta^2)$    &0.00936 & 0.00911 & 0.00985  & 0.0114& 0.0104\\
$v^f_{\infty}$ (km$/$s)             &0.0     &258.0   &250.3  &275.9     & 257.6 \\
\hline
\end{tabular}
\end{center}
\caption{\label{tab:UM4Dr0dependence} Summary of our results when fitted to
Eqs.~(\ref{eq:UM4Dr0dep}) and (\ref{eq:UM4DfitV}). The last column refers to PP results, as extrapolated 
from Lousto and Price \cite{Lousto:2004pr}. 
}
\end{table}

The total amount of energy radiated in gravitational waves during the
collision depends on the initial separation of the holes. As discussed
in Anninos {\em et al.} \cite{Anninos:1994gp}, two effects contribute
to increasing the
GW energy at larger initial separations; (i) there is more time
to radiate GWs during the infall and (ii) the infalling velocity
is larger. In practice, the second effect is found to be dominant.
Anninos {\em et al.} have accounted for both contributions by defining
\begin{align}
F_{L} = & \frac{\int_{L}^{2M}\dot{r}\ddot{r}^2dr}{\int_{\infty}^{2M}
        \lim_{L\to\infty}\dot{r}\ddot{r}^2dr}
%
\,,\quad\text{with}\quad
\dot{r} = \frac{(1-2M/r)\sqrt{2ML/r-2M}}{\sqrt{L-2M}}
\,.\nonumber
\end{align}
One can write the corrections to the radiation emission
\begin{align}
E^{\rm rad}_{L} = & F_{L}E^{\rm rad}_{\infty}
 = \left(1-\frac{40M}{9L}\right)E^{\rm rad}_{\infty}+{\cal O}\left(\frac{M^2}{L^2}\right)
\,.\nonumber
\end{align}
With the above as motivation, we have fitted our results to a
$1/L$ dependence, of the form
\begin{align}
\label{eq:UM4Dr0dep}
\frac{E^{\rm rad}(L)}{M\eta^2} = & \frac{E^{\rm rad}_{\infty}}{M\eta^2} \left(1+a_E\,M/L\right )
\,,
\end{align}
with $E^{\rm rad}_{\infty}$ the radiated energy for infinite initial
separation.  The results are summarized in Table~\ref{tab:UM4Dr0dependence}.
We remind the reader that $L$ stands for proper initial separation between
the holes.  We also note that the results in Table \ref{tab:UM4Dr0dependence}
are normalized by $\eta^2$.  For comparison, we also show in the
last entry of the table the results obtained in the PP limit, within
a linearized calculation. This study was done by Lousto and Price
\cite{Lousto:2004pr} using the same type of initial data; we have used
their Table I to obtain the behavior shown in Table~\ref{tab:UM4Dr0dependence}
above. We note that already for $q=1/10$ and $q=1/100$ our results are in good
agreement with PP calculations. We remind the reader, however, that
in the $q=1/10$ case there is a larger deviation in the $l=3$ modes.

With the extrapolation above one gets an estimate for the total radiation
of two BHs merging from infinite initial separation. A best fit
of this number as function of mass ratio yields
\begin{align}
\label{eq:UM4DfitA}
\frac{E^{\rm rad}_{\infty}}{M\eta^2} = & 0.0110 -0.0088 \eta 
\, 
\end{align}
In the PP limit, when $\eta \to 0$, this agrees with the classical PP
calculation, Eq.~(\ref{eq:UM4DDRPP}) to within $6\%$, i.e., within the numerical
uncertainties. Overall, the results in Table \ref{tab:UM4Dmodels} demonstrate
that we are able to accurately measure amounts of order $E^{\rm rad}\sim
10^{-6}M$ in these fully nonlinear evolutions.


The amount of spurious radiation in the initial data is also
consistent with predictions from linearized gravity. Lousto and Price
performed a detailed analysis of the amount of spurious radiation in the
infall of PPs into massive BHs, using the same type of initial
data \cite{Lousto:2004pr}. Using their Table I for $L>11$,
we find that the amount of spurious radiation varies with $L$ according
to $E_{\rm rad}/(M\eta^2) \sim 0.15(L/M)^{-2.5}$.
For $q=1/100$, for instance, we obtain
$E_{\rm rad}/(M\eta^2)=0.26(L/M)^{-2.55}$.
Thus, we find good agreement in the decay power (roughly $-2.5$) and
also in the proportionality coefficient.

If two BHs with different masses collide head-on, the remnant BH will recoil 
with respect to the center-of-mass frame, due to the emission of energy and 
momentum carried by gravitational waves. Based on PN tools, we have fit
our results to \cite{Fitchett:1983}
\begin{align}
\label{eq:UM4DfitV}
  v_{\rm recoil} = & v^f_{\infty}\, \frac{q^2(1-q)}{(1+q)^5}
    \left(1+b_E\,M/L\right)
\,,
\end{align}
where $v^f_{\infty}$ is a normalized recoil velocity for infinite initial
separation.  The normalized recoil velocity $v^f_{\infty}$ is shown in
Table \ref{tab:UM4Dr0dependence}.  The point particle limit was considered in
Ref.~\cite{Nakamura:1983hk}, who obtained $v^f_{\infty}=263 {\rm km/s}$ \footnote{note the slight disagreement with the extrapolation of Lousto
and Price's results, shown in Table \ref{tab:UM4Dr0dependence}}.  We note this
is not a trivial agreement: unlike energy calculations, momentum involves
interference with higher (typically highly suppressed) multipoles.
Overall, our results agree well in the limit of small mass-ratios with the
point particle limit. It is interesting to note in this context that for
both, radiated energy and linear momentum, the numerical results exceed
those obtained from the point particle limit by about $6~\%$. This value
agrees in sign and magnitude with the discretization error obtained for
the $q=1/100$ simulation in Sec.~\ref{sec:UM4Dsims}. We therefore consider
the discretization error the dominant source of the remaining
discrepancies.

\vspace{0.2cm}
\section{Conclusions}\label{sec:UM4Dconclusion}

The simulation of dynamical, interacting BHs has a tremendous
potential to provide answers to some of the most fundamental questions
in physics. Recent developments in experimental and theoretical
physics make this a pressing issue. We refer, in particular, to the
prominent role of BHs in the gauge-gravity duality, in TeV-scale
gravity or even on their own as solutions of the field equations
\cite{Zilhao:2010sr}. Recent work along these lines includes the
successful simulation and understanding of the collision of two
BHs at close to the speed of light in four-dimensional spacetime
\cite{Sperhake:2008ga,Sperhake:2009jz,Sperhake:2010uv,Shibata:2008rq},
the low energy collisions in higher spacetime dimensions
\cite{Zilhao:2010sr,Witek:2010xi,Witek:2010az}, BH scattering in five
dimensions \cite{Okawa:2011fv}, stability studies in
higher dimensions \cite{Shibata:2009ad, Shibata:2010wz,Lehner:2010pn} and
BH evolutions in non asymptotically flat spacetimes 
\cite{Bantilan:2012vu,Zilhao:2012bb,Witek:2010qc}.

We have shown here that NR is capable of simulating
dynamical BHs close to the regime of validity of linear
calculations, and to make contact with (semi-)analytic
approximation techniques. For this purpose we have evolved head-on
collisions of non-spinning BH binaries over a range of
mass ratios from $q=1$ to $q=1/100$.
We obtain radiated energies decreasing from about $5.5 \times 10^{-4}$
for $q=1$ to $10^{-6}$ for $q=1/100$. The recoil reaches
a maximum of about $4~{\rm km/s}$ near $q=3$ and decreases towards
$26~{\rm m/s}$ for $q=1/100$. In the limit of small mass ratios and
extrapolating our results to infinite initial separation, we find the
numerical values for radiated energy and linear momentum to be
$\approx 6~\%$ larger than the point-particle predictions. This
discrepancy agrees rather well in sign and magnitude with the
discretization error obtained from a convergence study of our
$q=1/100$ simulations. It thus appears likely that a significant
part of the remaining differences can be attributed to the
discretization error which mirrors the computational demands
of numerical BH binary simulations with such small
mass ratios.

With regard to the waveforms, the most remarkable result is the
suppression of odd $l$ multipoles. While we observe good
agreement between numerical and point-particle results for the
$l=2$ mode, already for $q=1/10$, the numerically calculated
$l=3$ multipole is visibly
suppressed for this case and only agrees well with the PP limit
for $q=1/100$.

Overall, the good agreement for waveforms and radiated energy and
momenta for the case $q=1/100$ demonstrates that numerical techniques
are capable of bridging the gap between linear analysis and the
fully non-linear regime of general relativity.


\chapter{Black hole collisions in higher dimensional spacetimes -- Framework}
\label{chapter:HigherDimF}
\section{Introduction}\label{sec:HDFIntro}
Black hole physics in higher dimensional spacetimes exhibit a plethora of fascinating 
phenomena, ranging from potential applications to high energy 
physics in the context of TeV scale gravity scenarios
\cite{Antoniadis:1990ew,ArkaniHamed:1998rs,Antoniadis:1998ig,Randall:1999ee,Randall:1999vf,
Argyres:1998qn,Giddings:2001bu,Dimopoulos:2001hw,Ahn:2002mj,
Chamblin:2004zg,Feng:2001ib,Ahn:2003qn,Cardoso:2004zi}
to fundamental questions about the stability of higher dimensional black objects and their phase diagrams
\cite{Emparan:2008eg,Emparan:2003sy,Shibata:2009ad,Cardoso:2006sj,Cardoso:2009bv,
Cardoso:2009nz,Gregory:1993vy,Choptuik:2003qd,Dias:2011jg,Dias:2010maa,Dias:2010eu,Lehner:2010pn,Lehner:2011wc,
Reall:2012it}.
Of particular interest are the dynamics of BHs in higher dimensional spacetimes 
-- a quite recent research branch in NR.
Perturbative methods are not sufficient to tackle most outstanding issues in the strong-field regime, 
and the employment of numerical relativity methods is inevitable.
Over the past few years, our group -- among few others worldwide -- has started to explore 
highly non-linear, dynamical scenarios, such as (head-on) collisions of BHs in
$D\ge5$ dimensional spacetimes and calculated the associated emission of gravitational radiation.
These calculations provide improved estimates for the amount of energy and angular momentum
emitted in form of gravitational waves and allows to compute 
(by energy balance arguments) the mass and spin of the newly created BH. In the context of TeV gravity
scenarios this information is of utmost importance for BH event generators, such as 
{\textsc{Blackmax}}, {\textsc{Catfish}}, {\textsc{Charybdis2}} or {\textsc{TrueNoir}} 
\cite{Dimopoulos:2001hw,Cavaglia:2006uk,Frost:2009cf,Dai:2007ki,Dai:2009by}.

Conceptually, such a NR code is based either on an extension of the GHG formulation 
or on the explicit splitting of space and time, 
thus evolving the dynamical system in time
(see e.g.\cite{Cardoso:2012qm,Yoshino:2011zz,Yoshino:2011zza,Sperhake:2011xk,Sperhake:2013qa}
for reviews on recent progress in higher dimensional NR).
However, a full-blown $(D-1)+1$-approach, analogous to the $3+1$-approach
presented in Chapter~\ref{chapter:NRframework}, 
is computationally too expensive for currently available computational resources.
Therefore, the problem of dynamically evolving higher dimensional spacetimes has been 
simplified to axissymmetric or effectively $3+1$-dimensional setups
\cite{Yoshino:2009xp,Shibata:2009ad,Shibata:2010wz,Okawa:2011fv,Lehner:2010pn,Lehner:2011wc,Dennison:2010wd,
Sorkin:2009wh,Sorkin:2009bc,Witek:2010xi,Zilhao:2010sr}.

In this chapter I will review the formalism and framework (developed by our group) 
adopted to perform numerical simulations
of BH collisions in higher dimensional spacetimes. 
For a summary of the numerical results
I refer the reader to Chapter~\ref{chapter:HigherDimR}. 
The presented summary is based on the publications 
\cite{Witek:2010xi,Witek:2010az,Zilhao:2010sr,Zilhao:2011yc}
as well as on as yet unpublished work in progress
\cite{Hilditch2012}. 

The starting point for our approach are $D$-dimensional spacetimes with an isometry group.
If this isometry group is sufficiently large, it permits the dimensional reduction of the problem 
to a $4$-dimensional model.
Specifically, I obtain $4$-dimensional GR coupled, in the most general case, 
to scalar and gauge fields.
The additional $(D-4)$ spatial dimensions manifest themselves only in these additional
fields emerging from the dimensional reduction.
We focus on two types of models in particular, which are generalisations of axial symmetry
to higher dimensional spacetimes: 
(i) $D\ge5$ dimensional vacuum spacetimes with an $SO(D-2)$ isometry group, 
and (ii) $D\ge6$ dimensional vacuum spacetimes with an $SO(D-3)$ isometry group. 
In order to end up with an effectively $4$-dimensional model I employ only part of this symmetry: 
Specifically, I perform a dimensional reduction by isometry on a $(D-4)$-sphere 
that has an $SO(D-3) \subset SO(D-2)$ isometry group
which allows us to investigate head-on collisions of BHs in $D\ge5$ as well as 
BH collisions with impact parameter and with spinning BHs in $D\ge6$, 
as long as all the dynamics are restricted to a single plane.
These classes include the most interesting physical configurations relevant to
accelerator and cosmic ray physics in the context of TeV-scale gravity,
and to fundamental questions concerning the properties of higher-dimensional black objects 
such as their stability and phase diagrams.

Subsequently, I cast the effective $4$-dimensional model into a time evolution problem by employing the 
$3+1$-split, discussed in Sec.~\ref{sec:NRdecomp}.
The resulting ADM-like equations are then reformulated in the generalized
BSSN evolution scheme \cite{Shibata:1995we,Baumgarte:1998te,Witek:2010es}, 
as described in Sec.~\ref{sec:NRgBSSN}.
The system is completed by specifying the gauge using the moving puncture approach \cite{Campanelli:2005dd,Baker:2005vv}
adapted to our setup.

In the following sections I will present the framework for NR simulations of higher dimensional
spacetimes, including the dimensional reduction to an effectively $4$-dimensional model in Sec.~\ref{sec:HDFDimRed},
the specific formulation of the time evolution problem in Sec.~\ref{sec:HDFevol},
the setup of initial data in Sec.~\ref{sec:HDFInitData} and the technique used to extract
gravitational radiation in Sec.~\ref{sec:HDFWaveExtr}.
Additionally, I show that the generalized BSSN formulation of the evolution equations 
of the effectively $3+1$-dimensional GR model
coupled to a scalar field together with the modified moving puncture gauge 
exhibits a strongly hyperbolic PDE system in Sec.~\ref{sec:HDFHyperbolicity}.
The presented formalism has been implemented in the {\textsc{HD-Lean}} code 
\cite{Zilhao:2010sr,Witek:2010xi} which is an extension of Sperhake's original {\textsc{Lean}} code 
\cite{Sperhake:2006cy}.

\section{$4+(D-4)$-form and dimensional reduction}\label{sec:HDFDimRed}
%
The key ingredient of our NR framework for higher dimensional spacetimes is 
the reduction from $D$-dimensional vacuum GR to an effectively $4$-dimensional model. 
The isometry group of $D$-dimensional Minkowski spacetime is $ISO(1,D-1)$ while BH
solutions of GR 
generically break this symmetry into a subgroup. For instance, the isometry
group of a $D$-dimensional Schwarzschild-Tangherlini BH \cite{Tangherlini:1963bw}
is $SO(D-1)\times \mathbb{R}$, whereas the isometry group for a head-on collision of two
non-rotating BHs is $SO(D-2)$. 
In fact, neither the time direction nor the direction of the collision correspond to symmetries, 
but a rotation of the remaining $(D-2)$ spatial directions leaves the spacetime invariant. 
The spacetime can then be considered as the semi-direct product of a $3$-dimensional 
spacetime $\,^{(3)}\!\M$ with the sphere $S^{D-3}=SO(D-2)/SO(D-3)$
and we take advantage of this symmetry to reduce the spacetime dimensionality. 
This can be accomplished by writing the $D$-dimensional vacuum Einstein's equations 
\begin{align}
\label{eq:HDFdimEE}
\,^{(D)}\!G_{MN} = & \,^{(D)}\!R_{MN} - \frac{1}{2} g_{MN} \,^{(D)}\!R = 0
\,,
\end{align}
where $M,N=0,...,(D-1)$ are spacetime indices,
in a coordinate system 
which makes the symmetry manifest and decomposing Einstein's equations with respect to this symmetry,
thus providing a lower dimensional interpretation 
(in the spirit of Kaluza-Klein reduction). 
Note, however, that we are not performing a compactification but a 
dimensional reduction by isometry, as first proposed by Geroch \cite{Geroch:1970nt}. 
Then, the extra dimensions appear as source terms of the lower
dimensional Einstein's equations, 
defined on the lower dimensional manifold.
Specifically, we use the $SO(D-3)$ symmetry, thus reducing the spacetime
on a $(D-4)$-sphere which yields a $4$-dimensional manifold.
The original proposal by Geroch \cite{Geroch:1970nt} considered the symmetry space $SO(2)$.  
A $5$-dimensional extension with the same symmetry space has been derived in \cite{Chiang:1985rk}.  
A generalisation to coset manifolds (like the sphere $S^n$) was given by Cho in \cite{Cho:1986wk,Cho:1987jf}.
We now describe in detail the reduction from a $D$-~to a $4$-dimensional spacetime.  
In order to highlight the particular classes of BH binaries we are able to study with this
framework, it is convenient to begin this discussion with the isometry group of
the $S^{D-3}$ sphere, i.e., with the $3+(D-3)$-split.
The general $D$-dimensional spacetime metric can be written in the form
\begin{align}
ds^2 = & g_{MN} dx^M dx^N
\,\non\\
 = & g_{\bar{\mu}\bar{\nu}} (x^M) dx^{\bar{\mu}} dx^{\bar{\nu}}
   + \Omega_{\bar{i}\bar{j}}(x^M) 
     \left( dx^{\bar{i}} - A_{\bar{\mu}}^{\bar{i}}(x^M) dx^{\bar{\mu}} \right)
     \left( dx^{\bar{j}} - A_{\bar{\nu}}^{\bar{j}}(x^M) dx^{\bar{\nu}}\right) 
\,,
\end{align}
where we have split the spacetime coordinates as $x^M=(x^{\bar{\mu}},x^{\bar{i}})$. 
$\bar{\mu},\bar{\nu} = 0,1,2$ are $3$-dimensional indices and
$\bar{i},\bar{j}=3,\dots,(D-1)$ are indices in the remaining $(D-3)$-dimensions. 
We may think of the spacetime as a fibre bundle, where $\{x^{\bar{i}}\}$ are coordinates
along the fibre and $\{x^{\bar{\mu}}\}$ are coordinates on the base space.
We intend to investigate $D$-dimensional spacetimes with an $SO(D-2)$
isometry group, appropriate to describe head-on collisions of non-rotating BHs. 
We furthermore assume that $\xi_{\ba}$, $\ba=1,\dots, (D-3)(D-2)/2$, are Killing vector fields,
\begin{align}
\label{eq:HDFkvf}
\mathcal{L}_{\xi_{\ba}} g_{MN} = & 0 
\,,
\end{align}
with Lie algebra
\begin{align}
\label{eq:HDFalgebra}
\left[ \xi_{\ba}, \xi_{\bar{b}} \right] = & \eps_{\ba\bar{b}}{}^{\bar{c}}\xi_{\bar{c}} 
\,,
\end{align}
where $\eps_{\ba\bar{b}}{}^{\bar{c}}$ are the structure constants of $SO(D-2)$. 
Because the fibre has the minimal dimension necessary to accommodate $(D-3)(D-2)/2$
independent Killing vector fields, we assume, without loss of generality, 
that the Killing vector fields have components exclusively along the fibre:
$\xi_{\ba} = \xi_{\ba}^{\bar{i}}\p_{\bar{i}}$.  
Furthermore, we normalise the Killing vectors so that they only depend on the coordinates of the fibre,
i.e. $\p_{\bar\mu}\xi_{\ba}^{\bar i}=0$. 
Then, Eq.~\eqref{eq:HDFkvf} yields the conditions
\begin{subequations}
\label{eq:HDFDRcond}
\begin{align}
\label{eq:HDFOmega}
\mathcal{L}_{\xi_{\ba}} \Omega_{\bar{i}\bar{j}} & = 0 
\,, \\
\label{eq:HDFcomm}
\mathcal{L}_{\xi_{\ba}} A_{\bar{\mu}}^{\bar{i}} & =0 
\,, \\
\label{eq:HDFgg}
\mathcal{L}_{\xi_{\ba}} g_{\bar{\mu}\bar{\nu}} & =0 
\,.
\end{align}
\end{subequations}
These expressions can be interpreted either as Lie derivatives of
rank-$2$ tensors defined on the $D$-dimensional spacetime, or as Lie
derivatives of a rank-$2$ tensor, a vector and a scalar, which are
defined on $S^{D-3}$.
Condition \eqref{eq:HDFOmega} implies
\begin{align}
\Omega_{\bar{i}\bar{j}} = &
f(x^{\bar{\mu}})h_{\bar{i}\bar{j}}^{S^{D-3}} 
\,,
\end{align}
because $\Omega_{\bar{i}\bar{j}}$ admits the maximal number 
of Killing vector fields and thus must be the metric on a maximally symmetric space 
at each $x^{\bar{\mu}}$. Due to \eqref{eq:HDFalgebra} this space must be the $S^{D-3}$ sphere.
$h_{\bar{i}\bar{j}}^{S^{D-3}}$ denotes the metric on an $S^{D-3}$ sphere with unit radius.
Eq.~\eqref{eq:HDFgg} implies
\begin{align}
g_{\bar{\mu}\bar{\nu}} = &
g_{\bar{\mu}\bar{\nu}}(x^{\bar{\mu}}) 
\,,
\end{align}
because the Killing vector fields $\xi_{\ba}$ act transitively on the fibre
and therefore the base space metric must be independent of the fibre coordinates.
Because Eq.~\eqref{eq:HDFcomm} is equivalent to 
\begin{align}
\label{eq:HDFcomm1} 
\left[ \xi_{\ba},A_{\bar\mu} \right]= & 0
\,,
\end{align}
it follows that
\begin{align}
\label{eq:HDFnovectors}
A_{\bar{\mu}}^{\bar{i}} = & 0
\,,
\end{align}
and there exist no non-trivial vector fields on $S^{D-3}$ for $D\ge 5$
that commute with all Killing vector fields on the sphere.

\begin{figure}
\begin{center}
\includegraphics[width=0.75\textwidth]{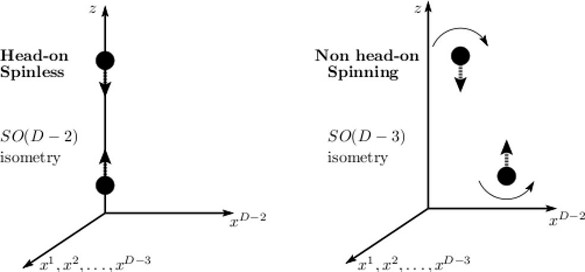}
\end{center}
\caption{\label{fig:HDFDimReduction} 
$D$-dimensional representation of two types of BH collisions, 
using coordinates $(t, x^1,x^2,\dots,x^{D-3},x^{D-2},z)$.
Left: head-on collision of spinless BHs, for which the isometry group is $SO(D-2)$. 
Right: non head-on BH system, with motion on a {\textit{single}} $2$-plane, 
       for BHs spinning only in the {\textit{same}} plane, 
       for which the isometry group is $SO(D-3)$. 
}
\end{figure}

We remark that \eqref{eq:HDFcomm1} corresponds to the statement, expressed in
\cite{Cho:1986wk} in group theoretical language, that the gauge group for a
theory reduced on a coset space $G/H$ is the normaliser of $H$ in $G$; in the
case of a sphere, where $G=SO(D-2)$ and $H=SO(D-3)$, the normaliser vanishes and
then there are no ``gauge vectors'', i.e., no non-vanishing metric
components $g_{\bar\mu\bar i}$. If the normaliser of $H$ in $G$ is
non-vanishing, such metric components appear, and with the dimensional reduction
they yield gauge fields in the reduced theory.  
For example, in the case of head-on collision in $D=4$ the isometry space is $SO(2)$ 
and scalar as well as vector fields emerge in the reduced theory \cite{Geroch:1970nt}.
In $D>4$, the isometry space is $SO(D-2)/SO(D-3)$
and the dimensional reduction only yields a single scalar field.
In the remainder of this work we focus on this subclass of spacetimes.
Because we intend to employ and modify an existing $3+1$-NR code, we now 
perform the $4+(D-4)$-split of the $D$-dimensional spacetime.
The metric on a unit sphere $S^{D-3}$ can always be written in terms of the line element $d\Omega_{D-4}$ 
on a unit sphere $S^{D-4}$ 
\begin{align}
h_{\bar{i}\bar{j}}^{S^{D-3}} dx^{\bar{i}}dx^{\bar{j}} = &
d\theta^2+\sin^2\theta d\Omega_{D-4} 
\,,
\end{align}
where $\theta\in [0,\pi]$ is a polar-like coordinate. 
Now we introduce $4$-dimensional coordinates, $x^\mu=(x^{\bar{\mu}},\theta)$,
$\mu=0,\dots,3$, and define a $4$-dimensional metric
\begin{align}
\label{eq:NR4dm}
g_{\mu\nu}dx^\mu dx^\nu = & 
g_{\bar{\mu}\bar{\nu}}dx^{\bar{\mu}}dx^{\bar{\nu}} + f(x^{\bar{\mu}}) d\theta^2 
\,,
\end{align}
as well as a new (conformal) factor
\begin{align}
\label{eq:HDFdeflam}
\la(x^{\mu}) = \sin^2\theta g_{\theta\theta} = y^2 
\,.
\end{align}
The last equality holds for transformations into coordinates adapted to axial symmetry.
Then, the most general $D\ge5$-dimensional metric compatible with $SO(D-2)$
isometry is
\begin{align}
\label{eq:HDFmetricansatz}
d s^2 = & g_{\mu\nu}dx^\mu dx^\nu + \la(x^{\mu}) d \Omega_{D-4} 
\,.
\end{align}
The geometry \eqref{eq:HDFmetricansatz} has only a manifest $SO(D-3)$ symmetry. 
Note, that the additional scalar field $\la\sim y^2$ 
(in coordinates adapted to the axial symmetry) 
arouses problems for the numerical implementation when $y=0$.
Therefore we will introduce regular variables in Sec.~\ref{ssec:HDFBSSNform}. 
The treatment of potentially pathogolical terms is presented in Appendix~\ref{asec:troubleterms}.
Performing the dimensional reduction of the $D$-dimensional vacuum Einstein's equations~\eqref{eq:HDFdimEE} on a 
$(D-4)$-sphere and considering the discussed symmetries leads to the $3+1$-dimensional Einstein's equations coupled
to the scalar field $\la$.
If we consider $SO(D-2)$ as the full isometry group, the emergent scalar field is completely determined by
the $3+1$-dimensional geometry via Eq.~\eqref{eq:HDFdeflam}.
Instead, considering the full isometry group $SO(D-3)$ allows us to explore
BH collisions with impact parameter and with spin as long as it is restricted to the same $2$-plane.
We depict both considered scenarios in Fig.~\ref{fig:HDFDimReduction}.
Taking metric \eqref{eq:HDFmetricansatz}, which describes general spacetimes with $SO(D-3)$ isometry in $D\ge6$
as follows from the discussion of \eqref{eq:HDFnovectors},
the $D$-dimensional vacuum Einstein's equations~\eqref{eq:HDFdimEE} reduce to a
$4$-dimensional theory of gravity coupled to a scalar field $\la(x^\mu)$. 
We remark that in this theory $\la$ and $g_{\mu\nu}$ are viewed as independent degrees of freedom.
The $D$-dimensional Einstein-Hilbert action reduces to
\begin{align}
\label{eq:HDFEHaction}
\mathcal{S} = & \frac{1}{16\pi G_4}\int d^4x\sqrt{-g} \la^{\frac{D-4}{2}}
\,\non\\ & \times
 \left[ \,^{(4)}\!R + (D-4) \left(  (D-5) \frac{1}{\la} - \frac{1}{\la} \Box \la
                        - \frac{D-7}{4\la^{2}} \p_\mu \la \p^\mu \la \right)
 \right] 
\,,
\end{align}
where the $D$-dimensional Newton's constant $G_D$ is related to the $4$-dimensional 
one $G_4$ by the area of the unit ${D-4}$-dimensional sphere $G_4=G_D/ \A^{S^{D-4}}$. 
The resulting equations of motion are 
\begin{subequations}
\label{eq:HDF4dEoM}
\begin{align}
\label{eq:HDFscalarEoM}
\na^{\mu}\na_{\mu} \la = & 2(D-5) - \frac{D-6}{2\la}\na^{\mu}\la \na_{\mu}\la
\,,\\
\label{eq:HDFEE}
\,^{(4)}\!G_{\mu\nu} = & \,^{(4)}\!R_{\mu\nu} - \frac{1}{2}g_{\mu\nu}\,^{(4)}\!R = 8\pi T_{\mu\nu}
\,,
\end{align}
\end{subequations}
where $\na_{\mu}$ is the covariant derivative with respect to the $4$-metric $g_{\mu\nu}$.
The energy-momentum tensor is given by
\begin{align}
\label{eq:HDFEMT}
T_{\mu\nu} = &
        \frac{D-4}{16\pi\la} \left( \na_{\mu}\na_{\nu}\la - \frac{1}{2\la}\na_{\mu}\la \na_{\nu}\la
        - (D-5) g_{\mu\nu} + \frac{D-5}{4\la}g_{\mu\nu}\na^{\rho}\la \na_{\rho}\la \right)
\,.
\end{align}

\section{Formulation as time evolution problem}\label{sec:HDFevol}
\subsection{ADM formulation}\label{ssec:HDFADMform}
As we have seen in the previous section, 
the dimensional reduction on a $(D-4)$-sphere of $D$-dimensional vacuum GR
results in the $4$-dimensional Einstein's equations coupled to a scalar field $\la$,
where the EoMs are given by Eqs.~\eqref{eq:HDF4dEoM}.
We wish to evolve this system numerically and therefore 
rewrite Eqs.~\eqref{eq:HDF4dEoM} as time evolution problem
by adopting the $3+1$-decomposition
\cite{York1979,Arnowitt:1962hi}, discussed in detail in Sec.~\ref{sec:NRdecomp}.
In order to close the system we introduce the conjugated momentum $K_{\la}$ related to
the scalar field $\la$ via
\begin{align}
\label{eq:HDFKlam}
K_{\la} = & -\frac{1}{2}\Lie_{n}\la
\,,
\end{align}
analogous to the definition of the extrinsic curvature, Eq.~\eqref{eq:NRevolga0}.
Then, the evolution equations for the $3$-metric $\ga_{ij}$ and scalar field $\la$ are given by
\begin{subequations}
\label{eq:HDFevol1}
\begin{align}
\label{eq:HDFevolgam}
(\p_t - \Lie_{\be}) \ga_{ij} = & -2\al K_{ij}
\,,\\
\label{eq:HDFevollam}
(\p_t - \Lie_{\be}) \la = &  -2\al K_{\la}
\,.
\end{align}
\end{subequations}
The evolution equation for $K_{\la}$ is obtained from the EoM 
of the scalar field, Eq.~\eqref{eq:HDFscalarEoM}.
The $3+1$-split yields
\begin{align}
\label{eq:HDFevolKlam}
(\p_t - \Lie_{\be}) K_{\la} = &
\al\left( (D-5) + K K_{\la} + \frac{D-6}{\la} K^2_{\la} - \frac{1}{2}D^i D_i \la - \frac{D-6}{4\la} D^i\la D_i\la \right)
\,\non\\ &
- \frac{1}{2} D^i\al D_i\la
\,,
\end{align}
where $D_i$ is the covariant derivative associated to the $3$-metric $\ga_{ij}$.
Next, we consider the various projections of the Einstein's equations~\eqref{eq:HDFEE},
as outlined in Sec.~\ref{sec:NRdecomp},
with the energy-momentum tensor specified in Eq.~\eqref{eq:HDFEMT}.
In particular, the projection of Eq.~\eqref{eq:HDFEE} onto the spatial hypersurface $\Si_t$
provides an evolution equation for the extrinsic curvature $K_{ij}$
\begin{align}
\label{eq:HDFevolKij}
(\p_t - \Lie_{\be}) K_{ij} = & - D_iD_j\al
+ \al\left( R_{ij} - 2 K^{k}{}_{i} K_{kj} + K K_{ij} \right)
\non\\ &
- \al \frac{D-4}{2\la}\left( D_iD_j\la -\frac{1}{2\la} D_i\la D_j\la - 2K_{\la} K_{ij} \right)
\,,
\end{align}
where $R_{ij}$ is the Ricci tensor associated with the $3$-metric $\ga_{ij}$.
The Lie derivatives of the dynamical variables along the shift vector $\be^i$ are given by
\begin{subequations}
\label{eqHDFLieADM}
\begin{align}
\Lie_{\be} \ga_{ij} = & D_i \be_j + D_j \be_i
\,,\\
\Lie_{\be} \la      = & \be^k \p_k \la
\,,\\
\Lie_{\be} K_{ij}   = & \be^k \p_k K_{ij} + K_{ik}\p_j \be^k + K_{kj} \p_j \be^k
\,,\\
\Lie_{\be} K_{\la}  = & \be^k \p_k K_{\la}
\,.
\end{align}
\end{subequations}
The physical constraints are obtained in an analogous manner as in the ``standard'' $4$-dimensional case,
see Eqs.~\eqref{eq:NRHamiltonian} and~\eqref{eq:NRMomentum}.
In particular the full contraction of Eq.~\eqref{eq:HDFEE} with the vector $n^{\mu}$ normal to the hypersurface $\Si_t$
yields the Hamiltonian constraint
\begin{align}
\label{eq:HDFHamADM}
\H = &
R - K_{ij} K^{ij} + K^2
\non\\ &
-\frac{D-4}{\la}\left(- (D-5) + D^i D_i\la + \frac{D-7}{4\la} D^i\la D_i \la
                      - (D-5)\frac{K^2_{\la}}{\la} - 2 K K_{\la} \right)
= 0
\,.
\end{align}
The mixed projection of Eq.~\eqref{eq:HDFEE} once onto the hypersurface and once perpendicular to it
results in the momentum constraint
\begin{align}
\label{eq:HDFMomADM}
\M_i = & D_j K^{j}{}_i - D_i K
-\frac{D-4}{2\la} \left( D_i K_{\la} - K^{j}{}_{i} D_j\la - \frac{K_{\la}}{\la}D_i\la \right)
= 0
\,.
\end{align}

In our framework 
the complete set of evolution equations 
for the $3$-metric $\ga_{ij}$, the scalar field $\la$, the extrinsic curvature $K_{ij}$ and the
scalar curvature $K_{\la}$ 
in ADM form are given by Eqs.~\eqref{eq:HDFevol1},~\eqref{eq:HDFevolKij} 
and~\eqref{eq:HDFevolKlam}, respectively.
The performed $3+1$-decomposition gives rise to the Hamiltonian and momentum constraint,
Eqs.~\eqref{eq:HDFHamADM} and~\eqref{eq:HDFMomADM}.
%
\newpage
\subsection{Generalized BSSN formulation}\label{ssec:HDFBSSNform}
%
In order to obtain a hyperbolic formulation of the ADM evolution Eqs.~\eqref{eq:HDFevol1},~\eqref{eq:HDFevolKlam}
and~\eqref{eq:HDFevolKij},
we reformulate them in terms of the generalized BSSN system \cite{Baumgarte:1998te,Shibata:1995we,Witek:2010es}
(cf. Sec.~\ref{sec:NRgBSSN}).
Employing the approach discussed in Sec.~\ref{sec:NRgBSSN},
we alter the characteristics of the PDE system by 
adding the definition-differential constraint $G_i$, given in Eq.~\eqref{eq:NRGi},
and the Hamiltonian and momentum constraint, Eqs.~\eqref{eq:HDFHamADM} and~\eqref{eq:HDFMomADM}.
Then the system of evolution equations is modified according to 
\begin{subequations}
\label{eq:HDFgBSSNconstradd}
\begin{align}
\p_t \ga_{ij} = & [\textrm{ADMHD}]
\,,\\
\p_t \la      = & [\textrm{ADMHD}]
\,,\\
\p_t K_{ij}   = & [\textrm{ADMHD}] + \al \left( \frac{1}{2}\p_i G_j + \frac{1}{2}\p_j G_i - \frac{1}{3}\ga_{ij} \p^k G_k
                                       - \frac{1}{3} \ga_{ij} \H \right)
\,,\\
\p_t K_{\la}  = & [\textrm{ADMHD}]
\,,\\
\p_t G_i      = & 2\al \M_i + \be^j\p_j G_i + \ga^{1/3} G^j \left( \p_t - \be^k \p_k \right)\left( \ga^{-1/3} \ga_{ij}\right)
\,,
\end{align}
\end{subequations}
where $[\textrm{ADMHD}]$ denotes the source terms of the ADM-like Eqs.~\eqref{eq:HDFevol1},~\eqref{eq:HDFevolKlam}
and~\eqref{eq:HDFevolKij}.
Next, we perform the conformal decomposition of the dynamical variables in order to
evolve quantities that are regular (everywhere in space).
Therefore we consider the conformal variables
\begin{subequations}
\label{eq:HDFgBSSNvars}
\begin{align}
\chi = & \ga^{-1/3}
\,,\quad
\tg_{ij} = \ga^{-1/3}\ga_{ij} = \chi \ga_{ij}
\,,\\
\tK = & \chi^{-3 \nk /2} K
\,,\quad
\tA_{ij} = \chi^{1-3 \nk /2} \left( K_{ij} - \frac{1}{3}\ga_{ij} K \right)
\,,\\
\tG^i = & \tg^{jk}\tG^{i}{}_{jk} = - \p_j\tg^{ij}
\,,\\
\ze = & \frac{\chi}{y^2} \la
\,,\quad
K_{\ze} = \frac{\chi}{y^2} K_{\la}
\,,\\
Q = & \chi^{3/2\nq}\al
\,,
\end{align}
\end{subequations}
where the densitization constants $\nk,\nq$ parametrize the 
change of variables\footnote{Note, that we employ a different re-scaling for the scalar momentum $K_{\la}$
as compared to Eq.~(4.3) in \cite{Zilhao:2010sr}.}.
The definition of the conformal variables introduces additional algebraic constraints
\begin{align}
T = & \tg^{ij} \tA_{ij} = 0
\,,\quad
D = \ln(\tg) = 0
\,,
\end{align}
representing the requirements that $\tA_{ij}$ remains tracefree
and the determinant of the conformal metric $\tg = 1$.
Under this change of variables the evolution equations~\eqref{eq:HDFgBSSNconstradd}
become 
\begin{subequations}
\label{eq:HDFgBSSNevol}
\begin{align}
\p_t \chi       = & [\textrm{gBSSN4D}]
\,,\\
\p_t \tg_{ij}   = & [\textrm{gBSSN4D}]
\,,\\
\p_t \tK        = & [\textrm{gBSSN4D}] + \al (D-4) \chi^{-3/2\nk} S^{\tK} 
\,,\\
\p_t \tA_{ij}   = & [\textrm{gBSSN4D}] + \al (D-4) \chi^{1-3/2\nk} S^{\tA}_{ij}
\,,\\
\p_t \tG_i      = & [\textrm{gBSSN4D}] + \al (D-4) \tg^{ij} S^{\tG}_j
\,,\\
\p_t \ze      = &  -2\al \left( K_\ze - \chi^{3/2 n_K} \ze \frac{\tK}{3} \right)
        +\frac{2}{y}\ze\be^y
        + \be^i\p_i\ze - \frac{2}{3}\ze\p_i\be^i
\,,\\
\p_t K_{\ze}  = & \be^i\p_i K_{\ze} - \frac{2}{3} K_{\ze} \p_i\be^i
        + \frac{1}{2} \al\ze \tD^i \tD_i\chi
        - \frac{1}{2}\al\chi \tD^i \tD_i\ze
        - \frac{D-6}{4\ze}\al\chi \p^i\ze \p_i\ze
\non\\ &
        + \frac{2D-7}{4} \al \p^i\chi \p_i\ze
        - \frac{(D-1)\al\ze}{4\chi}\p^i\chi \p_i\chi
        + \frac{1}{2} \p^i\al (\ze \p_i\chi - \chi \p_i\ze )
\non\\ &
        + \frac{5}{3} \al \chi^{3/2n_K} \tK K_{\ze}
        + (D-6) \al \frac{K_{\ze}^2}{\ze}
        - \chi \ze \frac{\p^y\al}{y}
        - (D-5)\al\chi\frac{\ze \tg^{yy} - 1}{y^2}
\non\\ &
        + \al\chi\ze \frac{\tG^y}{y}
        + 2 K_{\ze} \frac{\be^y}{y} 
        - (D-4)\al\chi \frac{\p^y\ze}{y} 
        + \frac{(2D-7)\al\ze}{2} \frac{\p^y\chi}{y} 
\,,
\end{align}
\end{subequations}
where $[\textrm{gBSSN4D}]$ denotes the $4$-dimensional vacuum evolution Eqs.~\eqref{eq:NRgBSSNevol}
and $\tD_i$ denotes the covariant derivative with respect to the conformal metric $\tg_{ij}$.
The coupling terms $S^{\tK}$, $S^{\tA}_{ij}$ and $S^{\tG}_j$ are given by
\begin{subequations}
\label{eq:HDFgBSSNcoupling}
\begin{align}
S^{\tK} = &  (D-5) \frac{\chi}{\ze} \frac{\ze\tg^{yy} - 1}{y^2} 
                 - (D-5) \frac{K_{\ze}^2}{\ze^2} 
                 - \chi^{3/2n_K} \tK \frac{K_{\ze}}{\ze}
\non\\ &
                 + \frac{\chi}{2\ze} \tD^i\tD_i\ze 
                 - \frac{1}{2} \tD^i \tD_i\chi 
                 + \frac{D-1}{4\chi} \p^i\chi\p_i\chi 
                 + \frac{(D-6)\chi}{4\ze^2} \p^i\ze \p_i\ze
\non\\ &
                 - \frac{2D-7}{4\ze} \p^i\chi \p_i\ze
                 - \frac{2D-7}{2} \frac{\p^y\chi}{y}
                 + \frac{(D-4)\chi}{\ze} \frac{\p^y\ze}{y}
                 - \chi \frac{\tG^y}{y} 
\,,\\
S^{\tA}_{ij} = & \chi^{3/2 n_K - 1} \tA_{ij} \frac{K_{\ze}}{\ze}
                + \frac{1}{2\chi} [\tD_i\tD_j\chi]^{tf}
                - \frac{1}{2\ze} [\tD_i\tD_j\ze]^{tf}
                - \frac{1}{4\chi^2} [\p_i\chi\p_j\chi]^{tf}
                + \frac{1}{4\ze^2} [\p_i\ze \p_j\ze]^{tf}
\non\\ &
                + \frac{1}{2\ze y} \big( 2\ze\tG^y_{ij} 
                        - \delta^y_i\p_j\ze - \delta^y_j\p_i\ze \big)
                + \frac{\tg_{ij}}{3\ze} \frac{\p^y\ze}{y}
                - \frac{\tg_{ij}}{3} \frac{\tG^y}{y}
\,,\\
S^{\tG}_i = & - \frac{2}{\ze} \p_i K_{\ze}
                 + \left( \frac{K_{\ze}}{\ze} - \chi^{3/2 n_K} 
                 \frac{\tK}{3} \right) \frac{\p_i\chi}{\chi}
                 + \left( \frac{K_{\ze}}{\ze} + \chi^{3/2 n_K} \frac{\tK}{3} \right) \frac{\p_i\ze}{\ze}
\non\\ &
                 + \frac{2}{y}\left( \chi^{3/2 n_K} \tA^y_i 
                        + \delta^y_i \left( \chi^{3/2n_K}\frac{\tK}{3} -\frac{K_{\ze}}{\ze}\right) \right)
                 + \chi^{3/2 n_K} \tA^k_i \left( \frac{1}{\ze} \p_k \ze - \frac{1}{\chi}\p_k\chi \right)   
\,.
\end{align}
\end{subequations}
The adopted treatment and the regularization of terms $\sim1/y$ or $\sim1/y^2$, 
which are potentially troublesome when they reach $y=0$ throughout a numerical simulation,
are summarized in Appendix~\ref{asec:troubleterms}.

In order to close the system of evolution equations we have to specify the gauge for 
the lapse function $\al$ and shift vector $\be^i$.
In an attempt to find gauge conditions well-suited to the numerical evolution of higher 
dimensional black holes we modify the standard moving puncture gauge \cite{Campanelli:2005dd,Baker:2005vv}
by terms that couple the lapse and shift to the scalar field and its momenta. 
For the lapse we choose a modification of the 
Bona-Mass\'o~\cite{Bona:1994dr} condition
\begin{align}
\label{eq:HDFslicingAlp}
\p_t \al = & \be^i\p_i\al - \al^2 \mu_L\left( \tK +(D-4)\mu_{\lambda}\frac{K_{\ze}}{\ze}\right)
\,.
\end{align}
In applications we always choose the $1+log$-slicing with $\mu_L=2/\al$
and, typically, $\mu_{\lambda}=1$. 
Then, the $1+log$-slicing condition translates into 
\begin{align}
\label{eq:HDFslicingQ}
\p_t Q = & \be^i\p_i Q - \nq Q \p_i \be^i + \chi^{\frac{3}{2}(\nk - \nq)} Q \tK \left(\nq Q -2\chi^{\frac{3}{2}\nq} \right)
- 2 (D-4) \mu_{\la} Q \frac{K_{\ze}}{\ze}
\,,
\end{align}
for the densitized lapse $Q$ considered as evolved variable in the generalized
BSSN scheme.
For the shift we employ the modified $\Ga$-driver condition
\begin{align}
\label{eq:HDFgammadriver}
\p_t\be^i = & \be^k\p_k\be^i
        - \eta_{\be} \be^i
        + \xi_{\Ga} \tilde{\Gamma}^i
        + \xi_{\la} \frac{D-4}{2}\frac{\p^i\zeta}{\zeta}
\end{align}
The free parameters in the gauge conditions that affect the 
principal part of the system are the scalar functions 
$(\mu_L,\mu_\la,\xi_{\Ga},\xi_{\la})$. 
The damping parameter~$\eta_{\be}$ is taken to be a constant in our applications. 

\section{Hyperbolicity of the evolution system}\label{sec:HDFHyperbolicity}
\subsection{Well-posedness of the initial value problem and strong hyperbolicity}
A system of partial differential equations is said to be 
well-posed if it admits a unique solution that depends 
continuously, in some appropriate norm, on given data. 
In physical applications well-posedness is a fundamental 
requirement both for a system to have predictive power, and 
for its treatment by numerical methods.

Consider the linear, constant coefficient, first order in time, 
second order in space evolution system
\begin{subequations}\label{eqn:FOITSOIS}
\begin{align}
\p_tv&=M^{-1}w+F^i\p_iv+S_v,\\
\p_tw&=V^{ij}\p_i\p_jv+G^i\p_iw+S_w,
\end{align}
\end{subequations}
where $S_v,S_w$ denote lower order derivatives. 
We denote $s^i$ as a unit spatial vector and we use the subscript $s$ to denote 
contractions of the partial derivative with $s^i$, 
A linear combination $u$ of components of $\p_sv$ and $w$, 
is called a 
characteristic variable with speed $\lambda$ in the $s^i$ 
direction if it satisfies 
\begin{align}
\p_tu&=\lambda\p_su+\textrm{transverse derivatives}.
\end{align}
The evolution system~\eqref{eqn:FOITSOIS} is called strongly 
hyperbolic if it admits a complete set of characteristic 
variables that depend continuously on $s^i$. The evolution 
system~\eqref{eqn:FOITSOIS} admits a well-posed initial value 
problem if and only if it is strongly hyperbolic. These 
definitions are extended to non-linear systems with variable 
coefficients by linearizing them and working in the frozen 
coefficient approximation.

\subsection{Hyperbolicity of the dimensionally reduced BSSN system} 
Strong hyperbolicity for quasi-linear and variable 
coefficient problems is defined by linearizing around a given 
solution and working in the frozen coefficient approximation. 
We perform a $2+1$ split in space against the unit spatial 
vector~$s^i$ by defining the standard projection operator
\begin{align}
q^i{}_j=&\gamma^i{}_j-s^is_j
\,,\qquad
s^{i} s_{i} = 1
\,,
\end{align}
Under this split a convenient choice for the scalar variables 
is
\begin{align}
\alpha,&&\quad\beta^{s}&=s_i\beta^i,\non\\
\gamma_{ss}&=s^is^j\gamma_{ij},&\quad\hat{\gamma}_{qq}&=q^{ij}\gamma_{ij}
+(D-4)\ln\lambda,\non\\
\ln\lambda,&&\quad G^s&=s_iG^i,\non\\
\hat{K}_\lambda&=\frac{1}{\lambda}K_\lambda,&
\quad\hat{K}_{qq}&=q^{ij}K_{ij}+(D-4)\hat{K}_\lambda,\non\\
K_{ss}&=s^is^jK_{ij}.&&\quad &
\end{align}
Note that~$\hat{\gamma}_{qq}$ and~$\hat{K}_{qq}$ are the natural choice 
of variables in the scalar sector because they correspond to the quantities 
that would be obtained by a $(D-2) + 1$ split against~$s^i$ in the higher 
dimensional BSSN formulation without the dimensional reduction by symmetry. 
In this section we use indices $A,B,...$ to denote projected quantities, so 
the vectors are written
\begin{subequations}
\begin{align}
\beta^A&=q^A{}_i\beta^i,&\quad\gamma_{sA}&=s^iq^j{}_A\gamma_{ij},\\
G^A&=q^A{}_iG^i,&K_{sA}&=s^iq^j{}_AK_{ij},
\end{align}
\end{subequations}
and the tensors are
\begin{subequations}
\begin{align}
\gamma_{AB}^{\textrm{TF}}&=
\big(q^i{}_Aq^j{}_B-\frac{1}{2}q_{AB}q^{ij}\big)\gamma_{ij},\\
K_{AB}^{\textrm{TF}}&=
\big(q^i{}_Aq^j{}_B-\frac{1}{2}q_{AB}q^{ij}\big)K_{ij}.
\end{align}
\end{subequations}
Up to derivatives transverse to~$s^i$ the system decomposes into scalar, 
vector and tensor blocks. We introduce the time 
derivative~$\p_0\simeq\frac{1}{\alpha}(\p_t-\beta^s\p_s)$.
Then, the principal symbol of the scalar block for the metric can be read off from Eqs.~\eqref{eq:HDFgBSSNconstradd}.
\begin{subequations}
\begin{align}
\p_0\gamma_{ss}&\simeq-2 K_{ss}+\frac{2}{\alpha}\p_s\beta_s,\\
\p_0\hat{\gamma}_{qq}&\simeq-2\hat{K}_{qq},\\
\p_0G_s&\simeq -2 \p_s\hat{K}_{qq},\\
\p_0K_{ss}&\simeq-\frac{1}{6}\p_s^2\hat{\gamma}_{qq}-\p_s^2\ln\alpha
+\frac{2}{3}\p_sG_s,\\
\p_0\hat{K}_{qq}&\simeq \frac{1}{6}\p_s^2\hat{\gamma}_{qq}-\frac{2}{3}\p_sG_s,
\end{align}
\end{subequations}
and for the scalar field
\begin{subequations}
\begin{align}
\p_0\lambda&\simeq -2 K_{\lambda},\\
\p_0\hat{K}_{\lambda}&\simeq-\frac{1}{2}\p_s^2\ln\lambda.
\end{align}
\end{subequations}
For the gauge conditions we consider a modification of the 
Bona-Mass\'o condition~\eqref{eq:HDFslicingAlp}
\begin{align}
\p_0\ln\alpha&\simeq-\mu_L\left(K_{ss}+\hat{K}_{qq}
+\bmu_{\lambda}\hat{K}_{\lambda}\right),
\end{align}
coupled to a parametrized modification of the standard~$3+1$ $\Ga$-driver 
condition~\eqref{eq:HDFgammadriver}\footnote{Note the difference in the last term. 
Here we consider $\sim\p\la/\la$ in contrast to $\sim\p\ze/\ze$.}.
\begin{align}
\p_0\beta_s&\simeq
\bar{\mu}_S\alpha\bigg(G_s+\frac{2}{3}\p_s\gamma_{ss}-\frac{1}{3}
\p_s\hat{\gamma}_{qq}+\bmu_\zeta\p_s\ln\lambda\bigg),\label{eqn:paramdriver}
\end{align}
where we have re-defined the gauge parameter employed in Eqs.~\eqref{eq:HDFslicingAlp} and~\eqref{eq:HDFgammadriver}
to
\begin{align}
\label{eq:HDFhypParDef}
\bmu_{\la} = & (D-4) \mu_{\la}
\,,\quad
\bmu_S = \ga^{1/3} \xi_{\Ga}
\,,\quad
\bmu_{\zeta} \bmu_S = \frac{D-4}{2}\xi_{\la}
\,.
\end{align}
We recover the $4$-dimensional puncture gauge and the standard $3+1$~principal
symbol of BSSN when $\bmu_{\la} = \bmu_{\ze} = 0$.
Then, the fields $(\ln\lambda)$ and~$\hat{K}_\lambda$ 
satisfy a decoupled wave equation in the principal symbol. 

The vector part of the system is
\begin{subequations}
\begin{align}
\p_0\beta_A&\simeq\bar{\mu}_S\alpha\big(G_A+\p_s\gamma_{sA}\big),\\
\p_0\gamma_{sA}&\simeq -2 K_{sA}+\frac{1}{\alpha}\p_s\beta_A,\\
\p_0G_A&\simeq 2\p_sK_{sA},\\
\p_0K_{sA}&\simeq\frac{1}{2}\p_sG_A
\,.
\end{align}
\end{subequations}
Finally, the tensor part of the system is given by
\begin{subequations}
\begin{align}
\p_0\gamma_{AB}^{\textrm{TF}}&\simeq - 2 K_{AB}^{\textrm{TF}},\\
\p_0K_{AB}^{\textrm{TF}}&\simeq
-\frac{1}{2}\p^2_s\gamma_{AB}^{\textrm{TF}}.
\end{align}
\end{subequations}

\paragraph{Hyperbolicity of the scalar block:} 
The scalar sector generically admits a complete set of characteristic variables,
which are given by 
\begin{subequations}
\begin{align}
u_{\alpha\pm\mu_L}&=\p_s\ln\alpha\pm\sqrt{\mu_L}\hat{K}
+\frac{\mu_L\bmu_\lambda}{\mu_L-1}
\left(\frac{1}{2}\p_s\ln\lambda\pm \sqrt{\mu_L}\hat{K}_\lambda\right),\\
u_{\lambda\pm1}&=\frac{1}{2}\p_s\ln\lambda\pm \hat{K}_{\lambda},\\
u_{\pm 1}&=\hat{K}_{qq}\pm\frac{2}{3}G_s
\mp\frac{1}{6}\p_s\hat{\gamma}_{qq},\\
u_{\pm \bar{\mu}_S}&=\p_s\beta_s\pm\frac{\sqrt{3\bar{\mu}_S}}{2}\alpha
        \left(G_s+\frac{2}{3}\p_s\gamma_{ss}-\frac{1}{3}\p_s\hat{\gamma}_{qq}\right)
+\frac{2\alpha\sqrt{\bar{\mu}_S}}{3\mu_L-4\bar{\mu}_S}\left(
        \frac{2}{\sqrt{3}}\sqrt{\bar{\mu}_S}\hat{K}\mp\p_s\alpha\right)
\nonumber\\ &
+\frac{3\alpha\mu_L\sqrt{\bar{\mu}_S}\bmu_\lambda}
        {(4\bar{\mu}_S-3)(3\mu_L-4\bar{\mu}_S)}
        \left(4\sqrt{\bar{\mu}_S}\hat{K}_\lambda
        \mp\sqrt{3}\p_s\ln\lambda\right)
\nonumber\\ &
+\frac{2\sqrt{3}\alpha\bar{\mu}_S\bmu_\zeta}{4\bar{\mu}_S-3}
        \left(\sqrt{3}\hat{K}_\lambda\mp\sqrt{\bar{\mu}_S}\p_s\ln\lambda\right)
\,,\\
u_0&=G_s-\p_s\hat{\gamma}_{qq}
\,,
\end{align}
\end{subequations}
with speeds
\begin{align}
& (\pm\sqrt{\mu_L},\pm1,\pm1,\pm\frac{2}{\sqrt{3}}\sqrt{\bar{\mu}_S},0)
\non
\end{align}
respectively, and where we employ the shorthand~$\hat{K}=K_{ss}+\hat{K}_{qq}$. 
In the special case that generically distinct characteristic speeds coincide, the 
system fails to be strongly hyperbolic.
In applications however, the speeds will typically intersect only on sets of measure zero. 
This scenario can be avoided by choosing the gauge 
parameters~$(\mu_L,\bmu_S,\bmu_\lambda,\bmu_\zeta)$ appropriately, such that 
either the characteristic speeds never coincide, or, if they do, 
the corresonding variables are decoupled in the principal symbol. 
For completeness we discuss the special cases in the following.
In the scalar sector we recover the standard~$D=4$ restrictions~\cite{Sarbach:2012pr,Gundlach:2006tw},
namely that the system is only weakly hyperbolic if~$\bmu_S=\ga^{1/3}\xi_{\Ga}=3/4$. 
This clash must be avoided by an appropriate choice of~$\bmu_S$. 
For the new parameters we have the additional restrictions that 
in case~$\mu_L=1$, we must choose $\bmu_\lambda=0$. In other words, 
if we choose harmonic slicing, corresponding to~$\mu_L=1$, it must be chosen in the 
full~$D$ dimensional spacetime to maintain strong hyperbolicity. 

\paragraph{Hyperbolicity of the vector block} 
The vector sector always admits a complete set of characteristic variables. 
They are given by
\begin{subequations}
\begin{align}
u_{A\pm1}&=\frac{1}{2}G_A \pm K_{sA}
\,,\\
u_{A\pm\bar{\mu}_S}&=\p_s\beta_A\pm\sqrt{\bar{\mu}_S}(G_A+\p_s\gamma_{sA})
\,,
\end{align}
\end{subequations}
with characteristic speeds $(\pm1,\pm\sqrt{\bar{\mu}_S})$. 

\paragraph{Hyperbolicity of the tensor block}
The tensor sector 
always admits a complete set of characteristic variables
\begin{align}
u_{AB\,\pm1}&=\p_s\gamma_{AB}^{\textrm{TF}}
\pm\frac{1}{2}K_{AB}^{\textrm{TF}},
\end{align}
with speeds~$\pm 1$. These characteristic variables correspond 
to the gravitational wave degrees of freedom in a linear, plane 
wave approximation to GR.

\section{Initial Data}\label{sec:HDFInitData}
In this section we will discuss our method to construct initial data for $D$-dimensional
BH spacetimes \cite{Zilhao:2011yc}, which are then evolved by the effective $3+1$-approach described in 
Secs.~\ref{sec:HDFDimRed} and~\ref{sec:HDFevol}.
In particular, we will focus on the preparation of initial configurations
describing head-on collisions of two BHs starting 
either from rest, which is realized by modified Brill-Lindquist type initial data,
or with non-zero boost, which is provided by generalized Bowen-York type initial data.
We start by considering a $(D-1)$-dimensional spatial hypersurface $\bar{\Si}_t$ with the induced 
$(D-1)$-dimensional metric $\bga_{ab}$ and extrinisic curvature $\bK_{ab}$
embedded in a $D$-dimensional spacetime,
where we use $a,b = 1,...,(D-1)$ throughout this section.
The $D$-dimensional spacetime metric $g_{MN}$ is given by the line element
\begin{align}
ds^2 = & g_{MN} dx^M dx^N 
 =  -\al^2 dt^2 + \bga_{ab} \left(dx^a + \be^a dt \right) \left(dx^b + \be^b dt \right) 
\,.
\end{align}
Performing a $(D-1)+1$-splitting of the $D$-dimensional vacuum Einstein's equations \eqref{eq:HDFdimEE}
yields the constraints
\begin{subequations}
\label{eq:HDFdm1constraints0}
\begin{align}
\label{eq:HDFdm1Ham}
\bar{\H} = & \bR + \bK^2 - \bK_{ab} \bK^{ab} = 0
\,,\\
\label{eq:HDFdm1mom}
\bar{\M}_a = & \bD^b \bK_{ab} - \bD_a \bK = 0
\,,
\end{align}
\end{subequations}
where $\bR$ and $\bD_a$ are the Ricci scalar and covariant derivative associated with the 
$(D-1)$-dimensional spatial metric $\bga_{ab}$.
Next, we conformally decompose the spatial metric and extrinsic curvature 
\begin{subequations}
\label{eq:HDFinitdataconfmetric}
\begin{align}
\bga_{ab} = & \psi^{\frac{4}{D-3}} \hga_{ab}
\,, \\
\bK_{ab}  = & \psi^{-2} \hA_{ab} + \frac{1}{D-1} \bga_{ab} \bK
\,,
\end{align}
\end{subequations}
which generalizes the $3+1$-dimensional conformal decompostion, Eqs.~\eqref{eq:NRconfmetric0} and~\eqref{eq:NRKsplit},
with 
\begin{align}
\bA^{ab} = & \psi^{-2\frac{D+1}{D-3}} \hA^{ab}
\,,\quad
\bA_{ab} = \bga_{ac}\bga_{bd} \bA^{cd} = \psi^{-2} \hA_{ab}
\,.
\end{align}
Furthermore, we assume a conformally flat space, i.e. $\hga_{ab} = \eta_{ab}$,
and impose the maximal slicing condition $\bK = 0$.
With these choices, the $(D-1)$-dimensional Hamiltonian and momentum constraints in vacuum,
Eqs.~\eqref{eq:HDFdm1constraints0}, become~\cite{Yoshino:2005ps,Yoshino:2006kc}
\begin{subequations}
\label{eq:HDFinitdataconstraints}
\begin{align}
\label{eq:HDFinitdataconfHam}
\bar{\H} = &  \hat \triangle \psi + \frac{D-3}{4(D-2)} \psi^{- \frac{3D -5}{D-3} } \hA^{ab} \hA_{ab} = 0
\,,\\
\label{eq:HDFinitdataconfMom}
\bar{\M}^a = &  \partial_b \hA^{ab} = 0 
\,,
\end{align}
\end{subequations}
where $\hat \triangle \equiv \partial_a \partial^a $ is the flat space
Laplace operator.

We employ a coordinate system which is adapted to the
generalized axial symmetry $SO(D-2)$ in $D=5$ dimensions and $SO(D-3)$ in $D\ge6$ dimensions,
as discussed in Sec.~\ref{sec:HDFDimRed}.
Therefore, we first consider the (flat) conformal metric in cylindrical coordinates
\begin{align}
\label{eq:HDFinitdatacylcoord}
\hga_{ab} dx^a dx^b = & dz^2 + d\rho^2 + \rho^2\left( d\varphi^2 + \sin^2\varphi d\Omega_{D-4}\right) 
\,,
\end{align}
where $d\Omega_{D-4}$ is the metric on the $(D-4)$-sphere.
Note, that $\varphi\in [0,\pi]$ is a polar rather than an azimuthal coordinate.
Next we introduce ``incomplete'' Cartesian coordinates as
\begin{align}
\label{eq:HDFinitdatacartcoord}
x = & \rho \cos\varphi 
\,,\quad
y = \rho \sin\varphi 
\,,
\end{align}
where $-\infty<x<+\infty$ and $0\le y<+\infty$.
Then, the $(D-1)$-dimensional metric is conformally decomposed according to
\begin{align}
\label{eq:HDFinitdataConfMetric}
\bga_{ab} dx^a dx^b = & \psi^{\frac{4}{D-3}} \left( dx^2 + dy^2 + dz^2 + y^2 d\Omega_{D-4} \right)
\,.
\end{align}

The ADM-like variables describing the dynamics of the $D$-dimensional
spacetime with $SO(D-2)$ or $SO(D-3)$ isometry
are the $3$-metric $\ga_{ij}$ and extrinisic curvature $K_{ij}$, the scalar field $\la$
and its conjugate momentum $K_{\la}$.
The goal of this section is the construction of initial data for these quantities.
Note, that $\ga_{ij}$, $K_{ij}$, $\la$ and $K_{\la}$ are related to the $(D-1)$-dimensional
quantities by
\begin{align}
\label{eq:HDFinitdatadimred}
\bga_{ij} = & \ga_{ij}
\,,\quad
\bga_{AB} = \la h_{AB}
\,,\quad
\bga_{iA} = 0
\,, \\
\bK_{ij} = & K_{ij}
\,,\quad
\bK_{AB} = \frac{1}{2} K_\la h_{AB}
\,,\quad
\bK_{iA} = 0
\,,\quad
\bK =  K + \frac{D-4}{2} \frac{K_\la}{\la}
\,,
\end{align}
where $A,B=4,...,(D-1)$ and $h_{AB}$ is the metric on the $(D-4)$-sphere.
In the following we summarize the construction of modified
Brill-Lindquist and Bowen-York type initial data \cite{Zilhao:2011yc}, setting up the initial configuration
for head-on collisions of two BHs in higher dimensional spacetimes.
%

\newpage
\subsection{Brill-Lindquist initial data}\label{ssec:HDFBLID}
%
First, we consider time symmetric initial data, i.e., $\bK_{ab} = 0$, which results in
{\textit{Brill-Lindquist}} type initial data.
Then, the momentum constraint~\eqref{eq:HDFinitdataconfMom} is trivially satisfied, whereas the Hamiltonian 
constraint~\eqref{eq:HDFinitdataconfHam} reduces to the
$(D-1)$-dimensional flat space Laplace equation,
\begin{align}
\label{eq:HDFinitdataFL}
\hat{\triangle} \psi = & 0
\,.
\end{align}
For asymptotically flat spacetimes, the conformal factor satisfies the
boundary condition $\lim_{ r \to \infty} \psi = 1$
and a solution $\psi_0$ to Eq.~\eqref{eq:HDFinitdataFL} is given by
\begin{align}
\label{eq:HDFinitdataBLpsi}
\psi_0 = & 1 + \sum_{i=1}^{N} \frac{\mu_{(i)}}{4 |r - r_{(i)}|^{D-3}}
\,,
\end{align}
where $r_{(i)}$ is the coordinate location of the $i^{\rm th}$ puncture.
The mass parameter $\mu_{(i)}$ is related to the horizon radius
$r_{S_{(i)}}$ and the ADM mass $M_{(i)}$ of the $i^{\rm th}$ BH by
\begin{align}
\mu_{(i)} \equiv r_{S_{(i)}}^{D-3} \equiv \frac{16\pi M_{(i)}}{\A_{D-2}(D-2)} 
\,,
\end{align}
where $\A_{D-2}$ is the area
of the unit $(D-2)$-sphere and we have set the $D$ dimensional Newton
constant to unity.

By inserting our assumptions of conformal flatness, maximal slicing 
and time symmetry into Eqs.~\eqref{eq:HDFinitdatadimred} the $3$-dimensional
variables describing Brill-Lindquist initial data become
\begin{subequations}
\label{eq:HDFinitdataBL}
\begin{align}
\ga_{ij} = & \psi^{\frac{4}{D-3}}\eta_{ij}
\,,\quad
K_{ij} = 0
\,,\\
\la = & \psi^{\frac{4}{D-3}} y^2
\,,\quad
K_{\la} = 0
\,,
\end{align}
\end{subequations}
where $\psi=\psi_0$ is given by Eq.~\eqref{eq:HDFinitdataBLpsi}.
These closed-form analytic data are the $D$-dimensional generalization of
Brill-Lindquist data \cite{Brill:1963yv}, adjusted to our approach of the (evolution) formalism,
and describe a spacetime containing
multiple non-spinning BHs at the moment of time symmetry, i.e., with
vanishing linear momentum.

\newpage
\subsection{Bowen-York initial data}\label{ssec:HDFBY}
%
In order to numerically evolve BHs with non-zero boost, we 
consider the generalization of {\textit{Bowen-York}} initial data to higher dimensional spacetimes.
As shown by Yoshino {\em et al.} \cite{Yoshino:2006kc}, we can write a solution
of the momentum constraint~\eqref{eq:HDFinitdataconfMom} describing a spacetime of arbitrary
dimensionality $D$ containing $N$ BHs in the form
\begin{align}
\label{eq:HDFinitdataBYA0}
\hA^{ab}_P = & \sum_{i = 1}^N \hat A_{P(i)}^{ab}
\,,
\end{align}
with
\begin{align}
\label{eq:HDFinitdataAAP0}
\hA^{ab}_{P(i)} = & \frac{4 \pi (D-1)}{(D-2) \A_{D-2} } \frac{1}{r^{D-2}_{(i)} }
\,\non\\ & \times
  \left(
    n^a_{(i)} P^b_{(i)} + n^b_{(i)} P^a_{(i)} 
    - (n_{(i)})_c P^c_{(i)} \hga^{ab} + (D-3) n^a_{(i)} n^b_{(i)} P^c_{(i)} (n_{(i)})_c
  \right)
\,,
\end{align}
where $P^a_{(i)}$ corresponds to the ADM momentum of the
$i^{\rm th}$ BH in the limit of large separation from all other
BHs.
Next, we consider the Hamiltonian constraint~\eqref{eq:HDFinitdataconfHam},
which reduces to an elliptic equation for the conformal factor $\psi$ where $\hA_{ab}$
is prescribed by Eq.~\eqref{eq:HDFinitdataBYA0}.
A solution $\psi_0$ of the conformal factor is given by
\begin{align}
\label{eq:HDFinitdataBYansatz}
\psi_0 = & u + \psi_{BL}
\,,\quad\text{with}\quad
\psi_{BL} = \sum_{i=1}^N \frac{\mu_{(i)}}{4|r - r_{(i)}|^{D-3}}
\,,
\end{align}
where $r_{(i)}$ and $\mu_{(i)}$ are the (coordinate) position and mass parameter of the $i^{th}$ BH
and $u$ is a regular function.
Inserting this ansatz into the Hamiltonian constraint~\eqref{eq:HDFinitdataconfHam} yields
\begin{align}
\label{eq:HDFinitdataU}
\hat{\triangle} u + \frac{D-3}{4(D-2)} \hA^{ab} \hA_{ab} \psi^{-\frac{3D-5}{D-3}} = 0
\,.
\end{align}
As in $D=4$, the higher dimensional extension of Bowen-York extrinsic
curvature data can also accommodate angular momentum of the BHs.
Here, however, we shall focus on initial data for
non-spinning, boosted BHs only.

Without loss of generality, we can always choose coordinates such that the
BHs are initially located on the $z$-axis at $z_1$ and $z_2$
and have momenta of
equal magnitude in opposite directions
$P_{(1)}^a = -P_{(2)}^a$.
Inserting the momenta into Eq.~\eqref{eq:HDFinitdataAAP0} then provides the conformal
traceless extrinsic cuvature and the differential equation \eqref{eq:HDFinitdataU}
which is solved numerically for $u$.
The class of symmetries covered by the formalism 
presented in Sec.~\ref{sec:HDFDimRed} includes head-on and grazing
collisions of non-spinning BHs with initial position 
$x_{(1)}^a = (0, 0, z_1,0,\dots, 0)\,,x_{(2)}^a = (0, 0, z_2,0,\dots, 0)$
and momenta
$P_{(1)}^a = (P^x, 0, P^z, 0, \dots, 0) = -P_{(2)}^a$,
whereas a non-zero $P^y$ is not compatible with the assumed symmetries.
However, the $x$-axis can always be oriented such that the
collision takes place in the $xz$-plane. Our formalism therefore
covers general grazing collisions of non-spinning BH binaries
in $D$ dimensions.
For now, we will focus on the case of BHs with momenta in the $z$ direction,
so that $P^x=0$ and the linear momenta are given by
\begin{align}
\label{eq:HDFinitdataPheadon}
P^a_{(1)} = & (0,0,P^z,0,\ldots ,0) = -P^a_{(2)}
\,.
\end{align}
The rescaled trace-free part of the extrinsic curvature for such a
configuration is
\begin{align}
\label{eq:HDFinitdata4}
\hat A_{ab} = & \hat A_{ab}^{(1)} + \hat A_{ab}^{(2)} \ ,
\end{align}
where $\hat A_{ab}^{(1)}$ and $\hat A_{ab}^{(2)}$ are given by
Eq.~\eqref{eq:HDFinitdataAAP0} with the linear momentum~\eqref{eq:HDFinitdataPheadon}.
In a coordinate system adapted to the spacetime symmetry we, thus, obtain
\begin{equation}
\label{eq:HDFinitdata8}
\hat A_{ab}^{(1)} =
  \displaystyle{
    \frac{4 \pi (D-1) P^z}{(D-2) \A_{D-2}
    (x^2 + y^2 + (z-z_1)^2)^{\frac{D+1}{2}} }
  }
\left(
\begin{array}{c|c}
 \hat a_{i j}^{(1)} & 0  \\
 \hline
 0 & \hat a_{AB}^{(1)}
\end{array}
\right),
\end{equation}
with
\begin{equation}
\label{eq:HDFinitdataA}
\hat a_{ij}^{(1)} =
\left(
\begin{smallmatrix}
   \left[ (D-4)x^2-y^2-(z-z_1)^2\right] (z-z_1) & (D-3)  x y
   (z-z_1) &  x \left[x^2+y^2+ (D-2) (z-z_1)^2\right] \\
 (D-3) x y (z-z_1) & 
  \left[(D-4)y^2 -x^2 -(z-z_1)^2\right]
   (z-z_1) &  y \left[x^2+y^2 + (D-2) (z-z_1)^2\right] \\
  x \left[x^2+y^2+ (D-2) (z-z_1)^2\right] &  y \left[x^2+y^2+ (D-2)
   (z-z_1)^2\right] &  \left[x^2+y^2 + (D-2) (z-z_1)^2\right] (z-z_1)
\end{smallmatrix}
\right)
\end{equation}
and
\begin{align}
\label{eq:HDFinitdata9}
\hat a_{AB}^{(1)} = & - y^2 (z-z_1) \left[ x^2 + y^2
                      + (z-z_1)^2 \right] h_{AB}\ ,
\end{align}
where $h_{AB}$ is the metric on the $(D-4)$-sphere.
The expression for $\hat A^{(2)}_{ab}$ is analogous, but with $z_2$ in place of $z_1$ and
$-P^z$ in place of $P^z$ in Eq.~\eqref{eq:HDFinitdata8}.

Employing the relations~\eqref{eq:HDFinitdatadimred} and the $4$-metric~\eqref{eq:HDFmetricansatz}
results in the Bowen-York type initial data for the $3$-dimensional dynamical quantities 
prescribed by
\begin{subequations}
\label{eq:HDFinitdataFinal}
\begin{align}
\ga_{ij} = & \psi^{\frac{4}{D-3}} \de_{ij}
\,,\quad 
\la = \psi^{\frac{4}{D-3}} y^2
\,, \\
K_{ij} = & \psi^{-2}(\hat A_{ij}^{(1)} + \hat A_{ij}^{(2)})
\,, \quad
K_\la = 2 \psi^{-2} y^2 ( P^{+} + P^{-})
\,,\quad
K  = - \frac{(D-4)K_{\la}}{2\la} 
\,,
\end{align}
\end{subequations}
where
\begin{subequations}
\begin{align}
P^{+} = & -\frac{4 \pi (D-1) P^z (z-z_1)}{(D-2) \A_{D-2} (x^2 + y^2+ (z-z_1)^2)^{\frac{D-1}{2}}  }
\,,\\
P^{-} = & \frac{4 \pi (D-1) P^z (z-z_2)}{(D-2) \A_{D-2} (x^2 + y^2+ (z-z_2)^2)^{\frac{D-1}{2}}  }
\,.
\end{align}
\end{subequations}
The elliptic equation for $u$, Eq.~\eqref{eq:HDFinitdataU}, writes
\begin{align}
\label{eq:HDFinitdataUfinal}
\left( \partial_{\rho\rho} + \partial_{zz} + \frac{D-3}{\rho}
       \partial_{\rho} \right) u
 = & \frac{3-D}{4(D-2)} \hat A^{ab} \hat A_{ab} \psi^{ -\frac{3D-5}{D-3} } \ ,
\end{align}
where
\begin{align}
\hat A^{ab} \hat A_{ab} = & 
(\hat A_{ij}^{(1)} + \hat A_{ij}^{(2)}) ( \hat A^{ij}{}^{(1)}+\hat A^{ij}{}^{(2)}) 
+ (D-4) ( P^{+} + P^{-})^2
\,.
\end{align}
Finally, Eq.~\eqref{eq:HDFinitdataUfinal} will be solved numerical by an extension
of the {\textsc{TwoPunctures}} spectral solver \cite{Ansorg:2004ds}
presented in \cite{Zilhao:2011yc}.

\section{Wave Extraction}\label{sec:HDFWaveExtr}
In the previous sections we have described our framework for numerical simulations
of higher dimensional BH spacetimes, providing the initial data setup as well as
a formulation of the evolution system to model the dynamics of these system.
The missing ingredient, namely the formalism that enables us to extract information
about gravitational radiation and energy emitted throughout the 
collision of two (higher dimensional) BHs,
will be presented in this section.

\subsection{Coordinate frames}
As discussed in Sec.~\ref{sec:HDFDimRed}, in our approach we perform a dimensional reduction
of the $D$-dimensional spacetime to an effectively $4$-dimensional gravity model and we consider
different generalizations of ``axial symmetries'' to higher dimensions: 
either $D\ge5$ dimensional spacetimes with $SO(D-2)$ isometry group, 
or $D\ge6$ dimensional spacetimes with $SO(D-3)$ isometry group. 
Here, we focus only on the former case, which allows us to model head-on collisions of non-spinning BHs.
Hereafter, we dub these spacetimes as \textit{axially symmetric}. 
Although the corresponding symmetry manifold is the $(D-3)$-sphere $S^{D-3}$, 
the quotient manifold in our dimensional reduction is its submanifold $S^{D-4}$. 
The coordinate frame in which the numerical simulations are performed is
\begin{align}
\label{eq:HDWEframe1}
(x^\mu,\phi^1,\dots,\phi^{D-4}) = & (t,x,y,z,\phi^1,\dots,\phi^{D-4}) 
\,,
\end{align}
where the angles $\phi^1,\dots,\phi^{D-4}$ describe the quotient manifold $S^{D-4}$ 
and do not appear explicitly in the simulations.
Here, $z$ is the symmetry axis, i.e. the collision line.

In the frame \eqref{eq:HDWEframe1}, the spacetime metric has the form 
(cf. Eqs.~\eqref{eq:HDFmetricansatz} and~\eqref{eq:NRlineeleg})
\begin{align}
\label{eq:HDWEtotalmetric}
ds^2 = &g_{\mu\nu}(x^\alpha)dx^\mu dx^\nu+\lambda(x^\mu)d\Omega_{D-4} 
\non\\
     = & -\alpha^2dt^2+\gamma_{ij}(dx^i+\beta^idt)(dx^j+\beta^jdt)
         +\lambda(x^\mu)d\Omega_{D-4} 
\,,
\end{align}
where $x^\mu=(t,x^i)$, $\lambda(x^\mu)$ is the scalar field and $\alpha,\beta^i$ are the lapse function and the shift
vector, respectively. It is worth noting that, although in $D=4$ a general axially symmetric spacetime has
non-vanishing mixed components of the metric, such as $g_{t\phi}$, in $D\ge5$ these components vanish in an appropriate
coordinate frame.
With an appropriate transformation of the four dimensional coordinates $x^\mu$, the residual symmetry left after the
dimensional reduction on $S^{D-4}$ can be made manifest: 
$x^\mu\rightarrow (x^{\bar\mu},\theta)$ ($\bar\mu=0,1,2$),
\begin{subequations}
\begin{align}
\label{eq:HDWEthetametric}
g_{\mu\nu}(x^\alpha)dx^\mu dx^\nu = & 
g_{\bar\mu\bar\nu}(x^{\bar\alpha})dx^{\bar\mu}dx^{\bar\nu}+g_{\theta\theta}(x^{\bar\alpha})d\theta^2
\,,\\
\label{eq:HDWElambdatheta}
\lambda(x^\mu) = & \sin^2\theta g_{\theta\theta}(x^{\bar\alpha})
\,,
\end{align}
\end{subequations}
so that Eq.~\eqref{eq:HDWEtotalmetric} takes the form 
$ds^2=g_{\bar\mu\bar\nu}dx^{\bar\mu}dx^{\bar\nu}+g_{\theta\theta}d\Omega_{D-3}$.

To extract gravitational waves far away from the symmetry axis
we employ the Kodama-Ishibashi (KI) formalism
\cite{Kodama:2003jz}, which generalizes the Regge-Wheeler-Zerilli \cite{Regge:1957td,Zerilli:1970se} 
approach to higher dimensions. 
We require that the spacetime, far away from the BHs, is approximately
spherically symmetric. Note, that spherical symmetry in $D$~dimensions means symmetry with respect to rotations on
$S^{D-2}$; this is an approximate symmetry which holds asymptotically
far away from the axis
and which is manifest in the coordinate frame:
\begin{align}
\label{eq:HDWEframe2}
(x^{a},\bar\theta,\theta,\phi^1,\dots,\phi^{D-4}) = & (t,r,\bar\theta,\theta,\phi^1,\dots,\phi^{D-4}) 
\,.
\end{align}
Note, that throughout this section $x^{a}=(t,r)$ and that we have 
introduced polar-like coordinates $\bar\theta,\theta\in[0,\pi]$ 
to ``build up'' the manifold $S^{D-2}$ in the background, together with a radial spherical coordinate $r$, which
is the areal coordinate in the background.

The coordinate frame \eqref{eq:HDWEframe2} is defined in such a way that the metric 
can be expressed as a stationary background $(ds^{(0)})^2$ (i.e., the Tangherlini metric) 
plus a perturbation $(ds^{(1)})^2$ which decays faster than $1/r^{D-3}$ for large $r$:
\begin{subequations}
\begin{align}
(ds^{(0)})^2 = & g^{(0)}_{ab}dx^{a}dx^{b}+r^2d\Omega_{D-2} 
\non\\
             =&  g_{tt}^{(0)}dt^2 + g_{rr}^{(0)}dr^2 
                + r^2\left(d\bar\theta^2+\sin^2\bar\theta d\Omega_{D-3}\right)
\non\\
\label{eq:HDWEds0}
             = & - \left(1-\frac{r_S^{D-3}}{r^{D-3}}\right)dt^2
                 + \left(1-\frac{r_S^{D-3}}{r^{D-3}}\right)^{-1}dr^2 
\non\\ &
                 + r^2\left[d\bar\theta^2+\sin^2\bar\theta\left(d\theta^2+\sin^2\theta d\Omega_{D-4}\right)\right] 
\,,\\
\label{eq:HDWEds1}
(ds^{(1)})^2 = &  h_{ab}dx^{a}dx^{b}+h_{a\bar\theta}dx^{a}d\bar\theta
                + h_{\bar\theta\bar\theta}d\bar\theta^2 
                + h_{\theta\theta}d\Omega_{D-3}
\,.
\end{align}
\end{subequations}
Here, the Schwarzschild radius $r_S$ is related to the ADM mass $M$ by
\begin{align}
\label{eq:HDWEdefadm}
r_S^{D-3} = & \frac{16\pi M}{(D-2){\cal A}_{D-2}} 
\,,
\end{align}
where ${\cal A}_{D-2}$ is the area of the $(D-2)$-sphere (see Eq.~\eqref{eq:HDWEareasphere}). 
For instance, $r_S=2M$ in $D=4$, $r^2_S=8M/(3\pi)$ in $D=5$ and $r^3_S = 3M / (2\pi)$ in $D=6$.

When we define the coordinate frame~\eqref{eq:HDWEframe2}, we also require that the coordinate $\theta$ in this frame 
coincides with the coordinate  $\theta$ appearing in Eq.~\eqref{eq:HDWEthetametric}. With this choice, the axial symmetry of the spacetime implies that
\begin{align}
\label{eq:HDWEaxialsymh}
h_{a\theta} = & h_{\bar\theta\theta} = 0
\,,
\end{align}
as in Eq.~\eqref{eq:HDWEds1}, and $\lambda=\sin^2\theta g_{\theta\theta}$, i.e. Eq.~\eqref{eq:HDWElambdatheta}.
The transformation from Cartesian coordinates $x^\mu=(t,x,y,z)$, in which the numerical simulation is implemented,
to the coordinates $(x^{a},\bar\theta,\theta)=(t,r,\bar\theta,\theta)$,
in which the wave extraction is performed, is given by
\begin{subequations}
\label{eq:HDWEtransfc}
\begin{align} 
x = & R\sin\bar\theta\cos\theta 
\,, \\ 
y = & R\sin\bar\theta\sin\theta 
\,,\\ 
z = & R\cos\bar\theta 
\,,
\end{align}
\end{subequations}
where $R = \sqrt{x^2 + y^2 + z^2}$
and by the reparametrization of the radial coordinate
\begin{align}
\label{eq:HDWEtransfr}
R = & R(r) \,.
\end{align}

We assume that the transformation~\eqref{eq:HDWEtransfc}, \eqref{eq:HDWEtransfr} yields the
``Tangherlini+perturbation'' splitting~\eqref{eq:HDWEds0}, \eqref{eq:HDWEds1} during the entire evolution of the system.
This statement can be checked
numerically by verifying the relations (see Appendix~\ref{appintegrals}):
\begin{subequations}
\begin{align}
\label{eq:HDWEtestgtt}
\mathcal{G}_{tt} & \equiv \frac{1}{K^{0D}\pi} \int_0^{\pi} d\bar{\theta} \sin^{D-3}\bar{\theta}
  \int_0^{\pi} d \theta \left( g_{tt}(\bar{\theta},\theta) - g_{tt}^{(0)} \right) = 0\,, \\
\label{eq:HDWEtestgtR}
\mathcal{G}_{tr} & \equiv  \frac{1}{K^{0D}\pi} \int_0^{\pi} d\bar{\theta} \sin^{D-3}\bar{\theta} 
  \int_0^{\pi} d \theta  g_{tR}(\bar{\theta},\theta)  =0\,, \\
\label{eq:HDWEtestgRR}
\mathcal{G}_{rr} & \equiv  
\frac{1}{K^{0D}\pi} \int_0^{\pi} d\bar{\theta} \sin^{D-3}\bar{\theta} 
  \int_0^{\pi} d \theta \left( g_{RR}(\bar{\theta},\theta) - g_{rr}^{(0)} \right) = 0\,,
\end{align} 
\end{subequations}
where $K^{0D}=\int_0^\pi d\bar\theta(\sin\bar\theta)^{D-3}$, together with the axisymmetry
conditions~\eqref{eq:HDWElambdatheta}, \eqref{eq:HDWEaxialsymh}.
The preservation of the above identities during the numerical evolution,
as shown in Secs.~\ref{ssec:HDRCCWE4D} and~\ref{sec:HDRcoll5D},
justifies also the identification of the time coordinate in the numerical and wave extraction frames, 
and our use of the KI formalism. 
Finally, Eqs.~\eqref{eq:HDWEtotalmetric}, \eqref{eq:HDWEds0}, \eqref{eq:HDWEds1} yield the $3+1$-splitting
\begin{align}
ds^2 = & (ds^{(0)})^2 + (ds^{(1)})^2 
\non\\
     = & g_{\bar\mu\bar\nu}dx^{\bar\mu}dx^{\bar\nu}+(r^2\sin^2 \bar \theta+h_{\theta\theta}) 
         (d\theta^2+\sin^2\theta d\Omega_{D-4})\non\\
\label{eq:HDWElineelement}
    =&-\alpha^2 dt^2+\gamma_{ij}(dx^i+\beta^idt) 
        (dx^j+\beta^jdt)+\lambda d\Omega_{D-4} 
\,,
\end{align}
where $x^{\bar\mu}=(t,r,\bar\theta)$.
With the $3+1$-split, the axisymmetry conditions~\eqref{eq:HDWElambdatheta},
\eqref{eq:HDWEaxialsymh} take the form
\begin{align}
\label{eq:HDWEcheckax}
\lambda = & \gamma_{\theta\theta}\sin^2\theta
\,,\quad
\gamma_{R\theta}=\gamma_{\bar\theta\theta} =\beta^\theta=0
\,.
\end{align}
The variable $r$ can be determined from the angular components of the metric~\eqref{eq:HDWElineelement}, 
by averaging out $h_{\bar\theta\bar\theta}$, $h_{\theta\theta}$ (see Appendix \ref{appintegrals});
its explicit expression is given by
\begin{align}
\label{eq:HDWEarealradius}
(r(R))^2 = & \frac{1}{(D-2)K^{0D}}\int_0^\pi d\bar\theta
           \left[ \gamma_{\bar\theta\bar\theta}(\sin\bar\theta)^{D-3} 
                 +(D-3)\gamma_{\theta\theta}(\sin\bar\theta)^{D-5}\right]\,.
\end{align}
In Secs.~\ref{ssec:HDRCCWE4D} and~\ref{sec:HDRcoll5D}
we will varify that the areal radius $r$ indeed is very close to $R$ throughout the numerical evolution.

\subsection{Harmonic expansion}
In the KI formalism \cite{Kodama:2003jz} (see also \cite{Kodama:2000fa}), the  background  spacetime has the form
\eqref{eq:HDWEds0}
\begin{align}
(ds^{(0)})^2 =& g^{(0)}_{MN}dx^Mdx^N =g^{(0)}_{ab}dx^{a}dx^{b}+r^2d\Omega_{D-2} 
 = g^{(0)}_{ab}dx^{a}dx^{b}+r^2\gamma_{\bar i\bar j}d\phi^{\bar i}d\phi^{\bar j} 
\,,
\end{align}
i.e. is the Tangherlini metric, where the $x^M$ coordinates refer to the full spacetime. 
The spacetime perturbations can be decomposed into spherical harmonics on
the $(D-2)$-sphere $S^{D-2}$. They are functions of the $(D-2)$ angles $\phi^{\bar i}=(\bar\theta,\theta,\phi^1,\dots,\phi^{D-4})$.  We denote  
the metric of $S^{D-2}$ by $\gamma_{\bar i\bar j}$, and with a subscript $_{:\bar i}$ the covariant derivative with
respect to this metric. Finally, we denote the covariant derivative with respect to the
metric $g^{(0)}_{ab}$ with a subscript $_{|a}$.

As discussed in \cite{Kodama:2003jz}, there are three types of spherical harmonics:
\begin{itemize}
\item The scalar harmonics ${\cal S}(\phi^{\bar i})$, which are
solutions of
\begin{equation}
\Box{\cal S}=\gamma^{\bar i\bar j}{\cal S}_{:\bar i\bar j}=-k^2{\cal S}\label{eqscal} \, ,
\end{equation}
with $k^2=l(l+D-3)$, $l=0,1,2,\dots$. 
The scalar harmonics ${\cal S}$ depend on the integer $l$ and on other
indices; we leave such dependence implicit. We also define
\begin{align}
{\cal S}_{\bar i}=&-\frac{1}{k}{\cal S}_{,\bar i}\,,\,\,
{\cal S}_{\bar i\bar j}=\frac{1}{k^2}{\cal S}_{:\bar i\bar j}+\frac{1}{D-2}\gamma_{\bar i\bar j}{\cal S} \, .
\label{defharm}
\end{align}
Observe, that $\gamma^{\bar i\bar j}{\cal S}_{\bar i\bar j}=0$.
Each harmonic mode of the metric perturbation $\delta g_{MN}=h_{MN}$ can be decomposed as
\begin{subequations}
\begin{align}
\label{eq:HDWEmpert0} 
\delta g_{ab} = & h_{ab} =  f_{ab}{\cal S}
\,, \\
\label{eq:HDWEmpert1}
\delta g_{a\bar i} = & h_{a\bar i} = rf_{a}{\cal S}_{\bar i}
\,,\\
\label{eq:HDWEmpert2} 
\delta g_{\bar i\bar j} = & h_{\bar i\bar j}
= 2r^2(H_L\gamma_{\bar i\bar j}{\cal S}+H_T{\cal S}_{\bar i\bar j})
\,,
\end{align}
\end{subequations}
where $f_{ab}$, $f_{a}$, $H_L$, $H_T$ are functions of $x^{a}=(t,r)$.
Note, that in each of these expressions there is a sum over the indices of the harmonic. 

For $l>1$, the metric perturbations can be expressed in terms of the gauge-invariant variables
\cite{Kodama:2000fa}
\begin{subequations}
\begin{align}
\label{eq:HDWEfunc1}
F = & H_L+\frac{1}{D-2}H_T+\frac{1}{r}X_{a} r^{|a}
\,,\\
\label{eq:HDWEfunc2}
F_{ab} = & f_{ab}+X_{a|b}+X_{b|a} 
\,,
\end{align}
\end{subequations}
where we have defined
\begin{align}
\label{eq:HDWEgi2}
X_{a} = & \frac{r}{k}\left(f_{a}+\frac{r}{k}H_{T|a}\right)
\,.
\end{align}
\item The vector harmonics ${\cal V}_{\bar i}(\phi^{\bar i})$, which are solutions of 
\begin{equation}
\gamma^{\bar i\bar j}{\cal V}_{\bar k:\bar i\bar j}=-k_V^2{\cal V}_{\bar k} \, ,
\end{equation} 
with $k_V^2=l(l+D-3)-1$, $l=1,2,\dots$. These harmonics satisfy the relation
\begin{equation}
{\cal V}^{\bar i}_{:\bar i}=0\,.\label{divV}
\end{equation}
The harmonic expansion of the corresponding metric perturbations is given by Eqs.~\eqref{eq:HDWEmpert1}-\eqref{eq:HDWEmpert2},
with ${\cal S}_{\bar i}$ replaced by ${\cal V}_{\bar i}$, ${\cal S}_{\bar i\bar j}$ replaced by
\begin{equation}
{\cal V}_{\bar i\bar j}=-\frac{1}{2k_V}(V_{\bar i:\bar j}+V_{\bar j:\bar i})\,,
\end{equation}
and $H_L=0$.
\item The tensor harmonics ${\cal T}_{\bar i\bar j}(\phi^{\bar i})$, which are solutions of 
\begin{equation}
\gamma^{\bar i\bar j}{\cal T}_{\bar r\bar s:\bar i\bar j}=-k_T^2{\cal T}_{\bar r\bar s} \, ,
\end{equation} 
with $k_T^2=l(l+D-3)-2$, $l=1,2,\dots$. These harmonics satisfy,
\begin{equation}
\gamma^{\bar i\bar j}{\cal T}_{\bar i\bar j}=0\,,
~~~~~{\cal T}^{:\bar i\bar j}_{:\bar j}=0\,.\label{divT}
\end{equation}
In the $D=4$ case they vanish. The harmonic expansion of the corresponding metric perturbations is given by
\eqref{eq:HDWEmpert2}, with ${\cal S}_{\bar i\bar j}$ replaced by ${\cal T}_{\bar i\bar j}$ and $H_L=0$.
\end{itemize}
%
\subsection{Implementation of axisymmetry}\label{implax}
In an axially symmetric spacetime, the metric perturbations are symmetric with respect to $S^{D-3}$. Therefore, the
harmonics in the expansion of $h_{MN}$ depend only on the angle $\bar\theta$.
Furthermore, since there are no off-diagonal terms in the metric, the only non-vanishing
$g_{a\bar i}$ components are $g_{a\bar\theta}$; the only components $g_{\bar i\bar j}$ are either
proportional to $\gamma_{\bar i\bar j}$, or all vanishing but $g_{\bar\theta\bar\theta}$. This implies that only
scalar spherical harmonics can appear in the expansion of the metric perturbations. Indeed, if
\begin{equation}
{\cal V}^{\bar i}=({\cal V}^{\bar\theta},0,\dots,0)\,,~~~~~{\cal V}^{\bar i}={\cal V}^{\bar i}(\bar\theta) \, ,
\end{equation}
then Eq.~(\ref{divV}) gives
\begin{equation}
{\cal V}^{\bar i}_{:\bar i}={\cal V}^{\bar\theta}_{,\bar\theta}=0~~~\Rightarrow~~~
{\cal V}^{\bar\theta}=0~~\Rightarrow~~~{\cal V}^{\bar i}=0\,.
\end{equation}
Similarly, from Eq.~(\ref{divT}) we obtain ${\cal T}_{\bar i\bar j}=0$.

The scalar harmonics, solutions of Eq.~\eqref{eqscal} and depending only on the coordinate $\bar\theta$, are given by the
Gegenbauer polynomials $C_l^{(D-3)/2}$, as discussed in Refs.~\cite{Berti:2003si,Cardoso:2002ay,Yoshino:2005ps}; writing explicitly the index $l$, they take the form
\begin{equation}
{\cal S}_l(\bar\theta)=(K^{lD})^{-1/2}C_l^{(D-3)/2}(\cos\bar\theta) \, , \label{gp}
\end{equation} 
where the normalization $K^{lD}$ is chosen such that
\begin{equation}
\int d\Omega^{D-2}{\cal S}_l{\cal S}_{l'}=\delta_{ll'}\,,~~~
\int d\Omega^{D-2}{\cal S}_{l\,,\bar\theta}{\cal S}_{l'\,,\bar\theta}=\delta_{ll'}k^2 \ ,\label{normalK}
\end{equation}
and $k^2=l(l+D-3)$ (see Appendix \ref{appintegrals}). 
By computing ${\cal S}_{l\,\bar i}$, ${\cal S}_{l\,\bar i\bar j}$ from
Eqs.~\eqref{defharm} (using Eq.~\eqref{eqscal}) 
we find
\begin{subequations}
\begin{align}
{\cal S}_{l\,\bar\theta\bar\theta}&=\frac{D-3}{k^2(D-2)} {\cal W}_l \, , \\
{\cal S}_{l\,\theta\theta}&=-\frac{\sin^2\bar\theta}{k^2(D-2)}{\cal W}_l\,,
\end{align}
\end{subequations}
where we have defined
\begin{align}
{\cal W}_l(\bar\theta) = & {\cal S}_{l\,,\bar\theta\bar\theta}-\cot\bar\theta{\cal S}_{l\,,\bar\theta}\,.
\end{align}
Therefore, the metric perturbations are given by
\begin{subequations}
\label{eq:HDWEmpert}
\begin{align}
h_{ab} =& f^l_{ab}{\cal S}_l(\bar\theta)
\,, \\
h_{a\bar\theta} =& rf^l_{a}{\cal S}_l(\bar\theta)_{\bar\theta}=
-\frac{1}{k} rf^l_{a}{\cal S}_l(\bar\theta)_{,\bar\theta}
\,,\\
h_{\bar\theta\bar\theta} =& 2r^2(H^l_L {\cal S}_l(\bar\theta)
                           +H^l_T {\cal S}_l(\bar\theta)_{\bar\theta\bar\theta}) 
                         = 2r^2\left(H^l_L {\cal S}_l(\bar\theta)
                           +H^l_T \frac{D-3}{k^2(D-2)}{\cal W}_l(\bar\theta)\right)
\,, \\
h_{\theta\theta} =& 2r^2(H^l_L\sin^2\bar\theta {\cal S}_l(\bar\theta)
                   +H^l_T {\cal S}_l(\bar\theta)_{\theta\theta}) 
                 =  2r^2\sin^2\bar\theta\left(H^l_L {\cal S}_l(\bar\theta)
                   -H^l_T \frac{1}{k^2(D-2)}{\cal W}_l(\bar\theta)\right) 
\,.
\end{align}
\end{subequations}
The quantities $f_{ab}$, $f_{a}$, $H_L$, $H_T$ are (see Appendix \ref{appintegrals}):
\begin{subequations}
\label{eq:HDWEmetricrelation}
\begin{align}
f^l_{ab}(t,r) =& \frac{{\cal A}_{D-3}}{\sqrt{K^{lD}}}\int_0^{\pi}
                 d\bar\theta (\sin\bar\theta)^{D-3}h_{ab}  C_l^{(D-3)/2}
\,,\\
f_{a}(t,r) =&-\frac{1}{\sqrt{l(l+D-3)}r}\frac{{\cal A}_{D-3}}{\sqrt{K^{lD}}} 
              \int_0^{\pi}d\bar\theta (\sin\bar\theta)^{D-3}
                    h_{a \bar \theta }C_{l\,,\bar\theta}^{(D-3)/2}(\cos{\bar\theta})
\,, \\
H_L(t,r) =& \frac{1}{2(D-2)r^2}\frac{{\cal A}_{D-3}}{\sqrt{K^{lD}}}\int_0^{\pi}
            d\bar\theta (\sin\bar\theta)^{D-3} 
           \left[h_{\bar\theta\bar\theta}+\frac{D-3}{\sin^2\bar\theta}h_{\theta\theta}\right] 
            C_l^{(D-3)/2}(\cos{\bar\theta})
\,, \\
H_T(t,r) =& \frac{1}{2r^2(k^2-D+2)}\frac{{\cal A}_{D-3}}{\sqrt{K^{lD}}}
            \int_0^{\pi}d\bar\theta (\sin\bar\theta)^{D-3} 
            \left[h_{\bar\theta\bar\theta}-\frac{1}{\sin^2\bar\theta}h_{\theta\theta}\right]W_l(\bar\theta)
\,,
\end{align}
\end{subequations}
where $h_{ab} = h_{ab}(t,r,\bar\theta)$, $h_{a\bar\theta} = h_{a\bar\theta}(t,r,\bar\theta)$,
$h_{\bar\theta \bar\theta} = h_{\bar\theta \bar\theta}(t,r,\bar\theta)$, 
$h_{\theta\theta} = h_{\theta \theta}(t,r,\bar\theta)$ and 
$C_l^{(D-3)/2} = C_l^{(D-3)/2}(\cos{\bar\theta})$.
Using Eqs.~\eqref{eq:HDWEfunc2}, \eqref{eq:HDWEgi2}, we obtain the gauge-invariant quantities $F$, $F_{ab}$
in terms of these functions. 

As we have discussed above, this approach has been developed for $D>4$, since in $D=4$ the off-diagonal terms
$g_{t\phi}$, $g_{r\phi}$ are not vanishing in general axially symmetric spacetimes. However, we can extend our
framework to $D=4$ if we restrict ourselves to axially symmetric spacetimes with $g_{t\phi}=g_{r\phi}=0$. In this way,
we can test our formalism by comparing our results to the existing literature. For instance, we note that in $D=4$ the
perturbation functions are related to the expressions in Ref.~\cite{Sperhake:2005uf}, with the identifications
\begin{subequations}
\begin{align}
\label{eq:HDWEcompPF1}
f^l_{ab} = & H_0,\,H_1,\,H_2\,,\\
\label{eq:HDWEcompPF2}
-\frac{r}{k}f^l_{a} = & h_0,\,h_1\,,\\
\label{eq:HDWEcompPF3}
\frac{2H_T}{k^2} = & G\,,\\
\label{eq:HDWEcompPF4}
2H_L+H_T = & K\,.
\end{align}
\end{subequations}
We also remark that in the transverse-traceless (TT) gauge, only $H_T$ is non-vanishing, but in a generic gauge (like the one
used in the numerical simulations) all these quantities are in principle non-vanishing.
\subsection{Extracting gravitational waves at infinity}\label{master}
In the KI framework, the emitted gravitational waves are described by the master function $\Phi$. To compute $\Phi$ in
terms of the gauge-invariant quantities $F$, $F_{ab}$ one should perform a Fourier transform or a time
integration (see \cite{Kodama:2003jz}). This can be avoided if we compute directly $\Phi_{,t}$, given by\footnote{Note
  that there is a factor $r$ missing in Eq.~(3.15) of Ref.~\cite{Kodama:2003jz}.}
\begin{equation} 
\label{eq:HDWEKIwavefunction}
\Phi_{,t}=(D-2)r^{(D-4)/2}\frac{-F^r_{~t}+2 r F_{,t}}{k^2-D+2+\frac{(D-2)(D-1)}{2}\frac{r_S^{D-3}}{r^{D-3}}} \, ,
\end{equation}
where $k^2=l(l+D-3)$. In the TT-gauge, the gravitational perturbation is described by $H_T$, which decays as
$r^{(D-2)/2}$ with increasing $r$, whereas the other perturbation functions have a faster decay (see
\cite{Berti:2003si}). In this gauge, the asymptotic behaviour of the master function is
\begin{equation}
\Phi\simeq\frac{2r^{(D-2)/2}H_T}{k^2}\,  ,\label{propfinite}
\end{equation}
and tends to an oscillating function with constant amplitude as $r\rightarrow\infty$.  
Writing the index $l$ explicitly, the energy flux in each $l-$multipole is \cite{Berti:2003si}
\begin{equation}
\label{eq:HDWEenergyflux}
\frac{dE_l}{dt}=\frac{1}{32\pi}\frac{D-3}{D-2}k^2(k^2-D+2)(\Phi^l_{,t})^2\,.
\end{equation} 
The total energy emitted in the process is then
\begin{equation}
\label{eq:HDWEenergyrad}
E=\sum_{l=2}^\infty\int_{-\infty}^{+\infty}dt\frac{dE_l}{dt}\,.
\end{equation}
%

\section{Discussion}\label{sec:HDFConclusion}
In this chapter we have presented a framework that allows the generalisation of the
current generation of 3+1 numerical codes to evolve, with relatively minor
modifications, spacetimes with $SO(D-2)$ symmetry in $D\ge5$ dimensions and
$SO(D-3)$ symmetry in $D\ge 6$ dimensions.  The key idea is a dimensional
reduction of the problem along the lines of Geroch's \cite{Geroch:1970nt}
procedure that recasts the $D$-dimensional vacuum Einstein's equations 
into an effectively $4$-dimensional system coupled to a scalar field.
The resulting equations can be transformed straightforwardly into the (generalized) BSSN
formulation that has proven to be remarkably successful in numerical evolutions of BH
configurations in $3+1$ spacetimes.  
We have isolated several issues related to the regularisation of the variables 
used in our formulation and demonstrated how
all difficulties related to the coordinate singularity arising from the use of
a ``radius-like'' coordinate can be successfully addressed in a numerical
implementation. 
We have further illustrated how 
Brill-Lindquist and Bowen-York type initial data
can be adapted straightforwardly to the formalism presented in this
chapter.  
More generally, the class of problems that may be studied with our
framework includes head-on collisions in $D\ge 5$ and a subset of BH collisions
with impact parameter and spin in $D\ge 6$.

Finally, we have presented our formalism to extract gravitational wave observables.
The technique is based on the studies of perturbations of Tangherlini BHs by 
Kodama \& Ishibashi \cite{Kodama:2003jz}
and has been adapted to our framework.

The formalism presented in this chapter has been implemented in the {\textsc{HD-Lean}}
code \cite{Zilhao:2010sr,Witek:2010xi}, an extension to higher dimensional spacetimes
of the original {\textsc{Lean}} code by Sperhake \cite{Sperhake:2006cy} developed for the 
modelling of $4$-dimensional spacetimes.

Our numerical studies, 
including thorough checks of the implementation as well as head-on collision 
of BHs in $D=5$ and $D=6$ spacetime dimensions,
will be discussed in the following Chapter~\ref{chapter:HigherDimR}.


\chapter{Black hole collisions in higher dimensional spacetimes -- Results}
\label{chapter:HigherDimR}
\section{Introduction}\label{sec:HDFIntro}
In high-energy physics, BHs are a central piece of the gauge/gravity
duality \cite{Maldacena:1997re,Witten:1998qj}, and are the generic outcome of
particle collisions at center-of-mass energies above the Planck scale
\cite{Choptuik:2009ww}.  In this regime the particular nature of the
particles' structure should become irrelevant, as indicated by Thorne's
``hoop'' conjecture \cite{Thorne:1972ji}, and ``no-hair theorem''-type 
arguments. 
These arguments have been supported by numerical studies \cite{Choptuik:2009ww,East:2012mb} and
indicate that, in general, trans-Planckian
collisions of particles are well described by collisions of highly
boosted BHs. In this context, scenarios such as TeV-gravity are
especially interesting, as they lower the fundamental Planck scale to the level
at which BHs would be produced in cosmic ray interactions with the atmosphere and particle accelerators
\cite{Banks:1999gd,Giddings:2001bu,Dimopoulos:2001hw,Ahn:2002mj,
Ahn:2003qn,Feng:2001ib, Cardoso:2004zi,
Cavaglia:2002si,Kanti:2004nr,Solodukhin:2002ui}.  Thus, high-energy
BH collisions could be used to look for signatures of extra dimensions
and BH production in ground-based experiments in the forthcoming years.
Note, however, that so far no BH signatures have been observed at the LHC,
working now at $7-8TeV$~\cite{Chatrchyan:2012taa,Aad:2012ic,Chatrchyan:2013xva}.

The above arguments illustrate the necessity to understand accurately
dynamical BH spacetimes, and their potential across a wide variety of
fields. Because the full system of Einstein equations needs to be carefully understood, 
this is a monumental task, and typically requires numerical methods.
With these fundamental issues as motivation, long-term efforts to
understand dynamical BHs in generic spacetimes have been initiated
\cite{Yoshino:2009xp,Zilhao:2010sr,Witek:2010xi,Sorkin:2009bc,Sorkin:2009wh,
Dennison:2010wd,
Yoshino:2011zz,Yoshino:2011zza,Cardoso:2012qm},
ranging from the inspiral of BH binaries \cite{Pretorius:2007nq,Hinder:2010vn,Centrella:2010mx},
high-energy collisions of BHs in four
\cite{Sperhake:2008ga,Shibata:2008rq,Sperhake:2009jz} and 
BH collisions in higher dimensional spacetimes
\cite{Zilhao:2010sr,Witek:2010xi,Witek:2010az,Okawa:2011fv}, 
stability studies in higher dimensions
\cite{Shibata:2009ad, Shibata:2010wz,Lehner:2010pn,Lehner:2011wc} and BH evolutions
in non-asymptotically flat spacetimes \cite{Bantilan:2012vu,Chesler:2010bi,
Witek:2010qc,Zilhao:2012bb}.

In the present chapter we focus mainly on evolutions of low energy head-on collisions in 
$D=5$ and $D=6$ dimensions. 
We have performed numerical simulations with the {\textsc{HD-Lean}} code, which implements 
the framework and formalism discussed Chapter~\ref{chapter:HigherDimF}.
The code is an extension of the original {\textsc{Lean}}
code \cite{Sperhake:2006cy,Sperhake:2007gu}, which is based on the \textsc{Cactus}
computational toolkit \cite{cactus} and uses the \textsc{Carpet}
mesh refinement package \cite{Schnetter:2003rb, carpet},
the apparent horizon finder \textsc{AHFinderDirect}
\cite{Thornburg:1995cp,Thornburg:2003sf} and an extension to $D$-dimensional spacetimes
of the {\textsc{TwoPunctures}} 
spectral initial data solver \cite{Ansorg:2004ds,Zilhao:2011yc}.

This chapter is organized as follows:
In Sec.~\ref{sec:HDRCodeTest} we perform several tests in order to verify the implementation.
In particular, we present simulations of single BHs in $D=5$ and $D=6$ and compare numerical versus analytic
data in case of geodesic slicing of the spacetime. Additionally, we have evolved head-on collisions
of non-boosted BHs in $D=4$ dimensions in order to verify the employed KI wave extraction. 
This setup allows for a benchmark test of the KI formalism against the well established 
Newman-Penrose formalism \cite{Newman:1961qr}
as well as wave extraction based on the Regge-Wheeler-Zerilli 
perturbation theory \cite{Regge:1957td,Zerilli:1970se} in $4$-dimensional configurations.
In Secs.~\ref{sec:HDRcoll5D} and~\ref{sec:HDRcoll6D} we present head-on collisions of
non-boosted, equal-mass BHs in $D=5$ and $D=6$ dimensions.
We discuss the gravitational wave emission as well as the energy outflow radiated throughout the 
collision.
This task becomes more challenging with increasing dimension and
the fact that perturbations decrease with $\sim\tfrac{1}{r^{D-3}}$
has far reaching implications for numerical simulations:
As we will see, much finer meshes are necessary to properly resolve
the BH region as well as the wave extraction zone.
Sec.~\ref{sec:HDRum5D} focuses on head-on collision of (non-boosted) unequal-mass BHs
in $D=5$ dimensions.
We analyse the emission of gravitational radiation, its energy and linear momentum content and 
provide a comparison with point-particle calculations \cite{Berti:2010gx}.

The results presented in this Chapter are based on the publications 
\cite{Witek:2010xi,Witek:2010az,Zilhao:2010sr,Zilhao:2011yc}
as well as on as yet unpublished work in progress
\cite{Hilditch2012,HDFcoll2012}.

Note, that length and time are measured in terms of the Schwarzschild radius $r_S$, which is related to the 
ADM mass $M$ of the system via Eq.~\eqref{eq:HDWEdefadm}. The total radiated energy is presented in terms
of the ADM mass.

\section{Code test}\label{sec:HDRCodeTest}
\subsection{Evolution of a single black hole}\label{ssec:HDRCCSingle}
In order to check our numerical framework and implementation we have performed a number of tests
evolving a single BH in $D=5$ and $D=6$ spacetime dimensions.
\paragraph{Geodesic slicing}
First, we have evolved the system by employing the geodesic slicing condition, i.e.,
the lapse function $\al$ and shift vector $\be^a$ have been set to
\begin{align}
\label{eq:HDRCCgeodsl}
\al = & 1
\,,\quad
\be^a = 0
\,,
\end{align}
with $a=0,\dots,(D-1)$.
Although any numerical simulation using geodesic slicing is doomed to fail 
because geodesics hit the singularity at finite proper time,
it allows for a comparison to analytic data. Thus, this strategy provides an excellent benchmark test
for the implementation
prior to the break-down of the simulation.

We start by studying a single Schwarzschild-Tangherlini BH in $D=5$ spacetime dimensions.
We express the $D$-dimensional Tangherlini
solution in a coordinate system of type \eqref{eq:HDFmetricansatz}
with geodesic slicing.
This coordinate system may be achieved by setting a congruence of in-falling radial time-like geodesics, each
geodesic starting from rest at radial coordinate $r_0$, with $r_0$ spanning the
interval $[r_S,+\infty[$, and using their proper time $\tau$ and $r_0$ as
coordinates (instead of the standard $(t,r)$ Schwarzschild-like coordinates). A
detailed construction of the Tangherlini solution in $D=5$ 
in these coordinates is given in Appendix~\ref{geoslice}. 
Then, the ($5$-dimensional) line element becomes
\begin{align}
\label{eq:HDRCCgeodesicmetric}
ds^2 = & -d\tau^2 + \frac{\left(r_0(R)^2+\left(\frac{r_S}{r_0(R)}\right)^2
       \tau^2\right)^2}{r_0(R)^2-\left(\frac{r_S}{r_0(R)}\right)^2\tau^2}
      \frac{dR^2}{R^2}+\left(r_0(R)^2-\left(\frac{r_S}{r_0(R)}\right)^2
      \tau^2\right)d\Omega_3 
\,,
\end{align}
where $r_0(R)$ is given by Eq.~\eqref{rzerotor}.

\begin{figure}[htpb!]
\begin{center} 
\begin{tabular}{cc}
\subfloat[$\tilde{\gamma}_{xx}$ along the $x$-axis]
{
 \includegraphics[clip=true,width=0.45\textwidth]{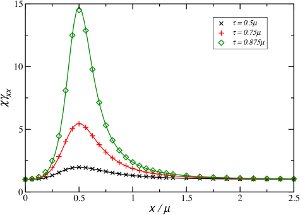}
} &
\subfloat[$\zeta/\chi=\lambda/y^2$ along the $y$-axis]
{
 \includegraphics[clip=true,width=0.45\textwidth]{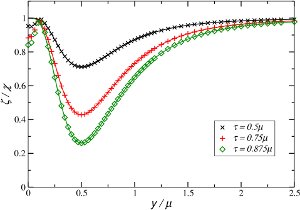}
}
\end{tabular}
\end{center}
\caption{\label{fig:HDRCCgammaxx}
Numerical values of $\tilde{\gamma}_{xx}$ and $\ze/\chi$ (points) versus the analytic solution (solid lines) 
plotted for various values of $\tau$, in the case of a single Tangherlini BH
in $D=5$ dimensions. 
}
\end{figure}

In Fig.~\ref{fig:HDRCCgammaxx} we plot the metric component
$\tilde{\gamma}_{xx}$ along the $x$ axis (left) and $\zeta/\chi$ (right), for
various values of $\tau$ using both the analytic solution and numerical
data. The agreement is excellent for $\tilde{\gamma}_{xx}$ and good for
$\zeta/\chi$. The latter shows some deviations very close to the puncture, but
we believe that it is not a problem for two reasons: \textit{(i)}~the agreement
improves for higher resolution; \textit{(ii)}~the mismatch does not propagate
outside of the horizon.

It is easy to interpret the behaviour observed for $\tilde{\gamma}_{xx}$. The
geodesic that starts from $r=r_0$ (in Schwarzschild-like coordinates) hits the
physical singularity of the Tangherlini solution within proper time
$\tau=r_0^2/r_S$. Moreover, this happens at
\begin{align}
\label{eq:HDRCChitssing}
R = & \frac{r_S}{2}\frac{1}{\sqrt{\tau/r_S}\pm\sqrt{\tau/r_S-1}} 
\,.
\end{align}
The earliest time at which the slicing hits the singularity is $\tau=r_S$, which
happens at $R=r_S/2$. On the $x$-axis $R=x$ and indeed one sees in
Fig.~\ref{fig:HDRCCgammaxx} that $\tilde{\gamma}_{xx}$ diverges at $x=r_S/2$. The
divergence then extends to both larger and smaller values of $x$, as expected
from \eqref{eq:HDRCChitssing}.

A quick glance at the evolution equations~\eqref{eq:HDFgBSSNevol} and~\eqref{eq:HDFgBSSNcoupling}
indicates that $D=5$ may be a special case. 
In all these expressions there exist terms which manifestly vanish for $D=5$.  

\begin{figure}
\begin{center}
\includegraphics[clip=true,width=0.5\textwidth]{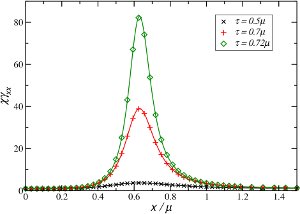}
\end{center}
\caption{\label{fig:HDRCCgeodesic_d6_hxx}
Numerical values versus the semi-analytic
expression of $\tilde{\gamma}_{xx}$ (cf.~Appendix~\ref{geoslice})
along the $x$-axis for the single Tangherlini BH in six dimensions.
}
\end{figure}

Therefore we next compare the numerical evolution of a single 
Tangherlini BH in $D=6$ spacetime dimensions with the analytic solution, using geodesic slicing. This
comparison has been more difficult than in case of $D=5$,
because the line element analogous to \eqref{eq:HDRCCgeodesicmetric} cannot be obtained
in a simple analytic form. In Appendix~\ref{geoslice} we demonstrate how a
semi-analytic solution can be obtained for the metric. In
Fig.~\ref{fig:HDRCCgeodesic_d6_hxx} we compare this expression with the 
numerical values at times $\tau=0.5~r_S$, $0.7~r_S$ and $0.72~r_S$. The
agreement is excellent and demonstrates that our implementation is correct.

\begin{figure}[htpb!]
\begin{center}
\begin{tabular}{cc}
\subfloat[Hamiltonian constraint]
{
\includegraphics[clip=true,width=0.45\textwidth]{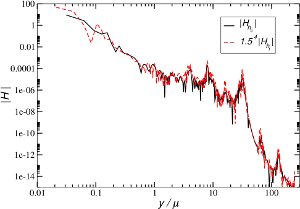}
} &
\subfloat[$y$-component of the  momentum constraint]
{
\includegraphics[clip=true,width=0.45\textwidth]{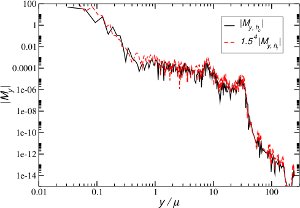}
}
\end{tabular}
\end{center}
\caption{
\label{fig:HDRCCconstraintsplot}
Constraints at time $t=28r_S$, for the evolution
of a single Tangherlini BH in $D=5$ dimensions.
}  
\end{figure}

\paragraph{Moving puncture gauge}
In the second part we evolve a single BH in $D=5$ and $D=6$ dimensions
employing the modified moving puncture gauge, Eqs.~\eqref{eq:HDFslicingQ} 
and~\eqref{eq:HDFgammadriver}, 
which is expected to yield long-term stable evolutions.
In particular, we perform a convergence analysis of the Hamiltonian 
and momentum constraint obtained from these simulations.
We choose Brill-Lindquist type initial data (cf. Sec.~\ref{ssec:HDFBLID}).
We employ the grid setup (cf.~Sec.~II E of Ref.~\cite{Sperhake:2006cy})
\begin{equation}
  \left\{(512,~256,~128,~64,~32,~16,~8,~4,~2) \times (),~h\right\} \, ,\nonumber
\end{equation}
in units of $r_S$ with resolutions $h_=r_S/32$ and $h_h=r_S/48$.  
In Fig.~\ref{fig:HDRCCconstraintsplot} we show the Hamiltonian constraint and the 
$y$-component of the momentum constraint 
resulting from the evolution of a single puncture in $D=5$
at evolution time $t=28~r_S$. 
By this time there are hardly any more gauge dynamics going on. 
We have re-scaled the high resolution run by $Q_4=1.5^4$, the 
factor anticipated for fourth order convergence \cite{Alcubierre:2008}.
One can see that there is some
noise, but the overall convergence is acceptable. For the Hamiltonian constraint
the convergence is essentially 4th order and for the momentum constraint it
decreases slightly towards 2nd or 3rd order in patches. From experience in 
$3+1$-dimensional numerical relativity this is perfectly acceptable, especially given
the fact that prolongation in time is second-order accurate.
%

Next, we have evolved a single puncture in $D=6$ spacetime dimensions.
Whereas we have been able to obtain long-term stable simulations
of single BHs in $D=5$ by modifying the
moving puncture gauge conditions in a very straight-forward manner, 
the $D=6$ case appeared to be much more difficult.
In fact, we could only accomplish this task by extensively studying various choices
of the gauge parameter $(\mu_L,\mu_\la,\xi_{\Ga},\xi_{\la})$ in 
Eqs.~\eqref{eq:HDFslicingQ} and~\eqref{eq:HDFgammadriver}, 
involving a large number of numerical experiments.

\begin{figure}[htpb!]
\begin{center}
\begin{tabular}{cc}
\subfloat[Hamiltonian constraint]{
 \includegraphics[clip=true,width=0.45\textwidth]{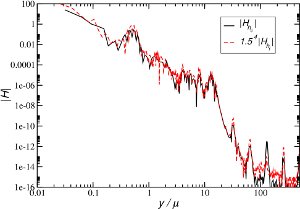}
} &
\subfloat[$y$-component of the momentum constraint]{
 \includegraphics[clip=true,width=0.45\textwidth]{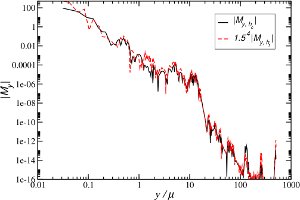}
}
\end{tabular}
\end{center}
\caption{\label{fig:HDRCCconstraintsd6}
Constraints at time $t=8r_S$, for the evolution
of a single Tangherlini BH in $D=6$ dimensions.
}  
\end{figure}

In Fig.~\ref{fig:HDRCCconstraintsd6} we show the Hamiltonian and the $y$-component of the
momentum constraint along the $y$-axis
obtained for a single puncture in $D=6$ at $t=8~r_S$. 
As for $D=5$, the high
resolution result is amplified by a factor $1.5^4$ expected for fourth order
convergence \cite{Alcubierre:2008}. While the convergence appears to be closer
to second order in some patches of the momentum constraint, the results are
clearly compatible with the numerical discretization.

The tests presented in this section demonstrate that our implementation of the higher dimensional framework
is correct and the code exhibits convergent results.

%

\subsection{Benchmark test - head-on collisions from rest in $D=4$}\label{ssec:HDRCCWE4D}
In order to test our implementation of the KI formalism, discussed in Sec.~\ref{sec:HDFWaveExtr},
we have simulated head-on collision
of an equal-mass, non-spinning BH binary initially at rest in $D=4$ spacetime dimensions.
Head-on collisions in $4$-dimensional spacetimes have been
studied extensively in the literature and provide valuable
opportunities to calibrate the wave extraction formalism.
In particular, we compare our results derived from the KI formalism
against those obtained from both,
the Regge-Wheeler-Zerilli wave extraction and the Newman-Penrose
framework; see, e.g., Refs.~\cite{Sperhake:2005uf,Sperhake:2006cy,
Lousto:2005xu,Reisswig:2010cd,Zenginoglu:2009ey} 
for corresponding literature studies.
The parameters used in the simulations presented in this section are shown in
Table \ref{tab:HDRWE4Dsetup}. 

In order to perform these tests, we need to relate our master
function $\Phi$ of Sec.~\ref{master} to the variables used in
traditional four dimensional studies. Specifically, a straightforward
calculation shows that the Zerilli wavefunction $\bar\Phi$
adopted in Ref.~\cite{Sperhake:2005uf} for $l=2$ multipoles
and the outgoing Weyl scalar $\Psi_4$ used in \cite{Sperhake:2006cy} 
can be expressed in terms of $\Phi$ according to
\begin{eqnarray}
\bar\Phi &=&6\Phi\,,\label{rel1}\\
r \Psi_4&=&\sqrt{6}\Phi_{,tt}\label{rel2}\,.
\end{eqnarray}
Note that the imaginary part of $\Psi_4$ vanishes in the case of a head on collision, due to symmetry.
The resolution is $h=r_S/96$ for all results reported in this section except for
the convergence study in Sec.~\ref{raden} which also uses the lower resolutions
$h_c=r_S/80$ and $h_m=r_S/88$.\footnote{In order to ensure that our fundamental
  unit is of physical dimension length for all values of spacetime dimension
  $D$, we believe it convenient to express our results in units of the radius
  $r_S$ (given by $r_S^{D-3} \equiv r^{D-3}_{S,1}+r^{D-3}_{S,2}$) of the
  ``total'' event horizon as opposed to the total BH mass $M$ commonly used in
  four dimensional numerical relativity. In $D=4$, of course, $r_S=2M$.}
Gravitational waves have been extracted at three different coordinate
  radii $R$ (cf. Eq.~\eqref{eq:HDWEtransfr}), which we denote by $R_{\rm ex} =
  30\,r_S\,,40\,r_S\,,50\,r_S$.

\begin{table*}[htpb!]
\begin{center}
\begin{tabular}{ccccc}
\hline
Run     & Grid Setup & $d/r_S$ & $L/r_S$ \\ \hline 
HD4$_c$ & $\{(128,64,32,16,8)\times(1,0.5,0.25),~h=r_S/80\}$ & $5.257$ & $7.154$ \\ 
HD4$_m$ & $\{(128,64,32,16,8)\times(1,0.5,0.25),~h=r_S/88\}$ & $5.257$ & $7.154$ \\ 
HD4$_f$ & $\{(128,64,32,16,8)\times(1,0.5,0.25),~h=r_S/96\}$ & $5.257$ & $7.154$ \\ 
\hline
\end{tabular}
\end{center}
\caption{\label{tab:HDRWE4Dsetup}
  Grid structure and initial parameters of the head-on collisions
  starting from rest in $D=4$.
  The grid setup is given in terms of the ``radii'' of the individual refinement levels, in units
  of $r_S$, as well as the resolution near
  the punctures $h$ (see Sec.~II E in \cite{Sperhake:2006cy} for details).
  $d$ is the initial coordinate  separation of the two punctures
  and $L$ denotes the proper initial separation.
}
\end{table*}
%
\paragraph{Tests on the numerical coordinates} 
The procedure described in Section \ref{sec:HDFWaveExtr} assumes that
the numerical spacetime consists of a small deviation from the
Schwarzschild-Tangherlini metric. In order to ensure that the
gravitational waves are extracted in
an appropriate coordinate system we perform a number of checks.
\begin{figure}[htpb!]
\begin{center}
\begin{tabular}{cc}
\includegraphics[width=0.5\textwidth]{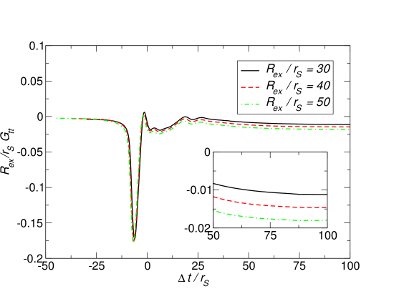} &
\includegraphics[width=0.5\textwidth]{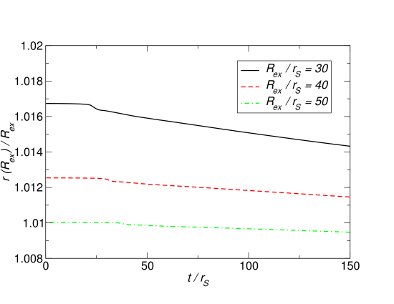}
\end{tabular}
\end{center}
\caption{\label{fig:HDRWE4Dcoordinates} 
Left panel: $\mathcal{G}_{tt}$ calculated from Eq.~\eqref{eq:HDWEtestgtt}
for $D=4$, at different extraction radii. This quantity has been shifted
in time to account for the different extraction radii and re-scaled
by the corresponding $R_{\rm ex}$. The late time behavior is shown in
the inset.
Right panel: time evolution of areal radius (cf. \eqref{eq:HDWEarealradius})  
re-scaled by the (coordinate) extraction radii
$R_{\rm ex} = 30r_S$ (black solid line), $R_{\rm ex} = 40r_S$ (red dashed line) and 
$R_{\rm ex} = 50r_S$ (green dashed-dotted line).
}
\end{figure}
First, we test the relations~\eqref{eq:HDWEtestgtt},~\eqref{eq:HDWEtestgtR}
and~\eqref{eq:HDWEtestgRR}. In Fig.~\ref{fig:HDRWE4Dcoordinates} we show
$\mathcal{G}_{tt}$, i.e., the difference between
the numerically calculated component $g_{tt}$, averaged over the extraction sphere
and the corresponding component of the assumed background metric.
Here we evaluate the background metric by assuming, as a first approximation,
that the Schwarzschild radius of the BH is $r_S=r_{S,1}+r_{S,2}$.

The deviation of the full 4-metric from the
Schwarzschild-Tangherlini background decreases as the extraction radius
increases. Indeed, a straightforward calculation shows that a deviation $\delta r_S$ of
the Schwarzschild radius from the background value leads to
$\mathcal{G}_{tt}\sim \delta r_S^{D-3}/r^{D-3}$,
i.e., $\mathcal{G}_{tt}\sim\delta r_S / r$ for $D=4$.
In the left panel of Fig.~\ref{fig:HDRWE4Dcoordinates} 
we therefore show the deviation $\mathcal{G}_{tt}$ re-scaled by $r$. We further apply a time shift
to account for the different propagation time of the wave to reach the extraction radii. As shown
in the figure, the deviation from the Schwarzschild line element {\it is} small and decreases
$\sim 1/r$ in accordance with our expectation. We also note that a deviation $\delta r_S$ represents a monopole
perturbation of the background which decouples from the quadrupole
wave signal at perturbative order, so that its impact on our results
is further reduced.

In summary, we can give an uncertainty estimate for the approximation $r_S = r_{S,1} + r_{S,2}$ for the Schwarzschild radius of the final BH,
which ignores the energy loss through gravitational radiation. As demonstrated by the left panel of
Fig.~\ref{fig:HDRWE4Dcoordinates}, at late times $|(R_{\rm ex}/r_S) \,{\cal G}_{tt}|\sim 0.01$, and, since
$r\simeq R_{\rm ex}$ (as we discuss below), we obtain the upper bound
\begin{equation}
\frac{\delta r_S}{r_S}\lesssim\frac{r}{r_S}{\cal G}_{tt}\sim 0.01\,.
\end{equation}
This crude analysis sets an upper bound of $\sim 1\%$ on the fraction of the center-of-mass energy radiated as gravitational waves.
We further note that the close agreement between $g_{tt}$ and its Tangherlini counterpart 
implies that the time coordinate employed in the numerical simulation and the Tangherlini coordinate time coincide. 
By analysing $\mathcal{G}_{tr}$ and $\mathcal{G}_{rr}$ in the same
manner, we find that relations~\eqref{eq:HDWEtestgtt}-\eqref{eq:HDWEtestgRR}
are satisfied with an accuracy of one part in $10^2$ throughout the evolution,
and one part in $10^3$ at late times, when the spacetime consists of a single distorted black hole.

In practice, gravitational waves are extracted on spherical shells
of constant coordinate radius. The significance of the areal
radius associated with such a coordinate sphere in the context
of extrapolation of GW signals has been studied
in detail in Ref.~\cite{Boyle:2009vi}. For our purposes, the most
important question is to what extent gauge effects change the
areal radius~\eqref{eq:HDWEarealradius} of our extraction spheres.
For this purpose, we show its time evolution in the right panel of Fig.~\ref{fig:HDRWE4Dcoordinates}
for different values of $R_{\rm ex}$. The reassuring result is that the areal radius
exceeds its coordinate counterpart by about $1~\%$ at $R_{\rm ex}=50~r_S$
and remains nearly constant in time. 

\begin{figure}[htpb!]
\begin{center}
\begin{tabular}{cc}
\includegraphics[width=0.5\textwidth]{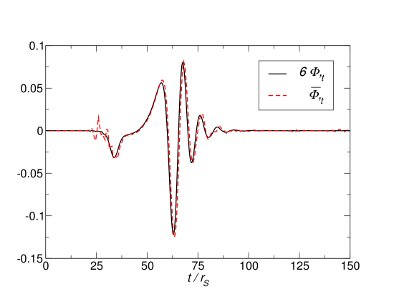} &
\includegraphics[width=0.5\textwidth]{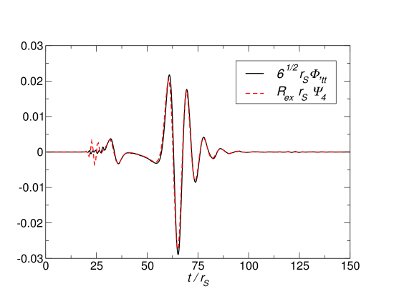}
\end{tabular}
\end{center}
\caption{\label{fig:HDRWE4Dcomparison1}  
  Left panel: Time derivatives of the $l=2$ modes of the KI
  function $\Phi$ (black solid line), and of the Zerilli function
  $\bar\Phi$ (red dashed line) extracted for model HD4$_f$
  at $R_{\rm ex}=30r_S$.
  The KI function has been re-scaled by a constant factor (cf. Eq.~\eqref{rel1}) 
  which accounts for the different normalizations of both formulations. 
  Right panel: comparison of the second time derivative $\Phi_{,tt}$
  with the outgoing Newman-Penrose scalar $\Psi_4$ for the same model. 
  The KI wavefunction has been re-scaled 
  according to Eq.~\eqref{rel2}.
}
\end{figure}

\paragraph{Waveforms}
As a benchmark for our wave extraction, we compare our results
obtained with independent wave extraction tools: 
(i) the explicitly four dimensional Zerilli formalism and 
(ii) the Newman-Penrose scalars.
For this purpose we have evolved model HD$4_f$ and extracted the
Zerilli function according to the procedure described in
\cite{Sperhake:2005uf}
(see also Eqs.~\eqref{eq:HDWEcompPF1}-\eqref{eq:HDWEcompPF4})
and the Newman Penrose scalar $\Psi_4$ as summarized in
\cite{Sperhake:2006cy}. These are compared
with the KI wave function $\Phi_{,t}$ and its time derivative
$\Phi_{,tt}$ in Fig.~\ref{fig:HDRWE4Dcomparison1}. Except for a small
amount of high frequency noise in the junk radiation at $t\approx 25 r_S$,
we observe excellent agreement between the different extraction methods.

\begin{figure}[htpb!]
\begin{center}
\begin{tabular}{cc}
\includegraphics[width=0.55\textwidth]{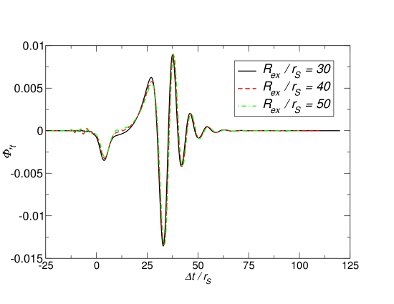}
\end{tabular}
\end{center}
\caption{\label{fig:HDRWE4DPhir}  The $l=2$ component of the
  KI wave function $\Phi_{,t}$ extracted
  at the radii $R_{\rm ex} = 30r_S$ (black solid line), $R_{\rm ex} = 40r_S$
  (red dashed line) and $R_{\rm ex} = 50r_S$ (green dashed-dotted line). They
  have been shifted in time by the corresponding $R_{\rm ex}$.}
\end{figure}

Next, we consider the dependence of the wave signal on the
extraction radius. 
In Fig.~\ref{fig:HDRWE4DPhir} we show the $l=2$ component of $\Phi_{,t}$
extracted at three different radii and shifted in time by $R_{\rm ex}$.
As is apparent from the figure, the wave function shows little variation
with $R_{\rm ex}$ at large distances, in agreement with expectations.

A further test of the wave signal arises from its late-time behaviour which is dominated by the BH
ringdown \cite{Berti:2009kk}, an exponentially damped sinusoid of the 
form $e^{-{\rm i}\omega t}$, with $\omega$ being a characteristic frequency 
called quasinormal mode (QNM) frequency. Using well-known methods \cite{Berti:2005ys,Berti:2007dg,Berti:2009kk}, 
we estimate this frequency to be $r_S\,\omega\sim 0.746\pm0.002-{\rm i}\,(0.176\pm0.002)$.  
This can be compared with theoretical predictions from a linearized approach,
yielding $r_S\,\omega=0.747344-{\rm i}\,0.177925$ \cite{Berti:2009kk}.

\begin{figure}[htpb!]
\begin{center}
\includegraphics[width=0.55\textwidth]{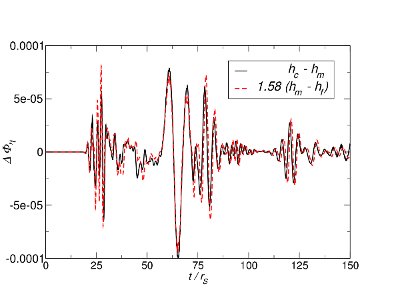}
\end{center}
\caption{\label{fig:HDRWE4Dconvergence}  
  Convergence analysis of the $l=2$ component of $\Phi_{,t}$ extracted at 
  $R_{\rm ex} = 30\,r_S$. We plot the differences between the low and medium resolution
  (black solid line) and medium and high resolution (red dashed line)
  run. The latter is re-scaled by the factor $Q_4=1.58$ expected for
  $4^{th}$ order convergence \cite{Witek:2010qc}. 
}
\end{figure}

Finally, we consider the numerical convergence of our results.
In Fig.~\ref{fig:HDRWE4Dconvergence}, we plot the differences obtained
for $\Phi_{,t}$ extracted at $R_{\rm ex} = 30\,r_S$, using the different 
resolutions of the three models HD$4$ listed in Table~\ref{tab:HDRWE4Dsetup}.
The differences thus obtained are consistent with $4^{th}$ order convergence. 
This implies a discretization error in the $l=2$ component of $\Phi_{,t}$ 
of about $4\%$ for the grid resolutions used in this work.


\begin{figure}[htpb!]
\begin{center}
\begin{tabular}{cc}
\includegraphics[width=0.5\textwidth]{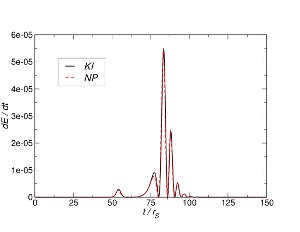} &
\includegraphics[width=0.5\textwidth]{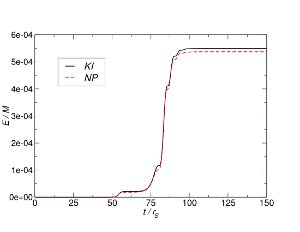} 
\end{tabular}
\end{center}
\caption{\label{fig:HDRWE4Denergyflux}  Energy flux (left panel)
  and radiated energy (right panel) for the $l=2$ mode extracted at $R_{\rm ex} = 50r_S$
  from the KI wave function $\Phi_{,t}$  (black solid curve)
  and the Newman Penrose scalar $\Psi_4$ (red dashed curve).
}
\end{figure}

\paragraph{Radiated energy}\label{raden}
Once the KI function $\Phi_{,t}$ is known, the energy flux
can be computed from Eq.~\eqref{eq:HDWEenergyflux}. For comparison, we have 
also determined the flux from the outgoing Newman Penrose scalar $\Psi_4$ according to Eq.~(22) 
in Ref.~\cite{Witek:2010qc}. The flux and energy radiated in 
the $l=2$ multipole, obtained with the two methods at $R_{\rm ex}=50\,r_S$
is shown in Fig.~\ref{fig:HDRWE4Denergyflux} and demonstrates agreement within the numerical uncertainties
of about $4~\%$ for either result.

We obtain an integrated energy of $5.5\times10^{-4}~M$
and $5.3\times10^{-4}~M$, respectively, for the gravitational wave energy
radiated in $l=2$, where $M$ denotes the center-of-mass energy.

The energy in the $l=2$ mode is known to contain
more than $99\%$ of the total radiated energy \cite{Sperhake:2006cy}.
Our analysis is compatible with this finding; while the energy in the $l=3$
mode is zero by symmetry, our result for the energy in the
$l=4$ mode obtained from the KI master function is three orders of magnitude smaller than that of the $l=2$
contribution.

\section{Head-on collisions from rest in $D=5$}\label{sec:HDRcoll5D}
Having tested the wave extraction formalism in $D=4$ dimensions in the previous section,
we now turn our attention to the results obtained for head-on collisions of BHs in 
$5$-dimensional spacetimes. As before, we consider nonspinning BH binaries initially at rest with 
coordinate separation $d/r_S$.
Note, that in five spacetime dimensions the Schwarzschild radius is 
related to the ADM mass $M$ via Eq.~\eqref{eq:HDWEdefadm},
\begin{align}
r_S^2=\frac{8M}{3\pi}\,.\label{HDRcoll5Ddefadm}
\end{align}
We therefore define the ``total'' Schwarzschild radius
$r_S$ such that $r_S^2=r_{S,1}^2 + r_{S,2}^2$. By using this definition,
$r_S$ has the physical dimension of a length and provides a suitable
unit for measuring both, results and grid setup.

\begin{table*}[htpb!]
\begin{center}
\begin{tabular}{ccccc}
\hline
Run &   Grid Setup & $d/r_S$ & $L/r_S$ \\ \hline 
HD5a &  $\{(256,128,64,32,16,8,4)\times(0.5,0.25),~h=r_S/84\}$ & $1.57$ & $1.42$ & \\ 
HD5b &  $\{(256,128,64,32,16,8,4)\times(0.5,0.25),~h=r_S/84\}$ & $1.99$ & $1.87$ \\ 
HD5c &  $\{(256,128,64,32,16,8,4)\times(1,0.5),~h=r_S/84\}$ & $2.51$ & $2.41$ \\ 
HD5d &  $\{(256,128,64,32,16,8,4)\times(1,0.5),~h=r_S/84\}$ & $3.17$ & $3.09$ \\ 
HD5e$_c$ & $\{(256,128,64,32,16,8)\times(2,1,0.5),~h=r_S/60\}$ & $6.37$ & $6.33$ \\ 
HD5e$_m$ & $\{(256,128,64,32,16,8)\times(2,1,0.5),~h=r_S/72\}$ & $6.37$ & $6.33$ \\ 
HD5e$_f$ & $\{(256,128,64,32,16,8)\times(2,1,0.5),~h=r_S/84\}$ & $6.37$ & $6.33$ \\ 
HD5f &  $\{(256,128,64,32,16,8)\times(2,1,0.5),~h=r_S/84\}$ & $10.37$ & $10.35$ \\ 
\hline
\end{tabular}
\end{center}
\caption{\label{tab:HDRcoll5Dsetup}
  Grid structure and initial parameters of the head-on collisions
  starting from rest in $D=5$.
  The grid setup is given in terms of the ``radii'' of the individual refinement levels, in units
  of $r_S$, as well as the resolution near
  the punctures $h$ (see Sec.~II E in \cite{Sperhake:2006cy} for details).
  $d/r_S$ is the initial coordinate  separation of the two punctures
  and $L/r_S$ denotes the proper initial separation.
}
\end{table*}

As summarized in Table \ref{tab:HDRcoll5Dsetup}, we consider a
sequence of BH binaries with initial coordinate
separation ranging from $d = 3.17\,r_{S}$ to $d = 10.37\,r_{S}$.
The table further lists the proper separation $L$ along the line of sight
between the BHs and the grid configurations used for the individual
simulations.

\subsection{Tests on the numerical coordinates}\label{sec:D5coordinatecheck}
In order to verify the assumptions underlying our formalism,
we have analysed the coordinate system in analogy to Sec.~\ref{ssec:HDRCCWE4D}. First, we have evaluated the
averaged areal radius on extraction spheres of constant coordinate
radius. 

\begin{figure}[htpb!]
\begin{center}
\begin{tabular}{cc}
\includegraphics[width=0.5\textwidth]{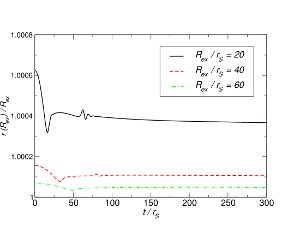} &
\includegraphics[width=0.5\textwidth]{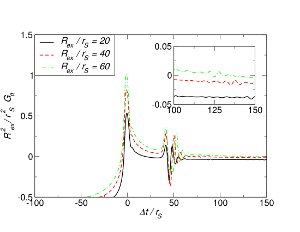} 
\end{tabular}
\end{center}
\caption{\label{fig:HDRcoll5DD5coordinates} 
Left panel:
     Time evolution of the areal radius $r$ in units of the extraction
     radius averaged over coordinate spheres at
     $R_{\rm ex}=20~r_S$ (black solid), $40~r_S$ (red dashed)
     and $60~r_S$ (green dash-dotted curve).
Right panel:
     Deviation of the metric component $R_{\rm ex}^2/r_S^2\mathcal{G}_{tt}$
     calculated from Eq.~\eqref{eq:HDWEtestgtt} at the same
     extraction radii and shifted in time to account for differences
     in the propagation time of the wave signal.
}
\end{figure}

The result shown in the left panel of Fig.~\ref{fig:HDRcoll5DD5coordinates} demonstrates that the coordinate and areal
radius agree within about 1 part in $10^4$ for $R_{\rm ex}\ge40~r_S$.
The Tangherlini coordinate $r$ equals by construction the areal
radius and our approximation of setting $r\approx R_{\rm ex}$ in the
wave extraction zone is satisfied with high precision.

Second, we evaluate the deviation of the metric components according to
Eqs.~\eqref{eq:HDWEtestgtt}-\eqref{eq:HDWEtestgRR}. From the discussion in Sec.~\ref{ssec:HDRCCWE4D} we
expect $\mathcal{G}_{tt}\sim r^2 /r_S^2$ in $D=5$. Our results in the right panel of Fig.~\ref{fig:HDRcoll5DD5coordinates} confirm 
this expectation and demonstrate that our spacetime is indeed perturbatively
close to that of a Tangherlini metric at sufficient distances from
the BHs; deviations in $\mathcal{G}_{tt}$ are well below 1
part in $10^3$ at $R_{\rm ex}=60~r_S$.
Furthermore, we can 
estimate the crudeness of the approximation $r^2_S = r^2_{S,1} + r^2_{S,2}$ for the Schwarzschild radius of the final BH:
as shown in the right panel of Fig.~\ref{fig:HDRcoll5DD5coordinates}, at late times
$|R_{ex}^2/r_S^2\mathcal{G}_{tt}|\sim 0.01$; 
this value gives an upper bound on the radiated energy.

For the third test, we recall that our higher dimensional implementation
does not employ the full isometry group of the $S^2$ sphere in
$D=5$ dimensions and axial symmetry manifests itself instead in the
conditions (\ref{eq:HDWEcheckax}) on the metric components and the scalar field.
We find these conditions to be satisfied within 1 part in $10^8$ and 1 part in $10^{16}$, respectively, in our numerical
simulations which thus represent axially symmetric configurations with high precision.

%

\subsection{Newtonian collision time}
\begin{figure}
\begin{center}
\begin{tabular}{cc}
\includegraphics[width=0.5\textwidth]{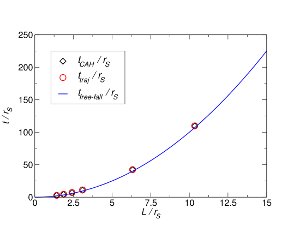} 
\end{tabular}
\end{center}
\caption{\label{fig:HDRcoll5DD5_NewtCollTime} 
 Estimates for the time it takes for two equal-mass BHs to
collide in $D=5$.  The first estimate is given by the time $t_{\text{CAH}}$
elapsed until a single common apparent horizon engulfs both BHs
(diamonds), the second estimate is obtained by using the trajectory of
the BHs, i.e., the time $t_{\text{traj}}$ at which their separation
has decreased below the Schwarzschild radius (circles). Finally, these
numerical results are compared against a simple Newtonian estimate,
given by Eq.~\eqref{eq:HDRcoll5D5dtime} (blue solid line).
}
\end{figure}

An estimate of the time at which the BHs ``collide'', can be obtained
by considering a Newtonian approximation to the kinematics of two point
particles in $D=5$. In the weak-field regime, Einstein's equations
reduce to ``Newton's law'' $a=-\nabla {\cal B}({\bf x})$, with $h_{00}=-2{\cal
B}({\bf  x})=r_S^{D-3}/2r^{D-3}$.
The Newtonian time it takes for two point-masses (with Schwarzschild
parameters $r_{S,1}$ and $r_{S,2}$) to collide from rest with initial
distance $L$ in $D$ dimensions is then given by
\begin{align}
\label{eq:HDRcoll5Dnewt-time}
\frac{t_{\text{free-fall}}}{r_S} = \frac{\mathcal{I}}{D-3}\left(\frac{L}{r_S}\right)^{\frac{D-1}{2}}\,,
\end{align}
where $r_S^{D-3} =r_{S,1}^{D-3} + r_{S,2}^{D-3}$ and
\begin{align}
\label{eq:HDRcoll5Dint}
\mathcal{I} = \int_0^1 \sqrt{\frac{z^{\frac{5-D}{D-3}}}{1-z}} dz
= \sqrt{\pi} \frac{\Gamma(\frac{1}{2} 
  + \frac{1}{D-3})}{\Gamma(1 + \frac{1}{D-3})}\,.
\end{align}
For $D=4$, one recovers the standard result
$t_{\text{free-fall}} = \pi \sqrt{L^3/r_S^3}r_S\,,$ whereas for $D=5$ we get
\begin{align}
t_{\text{free-fall}}=\left(L/r_S\right)^2r_S\,.\label{eq:HDRcoll5D5dtime}
\end{align}
In general relativity, BH trajectories
and merger times are intrinsically observer dependent quantities.
For our comparison with Newtonian estimates we have chosen relativistic
trajectories as viewed by observers adapted to the numerical coordinate
system. While the lack of fundamentally gauge invariant analogues in
general relativity prevents us from deriving rigorous conclusions,
we believe such a comparison to serve the intuitive interpretation
of results obtained within the moving puncture gauge.
Bearing in mind these caveats, we plot in Fig.~\ref{fig:HDRcoll5DD5_NewtCollTime} 
the analytical estimate of the Newtonian time of collision, together
with the numerically computed time of formation of a common apparent horizon.
Also shown in Fig.~\ref{fig:HDRcoll5DD5_NewtCollTime} is the time at
which the separation between the individual BH's puncture trajectory decreases below
the Schwarzschild parameter $r_S$. 
The remarkable agreement provides yet another example of how well
numerically successful gauge conditions appear to be adapted to the
BH kinematics. It is beyond the scope of this paper to
investigate whether this is coincidental or whether such agreement is
necessary or at least helpful for gauge conditions to ensure numerical
stability. Suffices it to say at this stage that
similar conclusions were reached by Anninos {\em et al.} \cite{Anninos:1994gp} and
Lovelace {\em et al.} \cite{Lovelace:2009dg} in similar four dimensional scenarios.

\subsection{Waveforms}\label{sec:D5waveforms}
%
\begin{figure}
\begin{center}
\begin{tabular}{cc}
\includegraphics[width=0.5\textwidth]{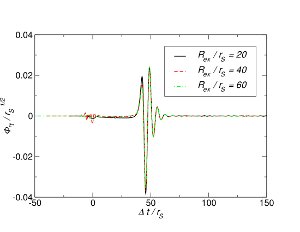} &
\includegraphics[width=0.5\textwidth]{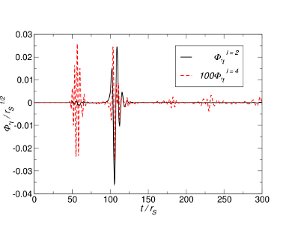} 
\end{tabular}
\end{center}
\caption{\label{fig:HDRcoll5DD5_Phir_hf84} 
Left panel: The $l=2$ component of the KI waveform for model HD5e$_f$  
extracted at radii $R_{ex} = 20, 40$ and $60~r_S$ and shifted in time by 
$R_{\rm ex}/r_S$. Right panel: The $l=2$ and $l=4$ mode of the KI function for the same 
simulation, extracted at $R_{ex}= 60~r_S$. For clarity, the $l=4$
component has been re-scaled by a factor of $100$.
}
\end{figure}
\begin{figure}
\begin{center}
\begin{tabular}{cc}
\includegraphics[width=0.5\textwidth]{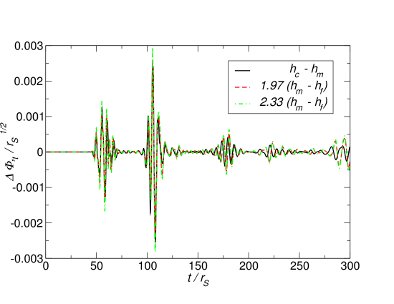} 
\end{tabular}
\end{center}
\caption{\label{fig:HDRcoll5DD5_convergence}  Convergence analysis
  of the $l=2$ component of the KI
  function generated by model HD5e  extracted at $R_{\rm ex}
  = 60r_{S}$.  The difference between the medium and high resolution
  waveforms has been amplified by the factors $Q_3=1.97$ (red dashed line)
  and $Q_4=2.33$ (green dashed-dotted line) indicating third and fourth
  order convergence. }
\end{figure}
We now discuss in detail the gravitational wave signal generated
by the head-on collision of two BHs in $D=5$ dimensions.
For this purpose, we plot in Fig.~\ref{fig:HDRcoll5DD5_Phir_hf84}
the $l=2$ multipole of the KI function $\Phi_{,t}$ for
model HD5e$_f$ obtained at different extraction radii.
Qualitatively, the signal looks similar to that
shown in the left panel of Fig.~\ref{fig:HDRWE4Dcomparison1} for $D=4$. A small spurious wavepulse
due to the initial data construction is visible at $\Delta t \approx 0$.
This so-called ``junk radiation'' increases in magnitude if the
simulation starts with smaller initial separation of the BHs.
We return to this issue further below, when we study the dependence of the
gravitational radiation on the initial BH separation.
The physical part of the waveform is dominated by the merger signal
around $\Delta t=50~r_S$, followed by the (exponentially damped) ringdown,
whereas the infall of the BHs before $\Delta t=40~r_S$ does not produce
a significant amount of gravitational waves.
Comparison of the waveforms extracted at different radii
demonstrates excellent agreement, in particular for those extracted at $R_{\rm ex}=40~r_S$ and
$60~r_S$. Extrapolation of the radiated energy
to infinite extraction radius yield a relative error of 5~\% at $R_{\rm ex}=60~r_S$,
indicating that such radii are adequate for the analysis presented in this work.

Due to symmetry, no gravitational waves are emitted in the $l=3$
multipole, so that $l=4$ represents the second strongest contribution
to the wave signal. As demonstrated in the right panel of
Fig.~\ref{fig:HDRWE4Dcomparison1}, however, its amplitude is two orders
of magnitude below that of the quadrupole.

A convergence analysis also using the lower resolution simulations
of models HD5e$_c$ and HD5e$_m$ is shown in Fig.~\ref{fig:HDRcoll5DD5_convergence}
and demonstrates overall convergence of third to fourth order, consistent with
the numerical implementation.
From this analysis we obtain a conservative estimate of
about $4\%$ for the discretization error in the waveform.

\begin{figure}[htpb!]
\begin{center}
\begin{tabular}{cc}
\includegraphics[width=0.5\textwidth]{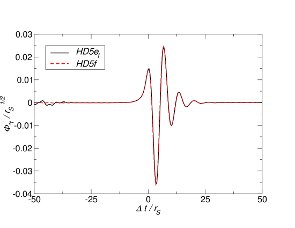} 
\end{tabular}
\end{center}
\caption{\label{fig:HDRcoll5DD5_Phit_models} 
The $l=2$ components of the KI function as generated by
a head-on collision of BHs with initial (coordinate) distance
$d=6.37\,r_S$ (black solid line) and $d=10.37\,r_S$ (red dashed line).
The wave functions have been shifted in time such that the formation of
a common apparent horizon corresponds to $\Delta t=0~r_S$ and taking into account
the time it takes for the waves to propagate up to the extraction radius
$R_{\rm ex} = 60\,r_{S}$.}
\end{figure}

In practice, numerical simulations will always start with a
finite separation of the two BHs. In order to assess how
accurately we are thus able to approximate an infall from infinity, we
have varied the initial separation for models HD5a to HD5f as
summarized in Table \ref{tab:HDRcoll5Dsetup}. 
For small $d$ we observe two
effects which make the physical interpretation of models HD5a$-$HD5c
difficult. First, the amplitude of the spurious initial
radiation increases and second, the shorter infall time causes an
overlap of this spurious radiation with the merger signal.
As demonstrated in Fig.~\ref{fig:HDRcoll5DD5_Phit_models} for models HD5e and HD5f,
however, we can safely neglect the spurious radiation as well
as the impact of a final initial separation, provided
we use a sufficiently large initial distance $d \gtrsim 6~r_S$
of the BH binary. Here, we
compare the radiation emitted during the head-on collision 
of BHs starting from rest with
initial separations $6.37~r_S$ and $10.37~r_S$.
The waveforms have been shifted in time by the extraction
radius $R_{\rm ex} = 60\,r_{S}$ and such that the formation of a common
apparent horizon occurs at $\Delta t=0$. The merger signal
starting around $\Delta t=0~r_S$ shows excellent agreement for the
two configurations and is not affected by the spurious signal
visible for HD5e at $\Delta t\approx -50~r_S$.

We conclude this discussion with two aspects of the post-merger
part of the gravitational radiation, the ringdown and the possibility
of GW tails. After formation of a common horizon,
the waveform is dominated by an exponentially damped
sinusoid, as the merged BH {\em rings down} into a stationary state.
By fitting our results with an exponentially damped sinusoid,
we obtain the characteristic frequency
\begin{align}
\label{eq:HDRcoll5Dqnm}
r_{S}\,\omega = & (0.955\pm 0.005) - i\,(0.255 \pm 0.005)
\,.
\end{align}
This value is in excellent agreement with perturbative calculations, which
predict a lowest quasinormal frequency $r_{S}\,
\omega=0.9477-i\,0.2561$ for $l=2$
\cite{Cardoso:2003qd,Yoshino:2005ps,Berti:2009kk}.

A well known feature in gravitational waveforms generated in
BH spacetimes with $D=4$ as well as $D>4$ are the so-called
{\it power-law tails} 
\cite{DeWitt:1960fc,Price:1971fb,Ching:1995tj, Cardoso:2003jf}.
In odd dimensional spacetimes an additional, different kind of late-time
power tails
arises, which does not depend on the presence of a BH. These
are due to a peculiar behavior of the wave-propagation
in {\it flat} odd dimensional spacetimes
because the Green's function has support inside the entire
light-cone \cite{Cardoso:2003jf}.
We have attempted to identify
such power-law tails in our signal at late times,
by subtracting a best-fit ringdown waveform. Unfortunately, we
cannot, at this stage, report
any evidence of such a power-law in our results, most likely because the
low amplitude tails are buried in numerical noise.

%
\subsection{Radiated energy}\label{sec:D5raden}
%
\begin{figure}
\begin{center}
\begin{tabular}{c}
\includegraphics[width=0.5\textwidth]{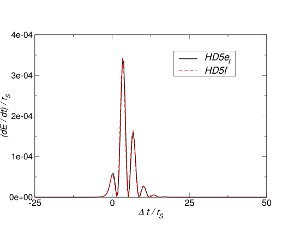} 
\end{tabular}
\end{center}
\caption{\label{fig:HDRcoll5DD5energy} 
 Energy flux in the $l=2$ component of the KI 
wave function   $\Phi_{,t}$, extracted at $R_{\rm ex} = 60\,r_{S}$, for models 
HD5e$_f$ (black solid line) and HD5f (red dashed line)
in Table~\ref{tab:HDRcoll5Dsetup}. The fluxes have been shifted in time by 
the extraction radius $R_{\rm ex} = 60\,r_{S}$ and the time $t_{\text{CAH}}$ 
at which the common apparent horizon forms.
}
\end{figure}
Comparison of Figs.~\ref{fig:HDRWE4DPhir} and \ref{fig:HDRcoll5DD5_Phit_models}
for the GW quadrupole in $D=4$ and $D=5$ shows a larger wave amplitude
in the five dimensional case and thus indicates that this case
may radiate more energy. We now investigate this question quantitatively
by calculating the energy flux from the KI master function via
Eq.~\eqref{eq:HDWEenergyflux}. The fluxes thus obtained for
the $l=2$ multipole of models HD5e$_f$ and HD5f
in Table~\ref{tab:HDRcoll5Dsetup}, extracted at $R_{\rm ex} = 60\,r_{S}$,
are shown in Fig.~\ref{fig:HDRcoll5DD5energy}.
As in the case of the KI master function in
Fig.~\ref{fig:HDRcoll5DD5_Phit_models}, we see no significant variation of the
flux for the two different initial separations.
The flux reaches a maximum value of $dE/dt\sim 3.4\times 10^{-4}\,r_S$,
and is then dominated by the ringdown flux. The energy flux from
the $l=4$ mode is typically four orders of magnitude smaller; this is consistent with the factor of 100 difference of the corresponding
wave multipoles observed in Fig.~\ref{fig:HDRcoll5DD5_Phir_hf84}, and the quadratic
dependence of the flux on the wave amplitude.

\begin{figure}
\begin{center}
\begin{tabular}{cc}
\includegraphics[width=0.5\textwidth]{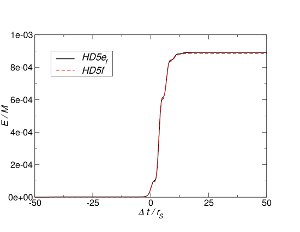} &
\includegraphics[width=0.5\textwidth]{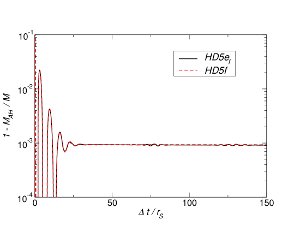}
\end{tabular}
\end{center}
\caption{\label{fig:HDRcoll5Dtotalenergy} 
Left panel: Fraction of the center-of-mass energy, $E_{rad}/M$, radiated
in the $l=2$ mode of the KI function shifted in time
such that the origin of the
time axis corresponds to the formation of a common apparent horizon.  %
Right panel: Fraction of the center-of-mass energy $1-M_{AH}/M$ radiated
during the collision, estimated using apparent horizon information.
The oscillations in this diagnostic quantity have a frequency comparable
to the $l=2$ quasinormal mode frequency.
}
\end{figure}
The total integrated energy emitted throughout the head-on collision
is presented in the left panel of Fig.~\ref{fig:HDRcoll5Dtotalenergy}. We find that a fraction of
$E_{rad}/M = (8.9\pm0.6)\times 10^{-4}$ of the center-of-mass energy
is emitted in the form of gravitational radiation. 
We have verified for these models that
the amount of energy contained in the spurious radiation is about
three orders of magnitude smaller than in the physical merger signal.

An independent estimate for the radiated energy can be obtained from the apparent
horizon area $A_4$ in the effective four dimensional spacetime
by using the spherical symmetry of the post-merger remnant BH. Energy balance then implies that the
energy $E$ radiated in the form of GWs is given by
\begin{align}
  \frac{E}{M} = 1-\frac{M_{AH}}{M} =
1-\frac{A_4}{4\pi r_S^2}\,,
\end{align}
where $M_{AH}$ is the apparent horizon mass. 
The estimate $E/M$ is shown in Fig.~\ref{fig:HDRcoll5Dtotalenergy} 
and reveals a behavior qualitatively similar to 
a damped sinusoid with constant offset. Indeed, by using
a least squares fit, we obtain a complex frequency
$r_{S}\,\omega\sim0.97-i\,0.29$, again similar to the fundamental $l=2$ quasinormal mode
frequency (see discussion around Eq.~\eqref{eq:HDRcoll5Dqnm}).
At late times, $1-M_{AH}/M$ asymptotes
to $1-M_{AH}/M\sim \left(9.3\pm 0.8\right)\times10^{-4}$
which agrees very well with the GW estimate, within the numerical uncertainties.

\section{Head-on collisions from rest in $D=6$}\label{sec:HDRcoll6D}
Here, I present evolutions of non-boosted, equal-mass BH binaries colliding head-on in $D=6$ spacetime dimensions.
This goal proved to be numerically more challenging than previous studies in $D=5$ 
and is subject to detailed investigations.
One reason might be the increasing fall-off with spacetime dimension which goes
as $\sim \tfrac{1}{r^{D-3}}$. Thus, not only do we need higher resolution grids to
properly resolve the BH region, but also gradients become steeper and might cause problems.
Another reason might be the formulation itself.
However, in Secs.~\ref{sec:HDFDimRed} and~\ref{sec:HDFevol}
I have shown that the employed formulation 
in the continuum limit
is strongly hyperbolic and therefore well-posed 
for generic spacetime dimension $D\ge5$ and most gauge parameters 
(cf. Sec.~\ref{sec:HDFHyperbolicity}).
Nevertheless, this is only a necessary but not sufficient condition for the numerical stability of the system.
In practice, I have only accomplished this task by extensively studying a wide range of parameter choices
$(\xi_{\Ga},\xi_{\la})$ of the $\Ga$-driver shift condition, Eq.~\eqref{eq:HDFgammadriver}.
The procedure has resulted in hundreds of numerical experiments.
Finally, the specification $\xi_{\Ga} = 0.25$ and $\xi_{\la} = 0.0$ has led to succesful, long-term stable 
simulations. 
Note, however, that there is no strong mathematical reason why this choice turns out to work better than others
and it seems to depend on the spacetime dimension in a non-trivial way.
Additionally, the particular setting of these parameters might depend sensitively on the 
particular configuration.
In this section, I present the thus obtained results.
The BHs, each with mass parameter $r^3_{S,i} = 0.5$,
start from rest with initial coordinate distance $d = 6.37~r_S$, $d=9.56~r_S$ or $d=12.74~r_S$.
The specifications of the simulations, such as the grid setup and resolution $h$ of the innermost
refinement level, the initial coordinate distance $d/r_S$ and the employed damping parameter
$\eta_{\be}$ in the $\Ga$-driver shift condition are summarized in 
Table~\ref{tab:HDR6Dsetup}.
I assume the (bare) total Schwarzschild radius $r_S^3 = r^3_{S,1} + r^3_{S,2} =1$ where I neglect radiation effects.
It is related to the ADM mass of the system via $r^3_S = \frac{3}{2\pi} M$ (cf. Eq.~\ref{eq:HDWEdefadm}).
\begin{table}[htpb!]
\begin{center}
\begin{tabular}{cccc}
\hline
Run      & Grid Setup & $d/r_S$ & $\eta_{\beta}$ \\ \hline 
HD6a$_c$ & $\{(256,128,80,48,16,8)\times(2,1,0.5),~h=r_S/84\}$  & $6.37$  & $1.75$ \\ 
HD6a$_m$ & $\{(256,128,80,48,16,8)\times(2,1,0.5),~h=r_S/88\}$  & $6.37$  & $1.75$ \\ 
HD6a$_f$ & $\{(256,128,80,48,16,8)\times(2,1,0.5),~h=r_S/92\}$  & $6.37$  & $1.75$ \\ 
HD6b     & $\{(256,128,80,48,12)\times(4,2,1,0.5),~h=r_S/92\}$  & $9.56$  & $1.75$ \\ 
HD6c     & $\{(256,128,80,48,16)\times(4,2,1,0.5),~h=r_S/92\}$  & $12.74$ & $1.50$ \\ 
\hline
\end{tabular}
\end{center}
\caption{\label{tab:HDR6Dsetup}
  Grid structure and initial parameters of the BH head-on collisions in $D=6$.
  The grid setup is given in terms of the ``radii'' of the individual 
  refinement levels, in units of $r_S$, as well as the resolution near
  the punctures $h$ (see Sec.~II E in \cite{Sperhake:2006cy} for details).
  $d/r_S$ is the initial coordinate  separation of the two punctures.
  $\eta_{\beta}$ is the damping parameter in the $\Ga$-driver shift condition, 
  Eq.~\eqref{eq:HDFgammadriver}.
}
\end{table}
%
\subsection{Waveforms}\label{ssec:HDR6Dwaveform}
\begin{figure}
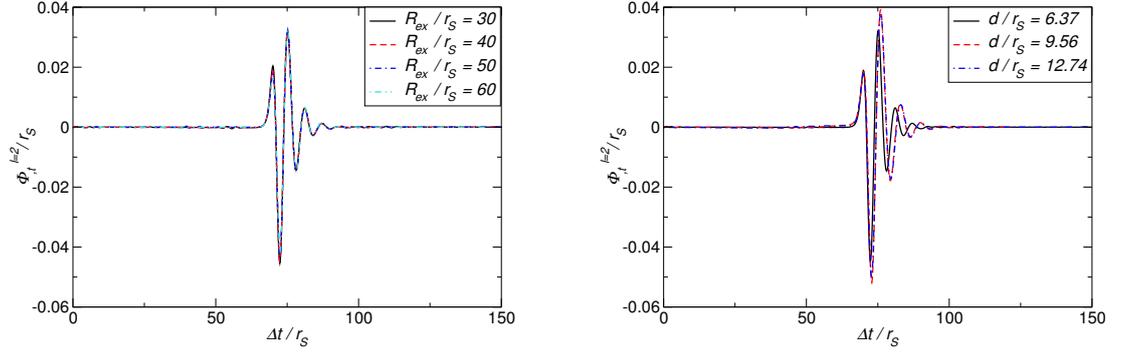

\begin{center}
\begin{tabular}{cc}
\includegraphics[width=0.5\textwidth]{HDR6Dfig1a} &
\includegraphics[width=0.5\textwidth]{HDR6Dfig1b} 
\end{tabular}
\end{center}
\caption{\label{fig:HDR6Dwaveforml2} 
Left panel:
$l=2$ multipole of the KI masterfunction $\Phi_{,t}$ for model HD6a$_f$ 
extracted at $R_{ex} = 30~r_S$ (black solid line), $R_{ex} = 40~r_S$ (red dashed line),
$R_{ex} = 50~r_S$ (blue dashed-dotted line) and $R_{ex} = 60~r_S$ (turquoise dashed-double-dotted line)
and shifted in time by $R_{ex}/r_S$.
Right panel:
$l=2$ multipole of the KI masterfunction $\Phi_{,t}$ for 
models HD6a$_f$ (black solid line), 
HD6b (red dashed line) and HD6c (blue dashed-dotted line).
The waveforms have been shifted in time by $R_{ex}/r_S$ and such that the first maxima coincide.
}
\end{figure}

First, I discuss the gravitational wave signal emitted throughout the head-on collision
of two BHs in $D=6$ dimensions.
In the left panel of Fig.~\ref{fig:HDR6Dwaveforml2}
I plot the $l=2$ multipole of the KI masterfunction $\Phi_{,t}$, Eq.~\eqref{eq:HDWEKIwavefunction},
for model HD6a$_f$, extracted at different radii $R_{ex} = 30,40,50,60~r_S$.
Qualitatively, the signal looks similar to those presented in the left panel of 
Figs.~\ref{fig:HDRWE4Dcomparison1} and~\ref{fig:HDRcoll5DD5_Phir_hf84} 
for $D=4$ and $D=5$ dimensions, respectively. 
A small spurious wavepulse, the so-called ``junk radiation'' which is
due to the initial data construction, would appear at $\Delta t \approx 0~r_S$.
However, the amplitude of the junk radiation is significantly less than in the previously studied cases
with comparable setups in $D=4$ and $D=5$ dimensions
and, in fact, not visible in Fig.~\ref{fig:HDR6Dwaveforml2}.
The physical part of the waveform is dominated by the merger signal
around $\Delta t = 75~r_S$, 
followed by the exponentially damped ringdown.
The comparison of the waveforms extracted at different radii demonstrates excellent agreement.
I will study the influence of the extraction at finite radii 
in more detail further below during the analysis of the radiated energy and give an estimate of the
thus induced error.

I further check our results by estimating the QNM ringdown frequency
of the $l=2$ and $l=4$ multipoles and compare them to the close-limit approximation 
\cite{VitorCompCL,Yoshino:2005ps}.
The QNM ringdown signal is an exponentially damped sinusoid 
which dominates the gravitational wave signal after the formation of a common horizon.
Therefore, I perform a fit of the numerically computed $l=2$ and $l=4$ waveforms 
to this functional form.
The resulting characteristic QNM frequencies are summarized in 
Table~\ref{tab:HDR6Dqnm}, where I list the dimensionless real and imaginary parts $r_S\,\omega_{R,l}$ and
$r_S\,\omega_{I,l}$.
A perturbative, close-limit analysis presented in \cite{VitorCompCL,Yoshino:2005ps} predicts the frequencies
\begin{align}
\label{eqHDR6DqnmCL}
r_S\,\omega_{l=2} = & 1.1369 - \imath\,0.3038 
\,,\non\\
r_S\,\omega_{l=4} = & 2.62337 - \imath\,0.438266 
\,,
\end{align}
and a complementary, point particle computation for the $l=2$ mode \cite{VitorCompCL,Berti:2009kk} gives
\begin{align}
\label{eqHDR6DqnmPP} 
r_S\,\omega_{l=2} = & 1.131 - \imath\,0.32
\,.
\end{align}
Whereas the values for simulation HD6a$_f$ with small initial separation of $d=6.37~r_S$ is in good agreement with 
the theoretical predictions \cite{VitorCompCL,Yoshino:2005ps,Berti:2009kk}, Eq.~\eqref{eqHDR6DqnmCL},
within about $6.5\%$ for $l=2$ and $9\%$ for $l=4$,
(i.e., within numerical errors),
the frequencies of the wave signal obtained in the case of larger separation $d=9.56, 12.74~r_S$ deviate
from the close-limit approximation \cite{VitorCompCL,Yoshino:2005ps,Berti:2009kk}
by about $20\%$ in both multipoles.

In the right panel of Fig.~\ref{fig:HDR6Dwaveforml2} I present the $l=2$ multipole of the 
KI masterfunction $\Phi_{,t}$ considering the initial
coordinate distances $d = 6.37~r_S$, $d=9.56~r_S$ and $d=12.74~r_S$.
Whereas the waveform for $d = 6.37~r_S$ has a significantly smaller amplitude and 
higher frequency, the wave signal for the cases with initial distance $d=9.56~r_S$ and $d=12.74~r_S$
basically coincide,
indicating that the BHs are initially far enough apart to mimic infinite separation.
However, the analysis of the QNM frequencies has revealed a significant deviation from the 
theoretically predicted value in the two latter cases.
Therefore, unless denoted otherwise, I will present results 
obtained for the high resolution run HD6a$_f$ with 
initial coordinate distance $d = 6.37~r_S$, extracted at $R_{ex} = 40~r_S$.

\begin{figure}
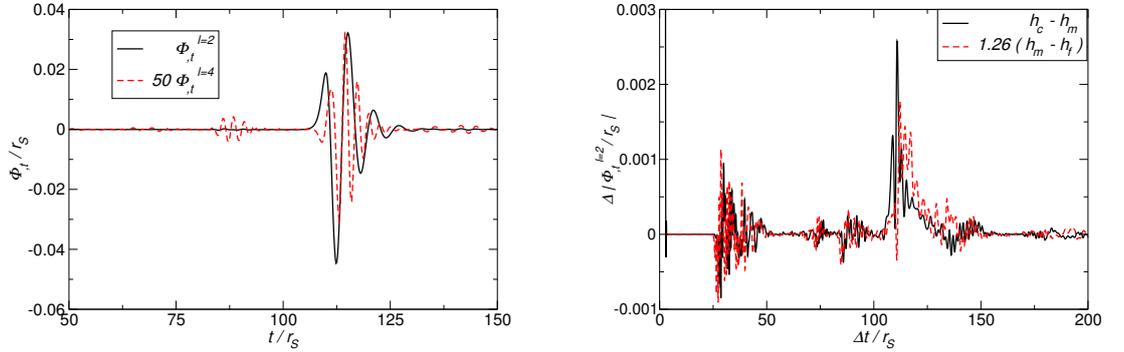

\begin{center}
\begin{tabular}{cc}
\includegraphics[width=0.5\textwidth]{HDR6Dfig2a} &
\includegraphics[width=0.5\textwidth]{HDR6Dfig2b} 
\end{tabular}
\end{center}
\caption{\label{fig:HDR6Dconvergence} 
Left panel:
Comparison of the $l=2$ (black solid line) and $l=4$ (red dashed line)
multipole of the KI masterfunction $\Phi_{,t}$ for model HD6a$_f$,
extracted at $R_{ex}=40~r_S$. The latter has been amplified by a factor $50$. 
Right panel:
Convergence analysis of the amplitude of the $l=2$ component of the KI function generated by model HD5a,
extracted at $R_{ex} = 40~r_S$.  The difference between the medium and high resolution
waveforms (red dashed line) has been amplified by the factor $Q_4=1.26$ 
indicating fourth order convergence. 
}
\end{figure}

In the left panel of Fig.~\ref{fig:HDR6Dconvergence} I present the $l=2$ and 
$l=4$ multipole of the KI function of run HD6a$_f$, where the latter has been re-scaled by a factor $50$.
The $l=4$ multipole represents the second strongest contribution to the wave signal, 
with an amplitude which is about two orders of magnitude smaller than the dominant $l=2$ mode,
whereas the $l=3$ multipole vanishes due to symmetry.

Furthermore, I perform a convergence test of the $l=2$ waveforms extracted at $R_{ex}=40~r_S$, 
presented in the right panel of
Fig.~\ref{fig:HDR6Dconvergence}, by employing model HD6a. 
Specifically, I have run the configuration with resolutions $h_c=r_S/84$, $h_m=r_S/88$ and 
$h=r_S/92$, which I denote as coarse, medium and high resolution.
I compare the differences of the amplitudes between the first two simulations 
with the difference between the medium and high resolution run.
The latter has been re-scaled by a factor $Q_4 = 1.26$ 
demonstrating fourth order convergence,
consistent with the numerical implementation.
The thus obtained discretization error in the amplitude of the waveform is about 
$\De|\Phi^{l=2}_{,t}|/|\Phi^{l=2}_{,t}| \le11\%$.
The analogous analysis performed for the phase $\phi$ reveals a discretization error up to $\De\phi^{l=2}/\phi^{l=2}\le 10\%$.

\begin{table}[htpb!]
\begin{center}
\begin{tabular}{ccccccc}
\hline
Run      & $d/r_S$ & $h/r_S$ & $r_S\,\omega_{R,l=2}$ & $-r_S\,\omega_{I,l=2}$ & $r_S\,\omega_{R,l=4}$ & $-r_S\,\omega_{I,l=4}$ 
\\ \hline 
HD6a$_c$ & $6.37$  & $1/84$  & $(1.07\pm0.02)$ & $(0.263\pm0.01)$ & $(2.32\pm0.1)$ & $(0.385\pm0.05)$ \\
HD6a$_m$ & $6.37$  & $1/88$  & $(1.04\pm0.03)$ & $(0.243\pm0.02)$ & $(2.39\pm0.1)$ & $(0.427\pm0.05)$ \\
HD6a$_f$ & $6.37$  & $1/92$  & $(1.06\pm0.01)$ & $(0.284\pm0.01)$ & $(2.39\pm0.1)$ & $(0.399\pm0.05)$ \\ 
HD6b     & $9.56$  & $1/92$  & $(0.91\pm0.02)$ & $(0.239\pm0.01)$ & $(2.11\pm0.1)$ & $(0.412\pm0.05)$ \\ 
HD6c     & $12.74$ & $1/92$  & $(0.93\pm0.05)$ & $(0.234\pm0.02)$ & $(2.09\pm0.1)$ & $(0.308\pm0.05)$ \\ 
\hline
\end{tabular}
\end{center}
\caption{\label{tab:HDR6Dqnm}
QNM frequencies $r_S\,\omega = r_S\,\omega_R + \imath\,r_S\,\omega_I$ of the $l=2$ and $l=4$ multipoles
of the KI wavefunction, 
resulting from a fit of the numerical waveform to
an exponentially damped sinusoid.
}
\end{table}

%
\subsection{Radiated energy}\label{ssec:HDR6Denergy}
\begin{figure}
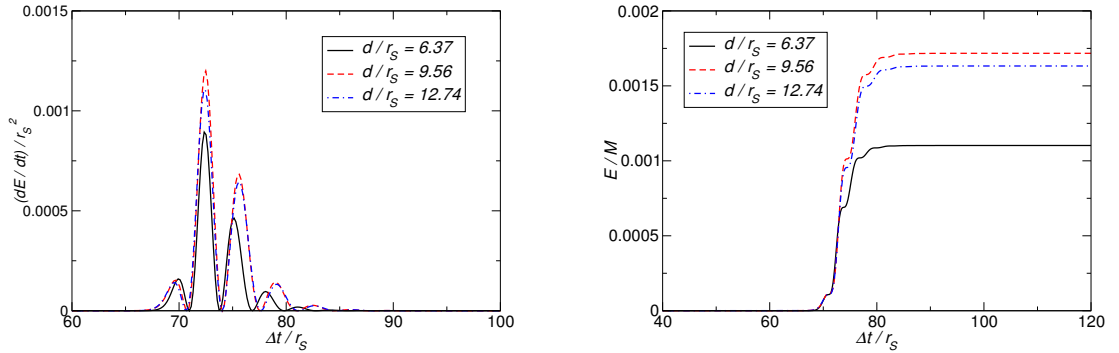

\begin{center}
\begin{tabular}{cc}
\includegraphics[width=0.5\textwidth]{HDR6Dfig3a} &
\includegraphics[width=0.5\textwidth]{HDR6Dfig3b} 
\end{tabular}
\end{center}
\caption{\label{fig:HDR6energyDist} 
Energy flux emitted in the $l=2$ multipole of the KI wavefunction (left panel) 
and total radiated energy (right panel), computed via Eq.~\eqref{eq:HDWEenergyrad},
for models HD6a$_f$ (black solid line), HD6b (red dashed line) and HD6c (blue dashed-dotted line).
I have shifted the signals by the extraction radius $R_{ex}=40~r_S$ and such that the maximum in the
energy flux overlap.
}
\end{figure}

I next investigate the emission of gravitational wave energy in more detail.
For this purpose, I first plot the energy flux emitted in the $l=2$ multipole of the KI wavefunction
as well as as the total radiated energy $E/M$ 
in Fig.~\ref{fig:HDR6energyDist}.
I present the radiated energy and flux, extracted at $R_{ex}=40~r_S$, 
for all three types of models 
corresponding to head-on collisions of BHs with initial (coordinate) separation $d=6.27, 9.56, 12.74~r_S$.
The signals shown in Fig.~\ref{fig:HDR6energyDist} have been shifted in time by the 
extraction radius $R_{ex}=40~r_S$ and such that the maxima in the energy flux overlap.
As I have previously observed in the analysis of the waveforms,
also the radiated energy and energy flux 
resulting from the HD6a$_f$ run on the one hand and the HD6b and HD6c runs on the other 
differ enormously.

\begin{figure}
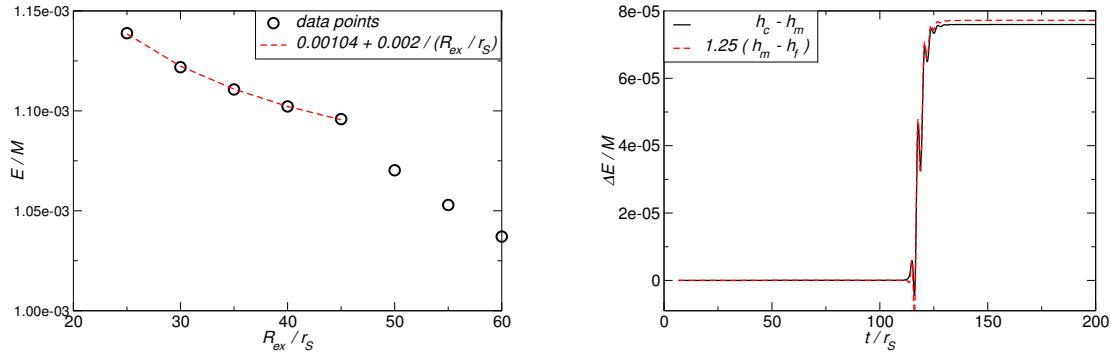

\begin{center}
\begin{tabular}{cc}
\includegraphics[width=0.5\textwidth]{HDR6Dfig4a} &
\includegraphics[width=0.5\textwidth]{HDR6Dfig4b} 
\end{tabular}
\end{center}
\caption{\label{fig:HDR6DenergyConv} 
Left:
Total radiated energy $E/M$ as function of the extraction radius $R_{ex}/r_S$ for model HD6a$_f$.
I show the numerical data (black cycles) together with the fitting function, 
Eq.~\eqref{eq:HDR6Denergyfit} (red dashed line).
Right:
Convergence plot for the total radiated energy $E/M$. Specifically, I show the differences between the 
coarse and medium resolution (black solid line) and medium and coarse resolution (red dashed line).
The latter has been amplified by a factor $Q_4=1.25$ indicating fourth order convergence.
}
\end{figure}

In the left panel of Fig.~\ref{fig:HDR6DenergyConv}
I present the radiated energy $E/M$ as function of the extraction radius 
$R_{ex} = 25,\dots,60~r_S$ for model HD6a$_f$.
Whereas the first five values, which have been extracted on the same refinement level, vary little,
the energy measured at $R_{ex}=50,55,60~r_S$, located at the next coarser refinement level,
shows significant deviations.
I find a similar behaviour also for models HD6b and HD6c.
Therefore, I consider only the total energy extracted at $R_{ex}=25,30,35,40,45~r_S$ and
perform a fit of the form
\begin{align}
\label{eq:HDR6Denergyfit}
\frac{E}{M} = & \frac{E_{\infty}}{M} + \frac{A_1}{(R_{ex}/r_S)}
\,.
\end{align}
The limit $R_{ex}/r_S \rightarrow \infty$ allows us to extrapolate the energy $E_{\infty}/M$
measured at infinite  radius. 
I summarize the extrapolated energy $E_{\infty}/M$
in Table~\ref{tab:HDR6Denergyinf}.
\begin{table}[htpb!]
\begin{center}
\begin{tabular}{ccc}
\hline
Run      & $d/r_S$ & $E_{\infty}/M$ \\
\hline
HD6a$_f$ & $6.37$  & $(0.1042\pm(0.0125 + 0.0009))\%$ \\
HD6b     & $9.56$  & $(0.1617\pm(0.0194 + 0.0019))\%$ \\
HD6c     & $12.74$ & $(0.1549\pm(0.0186 + 0.01))\%$ \\
\hline
\end{tabular}
\end{center}
\caption{\label{tab:HDR6Denergyinf}
Energy $E_{\infty}/M$ extrapolated to infinite extraction radius.
The deviation accommodates the discretization error of about $12\%$ as well as the
error arising from the measurment at finite extraction radii.
}
\end{table}

I account for the error arising from the measurment at finite extraction radii
by comparing $E_{\infty}/M$ with the value obtained from a second fit 
including the quadratic term $A_2 / (R_{ex}/r_S)^2$.
Then, this error is estimated to be about $\De E/E_{\infty} \sim 1\%, 1.2\%, 6\%$ 
for models HD6a$_f$, HD6b and HD6c, respectively.

In order to investigate the numerical accuracy 
I have performed a convergence analysis, presented in the right panel of Fig.~\ref{fig:HDR6DenergyConv}.
For this purpose, I have employed model HD6a 
representing a head-on collision of two BHs with initial distance $d=6.37~r_S$,
simulated with resolutions $h_c=r_S/84$, $h_m=r_S/88$ and $h_f=r_S/92$.
The corresponding convergence plot is presented in the right panel of Fig.~\ref{fig:HDR6DenergyConv}.
I depict 
the differences of the total radiated energy $E/M$, extracted at $R_{ex}=40~r_S$,
between the coarse-medium and medium-high resolutions simulations.
The latter difference has been re-scaled by the factor $Q_4=1.25$, thus demonstrating
convergence of fourth order consistent with the implementation.
The discretization error is estimated to be about $\De E/E \le 12\%$.
I have not performed convergence runs for the remaining models HD6b and HD6c and
assume that the discretization error is of the same order.
I list the error in the total radiated energy $E_{\infty}/M$
due to discretization and finite extraction in Table~\ref{tab:HDR6Denergyinf}.
Note, that the energy appears to depend non-monotonously on the initial distance. However,
the expected monotonous behaviour is still covered within the error estimates.
Nevertheless, because the extracted total energy $E_{\infty}/M$
appears to depend non-monotonously on the (initial) distance $d/r_S$, 
and because I only considered three different initial separations 
I have not extrapolated the 
radiated energy to infinite initial separation $d/r_S \rightarrow\infty$.
Instead, I consider model HD6a$_f$ which has exhibited waveforms with the correct QNM
frequencies. Then, I estimate the total radiated energy emitted throughout the 
head-on collision of two non-boosted, equal-mass BHs in $D=6$ spacetime dimensions to be
$E_{\infty}/M \sim (0.104\pm0.013)\%$.

Nevertheless, this result has to be taken with a grain of salt until ongoing investigations
conclusively answer the question arosen by the observed discrepancies.
Possible sources of error include
\noindent{(i)} lack of numerical accuracy or 
\noindent{(ii)} radiation effects.
In order to test the first possibility, simulations of model HD6c with higher resolutions are underway.
Because the radiated energy is expected to be small, I have assumed the total Schwarzschild radius
$r_S^3 = r^3_{S,1} + r^3_{S,2}=1$. However, the radiation effects might be non-negligible and are 
subject to further analysis.

%
\section{Head-on collsions of unequal mass black holes in $D=5$}\label{sec:HDRum5D}
In the previous sections we reported our results of head-on collisions of 
equal-mass BHs in $D=5$ and $D=6$ spacetime dimensions.
In the present section, we wish to extend that study to the case of
unequal mass BH binaries.  This is an interesting extension for several
reasons, perhaps the most important of which is the non-trivial comparison
with point-particle (PP) calculations in the linearized regime. 
We will compare radiated energy, momentum and multipolar dependence
of our full nonlinear results with results from linearized Einstein
equations. It turns out that the agreement is remarkable, providing
an outstanding consistency check on our codes and results.  A thorough
analysis of the linearized Einstein equations has been done 
by Berti {\textit{et al}} \cite{Berti:2010gx}.

Here, we have evolved BH binaries, colliding head-on from rest
with mass ratios
$q \equiv M_1/M_2 = r^{D-3}_{S,1} / r^{D-3}_{S,2} = 1,\,1/2,\,1/3,\,1/4$,
where $M_i$ is the mass of the $i$-th BH.  
The mass parameter $r^{D-3}_{S,1}/ r^{D-3}_S$ of the smaller BH is
given in Table~\ref{tab:HDRumsetup1} and we adapt the value of the second BH 
accordingly.
The initial coordinate separation of the two BHs is set to $d/r_S =
6.37$ which translates to a proper initial separation of $L/r_S = 6.33$.
Further details of the setup of the simulations are summarized in
Table~\ref{tab:HDRumsetup1} and~\ref{tab:HDRumsetup2}.  Unless denoted otherwise, our discussion will
always refer to the highest resolution runs with 
$h_f = r_S/84$, $h_f = r_S/102.9$, $h_f = r_S/118.8$ and 
$h_f = r_S/132.8$ for models R5a, R5b, R5c and R5d$_f$ 
in Table~\ref{tab:HDRumsetup2}, respectively.
The energy flux is computed according to Eq.~\eqref{eq:HDWEenergyflux}
(see Eq.~(21) in Ref.~\cite{Berti:2003si} for
the corresponding expression in Fourier space). 
The momentum flux can be obtained from
\begin{align}
\label{eq:HDRumMomFlux0}
\frac{dP^i}{dt} = & \int_{S_{\infty}}d\Omega \frac{d^2E}{dt d\Omega}n^i
\,,
\end{align}
with $n^i$ a unit radial vector on the sphere at infinity $S_{\infty}$. 
This results in an infinite series coupling different multipoles. 
Using only the first two terms in the series, we find, for instance,
that in $D=5$ the momentum flux in the collision direction is given by
\begin{align}
\label{eq:HDRumMomFlux}
\frac{dP}{dt} = & \frac{1}{4 \pi} \Phi^{l=3}_{,t}
                  \left( 5 \Phi^{l=2}_{,t} +21 \Phi^{l=4}_{,t} \right) 
\,.
\end{align}
Here, $\Phi^l_{,t}$ is the $l-$pole component of the KI gauge-invariant wavefunction, 
Eq.~\eqref{eq:HDWEKIwavefunction}, 
\cite{Kodama:2000fa,Kodama:2003jz,Witek:2010xi}.
From the momentum radiated, the recoil velocity of the system can be obtained as
\begin{align}
\label{eq:HDRumkickvelocity}
v_{\rm recoil} = & \left| \int^{\infty}_{-\infty} dt \frac{d P}{d t} \right|
\,.
\end{align}
\begin{table}
\begin{center}
\begin{tabular}{c|cccc}
\hline
Run & $q$ & $m_i$ 
& $z_1/r_S $ & $z_2 / r_S$ \\ 
\hline
R5a & $1$ & $0.5$  & $3.185$ & $-3.185$ \\
R5b & $1/2$ & $0.33$ & $4.247$ & $-2.123$ \\ 
R5c & $1/3$ & $0.25$ & $4.777$ & $-1.592$ \\ 
\hline
R5d$_c$ & $1/4$ & $0.2$ & $5.096$ & $-1.274$ \\
R5d$_m$ & $1/4$ & $0.2$ & $5.096$ & $-1.274$ \\
R5d$_f$ & $1/4$ & $0.2$ & $5.096$ & $-1.274$ \\
\hline
\end{tabular}
\end{center}
\caption{\label{tab:HDRumsetup1}
  Initial parameters of the head-on collisions from rest in $D=5$.
  We give 
  the mass of the smaller BH $m_i = r^{D-3}_{S,1} / r^{D-3}_{S}$ 
  and the quantity $q \equiv M_1/M_2= r^{D-3}_{S,1}/r^{D-3}_{S,2}$ 
  denotes the mass ratio.
  $z_{i}/r_S$ 
  is the initial position of $i$-th BH.
}
\end{table}
\begin{table}
\begin{center}
\begin{tabular}{c|cc}
\hline
Run & $q$  & Grid Setup \\ 
\hline
R5a & $1$ 
& $\{(256,128,64,32,16,8)\times(2,1,0.5),~h=1/84\}$ \\
R5b & $1/2$ 
& $\{(209,104.5,52.3,26.1,13.1,6.5)\times(1.6,0.8,0.4),~h=1/102.9\}$ \\
R5c & $1/3$ 
& $\{(181.0,90.5,45.3,22.6,11.3)\times(2.8,1.4,0.7,0.4),~h=1/118.8\}$ \\
\hline
R5d$_c$ & $1/4$ 
& $\{(161.9,80.9,40.5,20.2,10.1)\times(2.5,1.3,0.6,0.3),~h=1/113.8\}$ \\
R5d$_m$ & $1/4$ 
& $\{(161.9,80.9,40.5,20.2,10.1)\times(2.5,1.3,0.6,0.3),~h=1/123.3\}$ \\
R5d$_f$ & $1/4$ 
& $\{(161.9,80.9,40.5,20.2,10.1)\times(2.5,1.3,0.6,0.3),~h=1/132.8\}$ \\
\hline
\end{tabular}
\end{center}
\caption{\label{tab:HDRumsetup2}
  Grid structure for the simulations of head-on collisions from rest in $D=5$.
  The grid setup is given in terms of the ``radii'' of the individual refinement levels,
  as well as the resolution near the punctures $h$, 
  in units of $r_S$ (see Sec.~II E in \cite{Sperhake:2006cy} for details).
}
\end{table}

\subsection{Waveforms}
\begin{figure}
\begin{center}
\includegraphics[width=0.5\textwidth]{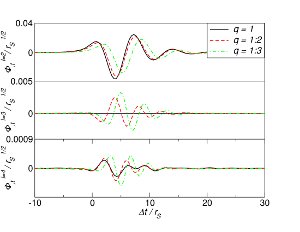}
\end{center}
\caption{\label{fig:HDRumwaveform} 
From top to bottom: $l=2$, $l=3$ and $l=4$ modes of the KI waveform
for the different mass ratios; $q = 1$ (black solid lines), $q = 1/2$
(red dashed lines) and $q = 1/3$ (green dash-dotted lines).  The curves
have been shifted in time such that the formation of the common apparent
horizon corresponds to $\Delta t / r_S = 0$ and taking into account the
time that it takes the
waves to propagate to the extraction radius.
}
\end{figure}
In Fig.~\ref{fig:HDRumwaveform} we show the $l=2,3,4$ waveforms for
different mass ratios, zoomed in around the time of the merger. 
The waveforms have been shifted in time such that
$\Delta t = (t - R_{ex} - t_{CAH})= 0~r_S$ corresponds to the
time $t_{CAH}$ at which the common apparent horizon forms and taking
into account the propagation time of the waves to the extraction
radius $R_{ex}= 60, 49, 42.4, 37.9~r_S$.  
The waveform is similar to previous four-dimensional
results (see, e.g. Ref.~\cite{Anninos:1998wt}, 
a more detailed study has been given by Sperhake {\textit{et al}} \cite{Sperhake:2011ik}). 
Although not shown in Fig.~\ref{fig:HDRumwaveform} we observe a small, spurious signal starting
around $(t - R_{ex})/r_S \approx 0$, which is an artifact of the initial data.

The actual physical part of the waveform is dominated by the merger signal
at $\Delta t / r_S \approx 0$ followed by the quasinormal ringdown.
We estimate that the different ringdown modes are given by
\begin{align}
\omega_{l=2} r_S = & (0.955 \pm 0.005) - i\, (0.255 \pm 0.005)
\,,\nonumber\\
\omega_{l=3} r_S = & (1.60 \pm 0.01) - i\, (0.31 \pm 0.01)
\,,\nonumber\\
\omega_{l=4} r_S = & (2.25 \pm 0.03) - i\, (0.35 \pm 0.05)
\,. 
\end{align}
These results agree well, and within uncertainties, with estimates from
linearized theory
\cite{Berti:2009kk,Cardoso:2003qd,Konoplya:2003dd,Berti:2003si}, providing
a strong consistency check on our results.
\begin{figure}
\begin{center}
\begin{tabular}{c}
\includegraphics[width=0.5\textwidth]{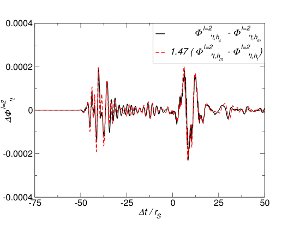} 
\end{tabular}
\end{center}
\caption{\label{fig:HDRumconvergence} 
Convergence analysis of the $l=2$ mode
for model ${\rm R5d}$ in Table~\ref{tab:HDRumsetup2}. We show the differences
between the coarse and medium resolution waveform (black solid line)
and the medium and high resolution waveform (red dashed line). The
latter has been amplified by the factor $Q_4=1.47$, indicating fourth
order convergence.  The curves have been shifted in time such that the
formation of the common apparent horizon corresponds to $\Delta t /
r_S = 0$ and taking into account the time that it takes the
waves to propagate to the extraction radius. 
}
\end{figure}
Finally, we consider numerical convergence of our waveforms. This study
is summarized in Fig.~\ref{fig:HDRumconvergence} for the $l=2$ mode of the KI
wavefunction, and for the most challenging mass ratio, $q=1/4$, model
${\rm R5d}$ in Table~\ref{tab:HDRumsetup2}.  We have evolved this setup at
three different resolutions, namely 
$h_c = r_S/113.8$, $h_m = r_S/123.3$ and $h_f = r_S/132.8$,
which we will refer to as ``coarse'', ``medium''
and ``high'' resolution in the following.  We show the difference between
the coarse and medium as well as between the medium and high resolution
waveforms. The latter has been amplified by the factor $Q_4=1.47$, which
indicates fourth order convergence. We obtain the same order of accuracy
for the higher modes. The discretization error in the waveforms is
estimated to be $\approx 1.5\%$.

\subsection{Radiated energy}
\begin{table}
\begin{center}
\begin{tabular}{c|ccccc}
\hline
$q$      & $ E^{\rm rad}/M (\%)$ & $E^{\rm rad}_{l=2}(\%)$ & $E^{\rm rad}_{l=3}(\%)$ & $E^{\rm rad}_{l=4}(\%)$& $v_{\rm recoil} (km/s)$ \\
\hline
1/1     & $0.089$ $(0.090)$ & $99.9$ & $0.0$ & $0.1$ & $0.00$  \\
1/2     & $0.073$ $(0.067)$ & $97.7$ & $2.2$ & $0.1$ & $11.37$ \\
1/3     & $0.054$ $(0.051)$ & $94.8$ & $4.8$ & $0.4$ & $12.64$ \\
1/4     & $0.040$ $(0.035)$ & $92.4$ & $7.0$ & $0.6$ & $11.38$ \\
\hline
\end{tabular}
\end{center}
\caption{\label{tab:HDRumD5EnergyModes}
Summary of our results concerning unequal-mass head-on collisions of
BHs in $D=5$.  We show the total radiated energy $E/M$ as measured from
the energy flux at $R_{ex}$, 
the quantity in parenthesis refers
to the estimate obtained using properties of the apparent horizon
 (see Sec.~\ref{sec:HDRcoll5D} for details).  The next three columns show the
fraction of energy $E^l$ excited in the $l$-th mode as compared to the
total radiated energy. The last column refers to the recoil velocity
$v_{\rm recoil}$ in $km/s$.
}
\end{table}
Table~\ref{tab:HDRumD5EnergyModes} lists some of the most important physical
quantities which characterize the head-on collision of BHs in $D=5$.
In particular, we show the radiated energy in units of total mass $M$,
and the recoil velocity of the final BH in ${\rm km/s}$. The maximum
amount of energy is emitted in the equal mass case 
($E^{\rm rad}/M = 0.089\%$ as presented in Sec.~\ref{sec:HDRcoll5D} \cite{Witek:2010xi}),
and it decreases for smaller mass ratios.  We estimate the error in the radiated
energy to be about $5~\%$.
These results have been obtained by integrating
the energy flux as given by the KI master wavefunction. We have also
estimated the radiated energy using properties of the apparent horizon
as described in Sec.~\ref{sec:HDRcoll5D}.
We estimate the discretization error to be about $\approx10~\%$ when using
this method.
The apparent horizon estimate for the total radiated
energy is shown in parenthesis in Table~\ref{tab:HDRumD5EnergyModes}, and
is consistent with the flux computation
within numerical uncertainties.
\begin{figure}
\begin{center}
\begin{tabular}{c}
\includegraphics[width=0.5\textwidth]{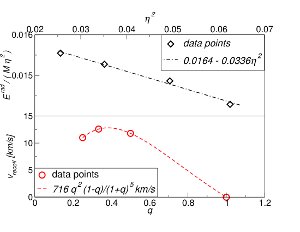}
\end{tabular}
\end{center}
\caption{\label{fig:HDRumintegratedenergy} 
Top:
Total integrated energy for different mass ratios, as function of
$\eta^2=[q/(1+q)^2]^2$.  The black diamonds denote the numerical data
and the black dashed-dotted line is the corresponding fitting function,
Eq.~\eqref{eq:HDRumfitA}.
Bottom:
Recoil velocity $v_{\rm recoil}$ in $km/s$ as function of the mass
ratio $q$.  The red circles denote the numerical data and the red dashed
line is the corresponding fitting function, Eq.~\eqref{eq:HDRumfitV}.
}
\end{figure}
Table~\ref{tab:HDRumD5EnergyModes} also shows the fraction of energy emitted in 
different multipoles.
Higher multipoles are clearly enhanced as the mass ratio decreases, 
in agreement with what we expect in the extreme case of a PP 
falling into a BH. In fact, we can make this statement more precise. 
Post-Newtonian arguments, which extend to generic $D$-dimensions, allow one 
to expect the functional dependence for the total radiated energy 
\cite{MadalenaThesis}, $E^{\rm rad}/M \propto \eta^2$, where $\eta=q/(1+q)^2$ is the 
dimensionless reduced mass. For clarity, we show the ratio $E^{\rm rad}/(M\eta^2)$
in the top panel of Fig.~\ref{fig:HDRumintegratedenergy}, which can be seen to 
depend very weakly on $\eta^2$. 
We can refine this argument by fitting our numerical results to an improved 
expression of the form $E^{\rm rad}/M \eta^2= A_0 + A_1 \eta^2$. We find
\begin{align}
\label{eq:HDRumfitA}
\frac{E^{\rm rad}}{M \eta^2} = & 0.0164 -0.0336 \eta^2 \,.
\end{align}
Moreover, the following expressions for the multipolar content 
provide a good fit to our numerical data,
\begin{align}
\frac{E^{\rm rad}_{l=2}}{E^{\rm rad}} = & 0.79 +0.83 \eta 
\,,\quad
\frac{E^{\rm rad}_{l=3}}{E^{\rm rad}} = 0.19 -0.77 \eta 
\,.
\end{align}
Linearized, PP calculations presented by Berti {\textit{et al}} \cite{Berti:2010gx} 
show that in the limit of zero mass ratio one obtains
\begin{align}
\frac{E^{\rm rad}_{\rm PP}}{M\eta^2} = & 0.0165 
\,,
\end{align}
which agrees with the extrapolation of our numerical results within less than $1\%$.
The multipole contents in the PP limit are 
\begin{align}
\frac{E^{\rm rad}_{l=2}}{E^{\rm rad}_{\rm PP}} = & 0.784  
\,,\quad
\frac{E^{\rm rad}_{l=3}}{E^{\rm rad}_{\rm PP}} =   0.167  
\,,
\end{align}
still in very good agreement with the extrapolation of our full numerical results 
to the zero mass ratio limit.
In fact, bearing in mind that we are extrapolating from mass ratios of $1/4$ 
down to the zero mass ratio limit, the agreement is impressive.
Finally, all these results are consistent with the fact that higher
multipoles contribute more to the radiation than in $D=4$, where for
instance the $l=3$ mode contributed roughly $10\%$ of the total energy in
the PP limit \cite{Davis:1971gg}. Linearized, point-particle calculations
show that the trend is consistent and continues in higher dimensions \cite{Berti:2010gx},
which might mean that accurate wave extraction will become extremely difficult, as higher
resolution is necessary to resolve higher-$l$ modes \cite{Berti:2007fi}.

\subsection{Radiated momentum}
\begin{figure}
\begin{center}
\includegraphics[width=0.5\textwidth]{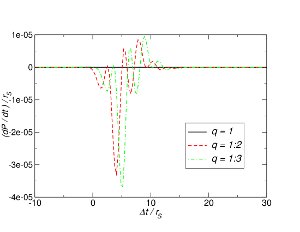} 
\end{center}
\caption{\label{fig:HDRummomentumflux} 
Momentum flux for different mass ratios.
The curves have been shifted in time such that the formation of the common apparent horizon
corresponds to $\Delta t = 0~r_S$ and taking into account the time that it takes the 
waves to propagate to the extraction radius.
}
\end{figure}
For unequal-mass collisions, the asymmetric emission of radiation along the collision axis causes a net momentum to be carried by
gravitational waves. As such, the final BH will ``recoil'', according to 
Eq.~\eqref{eq:HDRumkickvelocity}. Momentum fluxes and recoil velocity  
for different mass ratios are shown in Fig.~\ref{fig:HDRummomentumflux} and the bottom panel of Fig.~\ref{fig:HDRumintegratedenergy}, respectively.
We estimate the errors in the recoil velocity to be $\approx 5~\%$.
The general functional form for the dependence of momentum on the mass
parameters of the individual holes has been worked out by M. Lemos in
generic spacetime dimensions \cite{MadalenaThesis} and is the same
as in four dimensions
\begin{align}
\label{eq:HDRumfitV}
v_{\rm recoil} = &  C \frac{q^2(1-q)}{(1+q)^5}
\,.
\end{align}
By fitting this function to our numerical data, we obtain $C=716~{\rm km/s}$.
Observe that $v_{\rm recoil}$ reaches a maximum value at $q=2-\varphi\simeq 0.38$, where $\varphi$ is the golden ratio. The quality of the fit can be seen in the bottom panel of Fig.~\ref{fig:HDRumintegratedenergy}, 
where we overplot the numerical data points with the fitting function, 
Eq.~\eqref{eq:HDRumfitV}. This exercise is interesting because we can again extrapolate our results to the PP limit. 
Berti {\textit{et al.} \cite{Berti:2010gx} find
\begin{align}
v_{\rm recoil} = & 779 q^2 {\rm km/s}
\,,
\end{align}
in reasonably good agreement (better than 10$\%$) with our extrapolation. We note that momentum emission is given by a non-trivial
interference between different multipoles, so this is a non-trivial agreement.

\newpage
\section{Discussion}
In this chapter I have presented fully non-linear simulations of higher dimensional 
BH spacetimes.
These evolutions have been carried out with the {\textsc{HD-Lean}} code
(an extension of the original {\textsc{Lean}} code)
which employs the formalism discussed in Chapter~\ref{chapter:HigherDimF}.

First, I have tested our implementation by simulating single BH spacetimes.
Most importantly, I have demonstrated the internal consistency of our numerical 
framework in $D=5$ and $D=6$ dimensions by showing convergence of the 
Hamiltonian and momentum constraints as well as
comparing numerical results with (semi-)analytic expressions for a single
Tangherlini BH in geodesic slicing. 

In order to verify our wave extraction formalism I have evolved BH head-on 
collisions in $D=4$ dimensions as benchmark tests.
In this case a number of well tested, independent wave extractions techniques,
such as the Newman-Penrose or the Regge-Wheeler-Zerilli formalism,
are available.
I have compared the waveforms and radiated energy computed with the Kodama-Ishibashi formalism
to these alternative methods.
The excellent agreement of our results confirms the correctness and reliability
of our formalism and implementation.

I have accomplished successful numerical simulations of BH head-on collisions 
in $D=5$ and $D=6$ spacetime dimensions.
I have succeeded in monitoring the evolution throughout the collision,
resulting in the formation of a single BH,
and for a long time after the merger.
Along the way, I have extracted the corresponding waveforms and computed the total energy 
released during the collision.

I have performed a variety of further tests of the wave extraction
formalism in $D=5$. Besides testing the proximity of the
numerical coordinate system to the Tangherlini background spacetime,
I have demonstrated good agreement between the radiated energy
as derived directly from the KI master function with the values
obtained from the horizon area of the post-merger remnant hole.
Finally, the ringdown part of the waveform yields a quasinormal mode
frequency in excellent agreement with predictions from BH
perturbation theory.

I have shown that the total energy released in the form of gravitational
waves  is approximately $(0.089\pm 0.006)\%$ in $D=5$ 
and $(0.104\pm0.013)\%$ in $D=6$ 
of the initial center-of-mass energy of the system, for a head-on collision of two BHs starting from rest.
As a comparison, the analogous process in
$D=4$ releases a slightly smaller quantity: $(0.055\pm 0.006)\%$.
I summarize the main results for head-on collisions of two BHs starting from rest
in $D=4,5,6$ spacetime dimensions in Table \ref{tab:HDRcoll5Dsummary}.

The radiative efficiency $E^{\rm rad}/M$ in Table \ref{tab:HDRcoll5Dsummary}
shows that head-on collisions starting from rest in $D=5$ and $D=6$ dimensions
generate about $1.6$ and $1.9$ times 
as much GW energy as their four dimensional
counterparts. 
It will be very interesting to investigate to what
extent this observation holds for wider classes of BH collisions.
I can compare the radiation efficiency with the
upper limit derived by Hawking \cite{Hawking:1971} from the
requirement that the horizon area must not decrease in the collision.
This leads to the {\em area bound}
$\frac{E^{\rm area}}{M} \leq 1-2^{-\frac{1}{D-2}}\,.$
Evidently, this bound decreases with dimensionality, 
while in the present computation it increases with increasing dimensions.
As also shown in the table, the generation
of GWs in head-on collisions starting from rest is about 3 orders
of magnitude below this bound. 
In four dimensions it has already been demonstrated that there exist
more violent processes which release more radiation than the
head-on collisions considered in this work
\cite{Sperhake:2008ga, Shibata:2008rq, Sperhake:2009jz}.
A generalization to more generic configurations such as collisions of boosted
BHs is subject to ongoing investigations and will be presented elsewhere
\cite{HDFcoll2012}.

In the context of this work it is particulary interesting to point out
results in the point-particle approximation reported in \cite{Berti:2010gx,Berti:2003si}.
Although covering a complementary setup to our investigations it is interesting
to note that the {\textit{qualitative}} behaviour of both methods agrees well.
Specifically, the authors of \cite{Berti:2010gx} have shown that 
the gravitational radiation, emitted when a point particle 
without initial boost falls into a higher dimensional BH,
increases with the spacetime dimensionality.
\begin{table}
\begin{center}
\begin{tabular}{ccccc}
\hline
$D$ & $r_{S}\,\omega (l=2)$     & $E^{\rm rad}/M(\%)$   & $E^{\rm area}/M(\%)$  
& $E^{\rm rad}_{l=4}/E^{\rm rad}_{l=2}$\\ \hline
$4$ & $0.7473-\imath\,0.1779$   & $0.055$               & $29.3$ 
& $<10^{-3}$ \\
$5$ & $0.9477-\imath\,0.2561$   & $0.089$               & $20.6$ 
& $<10^{-4}$ \\
$6$ & $1.1369-\imath\,0.3038$   & $0.104$               & $15.9$
& $<10^{-3}$ \\
\hline
\end{tabular}
\end{center}
\caption{\label{tab:HDRcoll5Dsummary} Main results for head-on collisions
in $D=4$, $D=5$ and $D=6$ dimensions. I list the ringdown frequency $\omega$,
the total energy radiated in gravitational waves, the upper bound
$E^{\rm area}$ on the radiated energy obtained from Hawking's area theorem
and the fractional energy in the $l=4$ multipole relative to the
quadrupole radiation.}
\end{table}

I have, furthermore, achieved succesful evolutions of
unequal-mass BH binaries in higher
dimensions, by solving Einstein equations in the full nonlinear regime.
I have focused on head-on collisions in $D=5$ spacetime dimensions.
The gravitational waveforms exhibit similar features as their $4$-dimensional counterparts
\cite{Anninos:1998wt,Sperhake:2011ik},
and I were able to estimate the ringdown frequencies of the
lowest multipoles. 
I find good agreement with published values for
the quasinormal frequencies, extracted in a linearized formalism \cite{Berti:2009kk}.
When extrapolated to the zero-mass-ratio limit, our results agree with
linearized calculations \cite{Berti:2010gx} at the $\%$ level or better
for the energy and momentum radiated, as well as for the multipolar
dependence.
Our findings, supported by linearized analysis, indicate that the higher
multipoles become more important for larger $D$. This will certainly make
wave extraction at sufficiently large $D$ a more demanding task, since
higher resolutions are necessary to resolve these modes.  
Additionally, perturbations fall-off with a power $\tfrac{1}{r^{D-3}}$
resulting in steeper gradients and shorter length scales
and, thus, require finer meshes.
Finally, it would be very interesting to perform an exhaustive set
of simulations in higher $D$: our results, together with linearized
analysis \cite{Berti:2010gx}, suggest a qualitative change in radiation
emission for $D \leq 12-13$.
In fact, this change is required
by the fact that Hawking's area theorem forces the total amount of
gravitational radiation to decrease with $D$, at sufficiently large $D$
\cite{Witek:2010xi,Berti:2010gx}. Understanding the mechanism at play
requires extension of our results to arbitrary spacetime dimensions.

The results presented in this chapter focused on low-energy, i.e., non-boosted 
head-on collisions of BHs in $D=5$ and $D=6$ spacetime dimensions.

Of utmost importance for potential applications in high energy collider physics 
are fully dynamical, non-linear evolution of boosted BH collisions.
Results of PP calculations \cite{Berti:2010gx} of a particle with initial kinetic energy varying from zero
to the UV limit 
raise the expectation for astonishing results.
Counter-intuitively, they find that particles with intermediate large boost 
radiate {\textit{less}} energy than their non-boosted counterparts.
Also, while the energy released by the infall of a non-boosted particle increases 
with spacetime dimension, the trend in the UV limit is reversed.
Preliminary fully dynamical studies of head-on collisions of boosted BHs in $D=5$ and $D=6$ dimensions, 
which will be presented in detail elsewhere \cite{HDFcoll2012},
indicate a qualitatively similar trend.

Conclusively, I have started to explore a wide range of interesting phenomena 
and accomplished the numerical modelling of BH collision higher dimensional spacetimes.
This is but the tip of the iceberg and leaves plenty of room for exciting future investigations.



\chapter{Black holes in a box: Toward the numerical evolution of black holes in AdS}
\label{chapter:BBbox}
\section{Introduction}\label{sec:BoxIntro}
A powerful tool to explore strongly coupled gauge theories by investigating classical gravity 
(and vice versa) is provided by the gauge/gravity duality, 
originally proposed by Maldacena in 1998 \cite{Maldacena:1997re}.
In a nutshell, the gauge/gravity duality is a correspondence between a gravity theory in 
$D$-dimensional anti-de Sitter (AdS) spacetime (on the one hand) and
a strongly coupled gauge theory ``living'' on its $(D-1)$-dimensional boundary.
In many of these developments BHs play a crucial role,
as may be seen by the following list of examples: 
(i) the successful microscopic computations of the Bekenstein-Hawking entropy for extremal
BHs \cite{Strominger:1996sh}
and Hawking emission rates for near extremal BHs \cite{Callan:1996dv}
are now seen as applications of the correspondence; 
(ii) the confinement/deconfinement phase transition in QCD-like theories has been
identified \cite{Witten:1998qj} with the Hawking-Page phase transition for
AdS BHs \cite{Hawking:1982dh}; 
(iii)  moving away from thermal equilibrium, the quasinormal frequencies of AdS BHs
have been identified with the poles of retarded correlators describing
the relaxation back to equilibrium of a perturbed dual field theory
\cite{Horowitz:1999jd,Birmingham:2001pj}; 
(iv) in a large class of
gauge theories with a gravity dual, a universal behaviour was obtained for
the ratio of the strongly coupled medium's viscosity to entropy density,
by computing the absorption cross-section of low energy gravitons in
the dual BH (or black brane) geometry \cite{Kovtun:2004de}. The result
is in good agreement with experimental results from the Relativistic
Heavy Ion Collider (RHIC); 
(v) critical exponents, of the type
found in spherical gravitational collapse by Choptuik in $4$-dimensional
asymptotically flat spacetime \cite{Choptuik:1992jv}, have been
conjectured to be dual to the asymptotic value of the parton saturation
exponent for high energy scattering in QCD, in the Regge limit, at weak
coupling \cite{AlvarezGaume:2006dw}. 
These and other examples have built
expectations that the gauge/gravity correspondence will be a useful 
laboratory for exploring and understanding very difficult problems of both 
field theory and gravity, such as the confinement problem, 
the information loss paradox and the problem of singularities.

Given the potential of the correspondence exemplified above, a working
framework to solve Einstein's equations exactly for a broad range
of initial conditions in AdS spaces would, unquestionably, be very
useful for deepening the study of the correspondence, particularly in dynamical situations. 

In \cite{Bantilan:2012vu}, the authors presented (for the first time) 
fully numerical evolutions of AdS$_5$ spacetimes with $SO(3)$ symmetry 
and studied prompt BH formation.
Non-linear evolutions of scalar fields in AdS background \cite{Bizon:2011gg,Jalmuzna:2011qw,Buchel:2012uh}
and investigations of perturbations of AdS spacetimes \cite{Dias:2011ss} 
(to higher order in perturbation theory)
suggest that AdS is subject to the non-linear turbulent instability.
On the other hand, recent simulations of a wider class of real and complex scalars
in AdS revealed a threshold for BH formation~\cite{Buchel:2013uba}
and the evolution of time-periodic solution hint at stable islands in the
phase-space~\cite{Maliborski:2013jca}.

It has been our long-term goal to assemble a framework for fully dynamical, numerical
evolutions in AdS spacetimes
and the achievements presented in this chapter serve as the first step in this direction.
In order to accomplish this goal, one has to go beyond the 
standard methods of NR in, at least, two obvious points.
First of all, AdS spacetimes are not globally hyperbolic. In asymptotically
AdS spaces the boundary plays an  ``active role'' for the bulk
evolution. This is easily visualised in the Penrose diagram of
AdS, which has a timelike boundary. Physically, null geodesics
in AdS reach the boundary for a finite affine parameter. One
thus often refers to an asymptotically AdS space as a ``box'',
having in mind that AdS boundary conditions directly affect the
bulk physics \cite{Wald:1980jn,Ishibashi:2003jd,Ishibashi:2004wx}.
In the gauge/gravity correspondence, the choice of the
AdS boundary conditions is dictated by a holographic prescription
\cite{Son:2002sd,Herzog:2002pc,Skenderis:2008dh,Skenderis:2008dg}.
Secondly, from the viewpoint of the duality, $D$-dimensional AdS
spacetimes are relevant. 
Thus, we intend to develop a framework that could be used in AdS$_D$, in particular
for $D=5$, which is related by the correspondence to 4-dimensional gauge
theories. The latter issue has been recently addressed, e.g., in Refs.~\cite{Zilhao:2010sr,
Witek:2010xi,Witek:2010az,Zilhao:2011yc,Yoshino:2009xp,Sorkin:2009wh,Sorkin:2009bc,Nakao:2009dc,
Shibata:2009ad,Okawa:2011fv,Lehner:2010pn} and in recent
reviews on the topic~\cite{Yoshino:2011zz,Yoshino:2011zza,Cardoso:2012qm,Sperhake:2013qa}.
 
Here we shall focus
on the former issue: the active role of boundary conditions.
The dynamics of BHs in AdS, and especially the role of spatial infinity
(``the box'') is poorly understood. In contrast to the asymptotically
flat case, interesting new phenomena may occur in AdS backgrounds.
For instance, superradiance effects have been shown to make small (as
measured by the AdS radius)
rotating BHs unstable, through a sequence of
reflections at the boundary and amplifications close to the ergoregion
\cite{Hawking:1999dp,Cardoso:2004hs,Cardoso:2004nk,Cardoso:2006wa,
Kodama:2009rq,Murata:2008xr,Kodama:2007sf,Aliev:2008yk,Uchikata:2009zz}.
The final state of this instability could be a new non-axisymmetric BH
configuration, which is also supported by recent gravity/hydrodynamics
arguments \cite{Cardoso:2009bv,Cardoso:2009nz}. Notice that
non-axisymmetric BHs are strictly forbidden in asymptotically flat
spacetimes \cite{Hollands:2006rj,Hollands:2008wn}, so the boundary does
play an important role in the description of BHs.

In order to identify 
the influence of the boundary on the bulk evolution, 
we consider here a toy model for AdS. 
We set the cosmological constant to zero and impose mirror-like
boundary conditions on a box that contains the dynamical system. This
mimics the AdS global geometry, keeping the local geometry of vacuum
models. We choose the dynamical system to be a BH binary, starting
at some given distance, producing either a head-on collision or an
inspiralling merger. In the latter situation we consider the initial
BHs without intrinsic angular momentum. These are, by now,
very well tested systems when purely outgoing boundary conditions are
imposed. Thus we will be able to see clearly the modifications due to
the non-outgoing boundary conditions in systems with non-trivial dynamics.

For the post-merger dynamics the inspiralling binaries provide a more
interesting analysis than the head-on collision case, because the initial
center-of-mass energy transferred into gravitational radiation is by more 
than one order of magnitude larger in the former case as compared to
the latter. Immediately after the merger, the system will contain a 
single (spinning or non-spinning) BH plus gravitational
radiation. This radiation will then be (repeatedly) reflected off the
boundary and interact  with the BH.

The first non-trivial result is that we \textit{can} follow the
numerical evolution for up to two reflections off the boundary of the
gravitational radiation produced in the merger. \textit{A priori} it 
was not guaranteed that this could be achieved, since it is not known
whether the formulation of the Einstein equations that we use provides a
well defined initial value boundary problem together with the boundary
conditions we impose. Our simple setting actually provides a first attempt
to test the (numerical) stability of the initial boundary value problem in a
non-globally hyperbolic spacetime.  We find that our numerical results
are at least second-order convergent for at least two  reflections off
the wall,  after which we gradually lose convergence.
A deeper study of these issues is clearly needed, as
well as an exploration of how the convergence (and remaining results)
change for different boundary conditions.

During the window of numerical convergence, we study the properties of both
outgoing \textit{and ingoing} gravitational radiation. The usual studies of BH
binaries with outgoing boundary conditions, focus only on the Weyl scalar
$\Psi_4$, which describes outgoing gravitational waves. However, an equally
relevant quantity for the description of gravitational radiation is the scalar
$\Psi_0$, which encodes ingoing waves, but which is seldom discussed in the
literature. 
Due to our special boundary conditions and setup, we are able to verify 
certain relations between these two quantities in a numerical evolution 
for the first time. This also provides a test on the correctness and meaning 
of the boundary conditions we have imposed. 

By analysing the properties of the apparent horizon of the BH produced in the
merger and after each interaction with the gravitational wave packet, we
estimate the amount of energy and angular momentum that is transferred from the
radiation into the BH per interaction. 
In case of the inspiralling binary the boxed BH is spinning and we expect
superradiant scattering of the waves generated during merger to become
important; in fact, the back and forth bouncing of the waves at the reflecting
wall and their subsequent amplification by superradiance close to the ergoregion
are expected to turn the system into a BH
bomb~\cite{Press:1972zz,Cardoso:2004hs,Cardoso:2004nk}.
Thus, these simulations will be the
first attempt at a non-linear study of the BH bomb. An important open question
concerns the understanding of how the evolution proceeds and 
the end point of the instability. 
This can only be achieved through non-linear studies. A final
statement on this issue will, however, require further analysis than that
provided herein.

This chapter is organized as follows:
In Sec.~\ref{sec:formulation} we review the numerical framework 
with particular focus on the implementation
of the ``spherical'' boundary and the imposed boundary conditions.
The numerical results of our simulations are shown in
Sec.~\ref{sec:BBoxnum}.  
In Sec.~\ref{sec:conclusions} we close with a 
discussion of the results and prospects for the future. 
Appendix~\ref{ssec:appsnapshotsBox} exhibits some snapshots for
visualising the evolution of the system we have studied. 
Appendix~\ref{sec:appcubebox} describes the simulations with a cubic,
rather than spherical, box. 
The results presented in this Chapter are based on the publication \cite{Witek:2010qc}.
%
\section{Numerical framework\label{sec:formulation}}
%
In order to numerically generate a solution to the Einstein field equations,
it is most convenient to view the problem as a initial value problem. 
Therefore we employ the $3+1$-decomposition of the spacetime and evolve 
the resulting $3+1$-Einstein's equation in the (generalized) BSSN form 
together with the moving puncture approach
(see Chapter~\ref{chapter:NRframework} for a detailed discussion).
We evolve the system at hand with the {\textsc{Lean}} code \cite{Sperhake:2006cy,Sperhake:2007gu}.
The key ingredient in which our current numerical framework differs from
previous implementations of the {\textsc{Lean}} code and most other NR codes is
the outer boundary condition, which we will discuss in more detail in the
remainder of this section.

The vast majority of numerical simulations
of BH binaries has been concerned with asymptotically flat spacetimes
and consequently employed either of the following boundary treatments:
(i) outgoing Sommerfeld conditions on Cartesian grids of finite size,
as described for example in \cite{Alcubierre:2002kk}; (ii) outgoing radiation
\cite{Pollney:2009yz, Pollney:2009ut} with multipatch methods,
including {\em Cauchy characteristic} wave extraction
\cite{Reisswig:2009us,Reisswig:2009rx} and (iii)
constraint preserving boundary conditions combined with multidomain
methods~\cite{Rinne:2007ui,Pazos:2009vb}.

In contrast we will study the dynamics of BH spacetimes under
the influence of a reflective outer boundary. It is natural to use
for this purpose an outer boundary of spherical shape. Most importantly,
this avoids mixing of different gravitational wave multipoles as would
occur in the case of a reflective, cubic outer boundary. This is
discussed in more detail in Appendix~\ref{sec:appcubebox} where we
compare simulations using both types of boundary. Except for this
comparison, however, we will exclusively study spherical
outer boundaries or, rather, approximate these by using so-called {\em
Lego} spheres; cf.~Sec.~3 in \cite{Shoemaker:2003td}.
In the left panel of Fig.~\ref{fig:Boxfoliation} we sketch the foliation of the spacetime
under consideration, suppressing one spatial dimension for simplicity. 
The numerical domain, i.e., the Lego sphere is visualized by a dark (red) 
domain on each timeslice $\Sigma_{t+n\delta t}$. 
Their numerical implementation is illustrated in the right panel of Fig.~\ref{fig:Boxfoliation}
which schematically displays a computational
domain using four refinement levels with one or two components each. The
individual components are labelled $G^i_m$ where the indices $i$
and $m$ denote the refinement level and component number.
Note that one spatial dimension is suppressed for visualisation purposes.
In order to update a grid function at a particular vertex, we
require information from neighbouring points because of the discretization
of spatial derivatives
in the evolution equations. The exact number $n$ of neighbouring points
required in each direction depends on the finite difference stencils
employed. While $n=3$ for the $4^{\rm th}$ order accurate stencils
used in our simulations, we use $n=1$ for simplicity in our illustration in
the right panel of Fig.~\ref{fig:Boxfoliation}. Consider first the dark (blue) shaded area inside the
inner solid circle of radius $R_B$. Each point in this {\em regular}
domain can be updated straightforwardly
provided we also have valid data on the boundary points marked by $\times$
symbols. Points outside the circle of radius $R_B$ are not
required for updating regular points and are simply ignored in the
numerical evolution. 
The specific boundary condition is then determined
by the manner in which we update grid functions on the boundary points
marked as $\times$ in the figure.

In order to mimic the global structure of an Anti-de Sitter spacetime
we effectively enclose the BH binary inside a spherical mirror
and set
\begin{align}
\label{eq:Boxrefbc}
\frac{\partial}{\partial t} f = & 0
\,,
\end{align}
at each boundary point with $f$ denoting any of the BSSN variables.
The use of fourth-order stencils adds one complication to this picture:
the upgrade of a grid point requires two neighbors, so that points
right next to the boundary need special treatment. In practice, we have
achieved optimal stability properties by evolving these points with
second-order stencils.
Our implementation requires one further ingredient in order
to handle the spurious radiation inherent to numerically generated
initial data of BH binary systems; cf.~\cite{Bode:2007dv}.
In order to avoid contamination of our simulations by such spurious
radiation being trapped inside our reflective boundary we employ
standard outgoing radiation boundary conditions at early times
and only switch on our reflective condition at
\begin{align}
t_{\rm ref} = &  R_B +\Delta t_{\rm pulse}
\,.
\end{align}
In order to avoid a discontinuous jump from outgoing to reflective
boundary conditions, we gradually switch off the time derivative
$\partial f / \partial t$ using a weighting factor $w(t)$ which smoothly
decreases from 1 to 0 over an interval 
$\Delta t = 10~M$ and $\Delta t = 20~M$ for the head-on collision and inspiral, respectively.
The duration of the spurious wave pulse $\Delta t_{\rm pulse}$
is estimated from previous simulations of similar setups in
asymptotically flat spacetimes as for example presented in
Refs.~\cite{Baker:2006yw,Sperhake:2006cy,Berti:2007fi}.
The spurious radiation is thus given sufficient time to leave
the computational domain.
\begin{figure}[htpb!]
\begin{center}
\begin{tabular}{cc}
\includegraphics[width=0.5\textwidth]{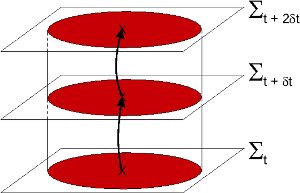} &
\includegraphics[width=0.4\textwidth]{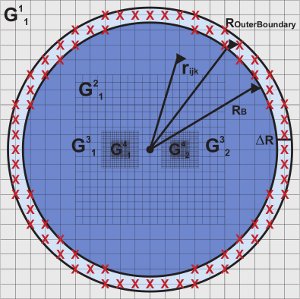}
\end{tabular}
\end{center}
\caption{\label{fig:Boxfoliation} 
Left:
Sketch of the foliation for the numerical
evolution of BH binaries in a (spherical)  box.  The location of the
considered numerical domain on each spatial hypersurface is shown as a
dark (red) sphere.
Right:
Illustration of a (Lego-)spherical outer boundary.
}
\end{figure}
%

\section{Numerical results}\label{sec:BBoxnum}
Our numerical study focuses on two types of binary BH
initial configurations; (i) head-on collisions of non-spinning
BHs starting from rest and (ii) quasicircular inspiral of
non-spinning holes.
In the remainder of this work we label these as HD and IN simulations.
The initial parameters of all our simulations
as well as the structure of the computational domain and the position of the
outer boundary $R_B$ are summarised in Table \ref{tab:Boxsphereruns}.
In all models the initial coordinate separations between the two BH has been set to 
$d=6.517~M$ and their irreducible mass to $M_{irr,i}=0.483~M$.
Unless denoted otherwise, the results presented refer to the highest
resolution available.

%
\begin{table*}
\begin{tabular}{cccccc}
\hline
Run  & Grid Setup   & $R_B/M$   & $P_{i}/M$  &
$J^{AH}_{\rm fin}/M^2$ & $J^{rad}_{\rm fin}/M^2$    
\\ \hline
IN1 & $\{(48,24,12,6)\times(1.5, 0.75),~1/56\}$ & $40$ & $\pm 0.133$ & $0.69$ & $0.70$  \\
IN2.1 & $\{(48,24,12,6)\times(1.5, 0.75),~1/48\}$ & $30$ & $\pm 0.133$ & $0.69$ & $0.65$  \\
IN2.2 & $\{(48,24,12,6)\times(1.5, 0.75),~1/52\}$ & $30$ & $\pm 0.133$ &  & $0.65$  \\
IN2.3 & $\{(48,24,12,6)\times(1.5, 0.75),~1/56\}$ & $30$ & $\pm 0.133$ &  & $0.65$  \\
HD1   & $\{(48,24,12,6)\times(1.5, 0.75),~1/60\}$ & $40$ & $0.0$ & $0.0$ & $0.0$ \\
VIS & $\{(48,24,12,6)\times(1.5, 0.75),~1/48\}$ & $48$ & $\pm 0.133$ & &  \\
\hline
\end{tabular}
\caption{\label{tab:Boxsphereruns} Grid structure, as well as initial and final
parameters of the simulated black holes. 
Their initial coordinate distance has been set to $d=6.517~M$ and the irreducible mass of
each BH is $M_{irr,i}=0.483~M$.
The grid setup is given in terms of the radii of the individual refinement levels as well as the
resolution near the punctures $h$ (see Sec.~II E in \cite{Sperhake:2006cy}
for details). The reflective outer boundary is located at radius
$R_B$. The table further shows the Bowen-York
\cite{Bowen:1980yu} parameter for initial linear momentum $P_i$
of the individual holes. $J^{AH}_{\rm fin}$ and $J^{rad}_{\rm fin}$
are the spin of the single hole after merger determined from the AH and
the merger radiation, respectively. All parameters are given in units of the 
ADM mass. We did not monitor the AH properties for all runs, therefore the spin of the
final BH is not determined (empty cells) for some cases.
The Weyl scalars have been extracted at $R_{\rm ex} = 35M$ (IN1, HD1) and
$R_{\rm ex} = 25M$ (IN2), respectively. Model VIS is used in
Appendix \ref{ssec:appsnapshotsBox} for visualisation.
}
\end{table*}
%

\subsection{\label{sec:convergence}Numerical convergence analysis}
%
Before we discuss in detail the physical properties of the BH
binary systems, we test the accuracy of our simulations by performing
a convergence analysis of model IN$2$.
Numerical simulations
based on the finite differencing method typically approximate
the continuum solution of differential equations with a
leading error
that has a polynomial dependence on the resolution, $f_{\rm cont}
= f_{\rm num} + \mathcal{O}(h^n)$. The order $n$ depends on the
specific numerical implementation. Consistency of the code
is tested by evolving the same configurations with low,
medium and high resolution $h_c$, $h_m$ and $h_f$. One straightforwardly
shows that the convergence factor is then given by
\begin{align}
Q \equiv &  \frac{f_{h_c} - f_{h_m}}{f_{h_m} - f_{h_f}}
          = \frac{h_c^n - h_m^n}{h_m^n - h_f^n},
\end{align}
where $f_h$ stands for any of the evolved variables obtained for
resolution $h$.
For the case of contemporary moving puncture codes, the
solution is complicated by the fact that the differential equations
are typically discretized with fourth- (or higher) order accurate
stencils but prolongation in time between different refinement levels
and implementation of outer boundary conditions
is only second-order accurate; see Sec.~IV in Ref.~\cite{Brugmann:2008zz}
and our discussion in Sec.~\ref{sec:formulation}.

Fig.~\ref{fig:BoxconvergencePsi} shows our convergence results for the variables
$\Psi_4$ and $\Psi_0$ obtained for resolutions $h_c=M/48$, $h_m=M/52$ and
$h_f=M/56$. Here the differences $f_{h_m} - f_{h_f}$ have been amplified by the
convergence factors $Q_2=1.26$ and $Q_4=1.47$ expected for second and fourth-order
convergence, respectively. The figure indicates fourth-order accuracy for the
first passage of the wave pulse and a gradual deterioration of convergence to
second-order accuracy afterwards. We believe this to be a consequence of the
different ingredients of the code as discussed above. At early stages, the
dominant error is the discretization of derivatives. As the pulse successively
passes across mesh refinement boundaries and is reflected off the outer
boundary, however, the second-order error in the
prolongation operation becomes dominant and reduces the order of convergence.
We also note, in this context, that well-posedness of the BSSN evolution system
with reflective boundary condition has so far not been demonstrated\footnote{To
  our knowledge, the well-posedness of the system of equations in combination
  with reflecting boundary conditions, as treated here, has not been studied
  yet. Some investigations of the wave equation with this type of boundary
  conditions suggest that it may be ill-posed \cite{Gustafsson1995,David}. These
  investigations also show that the wave equation with periodic boundary
  conditions is a well-posed initial boundary value problem
  \cite{Gustafsson1995,David}, pointing towards interesting future extension of
  our work.}. We can therefore not rule out adverse effects on the long-term
convergence properties due to potential ill-posedness of the continuum system of
equations. In the remainder of this discussion we will restrict ourselves to 2-3
passages of the wave pulse as covered in Fig.~\ref{fig:BoxconvergencePsi} during
which the relative uncertainties in $\Psi_4$ and $\Psi_0$ are $\le 5\%$.

\begin{figure}[htpb!]
\begin{center}
\begin{tabular}{cc}
\includegraphics[clip=true,width=0.5\textwidth]{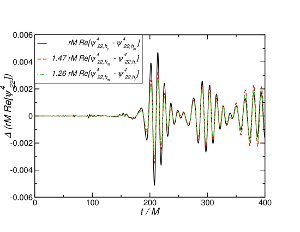} &
\includegraphics[clip=true,width=0.5\textwidth]{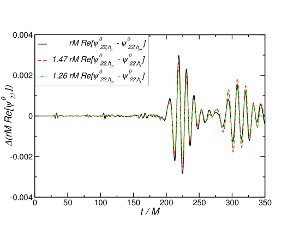}
\end{tabular}
\end{center}
\caption{\label{fig:BoxconvergencePsi} Convergence analysis of the outgoing
Weyl scalar $\Psi_4$ (left panel) and the ingoing Weyl scalar $\Psi_0$
(right panel) for the IN2 runs.  We show the differences of the
$l=m=2$ mode between the coarse and medium and the medium and fine resolution
run. The latter has been amplified by the factors $Q_4=1.47$ and $Q_2=1.26$.
We observe fourth-order convergence in the signal due to the merger whereas the first and
second after-merger cycles show only second-order convergence. The first
two reflected and ingoing wave pulses show second-order convergence.
}
\end{figure}
%

\subsection{\label{ssec:waveforms}Gravitational wave signal and black hole
dynamics}
To our knowledge, this work presents the first analysis of gravitational
waveforms with {\em both} outgoing ($\Psi_4$) and ingoing ($\Psi_0$)
contributions for long-term stable numerical simulations
of BH binaries. For this reason, we first illustrate the
general pattern of the wave signal obtained for model VIS of
Table \ref{tab:Boxsphereruns}. A series of snapshots of both
Newman-Penrose scalars are shown in Fig.~\ref{fig:Boxsnapshots}
in Appendix~\ref{ssec:appsnapshotsBox}
in superposed form.
\begin{figure}[htpb!]
\begin{center}
\includegraphics[width=0.5\textwidth]{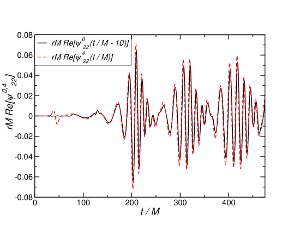}
\end{center}
\caption{\label{fig:BoxcomparePsi0Psi4_l2m2} 
Real part of the $l=m=2$ mode of
$rM\Psi_0$ and $rM\Psi_4$ of run IN1. The ingoing signal
$rM\Psi_0$ has been shifted in time by $\Delta t = 10M$ and in phase by $\pi$
(thus equivalent to an extra minus sign)
to account for the additional propagation time and the reflection.
}
\end{figure}

The gravitational wave signal is dominated by the quadrupole contributions
which we show in Fig.~\ref{fig:BoxcomparePsi0Psi4_l2m2}. For clarity, the
ingoing signal $\psi^0_{22}$ has been shifted in time by $\Delta t=10~M$
in order to compensate for the additional propagation time from the
extraction radius $R_{\rm ex}=35~M$ to the boundary $R_B=40~M$ and back
after reflection. The reflection introduces an additional phase shift
of $\Delta \phi = \pi$ which has also been taken into account in the
figure. Within numerical errors, we find
the resulting outgoing and subsequent ingoing pulses to overlap.

The first outgoing wave pulse, visible in Fig.~\ref{fig:BoxcomparePsi0Psi4_l2m2} around $150\le
t/M \le 250$, is generated during the inspiral, plunge and merger of
the binary and is similar to waveforms obtained for the inspiral of
non-spinning BH binaries in asymptotically flat spacetimes
(cf.~Fig.~1 in Refs.~\cite{Sperhake:2006cy,Brugmann:2008zz}).
Due to the reflecting boundary, however, this wave pulse does not escape
the computational domain. Instead it propagates inwards, interacts
with the post-merger remnant hole and eventually manifests itself as
a second wave pulse shifted by $\Delta t\approx 80~M$ relative to the
first. This process repeats itself many times, with the wave pulse being
presumably distorted (by absorption, superradiance and other curved spacetime effects on wave propagation) upon each interaction with the BH. We now investigate in detail these changes of the wave pulse upon interaction with the BH.
\paragraph{Interaction of the wave pulse with the remnant black hole}
As shown in Fig.~\ref{fig:BoxcomparePsi0Psi4_l2m2}, the outgoing
and subsequent ingoing wave pulses overlap within numerical uncertainties.
We therefore focus on the outgoing signal
in our study of subsequent wave pulses and the gradual changes
caused by successive scattering off the BH.
Changes in the wave pulse are best illustrated by considering the wave
amplitude as shown in Fig.~\ref{fig:Boxoverlapshifted}. Here we superpose
the $l=2$, $m=0$ mode for model HD1 and
the $l=2$, $m=2$ multipoles for models IN1 and IN2
of the first three successive outgoing
wave pulses by applying corresponding time shifts to the waveform.
Clearly, the wave pulses broaden after each scattering off the BH.
\begin{figure}[htpb!]
\begin{center}
\begin{tabular}{ccc}
\hspace{-0.5cm}
\includegraphics[clip=true,width=0.36\textwidth]{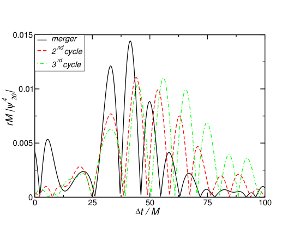} &
\hspace{-0.85cm}
\includegraphics[clip=true,width=0.36\textwidth]{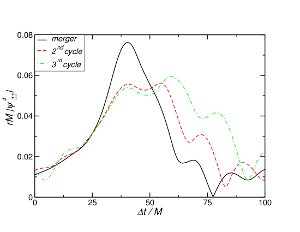} & 
\hspace{-0.85cm}
\includegraphics[clip=true,width=0.36\textwidth]{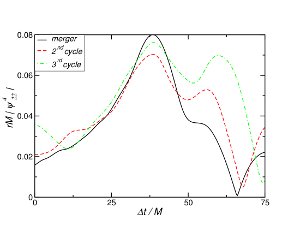} 
\end{tabular}
\end{center}
\caption{\label{fig:Boxoverlapshifted} 
Overlap of the amplitudes of
successive pulses of the same waveform; $l=2$,$m=0$ for the HD1 run (left),
$l=m=2$ for the IN1 (center) and IN2.3 (right panel) simulations,
obtained by time-shifting such that the maxima overlap.
}
\end{figure}
We emphasize that this distortion of the pulse is {\em not} an artifact
of the outer boundary condition as is demonstrated by the good overlap
between the ingoing and outgoing pulses in Fig.~\ref{fig:BoxcomparePsi0Psi4_l2m2}.
\begin{figure}[htpb!]
\begin{center}
\includegraphics[width=0.5\textwidth,clip=true]{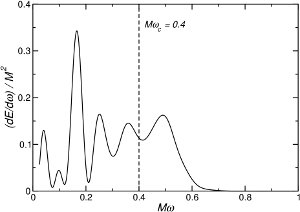}
\end{center}
\caption{\label{fig:Boxspectrum} The energy spectrum for the $l=m=2$
multipole of the outgoing scalar $\Psi_4$, for model IN2.3.
This spectrum corresponds solely to the initial wave packet, i.e.,
the waveform has been truncated immediately before the first reflection
off the boundary.  The vertical line marks the threshold frequency for
superradiance.
}
\end{figure}

One possible explanation for this distortion relies on strong field,
curved spacetime effects: massless waves in curved geometries do not
propagate {\it on} the light-cone.  Indeed, as shown in the classical
work by DeWitt and Brehme, the Green's function for a massless field
in a curved spacetime does not generally vanish inside the light cone
\cite{DeWitt:1960fc}. This gives rise to interesting effects.  It is
well-known that wave propagation in BH spacetimes shows that the signal
can roughly be divided in three parts: (i) the first part is the prompt
response, at very early times, whose form depends strongly on the initial
conditions. This is the most intuitive phase, being a counterpart of the
light cone propagation in flat spacetime; (ii) at intermediate times
the signal is dominated by an exponentially decaying ringing phase,
and corresponds to the excitation of the BH's characteristic modes of
vibration \cite{Berti:2009kk}; (iii) a late-time tail, usually a power
law falloff of the field \cite{Ching:1995tj,Cardoso:2003jf}.  Therefore,
a variety of possible distortions are possible.

Another possible explanation for the increasing width of the pulse arises in the
context of the {\em superradiance mechanism}.  It is well-known that the
scattering of a wave pulse off a rotating Kerr BH will result in superradiant
scattering -- amplification of the scattered wave packet via extraction of
rotational energy from the BH -- if the wave pulse satisfies
\begin{align}
\label{eq:Boxomthresh}
\omega < & m\Omega 
\,, 
\end{align}
where $\Omega \equiv j_{\rm fin}/(2r_+)$ is the BH angular velocity
\cite{Bardeen:1972fi,Teukolsky:1974yv,Cardoso:2005vk}. We note, however,
that Fig.~\ref{fig:Boxoverlapshifted} also indicates a broadening
of scattered wave pulses in the head-on case where no superradiance is
expected. While the observed broadening would be compatible with
superradiance, other effects appear to also influence the shape of
the pulse and our observations do not conclusively demonstrate its presence.

In order to investigate this dispersion in more detail, we plot in
Fig.~\ref{fig:Boxspectrum} the energy spectrum for the dominant
$l=m=2$ mode.
The vertical line in this figure denotes the threshold frequency
$m\Omega \approx 0.4/M$ corresponding to a final spin $j_{\rm fin} =0.69$
as obtained for the post-merger hole for configuration IN2;
cf.~Sec.~\ref{sec:AH}. The figure demonstrates that the $l=m=2$ mode 
does contain contributions which would be subject to
superradiance-induced amplification. These results then suggest that
the low-frequency component of the pulse is amplified due to superradiance,
while the high-frequency component is absorbed. A linear
analysis of superradiance in the Kerr geometry \cite{Teukolsky:1974yv}
shows that superradiant effects are always small, unless the hole is
rotating close to the extremal value. Thus, further studies,
including larger spins of the post-merger hole, are necessary to
comprehensively demonstrate superradiant wave amplification.

\paragraph{\label{ssec:ahdata}Black hole dynamics}
%
\begin{figure}[htpb!]
\begin{center}
\includegraphics[width=0.5\textwidth,clip=true]{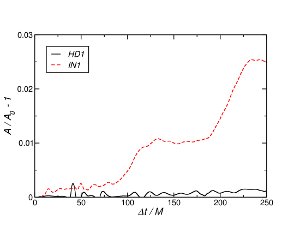}
\end{center}
\caption{\label{fig:BoxAHarea} 
Time evolution of the area of the apparent
horizon for the head-on and inspiral simulation HD1 (solid curve) and IN1 (dashed curve).
The area of the rotating BH increases at regular intervals
corresponding to the propagation time of the pulse between the hole and
the reflective boundary at $R_{\rm ex}=35~M$.
Due to the small amount of radiation
generated during the plunge 
in the head-on case (HD1), the variation in the AH area
is buried in numerical noise.
}
\end{figure}
In contrast to the case of asymptotically flat spacetimes, a BH
binary under the influence of a reflective shell
does not settle down into a stationary configuration
soon after merger. This is due to the repeated interaction with the
wave pulse passing back and forth across the spatially
finite spacetime. The
prolonged dynamical state of the system manifests itself prominently in
the area $A_{\rm AH}$ of the apparent horizon. In Fig.~\ref{fig:BoxAHarea}
we show the fractional deviation $(A_{\rm AH}-A_0)/A_0$ of the apparent
horizon area from its value $A_0$ immediately after merger, i.e. the
first instance a common apparent horizon is found.
As expected, the horizon area remains nearly constant for the duration
of the first passage of the pulse to the reflective shell and back,
$\Delta t \approx 80~M$ and $60~M$, respectively, for models IN1 and
IN2.1\footnote{Small oscillations in
the horizon area are due to numerical uncertainties.}.
The subsequent increase in $A_{\rm AH}$ demonstrates that some fraction of the
gravitational wave pulse energy is absorbed by the hole. It follows another
period of approximately constant horizon area, a further increase
upon the second scattering of the pulse and so on.
Unfortunately, the radiation efficiency is almost 2 orders of magnitude
lower for
head-on collisions (see Table II in Ref.~\cite{Sperhake:2006cy}),
so that the increase in horizon area is buried in the numerical uncertainties.
The head-on case serves as a useful comparison, however, as it demonstrates
that the changes observed for the inspiralling configurations are significant
relative to numerical uncertainties.

The BH mass, defined in terms of the
equatorial radius of the horizon $C_e$ by \cite{Kiuchi:2009jt}
\begin{align}
M = & \frac{C_e}{4\pi}
\,,
\end{align}
shows a similar behaviour as the horizon area. In Fig.~\ref{fig:BoxAHmass}
we plot the fractional deviation $(M-M_0)/M_0$ of the mass
from its value immediately after merger together with the irreducible
mass and the BH spin $J$ of the hole obtained for model IN1.
The mass remains approximately constant until the pulse returns after
its first reflection, then increases, remains constant during the second
passage of the pulse and so on. In contrast, the spin shows
a significant increase only during the first scattering of the pulse off the BH.

We conclude that in each interaction with the gravitational radiation, the
final BH mass increases.  It is interesting to compare the increase in the
horizon mass with the amount of gravitational wave energy radiated during
the last stages of the inspiral, plunge and merger of a corresponding
binary system in an asymptotically flat spacetime which is about $3.5~\%$
of the total energy of the system \cite{Berti:2007fi,Sperhake:2006cy}.
For the IN1 run, we estimate that about $15\%$ of the energy emitted
during the merger is absorbed by the central spinning BH per interaction.
\begin{figure}[htpb!]
\begin{center}
\includegraphics[width=0.5\textwidth]{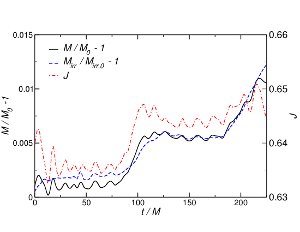}
\end{center}
\caption{\label{fig:BoxAHmass}
Time evolution of the (relative) mass of the BH (solid line) computed by $M =
C_e/4\pi$, the irreducible mass (dashed line) and the total
spin $J = j M^2$ (dashed-dotted line).
}
\end{figure}
Our results are consistent with total energy conservation. Moreover, they are {\it
not} incompatible with superradiant amplification:
typically, absorption of high-frequency waves is more effective than
superradiant amplification of low-frequency waves, such that the
net effect leads typically to absorption by the BH
\cite{Teukolsky:1974yv}.
The prolonged non-stationary character of the post-merger state is also
demonstrated by the time dependence of the BH's final spin.
Immediately after merger, we obtain $j_{\rm fin}= 0.69$ from
Eq.~\eqref{eq:BoxspinArea} in excellent
agreement with corresponding simulations using outgoing
radiation boundary conditions \cite{Berti:2007fi}. Successive interaction
with the reflected wave pulse, however, results in a small but significant
increase in the BH's spin as shown in Fig.~\ref{fig:BoxAHmass}.
As before, the first increase occurs about $\Delta t=80~M$ after merger,
when the pulse has returned to the BH. We estimate
the fractional increase in spin resulting from the
first scattering at about $5~\%$. For comparison, the total
angular momentum radiated in the case of an asymptotically flat spacetime
is reported as $J^{\rm rad}/M^2=0.246$ ($28~\%$ of the initial orbital angular momentum of the system) 
in Table I of Ref.~\cite{Berti:2007fi}.
It thus appears that a significant amount of angular momentum
remains in the form of gravitational waves.
Due to numerical uncertainties
it is not entirely clear whether later periods of interaction between
pulse and hole result in a further transfer of angular momentum
from the wave pulse to the hole or \textit{vice-versa}. Our results indicate,
however, that the amount of angular momentum exchanged in subsequent
interactions is significantly below $5~\%$.
\section{Discussion\label{sec:conclusions}}
The dynamics of BHs in generic spacetimes is a fascinating, yet extremely
challenging problem. The gauge/gravity duality, however, strongly
motivates us to solve dynamical problems with BHs on asymptotically AdS
backgrounds. In this work, we have studied a toy model that captures
one of the fundamental features of such backgrounds: the active role
played by the boundary conditions for the bulk evolution.

We have mimicked the global structure of an AdS background by
introducing a reflecting wall at some radius. Within this cavity we
evolved an inspiralling BH binary and a BH binary starting from rest at
a certain initial distance. Of course, these are very specific initial
configurations, and serve merely as tests for future, possibly more
complex, situations.

Perhaps the most important conclusion of the present work is that these
simulations {\it can} be done and represent the first step to a full numerical
evolution of BHs in AdS spacetimes. Indeed, as observed in Section
\ref{sec:convergence}, it is not known whether the BSSN evolution scheme
together with reflecting boundary conditions is a well-posed initial boundary value
problem. Thus, the convergence we have exhibited, which holds up to two reflections off
the boundary of the gravitational radiation produced in the merger, is the first
of our results.
Among the other results presented here we stress the following:
\begin{enumerate}
  \item
  For the first time, we were able to numerically study
  the scalar $\Psi_0$, describing ingoing waves, and check in the numerical
  data  the simple relations
  between $\Psi_0$ and $\Psi_4$ (cf. Eqs.~\eqref{eq:BoxPsi4_ddh_defn},
  \eqref{eq:BoxPsi4_ddh_defn2}, \eqref{asymptotics} and
  Fig.~\ref{fig:BoxcomparePsi0Psi4_l2m2}).
\item
  Our results are consistent with the intuitive
  expectations for a wave packet of radiation (generated during
  inspiral plus merger) travelling back and forth between the mirrorlike
  wall and the BH: part of this radiation is absorbed when
  interacting with the BH (especially high-frequencies). We estimate
  that about $15\%$ of the wave packet's energy is absorbed by the
  BH per interaction, at least during the first cycles.
\item
    The wave packet is clearly distorted upon interaction
  with the hole, which means some frequencies are absorbed
  more efficiently than others. In principle, we should observe a small but non-zero
  superradiance. Unfortunately, we have not obtained incontrovertible proof
  of non-linear superradiance from our numerical data. Presumably, the
  system will become unstable after a sufficiently long time, since the radiation should
  be exponentially amplified, once the high-frequency components had
  time to be completely absorbed by the BH. In future work
  we plan to investigate these instability studies further by considering
  a highly spinning, final BH produced by the inspiral of spinning
  BHs.
\end{enumerate}

One issue that we have not explored in this paper is the potential influence of
the box on the premerger dynamics. For sufficiently small size boxes, it is
plausible that the radiation produced in the inspiral may be reflected off the
boundary and interact with the binary \textit{before} the merger. This might
produce observable signatures in the premerger dynamics and even in the
properties of the remnant black hole. Whereas we seem to observe some hints of
this effect in our numerical data, a more exhaustive analysis is required to
produce some precise statements.


\chapter{Superradiant instabilities in astrophysical systems}
\label{chapter:Stability}
\section{Introduction}\label{sec:MFIntro}
Nowadays, BH physics has grown into a mature and fully developed branch of GR.
The theory has been worked out many decades ago, and tremendous progress has been made in actually observing
some of the fascinating general relativistic effects. 
From X-spectra on the inner edge of accretion disks, which probe the innermost stable circular orbit
of the geometry, to gravitational wave physics, 
``precision BH physics'' is a new and rapidly developing field 
\cite{Arvanitaki:2009fg, Arvanitaki:2010sy,McClintock:2009as,Antoniadis:2013pzd}.

One of the fundamental reasons why precision BH physics is possible at all, are the no-hair and uniqueness theorems: 
BHs in $4$-dimensional, asymptotically flat spacetimes must belong to the Kerr-Newman family
and are, thus, fully specified by three parameters only:
their mass, angular momentum and electric charge 
(see e.g. Ref.~\cite{Chrusciel:2012jk}, or Carter's contribution to Ref.~\cite{Hawking:1979ig}). 
In more colloquial terms,
this is commonly expressed by saying that BHs have no hair or, rather, have three hairs only. 
This simple yet powerful result has far reaching consequences:
Given some arbitrary perturbations with the same conserved charges, 
they must all decay to the same final state, namely one BH with those charges. 
By now, there are a plethora of studies, at the perturbative and fully non-linear level,
investigating the approach to the final stationary state (see, e.g., Ref.~\cite{Berti:2009kk} and references therein), 
which we summarize below.
The present chapter is part of a larger publication~\cite{Witek:2012tr} that appeared 
simultaneously with an accompanying work by Dolan~\cite{Dolan:2012yt}, exploring the dynamics of 
complex, massive scalar fields.
Additionally, in Ref.~\cite{Witek:2012tr} we have extended our study to investigate the time evolution
of massive vector fields, going up to very high spins (close to extremality) of the background
Kerr BH. A complentary study in the slow-rotation approximation has been 
presented by Pani et al~\cite{Pani:2012bp,Pani:2012vp}.
More recent reviews on the BH bomb mechanism and BH--scalar field systems can be found in 
Refs.~\cite{Berti:2013uda,Cardoso:2013zfa}.
%
\paragraph{Generic response of a BH spacetime to external perturbations}
The generic behavior of massless fields around a BH can be divided into three parts: 
\noindent(i) A prompt response, depending on the initial data, at early times, 
which is the counterpart to light-cone propagation in flat space; 
\noindent(ii) An exponentially decaying ``ringdown'' phase at intermediate times, where the BH is ringing 
with its characteristic quasinormal modes (QNMs). 
This stage typically dominates the signal, and its properties, such as vibration frequency and decay timescale,
depend solely on the parameters of the {\it final} BH \cite{Berti:2009kk}. 
The long-lasting oscillation of the lowest QNMs is the most important stage in the life of any
field around a BH. Its lifetime, or quality factor, depends solely on the BH spin \cite{Berti:2009kk}.
Specifically, the lifetime tends to increase with growing spin
and the decay timescale approaches zero for nearly extremal BHs.
Because of the no-hair theorem, the detection of QNMs allows to uniquely determine
the BH charges and provides tests of GR \cite{Berti:2009kk,Berti:2005ys,Kamaretsos:2011um};
\noindent (iii) At late times, the signal is dominated by a power-law fall-off, dubted ``late-time tail''
\cite{Price:1971fb,Leaver:1986gd,Ching:1995tj}. 
Tails are caused by backscattering off spacetime curvature 
and, as such, also appear in other situations where light propagation is not {\it on} 
the light cone \cite{Cardoso:2003jf}.

\paragraph{Superradiant effects}
%
Among the most exciting phenomena is the {\textit{superradiant effect}}
\cite{zeldovich1,zeldovich2,Bekenstein:1973mi}: 
In a scattering experiment off a BH the scattered wave will be amplified if its 
complex frequency $\omega=\omega_{R}+\imath\omega_{I}$ satisfies the superradiant condition
\begin{align}
\label{eq:MFSRcond}
\omega_{R} < & m\Omega_H
\,,
\end{align}
where $m$ is the azimuthal wave quantum number and $\Omega_H$ the angular velocity of the BH horizon.
Then, the excess energy is withdrawn from the object's rotational energy \cite{zeldovich1,zeldovich2}
and, in a dynamical scenario, the BH would spin down.
The effect can be attributed to the existence of negative-energy states in the ergo-region, 
and dissipation at the event horizon.

Superradiance is the chief cause of a number of exciting phenomena in BH physics:
\noindent{(i)}
Generic perturbations are damped away to infinity and across the event horizon.
Because rotating BHs amplify waves that fulfill the superradiant condition~\eqref{eq:MFSRcond}, 
the amplification factors as well as the quality factor, or lifetime, of these superradiant modes 
increase with rotation.

\noindent{(ii)}
Typically, any satellite around a BH spirals inwards as time goes by, due to
gravitational wave emission and energy conservation. 
Emission of radiation at infinity results in a larger binding energy of the particle.
Because superradiance implies the extraction of the BH's rotational energy, 
it is possible that the energy deficit comes entirely from the BH kinetic energy. 
In this way, satellites around rapidly spinning BHs can {\it in principle} orbit 
on a fixed radius -- on so-called floating orbits -- for a much longer time, tapping the BH's kinetic energy. 
In BH binaries, this effect can dominate in the presence of resonances \cite{Cardoso:2011xi,Yunes:2011aa,Cardoso:2012zn}.
Notice, that this phenomenon is analogous to tidal acceleration, e.g., 
in case of the Earth-moon system ~\cite{Hut:1981,Verbunt}. 

\noindent{(iii)}
A further interesting effect enters the game if we enclose the spinning BH inside 
a perfectly reflecting cavity.
As was recognized already by Zel'dovich \cite{zeldovich1,zeldovich2}, this way one would create an unstable system. 
Any initial perturbation will get successively amplified near the BH and reflected back at the mirror, 
thus creating an instability, which was termed the ``BH bomb'' \cite{Press:1972zz,Cardoso:2004nk}. 
Whereas the setup appears physically artifical at first glance, 
in Nature the role of the mirror can actually be realized by 
anti-de Sitter spatial infinity or massive fields.
In the former case, the BH bomb translates into a real, physical instability of (small) rotating BHs in 
asymptotically AdS spacetimes \cite{Cardoso:2004hs,Cardoso:2006wa,Kodama:2007sf,Uchikata:2009zz}.

\noindent{(iv)}
Finally, of direct interest for the present study is the fact that massive scalar fields 
around Kerr BHs are prone to a BH bomb-like instability, 
because the mass term effectively confines the field 
\cite{Damour:1976kh,Zouros:1979iw,Detweiler:1980uk,Furuhashi:2004jk,Cardoso:2005vk,Dolan:2007mj,Hod:2011zz}.
The instability is described by the time dependence of the field, 
$\Psi\sim e^{-i\omega t}$ with complex frequency $\omega$
For small coupling $M\mu_S\ll1$ between the BH mass $M$ and scalar field mass $\mu_S$, 
the characteristic frequency giving rise to the instability is
\cite{Pani:2012bp}
\footnote{Notice the difference of a factor $2$ to the original result \cite{Detweiler:1980uk}.}
\begin{align}
\label{eq:MFtsScaSmallMu}
M\omega_I = & \frac{a}{48M}(M\mu_S)^9
\,,
\end{align}
where the BH is spinning at angular momentum $J=aM$. 
In the opposite limit, i.e., for very large mass couplings $M\mu_S\gg1$, the instability time-scale is \cite{Zouros:1979iw}
\begin{align}
\label{eq:MFtsScaLargeMu}
\frac{\tau}{M} = \frac{1}{M\omega_I} = & 10^7\exp(1.84 M\mu_S)
\,.
\end{align}
The maximum (scalar field) growth rate of $M\omega_I\sim1.5\cdot10^{-7}$ has been found for the dipole mode
and the mass coupling $M\mu_S=0.42$ in the background of a Kerr BH with $a/M=0.99$ \cite{Dolan:2007mj,Cardoso:2005vk}.
These calculations have been performed in the linear regime, thus neglecting back-reaction effects such
as the BH spin-down or effects due to non-linear self-interaction of the scalar field.
Therefore, the final state of the superradiant instability in the fully dynamical regime is not known, 
partly because it requires the non-linear evolution of Einstein's equations for a timescale of order $10^6M$ 
(in geometric units $G=c=1$). 
A plausible evolution scenario consists on an exponentially growing scalar condensate outside the BH, 
extracting energy and angular momentum from the BH until the superradiant extraction stops, i.e., 
until condition~\eqref{eq:MFSRcond} is no longer satisfied. 
Further interesting new phenomena arise when we consider non-linear interaction terms 
such as the bosenova-type collapse recently reported in \cite{Kodama:2011zc,Yoshino:2012kn,Mocanu:2012fd}.
%

\paragraph{Superradiant instability in astrophysical systems}
We have argued beforehand that massive fields in the vicinity of BHs give rise to 
BH bomb-like, superradiant instabilities.
However, the effect is very weak for known standard model particles
in astrophysical environments:
For example, the mass coupling for the lightest known elementary scalar particle, the pion, 
around a solar mass BH is $M\mu_S\sim10^{18}$, resulting in an instability timescale much larger than the 
age of the universe.

Nevertheless, the superradiant instability around astrophysical BHs might become significant
if there exist fields with tiny, but non-vanishing mass.
One exciting possibility is provided by axions, ultralight bosonic states emerging
from string-theory compactifications, which have not been ruled out by current experiments.
In the ``axiverse'' scenario the existence of an entire 
landscape of ultra-light pseudo-scalar fields has been proposed,
covering a mass range from $10^{-33}eV\le\mu_S\le10^{-8}eV$
(see ~\cite{Arvanitaki:2009fg, Arvanitaki:2010sy,Kodama:2011zc} for recent overviews).
In fact, the existence of ultra-light axions leads to a plethora of possible observational implications and
signatures, such as 
modifications of the cosmic-microwave background polarization 
(for $10^{-33}eV\le\mu_S\le10^{-28}eV$).
They are also anticipated to make up a fraction of dark matter if $10^{-28}eV\le\mu_S\le10^{-22}eV$.
Of particular interest in the context of BH physics are axions in the mass range
$10^{-22}eV\le\mu_S\le10^{-10}eV$ \cite{Arvanitaki:2009fg, Arvanitaki:2010sy,Kodama:2011zc}.
Then, the time scales for the superradiant instability becomes astrophysically significant,
giving rise to a number of interesting effects:
\noindent{(i)}
A bosonic cloud bounded in the vicinity of a Kerr BH might create a ``gravitational atom'',
which can be de-excited by the emission of gravitons, thus carrying away BH angular momentum;
\noindent{(ii)}
If the accretion of bosons from this cloud is efficient, the rotation of the BH can be sustained 
and it might be turned into a GW pulsar;
\noindent{(iii)}
If, on the other hand, the accretion from the axionic cloud is not efficient enough,
the BH will eventually spin down, thus yielding gaps in the Regge plane 
(the phase-space spanned by mass and spin parameter of the BH).
Further possible effects have been discussed in 
Refs.~\cite{Arvanitaki:2009fg,Arvanitaki:2010sy,Cardoso:2011xi,
Yunes:2011aa,Alsing:2011er,Kodama:2011zc,Yoshino:2012kn,Mocanu:2012fd,Berti:2013uda,Cardoso:2013zfa}.

Similar superradiant instabilities are expected to occur for massive hidden $U(1)$ vector fields, 
which are also a generic feature of extensions of the standard model 
\cite{Goodsell:2009xc,Jaeckel:2010ni,Camara:2011jg,Goldhaber:2008xy}.  
While superradiant instabilities have been widely studied for massive scalar
fields~\cite{Press:1972zz,Damour:1976kh,Cardoso:2004nk,Cardoso:2005vk,Dolan:2007mj,Rosa:2009ei,Cardoso:2011xi}, 
the case of massive vector fields is still uncharted territory, 
the main reason being that the Proca equations seem to be non-separable
in the Kerr background. 
After the defence of this PhD thesis, investigations in the slow-rotation limit~\cite{Pani:2012vp,Pani:2012bp}
and full-blown numerical studies~\cite{Witek:2012tr}
have shed more light on the phase-space and time progression of Proca fields in Kerr.
Furthermore, so far most studies on the massive boson instability are done in Fourier space. 
As far as we are aware there has only been one attempt, 
besides the most recent studies in \cite{Yoshino:2012kn,Dolan:2012yt},
at studying the massive scalar field instability 
in the time domain, with generic initial conditions:
the work by Strafuss and Khanna \cite{Strafuss:2004qc}. 
We believe that some of the conclusions reported in that work are wrong; 
specifically the reported instability growth rate of $M\omega_I\sim2\cdot10^{-5}$, which
is two orders of magnitude larger than previous analytic results~\cite{Cardoso:2005vk,Dolan:2007mj}
and more recent numerical studies~\cite{Yoshino:2012kn}.
Here, we will attempt a correct explanation for these puzzling results.

In this chapter we will present the evolution of massive scalar fields in Schwarzschild and Kerr 
BH background in the time-domain, which is part of the publication~\cite{Witek:2012tr}.
We will provide generic Gaussian wave packets as well as quasi-bound states as initial configurations.
Investigations of massive vector fields around spinning BHs, 
which are expected to have a larger amplification factor than massive scalar fields
will be reported elsewhere~\cite{Witek:2012tr}.
The exploration of the dynamical coupling of the massive scalar to the BH spacetime as well as 
non-linear couplings and self-interaction terms are work in progress and subject to future investigations.

This chapter is organized as follows:
In Sec.~\ref{sec:MFFramework} we present the numerical framework, describing the formulation of the 
problem as Cauchy problem, the setup of initial configurations and the background spacetime as well as 
extraction of the scalar field.  
Sec.~\ref{sec:MFResScalar} is devoted to the numerical results of the (massive) scalar field evolution.
In particular, we present a number of benchmark tests to verify our implementation before studying more generic setups.
We will show that the evolution of a massive scalar has a non-trivial pattern,
which can be explained in terms of multi-mode excitation.
We believe that this pattern also describes the results reported by 
Strafuss and Khanna \cite{Strafuss:2004qc}.
Finally, we will finish this chapter with a summary of our results and some concluding remarks in 
Sec.~\ref{sec:MFconclusion}.

\section{Numerical framework}\label{sec:MFFramework}
I intend to explore the stability properties of Kerr BHs against massive 
scalar field perturbations
and  track their evolution in this background
in the time domain.
In the following I will describe the formalism and numerical framework that I have developed in order 
to achieve this goal.
\subsection{Action and equations of motion}\label{ssec:MFEoM}
I focus on the generic action involving massive scalar $\Psi$ and vector fields $A_{\mu}$ 
with mass $\mu_S$ and $\mu_V$, respectively,
\begin{align}
\label{eq:MFaction}
S = & \int d^4x \sqrt{-g} 
      \left( \frac{R}{k} - \frac{1}{4}F^{\mu\nu}F_{\mu\nu} - \frac{\mu_V^2}{2}A_{\nu}A^{\nu}
            -\frac{k_{\rm axion}}{2}\Psi \,{^*}F^{\mu\nu}F_{\mu\nu} 
\right.\non\\ & \left.
            -\frac{1}{2}g^{\mu\nu}\Psi_{,\mu}\Psi_{,\nu} - \frac{\mu_S^2}{2}\Psi^2-V(\Psi) \right)
\,,
\end{align}
where $k_{\rm axion}$ parametrizes the axion-like coupling, $F_{\mu\nu} \equiv \na_{\mu}A_{\nu} - \na_{\nu} A_{\mu}$
is the Maxwell tensor and $\,^{\ast}F^{\mu\nu} \equiv \frac{1}{2}\eps^{\mu\nu\rho\si}F_{\rho\si}$
is its dual.
Also $\epsilon^{\mu\nu\rho\si}\equiv \frac{1}{\sqrt{-g}}E^{\mu\nu\rho\si}$ and $E^{\mu\nu\rho\si}$ 
is the totally anti-symmetric Levi-Civita symbol with $E^{0123}=1$. 
\footnote{The identity $\nabla_{\nu} \,{^*}F^{\mu\nu}=0$ is useful to derive the equations of motion for the Chern-Simons term.}
The resulting equations of motion are
\begin{subequations}
\label{eq:MFEoMgen}
\begin{align}
\label{eq:MFEoMScalar}
& \left(\nabla_{\nu}\nabla^{\nu}-\mu_S^2\right)\Psi - \frac{k_{\rm axion}}{2}\,{^*}F^{\mu\nu}F_{\mu\nu}- V'(\Psi) = 0
\,,\\
\label{eq:MFEoMVector}
& \nabla_{\nu} F^{\mu\nu}+\mu_V^2A^\mu+2k_{\rm axion}\,{^*}F^{\mu\nu}\partial_{\nu}\Psi = 0
\,,\\
\label{eq:MFEoMTensor}
&  \frac{1}{k} \left(\frac{1}{2}g^{\mu\nu}R-R^{\mu\nu}\right) 
  + \frac{1}{8}F^{\alpha\beta}F_{\alpha\beta}g^{\mu\nu}
  -\frac{1}{2}F^{\mu}_{\,\,\alpha}F^{\nu\alpha} 
  + \frac{1}{4}\mu_V^2A_{\alpha}A^{\alpha}g^{\mu\nu}
  -\frac{\mu_V^2}{2}A^{\mu}A^{\nu}
\nonumber\\
& +\frac{1}{2}g^{\mu\nu}\left(\frac{1}{2}\Psi_{,\alpha}\Psi^{,\alpha}-\frac{\mu_S^2}{2}\Psi^2-V(\Psi)\right)
  -\frac{1}{2}\Psi^{,\mu}\Psi^{,\nu} = 0
\,.
\end{align}
\end{subequations}
I restrict myselve to vanishing axion-like coupling $k_{\rm axion}=0$
and investigate massive scalar fields only. 
Furthermore, I focus on weak fields throughout this study,
whereas non-linear effects and back-reaction on the metric will be studied elsewhere.
At linear order in the scalar field amplitude, the equation of motion 
for the metric, Eq.~\eqref{eq:MFEoMTensor}, is identical to Einstein's equations in vacuum. 
Therefore, I consider a Kerr background and focus on 
Eqs.~\eqref{eq:MFEoMScalar} 
in the remainder of the chapter.
In Boyer-Lindquist coordinates, the metric is described by
\begin{align}
\label{eq:MFKerrBL}
ds^2 = & - \left(1-\frac{2Mr}{\Sigma}\right) dt^2
         + \left(1+\frac{2Mr}{\Sigma}\right) dr^2 
         + \sin^2\theta\left(r^2+a^2+\frac{2Ma^2r\sin^2\theta}{\Sigma}\right) d\phi^2 
\nonumber \\ &
         + \Sigma d\theta^2
         + \left(\frac{4Mr}{\Sigma}\right) dtdr - \left(\frac{4Mra\sin^2\theta}{\Sigma}\right) dtd\phi
         - 2a\sin^2\theta\left(1+\frac{2Mr}{\Sigma}\right) drd\phi
\,, 
\end{align}
where
\begin{align}
\Sigma = & r^2+a^2\cos^2\theta, \quad
\Delta = r^2-2Mr+a^2
\,.
\end{align}
This geometry describes a rotating BH with mass $M$ and angular momentum $J=aM^2$. 
Note, that in order to ensure the regularity of the the spacetime, i.e. the existence of an event horizon,
the BH spin is constrained by the Kerr bound $a/M\leq 1$. 
I will set $M=1$ throughout our study.

Because I intend to solve the equations of motion, Eqs.~\eqref{eq:MFEoMScalar},
numerically it is convenient to reformulate them as time evolution problem.
For this purpose I employ the $3+1$-decomposition of the spacetime, discussed in Chapter~\ref{chapter:NRframework}
for the case of dynamical BH spacetimes.

\paragraph{Evolution equations for scalar fields} 
I focus on the setup for a massive scalar field without self-interaction, which 
implies $k_{\rm axion}=V=0$. Its dynamics are determined by the 
Klein-Gordon equation~\eqref{eq:MFEoMScalar}.
I write the equation of motions as a time evolution problem, i.e., 
as a set of first order in time, second order in space partial differential equations (PDEs).
Therefore, I introduce the conjugated momentum 
\begin{align}
\label{eq:MFdefPi}
\Pi = & -\frac{1}{\alpha}(\p_t - \Lie_{\beta}) \Psi
\,.
\end{align}
The definition~\eqref{eq:MFdefPi} provides evolution equations for the scalar field $\Psi$
\begin{align}
\label{eq:MFevolPsi}
\p_t \Psi = & \Lie_{\beta}\Psi - \alpha \Pi
\,,
\end{align}
where $\Lie_{\beta}\Psi = \beta^k \p_k\Psi$. 
By applying the $3+1$-split, the Klein-Gordon equation yields the evolution equations
for the momentum
\begin{align}
\label{eq:MFevolPi}
\p_t \Pi = & \Lie_{\beta}\Pi - D^i\alpha D_i \Psi + \alpha ( - D^i D_i \Psi + K \Pi + \mu_S^2 \Psi)
\,,
\end{align}
where $\Lie_{\beta}\Pi = \beta^k \p_k \Pi$ and I consider the background
spacetime in $3+1$-form given by Eq.~\eqref{eq:NRlineeleg}.
%

\subsection{Background in horizon penetrating coordinates}\label{ssec:MFbackground2}
%
For the purposes of our numerical implementation, I consider the Kerr spacetime 
in Cartesian Kerr-Schild coordinates $(t,x,y,z)$ as background spacetime. 
In particular, I employ the spacetime metric (\ref{eq:MFKerrBL}) in $3+1$-form 
given by Eq.~\eqref{eq:NRlineeleg}.
Then, considering a Kerr BH with angular momentum $aM$ in the $z$ direction, 
the lapse function $\al$, shift vector $\be^i$, $3$-metric $\ga_{ij}$ and extrinsic curvature $K_{ij}$ 
in Kerr-Schild coordinates $(t,x,y,z)$ are given by
\begin{subequations}
\label{eq:MFKSlapse}
\begin{align}
\alpha = & (1 + 2 H l^t l^t)^{-1/2} 
\,,\quad 
\beta^i = -\frac{2 H l^t l^i}{1 + 2 H l^t l^t} 
\,,\quad \gamma_{ij}=\delta_{ij} + 2 H l_i l_j 
\,,\\
K_{ij} = & - \frac{1}{\alpha} (l_i l_j \p_t H + 2 H l_{(i}\p_t l_{j)} ) 
\non\\ &
           - 2 \alpha 
               \left( \p_{(i} ( l_{j)} H l^t )+ 2 H^2 l^t l^k l_{(i}\p_{|k|}l_{j)}+ H l^t l_i l_j l^k \p_k H \right)
\,,
\end{align}
\end{subequations}
where  
\begin{align}
\label{eq:MFHfunc}
H = & \frac{M r_{BL}^3}{r_{BL}^4 + a^2 z^2}
\,,\quad
l_{\mu} = \left( 1, \frac{r_{BL} x + a y}{r_{BL}^2 + a^2}, 
                \frac{r_{BL} y - a x}{r_{BL}^2 + a^2}, \frac{z}{r_{BL}} \right)
\,.
\end{align}
The Boyer-Lindquist radial coordinate $r_{BL}$ is indirectly defined via
\begin{align}
\label{eq:MFdefrBL}
\frac{x^2 + y^2}{r_{BL}^2 + a^2} + \frac{z^2}{r_{BL}^2} = & 1
\,.
\end{align}
For our computations I will consider the root
\begin{align}
\label{eq:MFrBL}
r_{BL} = & \frac{1}{\sqrt{2}}\sqrt{R^2 + \sqrt{R^4 + 4 a^2 z^2}}
\,,
\end{align}
where I use the notation $R^2 = x^2 + y^2 + z^2 - a^2$.
Inserting the relations 
\begin{align*}
& l_t = 1\,,\quad l^t = -1\,,\quad l^i = l_i 
\,,\quad
\p_t H = \p_t l_{\mu} = 0
\,,
\end{align*}
into Eqs.\eqref{eq:MFKSlapse}, the spacetime is described explicitly by
\begin{subequations}
\label{eq:MFspacetime2}
\begin{align}
\label{eq:MFKSlapse2shift2}
\alpha = & \frac{1}{\sqrt{1 + 2 H}} 
\,,\quad
\beta^i =  \frac{2 H l^i}{1 + 2 H } 
\,,\quad
\gamma_{ij} =  \delta_{ij} + 2 H l_i l_j 
\,,\\
\label{eq:MFextrcurvature2}
K_{ij} = & \frac{2}{\sqrt{1+2H}}
           \left(   l_{(i} \p_{j)} H 
                  + H^2 ( l_i l^k \p_k l_j + l_j l^k \p_k l_i )
            + H \p_{(i} l_{j)}
                  + H l_i l_j l^k\p_k H \right)
\,.
\end{align}
\end{subequations}

\subsection{Initial data}\label{ssec:MFinitdata}
I will focus on generic-type initial configurations of Gaussian form, which gives rise to many interesting effects
as I will see in Sec.~\ref{sec:MFResScalar}.
Additionally, as a code test and a cleaner way to search for instabilities, I have implemented 
quasi-bound states as initial data. I describe each of these below.

\paragraph{Gaussian initial data}
 
In order to set up a generic Gaussian wave packet I choose initial data of the form
\begin{align}
\label{eq:MFinitdataQNM}
\Psi(t=0) = & 0
\,,\quad 
\Pi(t=0)=\exp{\left(-\frac{(r-r_0)^2}{w^2}\right)} {_0}\Sigma(\theta, \phi)
\,,
\end{align}
where $r_0$ is the center of the Gaussian and $w$ is its width. ${_0}\Sigma(\theta,\phi)$ is a superposition
of spherical harmonics $Y_{l m}(\theta,\phi)$. 
Because our evolutions are carried out on a Cartesian grid, I initialize the spherical harmonics {\it not} in
spherical coordinates $(r,\theta,\phi)$ but Cartesian coordinates $(x,y,z)$
\begin{align}
x = & r \sin\theta \cos\phi \,,\quad y=r \sin\theta \sin\phi \,,\quad z=r \cos\theta \,,
\end{align}
where $r^2 = x^2 + y^2 + z^2$ is the Kerr-Schild radial coordinate.
I list the spherical harmonics up to $l=2$ in Cartesian coordinates 
in Appendix~\ref{app:MFspherharm}.
\paragraph{Quasi-bound state initial data}
I consider a second, useful type of initial setup, which I term quasi-bound state initial data.
Perturbations of massive scalar fields around Schwarzschild or Kerr BHs give rise to long-lived modes, 
or quasi-bound states, which have been investigated in the frequency domain in 
the past \cite{Cardoso:2005vk,Konoplya:2006br,Dolan:2007mj,Berti:2009kk}.
These modes are interesting for our purposes because they prescribe a clean state, 
while other spurious effects are absent. 
For instance, a quasi-bound state consists of almost a single pure frequency, 
which can be superradiant. 
Because other modes are absent, no interference and beating of the kind described later will occur. 
Finally, the evolution of a pure mode serves as an additional benchmark test of the code \cite{Yoshino:2012kn}.

There is a powerful and simple method to compute these modes, either by direct 
numerical integration or via a continued fraction approach.
In order to construct the bound states for massive scalars I follow  
Leaver's method \cite{Leaver:1986gd,Cardoso:2005vk,Dolan:2007mj,Berti:2009kk}.
In Boyer-Lindquist coordinates, denoted by the subscript $BL$, the field reads
\begin{align}
\Psi_{lm} = &\exp(-\imath \omega t_{BL})\exp(-\imath m \phi_{BL}) S_{lm}(\theta) R_{lm}(r_{BL})
\,,
\end{align}
with
\begin{align}
R_{lm}(r_{BL}) = & (r_{BL}-r_{BL,+})^{-\imath \sigma} (r_{BL} - r_{BL,-})^{\imath \sigma + \chi -1} \exp(r_{BL} q)
                \sum_{n=0}^{\infty} a_n \left(\frac{r_{BL}-r_{BL,+}}{r_{BL}-r_{BL,-}}\right)^n
\,,
\end{align}
where
\begin{align}
\sigma = & \frac{2r_{BL,+} (\omega - \omega_c)}{r_{BL,+} - r_{BL,-}} 
\,,\quad 
q = \pm \sqrt{\mu_S^2 - \omega^2} 
\,,\quad 
\chi=\frac{\mu_S^2 - 2\omega^2}{q}
\,,
\end{align}
and $\omega_c = m\Omega_H$ is the critical frequency for superradiance.
All the remaining terms in this expression are known in closed form and 
the characteristic frequency $\omega$ can be obtained by solving a three-term recurrence 
relation \cite{Dolan:2007mj,Berti:2009kk}. 
The coefficients $a_n$ are given by, e.g., Eqs.~(35)-(48) of \cite{Dolan:2007mj}.
Because this solution is given in Brill-Lindquist coordinates I perform a coordinate transformation to
Kerr-Schild coordinates (see e.g.\cite{Alcubierre:2008}), for clarity denoted by the subscript $KS$,
\begin{align}
dt_{KS} = & dt_{BL} + \frac{2Mr}{\Delta} dr_{BL}
\,,\quad 
dr_{KS} = dr_{BL} 
\,,\quad 
d\theta_{KS} = d\theta_{BL} 
\,,\quad 
d\phi_{KS} = d\phi_{BL} + \frac{a}{\Delta} dr_{BL} 
\,.
\end{align}
Then, the bound state scalar field is given by (note that I drop the subscript $KS$ in the following)
\begin{align}
\label{eq:MFinitialBS}
\Psi_{lm} = & \exp(-\imath \omega t) (r-r_{+})^A (r-r_{-})^B 
              \left(\frac{r-r_{+}}{r-r_{-}}\right)^CY_{lm}(\theta,\phi) R_{lm} 
\,,
\end{align}
where $A = -\frac{2\imath \omega M r_{+}}{r_{-}-r_{+}}$, 
$B= \frac{2\imath \omega M r_{-}}{r_{-}-r_{+}}$, 
$C = \frac{\imath m a}{r_{-}-r_{+}}$.
The corresponding momenta $\Pi_{lm}$ are computed from Eq.~\eqref{eq:MFdefPi}. 
\subsection{Wave extraction and output}\label{sec:waveoutput}
There are several different quantities that one can evaluate and discuss. For scalar fields, I 
will measure the fields $\Psi$ and $\Pi$ on spheres with different discrete radius $r_{ex}$.
Therefore I interpolate these functions on a spherical grid at each time step
and decompose them into multipoles, with $r=r_{ex}$ fixed. Specifically,
\begin{subequations}
\label{eq:MFmodedecompScalar}
\begin{align}
\Psi_{lm}(t) = &\int d\Omega \Psi(t,\theta,\phi) Y^{\ast}_{lm}(\theta,\phi)
\,,\\
\Pi_{lm}(t)  = &\int d\Omega \Pi(t,\theta,\phi)  Y^{\ast}_{lm}(\theta,\phi)
\,.
\end{align}
\end{subequations}

\subsection{Numerical implementation}\label{ssec:MFImplementation}
It is our goal to investigate the dynamics of massive scalar fields in the background
of Kerr BHs in the time domain.
For this purpose I have developed the time evolution code 
{\textsc{Lin-Lean}}
which is based on the {\textsc{Cactus}} computational toolkit \cite{Goodale02a, cactus} and the
{\textsc{Carpet}} mesh refinement package \cite{Schnetter:2003rb, carpet}.
The evolution will be performed using the $3+1$ approach as explained in the previous sections 
and by employing  the method of lines (see e.g. \cite{Alcubierre:2008}).

Specifically, the evolution equations are integrated in time
using the fourth order Runge-Kutta scheme and spatial derivatives are discretized 
by second or fourth order centered finite difference (FD) stencils. 
Advection derivatives are discretized by lop-sided FDs.
Because I are considering a non-dynamical BH background,
I prescribe the spacetime in Kerr-Schild coordinates
and excise the 
region inside the event horizon $r_S = 2M$. 
In practice, I specify this region by a parameter $r_{exc}\leq2M$,
that I typically choose to be $r_{exc} = 1.0M$.
Inside the excised region, i.e., $r\leq r_{exc}$, I enforce flat spacetime,
i.e., $\alpha = 1$, $\beta^i=0$, $\gamma_{ij} = \delta_{ij}$ and $K_{ij} = 0$.
I realize the excision either by employing second order, one-sided, outwards pointing FD stencils 
or extrapolate the solution in the excision region.
For all practical applications, I choose the latter method and use second order FD 
stencils on the refinement level containing the excision region,
whereas I use fourth order FD stencils on all outer refinement levels.
%

\section{Evolutions of massive scalar fields}\label{sec:MFResScalar}
Here, we report on our investigation of massive scalar fields in Schwarzschild 
and Kerr backgrounds. 
The features of the evolved scalar field can be understood 
in terms of the proper oscillation modes of the BH 
in conjuction with late-time tails arising from curvature backscattering.
Whereas the massless scalar field clearly shows this quasinormal
ringdown signal followed by the expected power-law tail at late times,
massive fields exhibit a much richer structure.
The analysis of our results will focus mainly on the dipole mode, i.e., $l=m=1$,
although other modes are also presented.
In Table~\ref{tab:MFScaAnaModes} we summarize the frequencies of the two lowest lying
dipole modes for a number of mass parameters 
considering either a Schwarzschild or highly rotating Kerr BH with spin
parameters $a/M=0.0$ or $a/M=0.99$, respectively 
\cite{Berti:2005ys,Berti:2009kk}.
In case of massless perturbations, the definition of ``lowest lying'', or fundamental,
mode corresponds to the least damped mode. 
Massless perturbations are always short-lived, unless
the BH is nearly extremal.
In contrast, massive perturbation have a smaller decay rate.
Additionally, if the mass term is present, another family
of modes appears, which lives on the effective potential well,
and which we call quasi-bound states \cite{Dolan:2007mj}.
These are long-lived modes and we order them by {\textit{decreasing}} imaginary part:
the fundamental mode is the shortest-lived mode in this family.

At very late times, after the relaxation of the BH,
there is still scattering off the background curvature present.
This gives rise to late-time power-law tails 
\cite{Price:1971fb,Leaver:1986gd,Ching:1995tj}.
In case of massless perturbations they behave as 
\begin{align}
\label{eq:MFScaPLmassless}
\Psi \sim t^{-2l + 3}
\,.
\end{align} 

Instead, tails of massive perturbations exhibit two different stages, 
which strongly depent on the mass term $M\mu_S$ \cite{Koyama:2001ee,Koyama:2001qw,Burko:2004jn}

\begin{subequations}
\label{eq:MFScaPLmassive}
\begin{align}
\Psi\sim & t^{-l-3/2}\sin(\mu_S t)
\,,\quad {\textrm{at intermediate times}}
\,,\\
\Psi\sim & t^{-5/6}\sin(\mu_S t)
\,,\quad {\textrm{at very late times}}
\,.
\end{align}
\end{subequations}

\begin{table}
\begin{center}
\begin{tabular}{ccll}
\hline
$M \mu_S$& $a/M$   & $M \omega_{11}$ ($n=0$)     & $M \omega_{11}$ ($n=1$) \\ 
\hline
$0.00$ & $0.00$  & $0.2929-i0.09766$           & $0.2645-i0.3063$ \\ 
$0.00$ & $0.99$  & $0.4934-i0.03671$           & $0.4837-i0.09880$\\ 
\hline
$0.42$ & $0.00$  & $0.4075-i0.001026$          & $0.4147-i0.0004053$\\ 
$0.42$ & $0.99$  & $0.4088+i1.504\cdot10^{-7}$ & $0.4151+i5.364\cdot10^{-8}$ \\ 
\hline
$0.55$ & $0.00$  & $0.5207-i0.01076$           & $0.5376-i0.003743$ \\ 
$0.55$ & $0.99$  & $0.5358-i0.0007432$         & $0.5187-i0.0005984$ \\
\hline
\end{tabular}
\end{center}
\caption{\label{tab:MFScaAnaModes} 
Fundamental ($n=0$) and first overtone ($n=1$) 
modes for scalar perturbations. These modes have been computed with the continued fraction representation.
For massive perturbations, we list the two lowest order quasi-bound states, i.e, 
modes which are very long lived. These modes form a sequence
of {\it decreasing} imaginary frequency. 
We note that there are other modes which are highly damped, but which do not belong to the quasi-bound state sequence.
For $M\mu_S=0.55,\,a=0.99M$ for instance, we find at least one extra solution, 
at $M\omega=0.5233-i0.1059$, which we do not include in this family.
}
\end{table}
%

\subsection{Code test -- space dependent mass coupling}\label{ssec:MFScaMuVar}
As an initial code test, we take an unphysical space-dependent mass coupling
in the background of a Schwarzschild BH. In particular, we have chosen $\mu_S^2=-10M^2/r^4$.
This {\it{unphysical}} mass coupling quickly leads to an instability of the system, 
which provides a unique and fast test of the code. A mode analysis of the Klein-Gordon equation
is straightforward and shows the existence of at least one unstable mode, $\Psi\sim e^{0.071565 t}$.
We have evolved a spherically symmetric Gaussian profile centered at $r_0=12M$ and with width $w=2M$.
The grid setup is given by\\
$\{(1024,512,256,128,64,32,16,8,4,2),~h=M/40\}$, using the notation of Sec.~II E 
in \cite{Sperhake:2006cy}.
The results are shown in Fig.~\ref{fig:MFSca_a0muVar}, for which the scalar field is 
extracted as a function of time at $r_{ex} = 10M$.
Our numerical results are consistent with an exponential growth, $\Psi \sim e^{0.07161 t}$, which agrees to within
$0.06\%$ with the mode analysis.
\begin{figure}[htpb!]
\begin{center}
\includegraphics[width=0.50\textwidth]{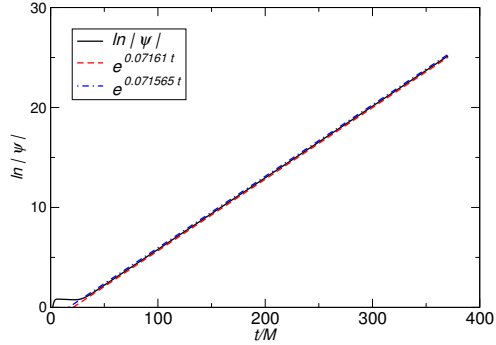}
\end{center}
\caption{\label{fig:MFSca_a0muVar} 
Evolution of a spherically symmetric scalar field with space dependent mass $\mu_S^2 = -10M^2/r^4$, 
around a non-rotating BH. 
The field has been extracted at $r_{ex} = 10M$.
}
\end{figure}
%

\subsection{Code test -- massless scalar fields}\label{ssec:MFScaMassless}
%
The main features resulting from the time evolution of massless fields around BHs have been known for 
decades (see e.g. \cite{Berti:2009kk} and references therein for an overview). 
Thus, they provide an excellent benchmark test for our code.
For this purpose we have evolved massless scalar fields around a non-rotating and highly
spinning BH with $a/M=0.99$. The field has been initiated as a Gaussian with width $w=2M$ located at $r_0=12M$.
The specifications of the setup are summarized in Table~\ref{tab:MFScaMLSetup}.
In Fig.~\ref{fig:MFScaMLwaveform} we display the evolution of the $l=m=0$ and $l=m=1$ modes of the scalar field
in the respective backgrounds.
The waveform exhibits all main features discussed in the Introduction: a prompt signal, 
followed by an exponentially decaying sinusoid and, at very late times, a power-law tail. 
A fit to the ringdown phase yields numerical QNM frequencies, which are summarized in Table~\ref{tab:MFScaMLRes}.
They are in excellent agreement within less than $2\%$ with tabulated values \cite{Berti:2005ys,Berti:2009kk}. 
The agreement increases for higher resolution.
The power-law tail is of the form $t^{p}$, with $p=-3.08$ ($a/M=0$) and $p=-3.07$ ($a/M=0.99$) 
for the $l=m=0$ mode, 
in agreement with a low frequency expansion of the wave equation which yields
$p=-3$, see Eq.~\eqref{eq:MFScaPLmassless}.
Additionally, we have performed a convergence analysis for the more challenging case of a highly
rotating BH background with $a/M=0.99$. Therefore we have evolved the massless scalar field with
three different resolutions $h_c=M/100$, $h_m=M/104$ and $h_f=M/108$.
In Fig.~\ref{fig:MFScaMLConvergence} we plot the differences of the $l=m=1$ mode 
between the coarse and medium resolution together
with the difference between the medium and high resolution runs. The latter has been rescaled by the
factor $Q_3=1.16$ demonstrating third order convergence.
This analysis allows us to estimate the discretization error in the signal which is about
$\De\Psi_{11}/\Psi_{11}\le8\%$ for the $l=m=1$ mode at late times of the evolution
and about $\De\Psi_{00}/\Psi_{00}\le3\%$ for $l=m=0$.
The performed checks confirm that our implementation works well and 
yields consistent results.

\begin{table}
\begin{center}
\begin{tabular}{cccc}
\hline
Run             & $a/M$  & $\,_0\Sigma(\theta,\phi)$    & Grid Setup   
\\ 
\hline
S\_mu0\_a0$_1$  & $0.0$  & $Y_{00,1-1,11}$      & $\{(384,192,96,48,24,12,6,3,1.5),~h=M/100\}$ \\
S\_mu0\_a0$_2$  & $0.0$  & $Y_{00,1-1,11}$      & $\{(192,96,48,24,12,6,3,1.5),~h=M/144\}$ \\
\hline
S\_mu0\_a99$_c$ & $0.99$ & $Y_{00,1-1,11}$      & $\{(384,192,96,48,24,12,6,3,1.5),~h=M/100\}$ \\
S\_mu0\_a99$_m$ & $0.99$ & $Y_{00,1-1,11}$      & $\{(384,192,96,48,24,12,6,3,1.5),~h=M/104\}$ \\
S\_mu0\_a99$_f$ & $0.99$ & $Y_{00,1-1,11}$      & $\{(384,192,96,48,24,12,6,3,1.5),~h=M/108\}$ \\
\hline
\end{tabular}
\end{center}
\caption{\label{tab:MFScaMLSetup} 
Initial setup for simulations of a massless scalar field with Gaussian initial data 
located at $r_0 = 12~M$ and with width $w = 2~M$ in Schwarzschild or Kerr background with
$a/M=0.99$.
We denote the modes of the initial pulse $\,_0\Sigma(\theta,\phi)$
and the specific grid setup, in units of the BH mass $M$, 
following the notation of Sec.~II E in \cite{Sperhake:2006cy}.
}
\end{table}
\begin{table}
\begin{center}
\begin{tabular}{cccc}
\hline
Run              & $a/M$ & $M\omega_R$ & $M\omega_I$ 
\\ 
\hline
S\_mu0\_a0$_1$  & $0.0$  & $0.291$     & $-0.097$ \\
S\_mu0\_a0$_2$  & $0.0$  & $0.294$     & $-0.096$ \\
\hline
S\_mu0\_a99$_c$ & $0.99$ & $0.497$     & $-0.0371$ \\
S\_mu0\_a99$_m$ & $0.99$ & $0.494$     & $-0.0370$ \\
S\_mu0\_a99$_f$ & $0.99$ & $0.493$     & $-0.0368$ \\
\hline
\end{tabular}
\end{center}
\caption{\label{tab:MFScaMLRes} 
Summary of results for a massless scalar field in Schwarzschild or Kerr BH background with $a/M = 0.99$.
We denote the real and imaginary part of the quasinormal frequency $M\omega_R$ and $M\omega_I$ of
the $l=m=1$ mode. 
}
\end{table}
\begin{figure}
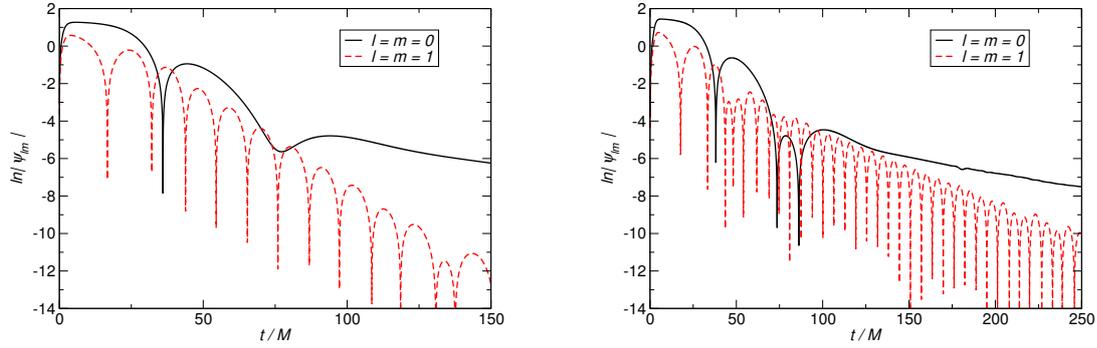

\begin{center}
\begin{tabular}{cc}
\includegraphics[width=0.50\textwidth]{SFMLfig1a} &
\includegraphics[width=0.50\textwidth]{SFMLfig1b} 
\end{tabular}
\end{center}
\caption{\label{fig:MFScaMLwaveform} 
Evolution of a Gaussian profile of a massless scalar field with width $w=2M$ centered at $r_0=12M$ 
around a Schwarzschild BH (left panel) and a Kerr BH with $a/M = 0.99$ (right panel).
The solid black line refers to the $l=m=0$ mode and the red dashed line to $l=m=1$. 
A prompt signal gives way to ringdown and a visible power-law tail for $l=m=0$. 
}
\end{figure}

\begin{figure}
\begin{center}
\includegraphics[width=0.50\textwidth]{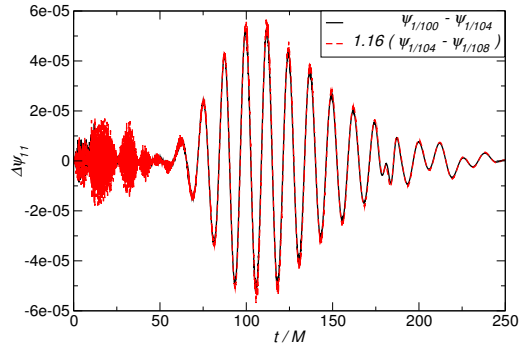}
\end{center}
\caption{\label{fig:MFScaMLConvergence} 
Convergence plot of the $l=m=1$ mode of the massless scalar field in Kerr background with $a/M=0.99$.
The differences between the coarse-medium (black solid line) and medium-high 
(red dashed line) resolution runs are shown.
The latter difference has been rescaled by $Q_3=1.16$ indicating third order convergence.
}
\end{figure}

\subsection{Massive scalar fields: quasi-bound states}\label{ssec:MFScaBS}
The remainder of this section is devoted to the investigation of massive scalar fields.
A mass term introduces a new scale in the problem and, thus, raises the expectation 
that new features appear in the evolution.
One of these features is the existence of long-lived, quasi-bound states, whose form and construction have been discussed in 
Sec.~\ref{ssec:MFinitdata}. 
We have constructed quasi-bound states for fields with mass coupling $M\mu_S=0.42$, 
following \cite{Dolan:2007mj,Cardoso:2005vk,Berti:2009kk},
and evolved them in Schwarzschild or Kerr background with $a/M=0.99$ according to 
Eqs.~\ref{eq:MFdefPi} and~\ref{eq:MFinitialBS}.
We show the initial data, specifically $|\Psi|^2$ along the x-axis, in Fig.~\ref{fig:MFScaBSID} 
for the fundamental and first overtone mode (see Table~\ref{tab:MFScaAnaModes}).
In case of non-rotating BHs the fundamental quasi-bound state is localized around the origin,
but moves outwards when rotation is included. 
Specifically, for a highly spinning BH with $a/M=0.99$, the fundamental mode is peaked at around $r\sim 12~M$. 
The first overtone on the other hand has a pronounced minimum or node around 
$r_{node}\sim 22.5~M$ and $r_{node}\sim26.5~M$ for $a/M=0$ and $a/M=0.99$, respectively.
This information will prove useful when trying to understand the interplay between
different modes during the time evolution of massive scalar fields
initially prescribed as generic Gaussian wave packet (see Sec.~\ref{ssec:MFScaMassiveSW} below).

Additionally, we have evolved the $l=m=1$ fundamental bound state mode  
of a scalar field with $M\mu_S = 0.42$ in Kerr background $a/M = 0.99$. 
The results are shown in the left panel of Fig.~\ref{fig:MFScaBSEv}, where we plot the variation of the
absolute value of $\Psi_{11}$ relative to its inital value at $t=0~M$.
By construction, the field should remain localized in the vicinity of the BH and its
absolute value $\Psi_{11}\Psi_{11}^{\ast} \sim \exp(-\imath \omega t)\exp(\imath \omega t) \sim const.$
should remain almost constant in time, with a small growth rate of 
$M\omega_I\sim1.5\cdot 10^{-7}$ \cite{Dolan:2007mj,Cardoso:2005vk}.
We have tested these properties numerically by evolving the field in time and extracting its $l=m=1$ mode
at radii $r_{ex}= 10, 20, 30, 40M$. 
The absolute value of the scalar field varies by less than $5\%$ 
($8\%$ for $r_{ex}= 40M$) until $t\sim 200M$ and by less than $1\%$ at late times.
Variations of this order of magnitude are expected for two reasons: first, the dependence of the quasi-bound state
was imposed to be a spherical harmonic, see \eqref{eq:MFinitialBS}. 
Secondly, finite resolution effects due to the numerical scheme play a crucial role 
and have been identified as a further cause for the slight discrepancy. 
This becomes evident in the right panel of Fig.~\ref{fig:MFScaBSEv}, 
in which we plot the relative absolute value of the scalar for the low and medium resolution run, 
$h=M/60$ and $h=M/72$, extracted at $r_{ex}=20M$.
In fact, the two curves agree very well until $t\sim100~M$, but differ by about $2\%$ at late times.

The quasi-bound states shown here should be unstable states with an instability time scale 
of $\sim10^{7}M$, as shown in Table~\ref{tab:MFScaAnaModes}. 
Naturally, with such a large instability timescale, 
such growth is completely buried in numerical noise.

%
\begin{figure}[htpb!]
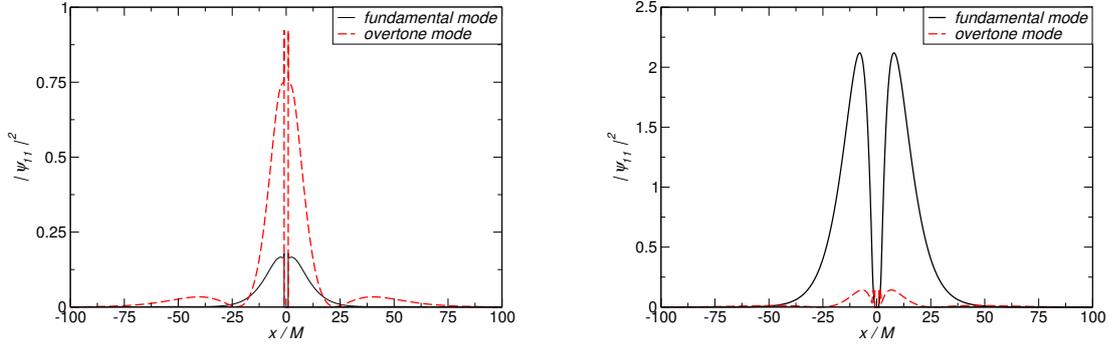

\begin{center}
\begin{tabular}{cc}
\includegraphics[width=0.50\textwidth]{SFBSfig1a} &
\includegraphics[width=0.50\textwidth]{SFBSfig1b}
\end{tabular}
\end{center}
\caption{\label{fig:MFScaBSID}
Absolute value of the bound state wave function of a massive scalar field with $M\mu_S=0.42$
at \textit{initial time}
in a Schwarzschild (left panel) and $a/M=0.99$ Kerr background (right panel). 
Black solid lines are the fundamental mode, red dashed lines denote the first overtone. 
}
\end{figure}
%
%
\begin{figure}[htpb!]
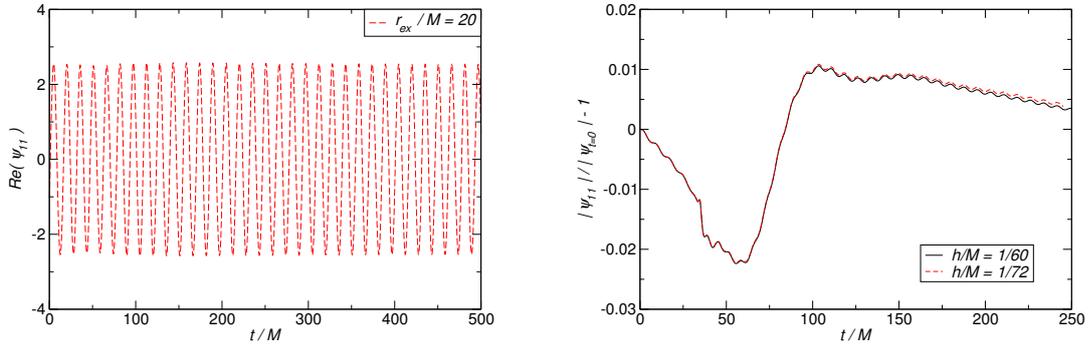

\begin{center}
\begin{tabular}{cc}
\includegraphics[width=0.5\textwidth]{SFBSfig2a} &
\includegraphics[width=0.5\textwidth]{SFBSfig2b}
\end{tabular}
\end{center}
\caption{\label{fig:MFScaBSEv}
Left: 
Evolution of a massive, $l=m=1$ field with $M\mu_S=0.42$ around a Kerr BH with $a/M=0.99$,
initialized as a quasi-bound state. The signal has been extracted at $r_{ex}=20M$.
The small bump at early times is presumably due to finite resolution and inaccurate spherical dependence 
(see main text for further discussion).
Right: 
Relative change of the absolute value of the scalar field $\Psi$, extracted at $r_{ex}=20M$
evolved with the numerical resolutions $h=M/60$ (black solid line) and
$h= M/72$ (red dashed line).
}
\end{figure}
\subsection{Massive scalar fields: ringdown and tails}
\label{ssec:MFScaMassiveSW}
After extensively testing the code, we now explore the dynamics of generic, massive scalar fields in the background
of a Schwarzschild or Kerr BH with $a/M=0.99$.
Therefore we initiate the field as a Gaussian wave packet, 
according to Eqs.~\eqref{eq:MFinitdataQNM},
typically with a width of $w=2M$ centered around $r_0=12M$.
The progression of scalar fields with mass coupling $M\mu_S$ is carried out by the time evolution 
system, Eqs.~\ref{eq:MFevolPsi} and~\ref{eq:MFevolPi}.
We summarize the specific configuration of our simulations, such as the mass coupling $M\mu_S$, 
the modes contributing to the initial Gaussian and the grid setup in Table~\ref{tab:MFScaMassiveSetup}.
Unless denoted otherwise, we set the excision radius to $r_{exc}=1.0M$.
In order to read off the results we interpolate the scalar field onto a sphere with fixed
radius $r_{ex}/M$ as seen by an observer at infinity
and decompose it into modes using spherical harmonics $Y_{lm}(\theta,\phi)$, 
according to Eq.~\eqref{eq:MFmodedecompScalar}.

In Table~\ref{tab:MFScaMassiveFreq} we summarize the results, 
including the quasinormal frequency $\omega$
and the fall-off of the tail $\Psi\sim t^{p}$ estimated from the numerical evolution.
We compute the real part of the QN frequency $\omega_R$ by considering various intervals, 
each about $2-5$ cycles long,
and performing a regression on the thus obtained values.
We obtain the imaginary part of the QN frequency $\omega_I$, which encodes information about 
the decay or growth rate $\tau = 1/\omega_I$, by performing a linear fit of the form
\begin{align}
\label{eq:MFScaFitIm}
\ln\Psi_{lm} = & A + \omega_{I,lm} t 
\,.
\end{align}
This dependence follows from $\Psi_{lm} \sim \exp(-\imath \omega t)$.
However, because in some cases the decay rate varies over time and because
of resonant or beating effects that we observe in some (other) cases, 
which we will discuss in more detail in Sec.~\ref{ssec:excitationcoefficients} below,
a clean fit of that form is not always possible. 
In order to estimate the power-law fall-off of the late-time tails 
$\Psi \sim t^p$
we employ the fit
\begin{align}
\label{eq:MFScaFitTail}
\ln|\Psi| \sim & p\,\ln t
\,.
\end{align}
%

We start by analysing massive scalar fields in a Schwarzschild BH background
considering a number of different mass couplings
varying from $M\mu_S=0.1,\dots,1.0$.

In Fig.~\ref{fig:waveformsSSm01} we present the logarithm of the $l=m=1$ mode
of the scalar field with mass coupling $M\mu_S=0.1$. 
We find the QN frequency $M \omega_{11} = 0.293 - \imath 0.036$.
In this case, we fit the tail to the oscillatory inverse power-law, Eq.~\eqref{eq:MFScaPLmassive},
expected for small mass couplings at intermediate late times.
Numerically we obtain $\Psi_{11} \sim t^{-2.543} \sin(0.1 t)$, 
which agrees with the theoretically expected fall-off
within less than $2\%$ \cite{Koyama:2001ee,Koyama:2001qw}.
The corresponding fit is presented in Fig.~\ref{fig:waveformsSSm01} (red-dashed line) 
which describes the numerical data well for $t\ge 100~M$.
%

Next, we consider a scalar field with mass coupling $M\mu_S = 0.42$.
In order to perform a convergence analysis of its $l=m=0$ mode, we have evolved the setup 
(run S\_m42\_a0 in Table~\ref{tab:MFScaMassiveSetup})
at three different resolutions $h_c=M/44$, $h_m=M/48$ and $h_f=M/52$. 
We show the corresponding convergence plot in Fig.~\ref{fig:convergenceSS042}.
In particular, we present the differences between the 
coarse - medium and medium - high resolution run. 
The latter has been amplified by the factor $Q_2 = 1.28$ demonstrating second order convergence.
We estimate the numerical error to be $\Delta\Psi_{00}/\Psi_{00} \sim 3.6\%$ at $t\sim1000~M$ which increases to
$\Delta\Psi_{00}/\Psi_{00} \sim 6.7\%$ at $t\sim1500~M$.

Additionally, we have simulated this field with a more generic wave packet for a longer time
by employing higher resolution $h=M/60$ and a larger outer boundary
located at $r_B=1536~M$.
The following results refer to this run.
In the left panel of Fig.~\ref{fig:waveformsSSm042_00} we present the logarithm of the $l=m=0$ mode of the 
scalar field extracted at different radii $r_{ex}=25,35,45,55M$.
In the right panel of Fig.~\ref{fig:waveformsSSm042_00} we show the log-log representation of the $l=m=0$ mode, 
extracted at the radii $r_{ex}=25M$, 
together with the fit to the late-time power-law tail, Eq.~\eqref{eq:MFScaFitTail}.
Numerically, we find the exponent $p=-0.880$ which agrees within $5.6\%$ 
with the theoretically expected late-time behaviour $p=-5/6$ 
\cite{Koyama:2001ee,Koyama:2001qw,Burko:2004jn}.

The $l=m=1$ mode of the scalar field with mass coupling $M\mu_S=0.42$, 
depicted in Fig.~\ref{fig:waveformsSSm042_11} for extraction radii 
$r_{ex}=22.5,30,40,50M$,
exhibits a much richer structure. 
In particular, we observe an overall modulation of the signal whose specific
features depend on the location of the measurement.
This behaviour can be understood as a result 
of beating between different overtone modes whose amplitude depends on the
specific location of the measurement.
We will discuss this effect in more detail in Sec.~\ref{ssec:excitationcoefficients} below.
Because of this beating effect we did not determine the decay rate or late-time tail of this mode.

These features of the beating effect appear also in the $l=m=1$ component of a scalar field with 
mass coupling $M\mu_S=0.55$, albeit not as pronounced.
In Fig.~\ref{fig:waveformsSSm055} we depict the logarithm of the $l=m=1$ mode, 
extracted at $r_{ex}=13, 20, 30, 40M$.
The first extraction radius corresponds to the node of the first overtone of the 
{\textit{bound state}} of a scalar field with $M\mu_S=0.55$ in Schwarzschild 
background. 
The behaviour of the $l=m=0$ mode is similar to the case with $M\mu_S=0.42$
and we summarize the (numerically computed) 
QN frequency and late-time tail in Table~\ref{tab:MFScaMassiveFreq}.

Finally, we have studied a scalar field with larger mass parameter $M\mu_S=1.0$ 
evolving in the background of a Schwarzschild BH.
In the left panel of Fig.~\ref{fig:waveformsSSm1} we show the logarithm of the $l=m=1$ mode,
extracted at radii $r_{ex}=10,20,30,40M$.
We numerically find the QN frequency $M \omega_{11} = 0.965 - \imath 0.0046$.
We show a log-log representation of the $l=m=1$ mode, measured at $r_{ex}=20M$, 
together with a fit to the late-time tail in the right panel of Fig.~\ref{fig:waveformsSSm1}.
We estimate the fall-off exponent to be $p=-0.873$ which is in good agreement,
within less than $4.9\%$,
with the previous results presented in \cite{Burko:2004jn}.

The numerical results of massive scalar fields in a Schwarzschild BH background,
specifically the QNM frequencies and power-law tails, are in good agreement
with theoretical predictions and previous numerical results (see e.g.\cite{
Koyama:2001ee,Koyama:2001qw,Burko:2004jn,Berti:2009kk} and references therein).
Thus, we have verified that our implementation works well and delivers reliable results.

In the second part of this section we focus on the time evolution of a massive scalar field 
with mass couplings $M\mu_S=0.42$ or $M\mu_S=0.55$ in the background of a Kerr BH with
spin parameter $a/M=0.99$. 
Initially, we choose a Gaussian profile of the scalar field located at $r_0=12M$ with width $w=2M$.
The specific setup is summarized in Table~\ref{tab:MFScaMassiveSetup}.
The excision radius is set to $r_{exc}=1.0M$ in order to guarantee that it always lies within the 
BH event horizon.
In order to read off the results we interpolate the scalar field onto a sphere at 
extraction radii $r_{ex}= 20,26.5,30,35,40,45,50,60M$ as seen by an observer at infinity
and decompose it into modes using spherical harmonics $Y_{lm}(\theta,\phi)$, 
according to Eq.~\eqref{eq:MFmodedecompScalar}.

First, we analyse the $l=m=0$ multipoles 
of both scalar field configurations, which are depicted in 
Figs.~\ref{fig:waveformsKerrmu042_l0} and~\ref{fig:waveformsKerrmu055_l0}.
In particular, we present the logarithm of the waveforms measured at 
$r_{ex}=20,26.5,35,45M$ in the respective left panels of the figures.
We estimate the real part of the frequencies to be $M\omega_{R,00}=0.419\pm0.01$ and
$M\omega_{R,00}=0.549\pm0.01$ for the scalar field with mass coupling $M\mu_S=0.42$ and $M\mu_S=0.55$,
respectively.
A clean fit to the imaginary parts of the frequencies has not been procurable because of variations in time.
In the right panels of Figs.~\ref{fig:waveformsKerrmu042_l0} and~\ref{fig:waveformsKerrmu055_l0}
we show a log-log representation of the $l=m=0$ mode, extracted at $r_{ex}=20M$,
in order to visualize the late-time tail. 
Specifically, we present the numerical data together with a fit of the form 
$\psi\sim t^{p}$ with the power-law fall-off
$p=-0.861$ and $p=-0.847$ in the case $M\mu_S=0.42$ and $M\mu_S=0.55$, respectively.
These numerically computed values are in good agreement, within $\le3.5\%$ and $\le2\%$, 
with the prediction $p=-5/6$ 
\cite{Koyama:2001ee,Koyama:2001qw,Burko:2004jn}.
%

Additionally, we have performed a convergence analysis for the evolution of a massive scalar field with
coupling $M\mu_S=0.42$ in order to estimate the numerical accuracy of the waveforms.
Therefore, we have simulated the setup 
with three different resolutions $h_c = M/60$,
$h_m = M/72$ and $h_h = M/84$ (runs S\_m42\_a99$_c$, S\_m42\_a99$_m$ and S\_m42\_a99$_f$
in Table~\ref{tab:MFScaMassiveSetup}).
In Fig.~\ref{fig:convergenceKerr042} we present the corresponding convergence plot
for the $l=m=0$ (left panel) and $l=m=1$ (right panel) modes.
The numerically found convergence factor is $Q_2=1.66$, thus demonstrating second order convergence.
This procedure allows us to estimate the discretization error to be about 
$\De\psi/\psi \le 1\%$ in both modes.

In summary, the $l=m=0$ modes of the considered massive scalar fields with coupling
$M\mu_S=0.42$ and $M\mu_S=0.55$ in Kerr background with $a/M=0.99$
show the expected ringdown signal followed by a late-time power-law tail.
The numerically obtained properties of the waveforms, such as (real) part of the ringdown frequency
at late-time fall-off are in good agreement with theoretical predictions and previous numerical results
\cite{Koyama:2001ee,Koyama:2001qw,Burko:2004jn}.
However, the corresponding $l=m=1$ modes, shown in Figs.~\ref{fig:Kerrmu042_l1} and~\ref{fig:waveformsKerrmu055},
exhibit interesting features which can be explained by a beating phenomenon.
We analyse this behaviour in more detail in the following section.

%
\begin{table}
\begin{center}
\begin{tabular}{lcccc}
\hline
Run             & $M \mu_S$ & $a/M$  & $\,_0\Sigma(\theta,\phi)$ 
& Grid Setup   \\ 
\hline
S\_m01\_a0      & $0.1$   & $0.0$  & $Y_{11}$ 
& $\{(1024,512,256,128,64,32,8,4,2), 1/40\}$ \\
S\_m42\_a0$_c$ & $0.42$  & $0.0$  & $Y_{00}$ 
& $\{(1024,512,256,128,64,32,8,4,2), 1/44\}$ \\
S\_m42\_a0$_m$ & $0.42$  & $0.0$  & $Y_{00}$
& $\{(1024,512,256,128,64,32,8,4,2), 1/48\}$ \\
S\_m42\_a0$_h$ & $0.42$  & $0.0$  & $Y_{00}$  
& $\{(1024,512,256,128,64,32,8,4,2), 1/52\}$ \\
S\_m42\_a0     & $0.42$  & $0.0$  & $Y_{10,11,20,22}$ 
& $\{(1536,384,192,96,48,24,12,6,3,1.5), 1/60\}$ \\
S\_m55\_a0     & $0.55$  & $0.0$  & $Y_{10,11,20,22}$ 
& $\{(1536,384,192,96,48,24,12,6,3,1.5), 1/60\}$ \\
S\_m1\_a0       & $1.0$   & $0.0$  & $Y_{11}$
& $\{(1024,512,256,128,64,32,8,4,2), 1/40\}$ \\
\hline
S\_m42\_a99$_c$ & $0.42$ & $0.99$ & $Y_{00,10,11,1-1}$
& $\{(1536,384,192,96,48,24,12,6,3,1.5), 1/60\}$ \\
S\_m42\_a99$_m$ & $0.42$ & $0.99$ & $Y_{11,1-1}$
& $\{(1536,384,192,96,48,24,12,6,3,1.5), 1/72\}$ \\
S\_m42\_a99$_f$ & $0.42$ & $0.99$ & $Y_{11,1-1}$
& $\{(1536,384,192,96,48,24,12,6,3,1.5), 1/84\}$ \\
S\_m55\_a99    & $0.55$  & $0.99$ & $Y_{00,10,11,1-1}$ 
& $\{(1536,384,192,96,48,24,12,6,3,1.5), 1/60\}$ \\
\hline
\end{tabular}
\end{center}
\caption{\label{tab:MFScaMassiveSetup} 
Initial setup for simulations of a massive scalar field with Gaussian initial data 
located at $r_0=12M$ and with width $w=2M$ in a Schwarzschild or Kerr background with 
$a/M=0.99$.
In particular we denote the mass parameter $M\mu_S$, the modes of the initial pulse $\,_0\Sigma(\theta,\phi)$
and the specific grid setup, measured in units of the BH mass $M$, 
following the notation of Sec.~II E in \cite{Sperhake:2006cy}.
}
\end{table}
%
%
\begin{table}
\begin{center}
\begin{tabular}{lcccccc}
\hline
Run              & $M \mu_S$ & $a/M$ & $(lm)$ & $M\omega_R$       & $M\omega_I$   & tail 
\\ 
\hline
S\_m01\_a0      & $0.1$   & $0.0$  & $(11)$ & $0.293$           & $-0.036$      & $t^{-2.543} \sin(0.1\, t)$
\\
S\_m42\_a0$_c$ & $0.42$  & $0.0$  & $(00)$ & $0.417$           & n.a.          & $t^{-0.8856}$ 
\\
S\_m42\_a0$_m$ & $0.42$  & $0.0$  & $(00)$ & $0.416$           & n.a.          & $t^{-0.8837}$ 
\\
S\_m42\_a0$_h$ & $0.42$  & $0.0$  & $(00)$ & $0.417$           & n.a.          & $t^{-0.882}$ 
\\
S\_m42\_a0     & $0.42$  & $0.0$  & $(00)$ & $0.415$           & n.a.          & $t^{-0.880}$
\\
S\_m42\_a0     & $0.42$  & $0.0$  & $(11)$ & $0.407$           & n.a.          & n.a.
\\
S\_m55\_a0     & $0.55$  & $0.0$  & $(00)$ & $0.543$           & n.a.          & $t^{-0.877}$
\\
S\_m55\_a0     & $0.55$  & $0.0$  & $(11)$ & $0.542$           & n.a.          & n.a.
\\
S\_m1\_a0       & $1.0$   & $0.0$  & $(11)$ & $0.965$           & $-0.0046$     & $t^{-0.873}$ 
\\
\hline
S\_m42\_a99$_c$ & $0.42$ & $0.99$ & $(00)$ & $0.419$           & n.a.          & $t^{-0.861}$
\\
S\_m42\_a99$_m$ & $0.42$ & $0.99$ & $(00)$ & $0.415$           & n.a.          & $t^{-0.841}$
\\
S\_m42\_a99$_f$ & $0.42$ & $0.99$ & $(00)$ & $0.406$           & n.a.          & n.a.
\\
S\_m55\_a99    &  $0.55$ & $0.99$ & $(00)$ & $0.549$           & n.a.          & $t^{-0.847}$
\\
\hline
\end{tabular}
\end{center}
\caption{\label{tab:MFScaMassiveFreq} 
Summary of results for a massive scalar field with mass coupling $M\mu_S$ in the background
of a Schwarzschild and Kerr BH with $a/M=0.99$.
We denote the real and imaginary part of the quasinormal frequency $M\omega_R$ and $M\omega_I$ of 
a specific mode $(lm)$ as well as its late-time tail.
In some cases a clean fit to $M\omega_I$ or the tail has not been feasible.
The cases are denoted by {\textit{n.a.}}.
}
\end{table}
%
%
\begin{figure}
\begin{center}
\includegraphics[width=0.50\textwidth]{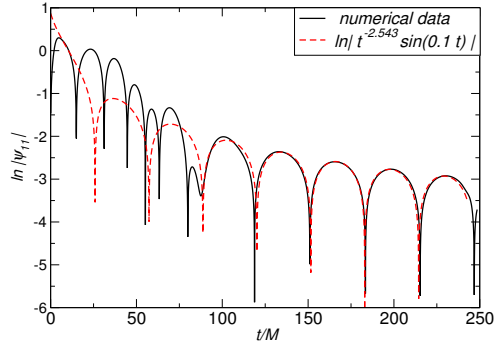}
\end{center}
\caption{\label{fig:waveformsSSm01}
Logarithm of the $l=m=1$ mode the massive scalar field with $M\mu_S = 0.1$.
We plot the numerical data (black solid line) as well as a fit to the tail
$\Psi_{11} \sim t^p \sin(\mu_S t)$ with $p=-2.543$ and $\mu_S=0.1$ (red dashed line).
}
\end{figure}
%
%
\begin{figure}
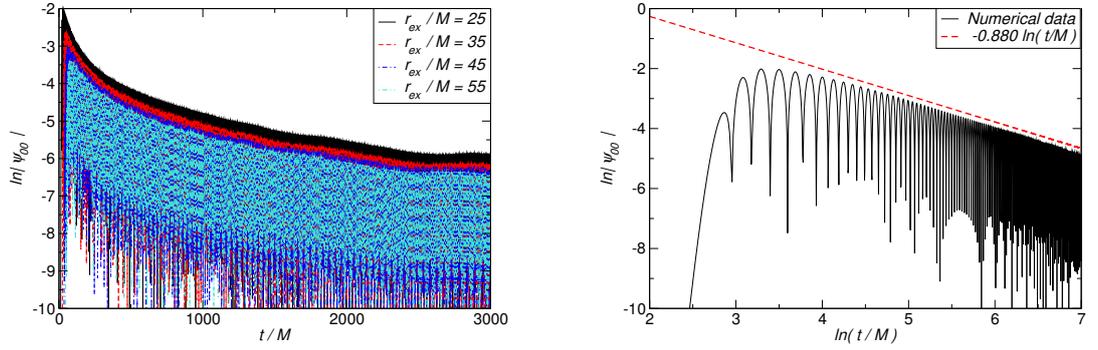

\begin{center}
\begin{tabular}{cc}
\includegraphics[width=0.50\textwidth]{SFMa0fig2a} &
\includegraphics[width=0.50\textwidth]{SFMa0fig2b} 
\end{tabular}
\end{center}
\caption{\label{fig:waveformsSSm042_00}
Logarithm of the $l=m=0$ mode of the scalar field with $M\mu_S = 0.42$ in Schwarzschild background
Left: as function of time and extracted at radii $r_{ex} = 25,35,45,55M$.
Right: as function of $ln(t/M)$. We plot the numerical data extracted at $r_{ex} = 25M$ (black solid line)
together with the fit of the tail $\Psi \sim t^{-0.880}$ (red dashed line).
}
\end{figure}
%
%
\begin{figure}
\begin{center}
\includegraphics[width=0.50\textwidth]{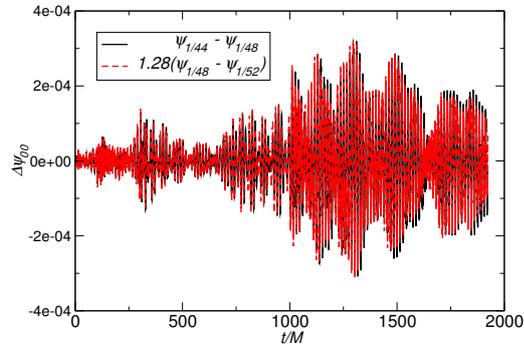}
\end{center}
\caption{\label{fig:convergenceSS042}
Convergence plot of the $l=m=0$ mode of the massive scalar field with $M\mu_S=0.42$.
We present the differences between the coarse-medium and medium-high resolution runs, where the
latter is amplified by $Q_2=1.28$ indicating second order convergence.
}
\end{figure}
%
%
\begin{figure}
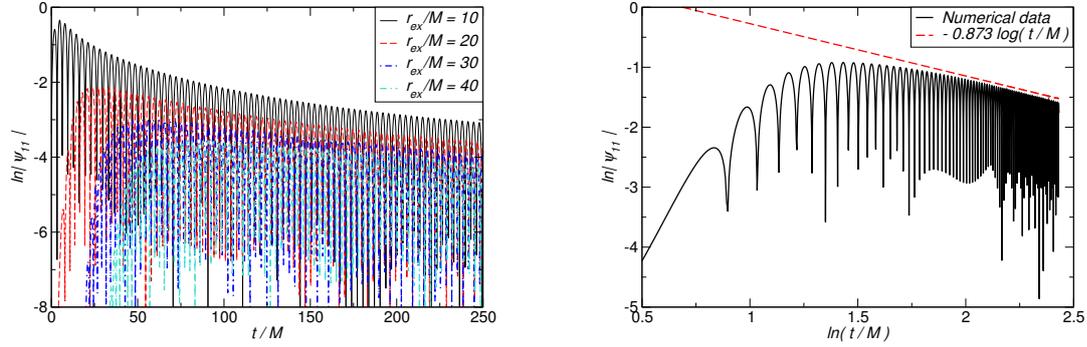

\begin{center}
\begin{tabular}{cc}
\includegraphics[width=0.50\textwidth]{SFMa0fig4a} & 
\includegraphics[width=0.50\textwidth]{SFMa0fig4b} 
\end{tabular}
\end{center}
\caption{\label{fig:waveformsSSm1}
Left:
Logarithm of the $l=m=1$ mode the massive scalar field with $M\mu_S = 1.0$
extracted at radii $r_{ex}= 10, 20, 30, 40M$ as function of time $t/M$.
Right: 
Log-log plot of the same mode.
We show the numerical data extracted at $r_{ex}=20M$ (black solid line)
together with the fit of the tail $\Psi \sim t^{-0.873}$ (red dashed line).
}
\end{figure}
%

%
\begin{figure}
\begin{center}
\begin{tabular}{cc}
\includegraphics[width=0.5\textwidth]{SFMa99fig1a} &
\includegraphics[width=0.5\textwidth]{SFMa99fig1b}
\end{tabular}
\end{center}
\caption{\label{fig:waveformsKerrmu042_l0}
Left: Logarithm of the $l=m=0$ mode of the massive scalar field with $M\mu_S=0.42$ in Kerr background
      with $a/M = 0.99$, extracted at $r_{ex}=20,26.5,35,45M$.
Right: Log-Log plot of the same mode, extracted at $r_{ex}= 20M$.
       We plot the numerical data (black solid line) together with the fit of the 
       late-time tail $\Psi\sim t^{-0.861}$.
}
\end{figure}
%
%
\begin{figure}
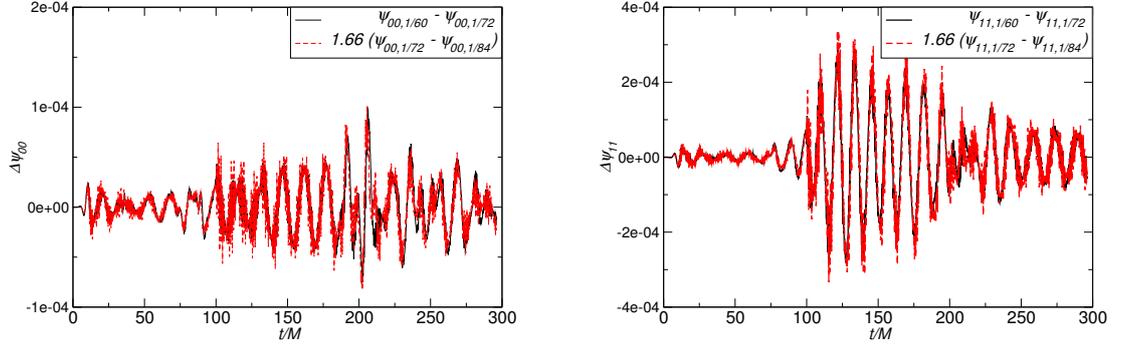

\begin{center}
\begin{tabular}{cc}
\includegraphics[width=0.50\textwidth]{SFMa99fig2a} &
\includegraphics[width=0.50\textwidth]{SFMa99fig2b} 
\end{tabular}
\end{center}
\caption{\label{fig:convergenceKerr042}
Convergence plot of the $l=m=0$ (left panel) and $l=m=1$ (right panel) modes 
of the scalar field with $M\mu_S=0.42$ in Kerr background with $a/M=0.99$.
We present the differences between the coarse-medium and medium-high resolution runs, where the
latter is amplified by $Q_2=1.66$ indicating second order convergence.
}
\end{figure}
%
%
\begin{figure}
\begin{center}
\begin{tabular}{cc}
\includegraphics[width=0.5\textwidth]{SFMa99fig3a} &
\includegraphics[width=0.5\textwidth]{SFMa99fig3b} 
\end{tabular}
\end{center}
\caption{\label{fig:waveformsKerrmu055_l0}
Left: Logarithm of the $l=m=0$ mode of the massive scalar field with $M\mu_S=0.55$ in Kerr background
      with $a/M = 0.99$, extracted at $r_{ex}=20,26.5,35,45M$.
Right: Log-Log plot of the same mode, extracted at $r_{ex}= 20M$.
       We plot the numerical data (black solid line) together with the fit of the 
       late-time tail $\Psi\sim t^{-0.847}$.
}
\end{figure}
%
\subsection{Mode excitation and beating }\label{ssec:excitationcoefficients}
We now discuss an exciting effect which we observe in the dipole modes of the scalar fields with
mass couplings $M\mu_S=0.42$ and $M\mu_S=0.55$ in Schwarzschild and Kerr background with $a/M=0.99$.
We present the waveforms, measured at different radii, in Figs.~\ref{fig:waveformsSSm042_11},
\ref{fig:waveformsSSm055}, \ref{fig:Kerrmu042_l1} and~\ref{fig:waveformsKerrmu055}.
Typically, we have chosen the extraction radii such that $r_{ex}<r_{node}$, 
$r_{ex}\sim r_{node}$, $r_{ex}>r_{node}$ and $r_{ex}\sim r_{max}$, where $r_{node}$ and $r_{max}$
denote the location of the node and local maximum ($r_{max}>r_{node}$) of the first overtone of 
the corresponding quasi-bound state. Although strictly speaking being a different state, 
the quasi-bound state serves as excellent guide for the numerical simulations of a scalar field 
with more generic initial profile.
By studying Figs.~\ref{fig:waveformsSSm042_11}, \ref{fig:waveformsSSm055}, 
\ref{fig:Kerrmu042_l1} and~\ref{fig:waveformsKerrmu055}, we observe two effects:
\noindent{(i)} 
a modulation of the $l=m=1$ waveform and
\noindent{(ii)} 
a dependence of this modulation on the location of measurement.
We further note, that these effects are more pronounced in the case of the scalar field
with mass coupling $M\mu_S=0.42$ than for $M\mu_S=0.55$.
Therefore, we will restrict ourselves to the analysis of the former case.
The two phenomena can be understood by a beating effect between the fundamental and overtone mode
and their space dependent excitation.
In the following, we will investigate both effects in more detail.

The observed modulation or envelope of the waveforms can be understood 
by considering the beating between the fundamental and $n$-th overtone mode with similar real 
parts of the frequencies, i.e., 
$\omega_0 = \omega_{R,0} + \imath \omega_{I,0}$ and 
$\omega_n = \omega_{R,n} + \imath \omega_{I,n}$
with $\omega_{n,R} = \omega_{0,R} + \delta$ and $\delta \ll 1$.
The imaginary parts are related via $\omega_{I,n} = \omega_{I,0} / (n+1)$.
Then, if we restrict ourselves to only two competing modes, 
the scalar field can be expressed as
\begin{align}
\label{eq:FitBeating}
\Psi \sim & A_0 \exp(-\imath \omega_0 t) + A_n \exp(-\imath \omega_n t)
\nonumber \\
        = & \exp(-\imath \omega_{R,0} t)
            \left(  A_0 \exp(\omega_{I,0} t)
                  + A_n \exp(-\imath\delta t) \exp\left(\frac{\omega_{I,0} t}{n+1}\right) 
            \right)
\,,
\end{align}
where $A_0$ and $A_n$ are the respective amplitudes of the fundamental and overtone mode.
In order to estimate the frequencies of the modes, their difference $\delta$ and amplitudes 
we fit the numerical data
to the real part of the scalar field which is (approximately) given by
\begin{align}
\label{eq:FitBeatingRe}
\Re(\Psi) \sim & A_0 \exp(\omega_{I,0} t) + A_1 \exp(\omega_{I,0} t/2) \cos(\delta t)
\,.
\end{align}
Here, we assume a beating between the fundamental and first overtone mode.
The results of this fit for the scalar field with $M\mu_S=0.42$ are summarized in Table~\ref{tab:fitsSFKerr}.
In some cases, the imaginary part of the frequency could not be fitted accurately because of numerical noise.
These cases are denoted by {\textit{n.a.}}.
We find good agreement between the fitted and theoretically predicted values 
of the real parts of the frequencies $\omega_{R,0}$ and $\omega_{R,1}$ 
and their difference $\delta$ (see Table~\ref{tab:MFScaAnaModes}),
within less than $7\%$.

Additionally, we plot the fitted curves together with the numerical ones in Figs.~\ref{fig:waveformsSSm042_11}
and~\ref{fig:Kerrmu042_l1} for specific values of the extraction radii.
We have focused on the cases $r_{ex}<r_{node}$, $r_{ex}\sim r_{node}$ and $r_{ex}\sim r_{max}>r_{node}$.
This is a reasonable choice because the generic scalar field eventually approaches the bound state,
as becomes evident in the top right panel of Fig.~\ref{fig:Kerrmu042_l1}.
The plot corresponds to the $l=m=1$ mode measured at $r_{ex}\sim r_{node}$, where the amplitude of the 
(quasi-bound state) overtone mode vanishes. 
Whereas the waveform shows a small modulation at early times, the signal flattens at late times,
indicating that the overtone amplitude decreases, thus indeed approaching the quasi-bound state.

Let's now focus on the scalar field with $M\mu_S=0.42$ in Kerr background, Fig.~\ref{fig:Kerrmu042_l1}.
The plots nicely illustrate the interplay between the fundamental and overtone mode
and the dependence of their amplitude on the location of measurement.
In the case $r_{ex}<r_{node}$ (top left panel of Fig.~\ref{fig:Kerrmu042_l1}), 
the amplitudes of the fundamental and overtone mode are both
present with $A_0>A_1$. The signal 
exhibits a clear, though not very strong modulation of the waveform.
In the case $r_{ex}\sim r_{node}$ (top right panel of Fig.~\ref{fig:Kerrmu042_l1})
we observe a slight modulation of the signal at early times
which flattens at late times. This indicates that the amplitude of the overtone mode 
decreases and eventually only the fundamental mode will be present.
In the case $r_{ex}\sim r_{max}$ we find that the amplitudes of the fundamental and overtone mode
are similar, $A_1\le A_0$. This results in a strong modulation of the 
scalar field as can be seen in the bottom left panel of Fig.~\ref{fig:Kerrmu042_l1}.

We have observed that different modes are excited to different amplitudes 
depending on initial data and the location where the mode is being measured.
In a seminal work, Leaver has established some rigorous results in this regard, 
by deforming the contour integral in the complex-frequency plane; 
for further details we refer the reader to the original work \cite{Leaver:1986gd} 
and to other comprehensive follow-ups \cite{Andersson:1995zk,Berti:2006wq,Berti:2009kk,Zhang:2013ksa}.
The upshot is that each quasinormal mode, which corresponds to a pole in the complex-frequency plane, 
is excited to a different degree depending on the initial data and on the mode in question. 
In the frequency-domain, it is useful to represent the scalar field by the Klein-Gordon equation in
the form
\begin{align}
\label{potbh1}
\frac{d^2\Psi(\omega\,,r)}{dr_*^2} +V\Psi(\omega\,,r) = & I(r)
\,,
\end{align}
where $V$ is a mass-dependent potential (see for instance \cite{Detweiler:1980uk} for details). 
The function $I(r)$ is a generic source term, which describes for instance
the initial data when $\Psi(\omega)$ is a Laplace transform of the time-domain wavefunction.
The QNM contribution can be isolated from other features of the signal, such
as the late-time tail, using the Green's function technique
\cite{Leaver:1986gd,Andersson:1995zk,Berti:2006wq}. First, one defines a solution of the homogeneous equation
having the correct behavior at the horizon (only in-going waves),
\begin{subequations}
\label{asrplus}
\begin{align}
& \lim_{r \to r_+} \Psi_{r_+}\sim e^{-i(\omega-m\Omega) r_*}
\,,\\
& \lim_{r \to \infty} \Psi_{r_+}\sim A_{\rm in}(\omega)e^{-q r}+A_{\rm out}(\omega)e^{qr}
\,, 
\end{align}
\end{subequations}
where $q=\pm \sqrt{\mu_S^2-\omega^2}$.
The excitation amplitude of each mode is characterized by the quantity
\begin{align}
\label{Cn}
C_n\propto & B_n{\hat \Psi}_{r_+}(\omega_n,r)\int I(r) {\hat \Psi}_{r_+}(\omega_n,r)dr
\,.
\end{align}
Here, the function ${\hat \Psi}_{r_+}(\omega_n,r)\equiv \Psi_{r_+}(\omega_n,r)/A_{\rm out}$ 
is the normalized wavefunction, to be evaluated {\it at} the quasinormal frequency $\omega_n$. 
Thus, two quantities are crucial to determine the excitation of a given mode: 
the convolution of the homogeneous solution with the initial data, 
and the {\it excitation factor} $B_n$ which is equal to
\begin{align}
B_n = & A_{\rm out}\left(\frac{dA_{\rm in}}{d\omega}\right)^{-1}
\,.
\end{align}
Thus, for instance, the relative amplitude between different modes depends strongly 
on the point where this amplitude is evaluated: if it is close to a node, 
the mode in question will have a very small amplitude (see Eq.~\eqref{Cn}): 
by definition a mode is not excited at its node. 
Likewise, Eq.~\eqref{Cn} implies that localized initial data close to the node of the mode does 
not excite the mode in question, a well-known result for closed systems \cite{Berti:2006wq}.

We have not attempted a complete description of mode excitation for this work, 
a preliminary analysis indicates that the excitation factors $B_n$ differ appreciable for
different quasi-bound states overtones, contrary to the usual quasinormal modes \cite{Berti:2006wq}. 
For instance, for $a=0$, the fundamental bound-state with $M\omega=0.407524-0.001026i$ has $B_n \sim 10^{-6}$ 
while the first overtone $M\omega=0.4146906 - 0.0004053i$ has $B_n\sim 10^{-9}$. 
On the other hand $N_n/A_{\rm out}^2$ is roughly of the same order of magnitude for both modes.

\begin{table*}
\begin{center}
\begin{tabular}{ccccccccc}
\hline
Run              & $M \mu_S$ & $a/M$  & $r_{ex}/M$ & $\delta$     & $M\omega_{R,0}$ & $M\omega_{R,1}$     
& $A_0$    & $A_1$ \\
\hline
S\_m042\_a0      & $0.42$  & $0.0$  & $22.5$     & $0.0077$     & $0.4077$        & $0.4154$           
& $0.15$ & $0.017$ \\
                 &         &        & $30.0$     & $0.0070$     & $0.4079$        & $0.4149$ 
& $0.08$ & $0.02$  \\
\hline
S\_m042\_a99$_c$ & $0.42$  & $0.99$ & $20.0$     & $0.0062$     & $0.408$          & $0.4143$ 
& $0.085$  & $0.013$ \\
                 &         &        & $26.5$     & $0.0069$     & $0.4085$         & $0.4154$ 
& $0.054$  & $0.0003$ \\
                 &         &        & $50.0$     & $0.0063$     & $0.4057$         & $0.412$ 
& $0.012$  & $0.008$  \\
\hline
\end{tabular}
\end{center}
\caption{\label{tab:fitsSFKerr} 
Parameters of the fit 
given by Eq.~\eqref{eq:FitBeatingRe},
modelling the $l=m=1$ mode of the massive scalar fields with $M\mu_S=0.42$ 
in Schwarzschild or Kerr background with $a/M=0.99$.
}
\end{table*}

\begin{figure}
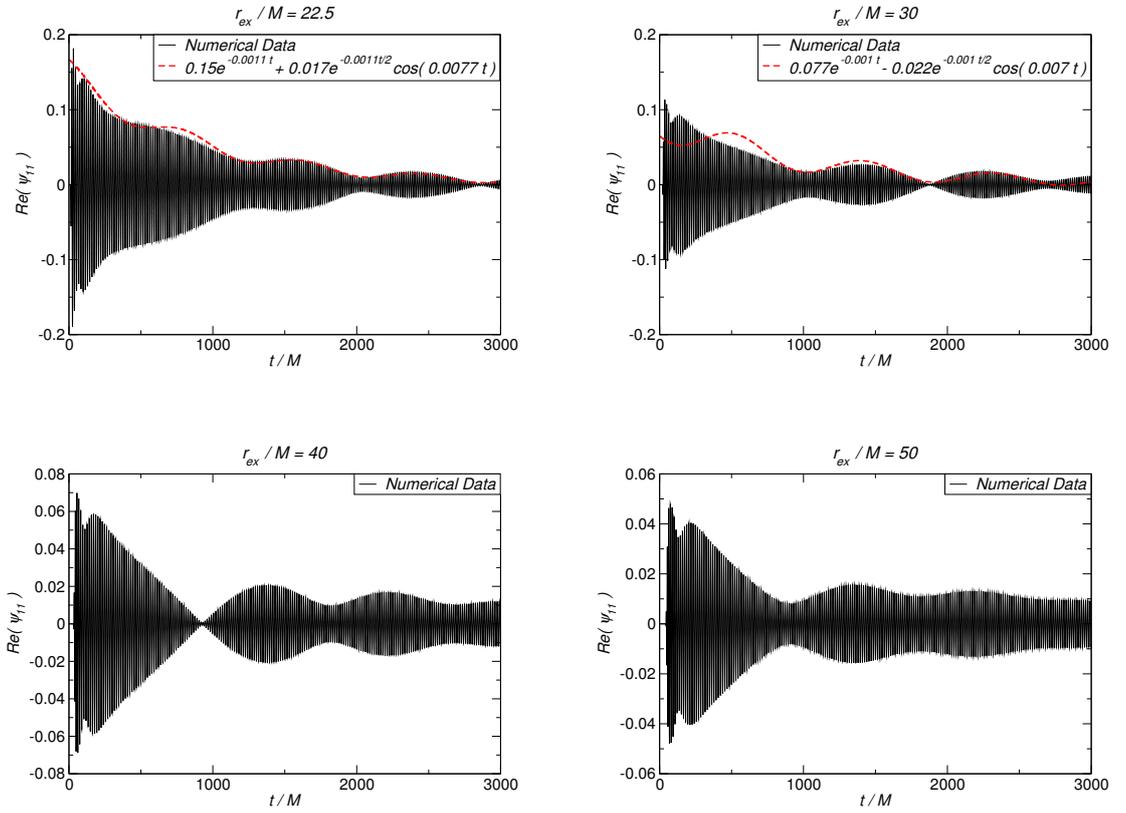

\begin{center}
\begin{tabular}{cc}
\includegraphics[width=0.50\textwidth]{SFBeatfig1a} & 
\includegraphics[width=0.50\textwidth]{SFBeatfig1b} \\ 
\includegraphics[width=0.50\textwidth]{SFBeatfig1c} & 
\includegraphics[width=0.50\textwidth]{SFBeatfig1d}  
\end{tabular}
\end{center}
\caption{\label{fig:waveformsSSm042_11}
Real part of the $l=m=1$ mode of the scalar field with $M\mu_S = 0.42$ in Schwarzschild background
as function of time and extracted at radii 
$r_{ex}=22.5~M$ (top left), $r_{ex}=30~M$ (top right), $r_{ex}=40~M$ (bottom left) and
$r_{ex}=50~M$ (bottom right). The first extraction radius marks the node position of the first overtone
of the corresponding quasi-bound state.
The red dashed line in the top panel are a fit of the form Eq.~\eqref{eq:FitBeatingRe}.
}
\end{figure}
%
\begin{figure}
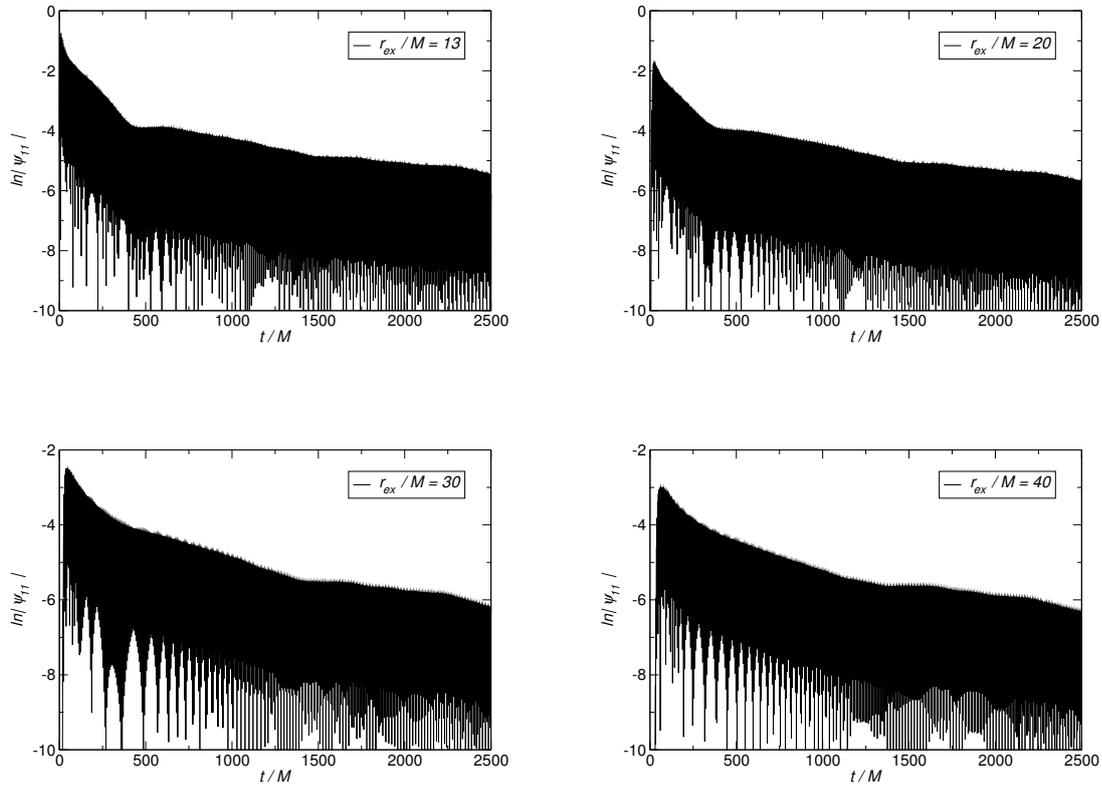

\begin{center}
\begin{tabular}{cc}
\includegraphics[width=0.50\textwidth]{SFBeatfig2a} &
\includegraphics[width=0.50\textwidth]{SFBeatfig2b} \\
\includegraphics[width=0.50\textwidth]{SFBeatfig2c} &
\includegraphics[width=0.50\textwidth]{SFBeatfig2d}
\end{tabular}
\end{center}
\caption{\label{fig:waveformsSSm055}
Logarithm of the $l=m=1$ mode of the real part of the massive scalar field with 
$M \mu_S = 0.55$ in Schwarzschild background. The waveforms have been extracted at 
$r_{ex}=13~M$ (top left), $r_{ex}=20~M$ (top right),
$r_{ex}=30~M$ (bottom left) and  $r_{ex}=40~M$ (bottom right).
The first extraction radius corresponds to the node position of the first overtone
of the corresponding quasi-bound state.
}
\end{figure}

\begin{figure}
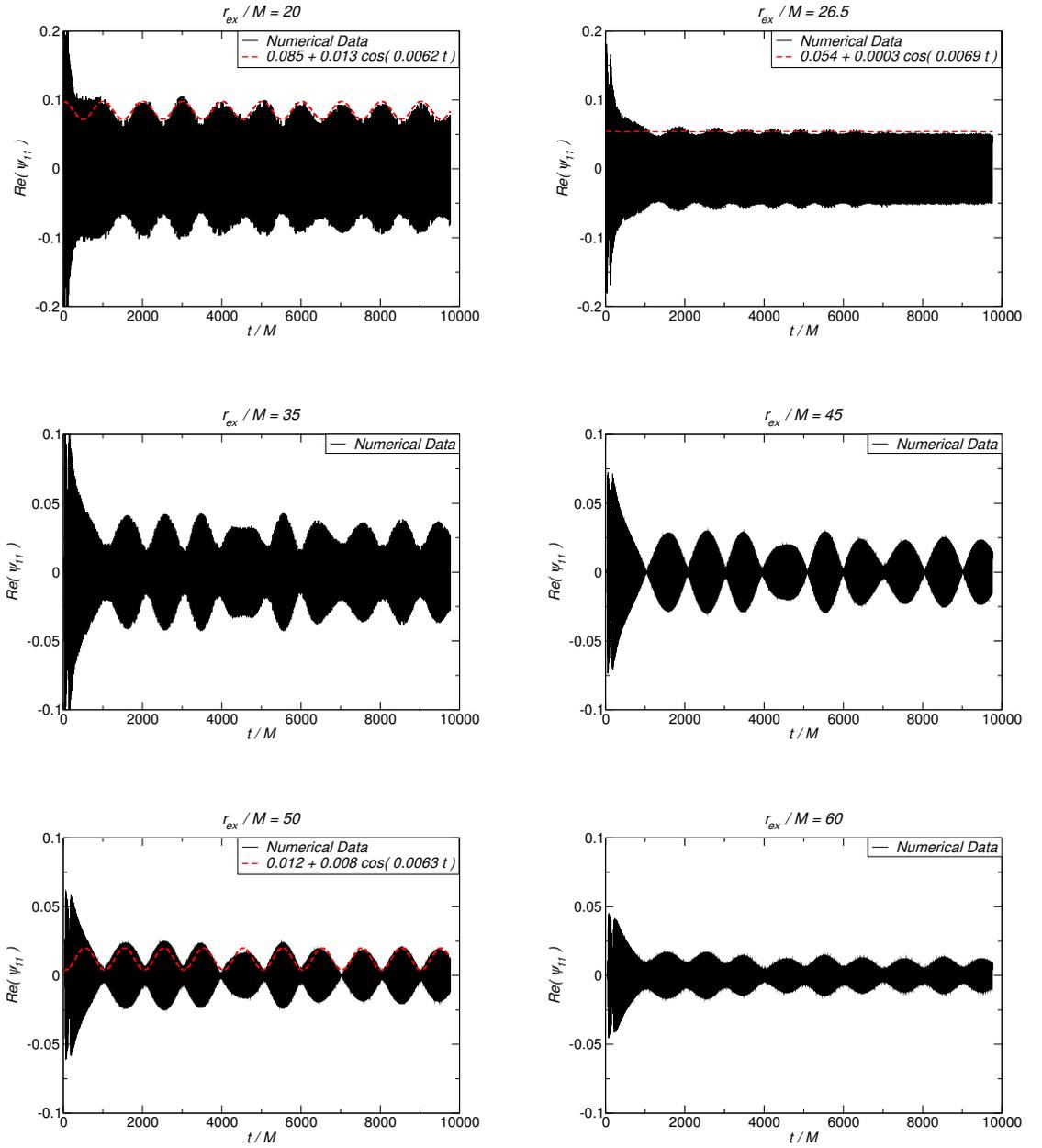

\begin{center}
\begin{tabular}{cc}
\includegraphics[width=0.5\textwidth]{SFBeatfig3a} & 
\includegraphics[width=0.5\textwidth]{SFBeatfig3b} \\ 
\includegraphics[width=0.5\textwidth]{SFBeatfig3c} & 
\includegraphics[width=0.5\textwidth]{SFBeatfig3d} \\
\includegraphics[width=0.5\textwidth]{SFBeatfig3e} & 
\includegraphics[width=0.5\textwidth]{SFBeatfig3f} \\
\end{tabular}
\end{center}
\caption{\label{fig:Kerrmu042_l1}
Real part of the $l=m=1$ mode of the massive scalar field with $M\mu_S=0.42$ in Kerr background
with $a/M = 0.99$, extracted at 
$r_{ex} = 20~M$ (top left), $r_{ex} = 26.5~M$ (top right),
$r_{ex} = 35~M$ (mid left), $r_{ex} = 45~M$ (mid right),
$r_{ex} = 50~M$ (bottom left) and $r_{ex}= 60~M$ (bottom right).
$r_{ex} = 26.5~M$ (top right) corresponds to the position of the node of the first overtone of the 
quasi-bound state, 
whereas $r_{ex}\sim50~M$ corresponds to its local maximum.
The red dashed lines indicate the fit, Eq.~\eqref{eq:FitBeatingRe}, 
modelling the envelope well at late times.
}
\end{figure}
\begin{figure}
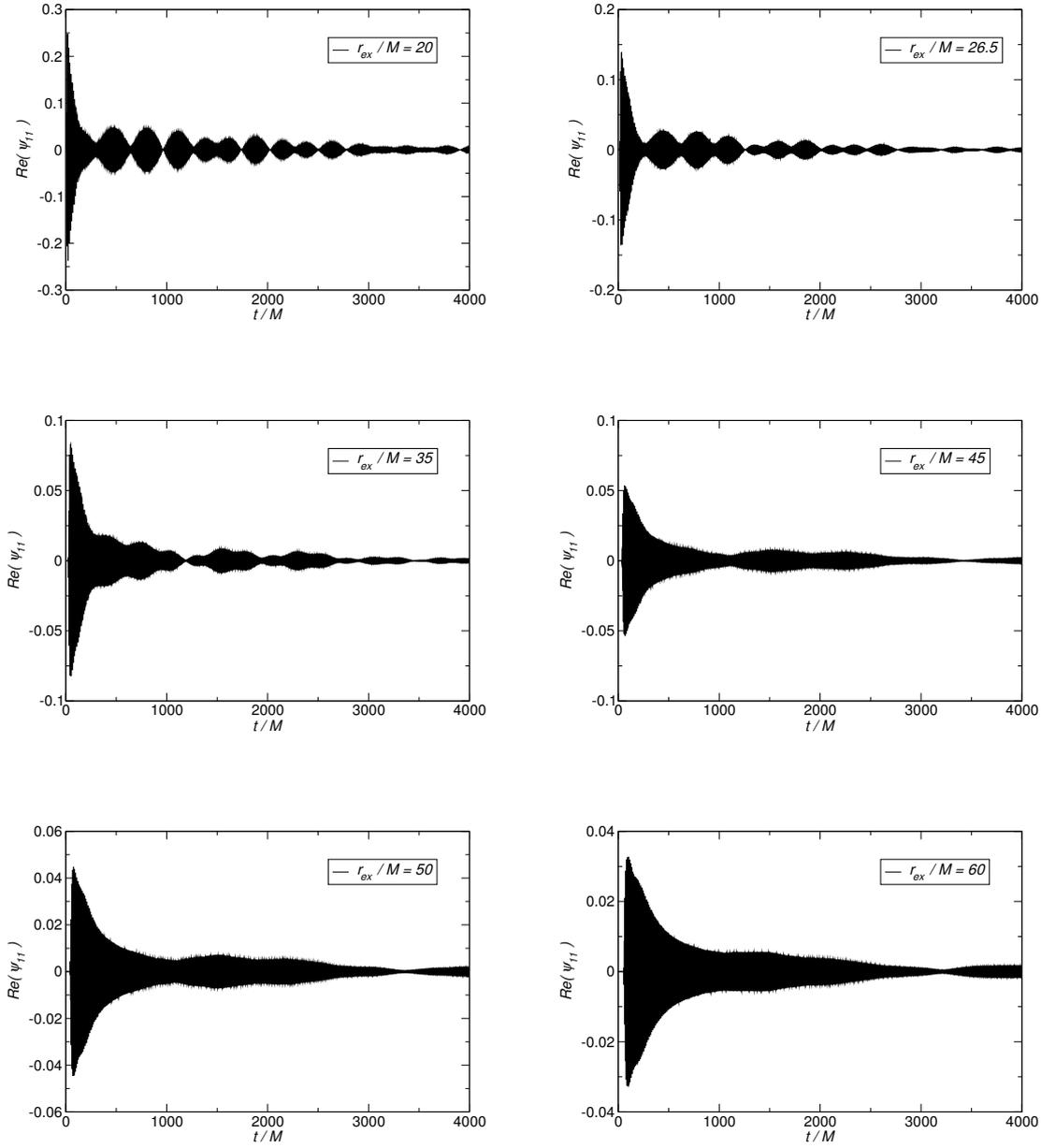

\begin{center}
\begin{tabular}{cc}
\includegraphics[width=0.5\textwidth]{SFBeatfig4a} &
\includegraphics[width=0.5\textwidth]{SFBeatfig4b} \\
\includegraphics[width=0.5\textwidth]{SFBeatfig4c} &
\includegraphics[width=0.5\textwidth]{SFBeatfig4d} \\
\includegraphics[width=0.5\textwidth]{SFBeatfig4e} &
\includegraphics[width=0.5\textwidth]{SFBeatfig4f} \\
\end{tabular}
\end{center}
\caption{\label{fig:waveformsKerrmu055}
Real part of the $l=m=1$ mode of the massive scalar field with $M\mu_S=0.55$ in Kerr background
with $a/M = 0.99$, extracted at 
$r_{ex} = 20~M$ (top left), $r_{ex} = 26.5~M$ (top right),
$r_{ex} = 35~M$ (mid left), $r_{ex} = 45~M$ (mid right),
$r_{ex} = 50~M$ (bottom left) and $r_{ex}= 60~M$ (bottom right).
}
\end{figure}
%

\clearpage
\newpage
\section{Conclusions}\label{sec:MFconclusion}
In this chapter we have studied in detail the evolution of massive scalar fields in the background
of Schwarzschild or Kerr BHs in the time-domain.
Therefore, we have presented a time evolution code {\textsc{Lin-Lean}}, which enables us to model the progression
of massive (scalar) fields in the background of BHs.
The motivation to revisit the evolution of massive scalar fields in the linear regime,
which has been widely studied in the literature 
\cite{Detweiler:1980uk,Zouros:1979iw,Damour:1976,Furuhashi:2004jk,
Dolan:2007mj,Cardoso:2005vk,Koyama:2001ee,Koyama:2001qw, Burko:2004jn,
Strafuss:2004qc,Yoshino:2012kn,Barranco:2012qs},
has been twofold:
\noindent{(i)} extensively testing and verifying our code in preparation for future applications; 
\noindent{(ii)} to shed some light onto the puzzling results reported by Strafuss \& Khanna \cite{Strafuss:2004qc}.
The authors found an instability growth rate $M\omega_I\sim2\cdot10^{-5}$ for a massive scalar field with $M\mu_S=0.25$ 
in Kerr background with $a/M=0.9999$ -- two orders of magnitude larger than the results of
frequency domain calculations for the maximum possible instability timescale\cite{Dolan:2007mj,Cardoso:2005vk}.

As initial configurations we have considered quasi-bound states, which are localized in the
vicinity of the BH and represent very clean states. The second class of initial data have been
Gaussian wave packets, representing more general and therefore possibly more realistic setups.

The first part has been devoted to check our implementation.
Therefore we have evolved a spherically symmetric scalar field with unphysical, spatial dependent mass coupling which 
yields an instability. The numerically computed growth rate agrees with the theoretical one within less than $1\%$.
As a second benchmark test we have evolved massless scalar fields in Schwarzschild and highly spinning Kerr 
background with $a/M=0.99$. The thus estimated QN frequencies and, if applicable, late time tails 
have been in excellent agreement with theoretical predictions \cite{Berti:2009kk}. 
Notice, that the more challenging case of highly rotating background has required
very high resolution of the numerical grid close to the BH as well as in the wave extraction zone. 

Our next step included the numerical construction of quasi-bound state initial data of a dipole
scalar field with mass coupling $M\mu_S=0.42$. 
As background we have choosen a Kerr BH with spin parameter $a/M=0.99$
which is known to give rise to the strongest instability \cite{Dolan:2007mj,Cardoso:2005vk}.
The thus computed fundamental and overtone modes of the quasi-bound state 
have been in excellent agreement with 
a comparative computation employing Leaver's method \cite{Leaver:1986gd,Dolan:2007mj,Cardoso:2005vk}.
Additionally, in a time evolution we have verified that the amplitude of the 
field hardly varied with time. However, we have not been able to observe the growth
of the instability because its expected value is tiny, $M\omega_I\sim1.5\cdot10^{-7}$,
and therefore completely buried in numerical noise.

Finally, we have evolved scalar fields with mass coupling $M\mu_S=0.1, 0.42, 0.55, 1.0$
in Schwarzschild and $M\mu_S=0.42, 0.55$ in highly rotating Kerr background with $a/M=0.99$.
As initial configuration we have chosen generic Gaussian wave packets.
We have first focused on 
the dipole modes of the field with small and large mass couplings $M\mu_S=0.1$ and $M\mu_S=1.0$
as well as the monopole modes for $M\mu_S=0.42$ and $M\mu_S=0.55$.
Their numerically computed QN frequencies and late-time tails have been in good agreement,
overall within less than $\sim5\%$, with previously reported results 
\cite{Dolan:2007mj,Cardoso:2005vk,Koyama:2001ee,Koyama:2001qw,Burko:2004jn}.
The time evolution of the dipole modes of the scalar field with mass couplings $M\mu_S=0.42$ and $M\mu_S=0.55$
have revealed an interesting phenomena: 
a modulation of the signal over time, where the amplitude of the envelope strongly depends on the location
of the measurement.
This behaviour can be understood by the beating effect between fundamental and overtone modes
present due to the generic initial data.
Indeed, the real parts of the fundamental and overtone frequncies as well as the 
beating frequncy $\delta$ estimated from the numerical data are in remarkable agreement
with theoretical predictions summarized in Table~\ref{tab:MFScaAnaModes}.

A further analysis and comparison to the corresponding quasi-bound states 
revealed that the excitation of the respective modes is space dependent.
For example, the amplitude of the overtone mode at its node vanishes by definition, 
thus suppressing any modulation.
Indeed, our numerical data match these expectations 
(as can be seen in the top right panel of Fig.~\ref{fig:Kerrmu042_l1}):
The dipole mode of the generic scalar field extracted at the overtone's node of the corresponding 
quasi-bound state shows little variation at early times and relaxes to an almost constant 
amplitude at late times.
Further evidence is given by the signal extracted at the local maximum of the overtone.
At this position the amplitudes of fundamental and overtone mode are similar, resulting
in a strong modulation of the waveform.
Indeed, we observe the expected behaviour in the bottom left panel of Fig.~\ref{fig:Kerrmu042_l1}.

The observed beating effect, supported by the findings in Ref.~\cite{Dolan:2012yt},
has the potential to explain the puzzling result reported
by Strafuss \& Khanna \cite{Strafuss:2004qc}.
As in our case, they have evolved a Gaussian wave packet which might 
excite not only the fundamental but also overtone modes.
A rough analysis employing Eq.~\eqref{eq:FitBeating} with their setup
reveals that beating should be present.
This results in a {\textit{local}} growth rate of $M\omega_I\sim 10^{-5}$
as has been observed in \cite{Strafuss:2004qc} and misinterpreted as overall instability growth of the
massive scalar field.

By developing and testing our $3+1$ time evolution code that simulates massive scalar fields in the background of
BH spacetimes we have performed first important step towards the exploration of more 
challenging scenarios:

\noindent{(1) Evolution of Proca fields in BH backgrounds:} 
A further exciting line of research is the investigation of massive vector fields in BH backgrounds.
They are expected to give rise to a much stronger superradiant instability than the scalar case.
Previous investigations have focused on massive vector fields in Schwarzschild 
\cite{Gal'tsov:1984nb, Konoplya:2005hr,Rosa:2011my} 
and Tangherlini \cite{Herdeiro:2011uu} backgrounds.
Especially, Rosa \& Dolan \cite{Rosa:2011my} have recently reported on the complete 
QNM and bound state spectrum of Proca fields in Schwarzschild backgrounds.
However, the Kerr case proves the be much more challenging because the equations of motion
appear to be non-separable for highly spinning BHs 
(but see \cite{Pani:2012vp,Pani:2012bp} for a computation in the slow rotation approximation).
Therefore, our code provides the ideal setup, with minor modifications, to address 
this challenging task. First results will be reported elsewhere \cite{Witek:2012tr}.

\noindent{(2) Self-interacting scalar fields:}
Another class of interesting problems is given by massive scalar fields whose dynamics are described by additional 
non-linear terms, modelling their self-interaction. 
This open issue has first been addressed by Yoshino \& Kodama \cite{Yoshino:2012kn}
who modelled the collapse of a so-called bosenova.

\noindent{(3) Backreaction effects:}
As far as we are aware, all studies involving massive scalar fields have been performed in the
linear regime, i.e., neglecting backreaction effects of the BH spacetime.
Therefore it is of utmost interest to explore the fully non-linear regime, 
which allows for the investigation of backreaction of the spacetime, such as the spin-down of the BH 
due to (subsequent) superradiant scattering.
This type of studies would enable us to throw a glance at the  
end-state of the superradiant instability or, possibly, equilibrium configurations.

There is an entire playground of exciting future applications of massive fields in
BH spacetimes.
Our code is perfectly suited to address these challenging problems, which are subject to ongoing work.


\appendix
\chapter{Appendix -- Electromagnetic decomposition of the Weyl tensor}
\label{sec:appWaveExtraction}
Because we analyse the Newman-Penrose scalar $\Psi_0$,
besides $\Psi_4$, which is uncommon in numerical works, we collect  
some useful results in this appendix. Following the sign convention in
\cite{Friedrich:1996hq, Sperhake:2006cy} the Newman-Penrose scalars
$\Psi_0$ and $\Psi_4$ are computed by Eqs.~\eqref{eq:BoxPsi0_Weyl} and \eqref{eq:BoxPsi4_Weyl}.
The vectors $\vec{k}, \vec{\ell}, \vec{m}, \bar{\vec{m}}$ form a null-tetrad. 
Their inner products vanish except for 
\begin{align}
-\vec{k}\cdot \vec{\ell} = & 1 = \vec{m}\cdot\bar{\vec{m}}
\,.
\end{align}
In practice, the vectors of the null-tetrad are constructed from a Cartesian
orthonormal basis $(u,v,w)$ in the spatial hypersurface
and the timelike orthonormal vector $\hat{\vec{n}}$ according to 
\begin{subequations}
\label{eq:BoxnulltetradCart}
\begin{align}
k^{\alpha} = & \frac{1}{\sqrt{2}} ( \hat{n}^{\alpha} + u^{\alpha})
\,, \quad
\ell^{\alpha} =  \frac{1}{\sqrt{2}} ( \hat{n}^{\alpha} - u^{\alpha} )
\,,\\
m^{\alpha} = & \frac{1}{\sqrt{2}} ( v^{\alpha} + i w^{\alpha} )
\,, \quad
\bar{m}^{\alpha} =  \frac{1}{\sqrt{2}} ( v^{\alpha} - i w^{\alpha} )
\,.
\end{align}
\end{subequations}
The orthonormal triad vectors are constructed via the Gram-Schmidt orthonormalization starting with
\begin{subequations}
\begin{align}
u^i  = & [x,y,z]
\,, \\
v^i  = & [xz, yz, -x^2-y^2]
\,, \\
w^i  = & \epsilon^i_{jk} u^j w^k
\,,
\end{align}
\end{subequations}
where $\epsilon^i_{jk}$ is the 3-dimensional Levi-Civita tensor.
Next, we decompose the Weyl tensor in terms of its electric and magnetic parts 
\cite{Friedrich:1996hq}
\begin{align}
\label{eq:BoxWeylDecomp}
C_{\alpha\beta\gamma\delta}  = &  
2\left(  l_{\alpha[\gamma}E_{\delta]\beta} - l_{\beta[\gamma}E_{\delta]\alpha} 
       - \hat{n}_{[\gamma}B_{\delta]\tau}\epsilon^{\tau}_{\alpha\beta} 
       - \hat{n}_{[\alpha}B_{\beta]\tau}\epsilon^{\tau}_{\gamma\delta}
  \right)
\,,
\end{align}
where $l_{\mu\nu} = \gamma_{\mu\nu} + \hat{n}_{\mu}\hat{n}_{\nu}$ and
$\epsilon_{\alpha\beta\gamma} = \epsilon_{\mu\nu\lambda\rho}\hat{n}^{\mu}
\ga^{\nu}_{\alpha}\ga^{\lambda}_{\beta}\ga^{\rho}_{\gamma}$.
The electric and magnetic part of the Weyl tensor are given by
\begin{subequations}
\begin{align}
E_{\alpha\beta} = & C_{\mu\nu\lambda\rho}\ga^{\mu}_{\alpha}\hat{n}^{\nu}\ga^{\lambda}_{\beta}\hat{n}^{\rho}
\,,\\
B_{\alpha\beta} = & ^{\ast}C_{\mu\nu\lambda\rho}\ga^{\mu}_{\alpha}\hat{n}^{\nu}\ga^{\lambda}_{\beta}\hat{n}^{\rho}
\,.
\end{align}
\end{subequations}
$\ga^{\mu}_{\nu}$ denotes the projection operator onto the hypersurface and ${^{\ast}}$ denotes the Hodge dual.
By using the Gauss-Codazzi equations we express the electromagnetic components
in terms of the ``3+1'' variables \cite{Sperhake:2006cy}
\begin{subequations}
\begin{align}
\label{eq:BoxelecWeylcomp3+1}
E_{ij} = & R_{ij} - \gamma^{kl}( K_{ij}K_{kl}-K_{ik}K_{jl} )
\,,\\
\label{eq:BoxmagnWeylcomp3+1}
B_{ij} = & \gamma_{ik}\epsilon^{klm}D_l K_{mj}
\,.
\end{align}
\end{subequations}
If we insert Eq. \eqref{eq:BoxWeylDecomp}, 
the definition of the null-tetrad \eqref{eq:BoxnulltetradCart} and the expressions 
\eqref{eq:BoxelecWeylcomp3+1}, \eqref{eq:BoxmagnWeylcomp3+1} into the definition of 
the Newman-Penrose scalars \eqref{eq:BoxPsi0_Weyl},~\eqref{eq:BoxPsi4_Weyl} we obtain
\begin{subequations}
\label{eq:BoxelmagnPsi}
\begin{align}
\Psi_0 = & \frac{1}{2} [ E_{kl} ( v^{k}v^{l} - w^{k}w^{l} )
                + B_{kl} ( v^{k} w^{l} + v^{l} w^{k} ] 
\non\\ &
          +\frac{i}{2} [ E_{kl} ( v^{k}w^{l} + v^{l}w^{k} )
                - B_{kl} ( v^{k}v^{l} - w^{k}w^{l} ) ]
\,, \\
\Psi_4 = & \frac{1}{2} [ E_{kl} ( v^{k}v^{l} - w^{k}w^{l} )
                - B_{kl} ( v^{k}w^{l} + v^{l}w^{k} ) ] 
\non\\ &
          -\frac{i}{2} [ E_{kl} ( v^{k}w^{l} + v^{l}w^{k} ) +
                B_{kl} ( v^{k}v^{l} - w^{k}w^{l} ) ] 
\,.
\end{align}
\end{subequations}
In the numerical code we use these relations in order to calculate 
$\Psi_0$ and $\Psi_4$ on the entire Cartesian grid. Then, they are 
interpolated onto coordinate spheres of various extraction radii $r_{\rm ex}$.
The Newman-Penrose scalars $\Psi_0$ and $\Psi_4$ are decomposed into 
spin-weighted spherical harmonics $_s Y_{lm}$ according to
\begin{subequations}
\begin{align}
\psi^0_{lm}(t) = & \int d\Omega\Psi_0(t,\theta,\phi  {_2}Y^{\ast}_{lm} \theta,\phi) 
        = (-)^{m+2}\int d\Omega \Psi_0(t,\theta,\phi) {_{-2}}Y_{lm}(\theta,\phi)
\,, \\
\psi^4_{lm}(t) = & \int d\Omega \Psi_4(t,\theta,\phi) {_{-2}}Y^{\ast}_{lm}(\theta,\phi)
\,.
\end{align}
\end{subequations}
In the first equation we have used the relation \cite{Goldberg:1966uu}
\begin{align}
 {_s} Y^{\ast}_{lm} = & (-)^{m+s} {_{-s}} Y_{lm}
\,.
\end{align}
Thus, in practice we implement $\psi^0_{lm}$ and $\psi^4_{lm}$ only 
in terms of the spherical harmonics ${_{-2}} Y_{lm}$ with spin-weight $-2$.

\chapter{Appendix -- List of spherical harmonics}\label{app:MFspherharm}
In this section we list the spherical harmonics up to $l=4$ 
in spherical coordinates $\{\theta,\phi\}$ and Cartesian coordinates 
\begin{subequations}
\begin{align}
\label{eq:CartCoords}
x = & r \sin\theta \cos\phi \,,\\
y = & r \sin\theta \sin\phi \,,\\
z = & r \cos\theta \,,
\end{align}
\end{subequations}
where $r^2 = x^2 + y^2 + z^2$ is the Kerr-Schild radial coordinate.
\begin{description}
\item [$l=0$] 
\begin{align}
Y^R_{00} = & \frac{1}{\sqrt{4\pi}}
\,,\quad
Y^I_{00} =   0
\end{align}
\item [$l=1$]
\begin{subequations}
\begin{align}
Y^R_{10} = & \sqrt{\frac{3}{4\pi}} \cos\theta
         =   \sqrt{\frac{3}{4\pi}} \frac{z}{r}
\,,\quad
Y^I_{10} = 0
\,,\\
Y^R_{11} = & -\sqrt{\frac{3}{8\pi}} \sin\theta \cos\phi
         =   -\sqrt{\frac{3}{8\pi}} \frac{x}{r}
\,,\quad
Y^I_{11} =   -\sqrt{\frac{3}{8\pi}} \sin\theta \sin\phi
         =   -\sqrt{\frac{3}{8\pi}} \frac{y}{r}
\,,\\
Y^R_{1-1} = & - Y^R_{11}
\,,\quad
Y^I_{1-1} =    Y^I_{11}
\end{align}
\end{subequations}
\item [$l=2$]
\begin{subequations}
\begin{align}
Y^R_{20} = & \sqrt{\frac{5}{16\pi}} (3\cos^2\theta - 1) 
         =   \sqrt{\frac{5}{16\pi}} \left(3\frac{z^2}{r^2} - 1\right)
\,,\quad
Y^I_{20} = 0
\,,\\
Y^R_{21} = & -\sqrt{\frac{5}{8\pi}} \cos\theta \sin\theta \cos\phi
         =   -\sqrt{\frac{5}{8\pi}} \frac{xz}{r^2}
\,,\quad
Y^I_{21} =   -\sqrt{\frac{5}{8\pi}} \cos\theta \sin\theta \sin\phi
         =   -\sqrt{\frac{5}{8\pi}} \frac{yz}{r^2}
\,,\\
Y^R_{22} = & \sqrt{\frac{15}{32\pi}} \sin^2\theta  \cos(2\phi)
         =   \sqrt{\frac{15}{32\pi}} \frac{x^2-y^2}{r^2}
\,,\quad
Y^I_{22} =   \sqrt{\frac{15}{32\pi}} \sin^2\theta \sin(2\phi)
         =   \sqrt{\frac{15}{8\pi}} \frac{xy}{r^2}
\,\\
Y^R_{2-1} = & - Y^R_{21}\,,\quad
Y^I_{2-1} =     Y^I_{21}\,,\\
Y^R_{2-2} = &   Y^R_{22}\,,\quad
Y^I_{2-2} =   - Y^I_{22}\,
\end{align}
\end{subequations}
\item [$l=3$]
\begin{subequations}
\begin{align}
Y^R_{30} = & \sqrt{\frac{7}{4\pi}} (2.5 \cos^3\theta - 1.5 \cos\theta)
\,,\quad
Y^I_{30} =   0
\,,\\
Y^R_{31} = & -\sqrt{\frac{21}{64\pi}} \sin\theta \cos\phi (5 \cos^2\theta - 1)
\,,\quad
Y^I_{31} =   -\sqrt{\frac{21}{64\pi}} \sin\theta \sin\phi (5 \cos^2\theta - 1) 
\,,\\
Y^R_{32} = & \sqrt{\frac{105}{32\pi}} \sin^2\theta \cos\theta \cos(2\phi)
\,,\quad
Y^I_{32} =   \sqrt{\frac{105}{32\pi}} \sin^2\theta \cos\theta \sin(2\phi)
\,,\\
Y^R_{33} = & -\sqrt{\frac{35}{64\pi}} \sin^3\theta \cos(3\phi)
\,,\quad
Y^I_{33} =   -\sqrt{\frac{35}{64\pi}} \sin^3\theta \sin(3\phi)
\,,\\
Y^R_{3-1} = & - Y^R_{31}\,,\quad
Y^I_{3-1} =     Y^I_{31}
\,,\\
Y^R_{3-2} = &   Y^R_{32}\,,\quad
Y^I_{3-2} =   - Y^I_{32}
\,,\\
Y^R_{3-3} = & - Y^R_{33}\,,\quad
Y^I_{3-3} =     Y^I_{33}
\end{align}
\end{subequations}
\end{description}
where $\cos(2\phi) = \cos^2\phi - \sin^2\phi$,
$\sin(2\phi) = 2\cos\phi \sin\phi$,
$\cos(3\phi) = 4 \cos^3\phi - 3 \cos\phi$
$\sin(3\phi) = 4 \cos^2\phi \sin\phi - \sin\phi$

\chapter{Appendix -- Black holes in higher dimensional spacetimes}
\section{Analysis of troublesome terms at $y=0$} \label{asec:troubleterms}
The right-hand sides of Eqs.~\eqref{eq:HDFgBSSNevol} and~\eqref{eq:HDFgBSSNcoupling}
contain various terms which cannot be evaluated directly at $y=0$ because they involve
explicit division by $y$.  Although these terms are regular by virtue of a
corresponding behaviour of the numerators, they need to be explicitly evaluated
in the numerical implementation. In this Appendix we outline how the regularity
of these terms can be implemented in a simple and efficient manner. For
convenience we use a special notation: late latin indices $i,~j,~\ldots$ run
from 1 to 3, covering $x$, $y$ and $z$, but early latin indices $a,~b,~\ldots$
take values 1 and 3 but not 2, i.e., they cover $x$ and $z$ but not $y$.

We begin this discussion by describing a simple manipulation which underlies
most of our regularisation procedure. Consider for this purpose a function $h$
which is linear in $y$ near $y=0$, i.e., its Taylor expansion is given by
$h(y) = h_1 y + \mathcal{O}(y^2)$. From this relation we directly obtain
\begin{equation}
  \lim_{y \rightarrow 0} \frac{h}{y} = h_1 = \partial_y h \, .
      \label{eq: yforderiv}
\end{equation}
This trading of divisions by $y$ for partial derivatives extends to higher
orders in a straightforward manner and will be used throughout the following
discussion.

Next, we consider the right-hand sides of
Eqs.~\eqref{eq:HDFgBSSNevol} and~\eqref{eq:HDFgBSSNcoupling}
and summarize the potentially troublesome terms as follows
\begin{align}
  &\frac{\beta^y}{y} \, , \qquad \frac{\tilde{\Gamma}^y}{y} \, ,
      \label{eq: irrbyoy} \\
  &\frac{\tilde{\gamma}^{ym}}{y}\partial_m f \, ,
      \label{eq: irrdf} \\
  &\frac{\tilde{\gamma}^{yy} \zeta - 1}{y^2} \, ,
      \label{eq: irrconsing} \\
  &\frac{1}{y} \left( \chi^{3/2\nk} \tA^{y}{}_{i} 
                + \de^y_i \left( \chi^{3/2\nk}\frac{\tK}{3} - \frac{K_{\ze}}{\ze}\right)
               \right)
  \,, \label{eq: irrdtconsing} \\
  &\frac{1}{y} \left( \delta_j{}^y \partial_i \zeta + \delta_i{}^y \partial_j
      \zeta - 2\zeta \tilde{\Gamma}^y_{ij} \right) \, .
      \label{eq: irrChris}
\end{align}
Here $f$ stands for either of the scalars or densities $\zeta$, $\chi$ and
$\alpha$.

Regularity of the terms \eqref{eq: irrbyoy} immediately follows from the
symmetry condition of the $y$-component of a vector
\begin{equation}
  \beta^y(-y) = -\beta^y(y) \, .
\end{equation}
We can therefore use the idea illustrated in Eq.~\eqref{eq: yforderiv}
and obtain
\begin{equation}
  \lim_{y\rightarrow 0} \frac{\beta^y}{y} = \partial_y \beta^y \, ,
\end{equation}
and likewise for $\tilde{\Gamma}^y/y$.
The terms \eqref{eq: irrdf} are treated in a similar manner because
the derivative of a scalar (density) behaves like a vector on our
Cartesian grid. We thus obtain
\begin{equation}
  \lim_{y \rightarrow 0} \left( \frac{\tilde{\gamma}^{ym}}{y}
        \partial_m f \right)
      = (\partial_y \tilde{\gamma}^{ya})(\partial_a f)
        + \tilde{\gamma}^{yy} \partial_y \partial_y f \, .
\end{equation}

Regularity of the expression \eqref{eq: irrconsing} is not immediately obvious
but can be shown to follow directly from the requirement that there should be no
conical singularity at $y=0$. Specifically, this condition implies that
$\tilde{\gamma}^{yy} \zeta = 1 + \mathcal{O}(y^2)$, so that
\begin{equation}
  \lim_{y \rightarrow 0} \left( \frac{\tilde{\gamma}^{yy} \zeta-1}{y^2} \right)
      = \frac{1}{2} \left( \zeta \partial_y \partial_y \tilde{\gamma}^{yy}
        + \tilde{\gamma}^{yy} \partial_y \partial_y \zeta \right) \, .
  \label{eq: regES_v3}
\end{equation}

The discussion of the term \eqref{eq: irrdtconsing} requires us to distinguish
between the cases $i=a\ne y$ and $i=y$. The former straightforwardly results in
\begin{equation}
  \lim_{y\rightarrow 0} \left( -\frac{\tilde{\gamma}^{ym}}{y} \tilde{A}_{ma}
       \right)
     = -\tilde{A}_{ba} \partial_y \tilde{\gamma}^{yb}
       - \tilde{\gamma}^{yy} \partial_y \tilde{A}_{ya} \, .
\end{equation}
For the case $i=y$, we first note that the limit $y \rightarrow0$ implies
$\tilde{\gamma}^{yy} = 1/\tilde{\gamma}_{yy} + \mathcal{O}(y^2)$, so that the
condition \eqref{eq: regES_v3}, i.e., no conical singularities, can be
written as
\begin{equation}
  \lim_{y \rightarrow 0} \left(\zeta - \tilde{\gamma}_{yy}\right)
      = \mathcal{O}(y^2) \, .
\end{equation}
Next we take the time derivative of this expression and obtain after some
manipulation
\begin{equation}
  \mathcal{O}(y^2) =
  \lim_{y \rightarrow 0} \partial_t(\zeta - \tilde{\gamma}_{yy})
     = -2\alpha \zeta \left( \frac{K_{\zeta}}{\zeta}
        - \chi^{3/2\nk} \frac{\tK}{3}
        - \chi^{3/2\nk} \tilde{\gamma}^{ym} \tilde{A}_{my} \right)
       + \mathcal{O}(y^2) \, ,
\end{equation}
and, consequently,
\begin{equation}
  \lim_{y \rightarrow 0} \left[ \frac{1}{y} \left( \frac{K_{\zeta}}{\zeta}
        - \chi^{3/2\nk} \frac{\tK}{3}
        - \chi^{3/2\nk} \tilde{\gamma}^{ym} \tilde{A}_{my} \right) \right]
      = 0 \, .
\end{equation}

Finally, we consider the term \eqref{eq: irrChris}. Expansion of the Christoffel
symbol, repeated use of the method illustrated in Eq.~\eqref{eq: yforderiv} and
the condition for avoiding a conical singularity enable us to regularise this
term for all combinations of the free indices $i$ and $j$. We thus obtain
\begin{align}
  \lim_{y \rightarrow 0} \left[
      \frac{1}{y} \left( 2\partial_y \zeta - 2\zeta \tilde{\Gamma}^y_{yy}
      \right) \right]
    & = 2\partial_y \partial_y \zeta - \zeta \tilde{\gamma}^{yy}
      \partial_y \partial_y \tilde{\gamma}_{yy}
      - \zeta (\partial_y \tilde{\gamma}^{yc}) (2\partial_y \tilde{\gamma}_{yc}
      - \partial_c \tilde{\gamma}_{yy}) \, , \\
  \lim_{y \rightarrow 0} \left[
      \frac{1}{y} \left( \partial_a \zeta - 2\zeta \tilde{\Gamma}^y_{ay}
      \right) \right]
    &=  0 \, , \\
    \begin{split}
      \lim_{y \rightarrow 0} \left[
      -2\frac{\zeta}{y} \tilde{\Gamma}^y_{ab} \right]
    & = -\zeta \tilde{\gamma}^{yy} (\partial_y \partial_a \tilde{\gamma}_{by}
      +\partial_y \partial_b \tilde{\gamma}_{ya} - \partial_y \partial_y
      \tilde{\gamma}_{ab}) 
      \\    
      & \quad 
      - \zeta (\partial_y \tilde{\gamma}^{yc}) (\partial_a \tilde{\gamma}_{bc}
      + \partial_b \tilde{\gamma}_{ac} - \partial_c \tilde{\gamma}_{ab}) \, .
    \end{split}
\end{align}
We conclude this discussion with a method to express derivatives of the inverse
metric in terms of derivatives of the metric. For this purpose we use the
condition that $\det \tilde{\gamma}_{ij}=1$ by construction and explicitly
invert the metric components as for example in
\begin{equation}
  \tilde{\gamma}^{xy} = \tilde{\gamma}_{xz}\tilde{\gamma}_{yz}
       - \tilde{\gamma}_{xy} \tilde{\gamma}_{zz} \, .
\end{equation}
A straightforward calculation gives us the derivatives of the inverse metric
components as follows
\begin{align}
  \partial_y \tilde{\gamma}^{xy}
    & = \tilde{\gamma}_{xz} \partial_y \tilde{\gamma}_{yz}
       - \tilde{\gamma}_{zz} \partial_y \tilde{\gamma}_{xy}
       + \mathcal{O}(y^2) \, , \\
  \partial_y \tilde{\gamma}^{yz} & = \tilde{\gamma}_{xz} \partial_y
      \tilde{\gamma}_{xy} - \tilde{\gamma}_{xx} \partial_y
      \tilde{\gamma}_{yz} + \mathcal{O}(y^2)\, , \\
  \partial_y \tilde{\gamma}^{yy} & = \tilde{\gamma}_{zz} \partial_y
      \tilde{\gamma}_{xx} + \tilde{\gamma}_{xx} \partial_y
      \tilde{\gamma}_{zz} - 2\tilde{\gamma}_{xz} \partial_y
      \tilde{\gamma}_{xz} \, , \\
  \partial_y \partial_y \tilde{\gamma}^{yy} &=
      \tilde{\gamma}_{zz} \partial_y \partial_y \tilde{\gamma}_{xx}
      + \tilde{\gamma}_{xx} \partial_y \partial_y \tilde{\gamma}_{zz}
      - 2\tilde{\gamma}_{xz} \partial_y \partial_y \tilde{\gamma}_{xz}
      + \mathcal{O}(y^2) \, .
      \label{eq:dydygammayy}
\end{align}
The benefit in using these expressions is purely numerical: we do not need to
store the inverse metric in grid functions which reduces the memory requirements
of the simulations.

\section{Geodesic slicing}
\label{geoslice}
In standard Schwarzschild-like coordinates, the Tangherlini metric reads
\begin{align}
  ds^2=-f(r)dt^2+\frac{dr^2}{f(r)}+r^2d\Omega_{D-2} \, ,
      \qquad f(r)=1-\left(\frac{r_S}{r}\right)^{D-3}  \, .
\end{align}
For a radially in-falling massive particle, starting from rest at $r=r_0$, the
energy per unit mass is $\sqrt{f(r_0)}$. The geodesic equation may then be
written as
\begin{align}
  \frac{dt}{d\tau}=\frac{\sqrt{f(r_0)}}{f(r)} \, ,
      \qquad \left(\frac{dr}{d\tau}\right)^2=f(r_0)-f(r) \, .
      \label{geodesics}
\end{align}
In four and five dimensions these equations have simple solutions. In
five dimensions the solutions are
\begin{align}
  t=\sqrt{f(r_0)}\tau +\frac{r_S}{2}\ln \left|\frac{\tau+\sqrt{f(r_0)}
       r_0^2/r_S}{\tau-\sqrt{f(r_0)} r_0^2/r_S}\right| \, , \qquad r^2=r_0^2
       -\left(\frac{r_S}{r_0}\right)^2\tau^2 \, .
\end{align}
Then, performing a coordinate transformation $(t,r)\rightarrow (\tau,
r_0)$ the line element becomes
\begin{align}
  ds^2=-d\tau^2+\frac{\left(r_0^2+\left(\frac{r_S}{r_0}\right)^2
      \tau^2\right)^2}{r_0^2-\left(\frac{r_S}{r_0}\right)^2
      \tau^2}\frac{dr_0^2}{r_0^2f(r_0)}+\left(r_0^2-\left(\frac{r_S}{r_0}
      \right)^2\tau^2\right)d\Omega_3 \, .
\end{align}
This coordinate system encodes a spacetime slicing with zero shift and constant
(unit) lapse, i.e., of type \eqref{eq:HDWEtotalmetric} with
$\alpha=1,\beta^a=0$ for \textit{all} times. To compare it with a numerical
evolution we must have the initial data for the spatial metric written in a
conformally flat form. Taking the initial hypersurface to be $\tau=0$ we see
that this is achieved by a coordinate transformation $r_0\rightarrow R$ with
\begin{align}
  \frac{dR}{R}=\frac{dr_0}{\sqrt{f(r_0)} r_0} \ \ \Rightarrow \ \ r_0(R)=
      R\left(1+\frac{r_S^2}{4R^2}\right) \, .
  \label{rzerotor}
\end{align}
This actually coincides with the standard coordinate transformation from
Schwarzschild to isotropic coordinates in five dimensions. The line element
finally reads \eqref{eq:HDRCCgeodesicmetric}. At the initial hypersurface $\tau=0$,
\begin{align}
ds^2_{\tau=0} = & \left(\frac{r_0(R)}{R}\right)^2\left(dR^2+R^2d\Omega_3\right)
\non\\ & 
        = \left(\frac{r_0(\sqrt{\rho^2+z^2})}{\sqrt{\rho^2+z^2}}\right)^2
          \left(dz^2+d\rho^2+\rho^2d\theta^2+\rho^2\sin^2\theta d\Omega_1 \right)\, ,
\end{align}
where we have used the metric on the $3$-sphere in the form
\begin{align}
  d\Omega_3=d\tilde{\theta}+\sin^2\tilde{\theta}(d\theta^2+\sin^2\theta
       d\Omega_1) \, , \
  \label{S3}
\end{align}
and performed the coordinate transformation $(R,\tilde{\theta})\rightarrow
(\rho,z)$ defined as
\begin{align}
  \rho=R\sin\tilde{\theta} \, , \ \ z=R\cos\tilde{\theta}  \, .
\end{align}
Using \eqref{eq:HDWEtransfc} we get
\begin{align}
  ds^2_{\tau=0}= \left(\frac{r_0(\sqrt{x^2+y^2+z^2})}{\sqrt{x^2+y^2+z^2}}
      \right)^2\left(dx^2+dy^2+dz^2+y^2d\Omega_1\right)\, .
\end{align}
Thus the coordinate transformation from the spherical coordinates $(R,\tilde{\theta},\theta)$
used in Eq.~\eqref{eq:HDRCCgeodesicmetric} to the ``incomplete''
Cartesian coordinates $(x,y,z)$ used in the numerical evolution  is
\begin{align}
  x=R\sin\tilde{\theta}\cos\theta \, , \ \ y=R\sin\tilde{\theta}
      \sin\theta \, , \ \ z=R\cos\tilde{\theta} \, ,
  \label{3dinccart}
\end{align}
which resembles the usual coordinate transformation from spherical polar
coordinates to Cartesian coordinates in $\mathbb{R}^3$; but note that
$\tilde{\theta}$ and $\theta$ are \textit{both} polar angles with range
$[0,\pi]$, which is the manifestation of the Cartesian coordinates
``incompleteness''.

The coordinate change \eqref{3dinccart} brings the five dimensional
Tan\-gher\-li\-ni metric in geode\-sic slicing to a conformally flat form at
$\tau=0$. This matches the initial data for the numerical evolution. One may
ask, however, if the coordinate transformation \textit{evolves}, in order to
compare the analytic form with the numerical evolution. This cannot be the case,
since the existence of $\tau$-dependent terms in the coordinate transformation
would imply a drift away from geodesic slicing. We are thus guaranteed that the
coordinate transformation \eqref{3dinccart} is valid for \textit{all} values of
$\tau$. Then, we can predict the value of the metric components that should be
obtained from the numerical evolution; say $\gamma_{xx}$ should be, at time
$\tau$
\begin{align}
  \gamma_{xx}(\tau,x,y,z)=\frac{x^2g_{RR}(\tau,R)}{R^2}+
      \frac{x^2z^2g_{\tilde{\theta}\tilde{\theta}}(\tau,R)}{R^4(x^2+y^2)}
      +\frac{y^2g_{\theta\theta}(\tau,R)}{(x^2+y^2)^2} \, ,
      \label{generalfig:HDRCCgammaxx}
\end{align}
where $R^2=x^2+y^2+z^2$ and
$g_{RR}(\tau,R),g_{\tilde{\theta}\tilde{\theta}}(\tau,R),
g_{\theta\theta}(\tau,R)$ are readily obtained from Eq.~\eqref{eq:HDRCCgeodesicmetric} with
\eqref{S3} and \eqref{3dinccart}. The result for $\tilde{\gamma}_{xx}$ along the
$x$-axis is plotted in Fig.~\ref{fig:HDRCCgammaxx} for various values of $\tau$.

For $D\ge 6$ the situation is more involved because equations \eqref{geodesics}
can no longer be integrated straightforwardly, but require a numerical
treatment.  First one notices that the coordinate transformation
$(t,r)\rightarrow (\tau,r_0)$, with initial conditions $t(\tau=0)=0$ and
$r(\tau=0)=r_0$, brings the $D$~dimensional Tangherlini metric to the form
\begin{align}
  ds^2=-d\tau^2+\left(\frac{\partial r(\tau,r_0)}{\partial r_0}\right)^2
    \frac{dr_0^2}{f(r_0)}+r^2(\tau,r_0)d\Omega_{D-2} \, .
\end{align}
Then, from the initial conditions, it follows that the coordinate
transformation to isotropic coordinates at $\tau=0$ is
\begin{align}
  \frac{dR}{R}=\frac{dr_0}{\sqrt{f(r_0)} r_0} \ \ \stackrel{D=6}{\Rightarrow} \ \ r_0(R)=\frac{
      R}{r_S}\left(1+\frac{r_S^3}{4R^3}\right)^{2/3} \, .
  \label{rzerotor6}
\end{align}
Writing the metric on the $(D-2)$-sphere as in Eq.~\eqref{S3} (replacing
$d\Omega_1\rightarrow d\Omega_{D-4}$), one concludes that the transformation to
``incomplete'' Cartesian coordinates is still \eqref{3dinccart}. Thus
\eqref{generalfig:HDRCCgammaxx} is still valid, which reduces to, along the $x$-axis
($R=x$):
\begin{align}
\gamma_{xx}(\tau,x,0,0)=g_{RR}(\tau,x)=\frac{r_0(x)^2}{x^2}
    \left(\frac{\partial r(\tau,r_0)}{\partial r_0}\right)^2_{r_0=r_0(x)} \, .
    \label{finalgs}
\end{align}
This expression is valid for any $D$. For $D=6$, $r_0(x)$ is explicitly given by
Eq.~\eqref{rzerotor6}. The derivative in Eq.~\eqref{finalgs} has to be computed
numerically. The result for $\tilde{\gamma}_{xx}$ is plotted, for various values
of $\tau$, in Fig.~\ref{fig:HDRCCgeodesic_d6_hxx}.

\newpage
\section{Coordinate transformation}\label{app:coordtrafo}
In order to extract gravitational radiation using the KI formalism one has to perform a coordinate transformation from 
Cartesian coordinates, which are used during the numerical evolution, to those adapted
for wave extraction.
The physical 3-metric $\gamma_{ij}$, the lapse function $\alpha$ and the shift vector $\beta^i$ computed on our
Cartesian grid are interpolated onto a Cartesian patch. In terms of these quantities we compute the 4-metric
$g_{\mu\nu}$ in Cartesian coordinates according to Eq.~\eqref{eq:HDWEtotalmetric}:
\begin{align}
g_{\mu\nu}dx^{\mu}dx^{\nu} =& (-\alpha^2 + \gamma_{ij} \beta^i\beta^j)dt^2
                             +\gamma_{ij}\beta^i dt dx^j 
                             + \gamma_{ij}\beta^j dt dx^i + \gamma_{ij} dx^i dx^j
\,.
\end{align}
Then, we transform the 4-metric in Cartesian coordinates into spherical coordinates, defined by Eq.~\eqref{eq:HDWEtransfc}
\begin{align}
x = & R\sin\bar\theta\cos\theta 
\,,\quad
y =   R\sin\bar\theta\sin\theta 
\,,\quad
z =   R\cos\bar\theta 
\,, 
\end{align}\label{eq:HDWEtransfc1}
where $\bar\theta,\theta\in[0,\pi]$ and $R=\sqrt{x^2+y^2+z^2}$.
If we denote the metric in spherical coordinates by $g_{\mu\nu}^S$
and define $\rho \equiv \sqrt{x^2 + y^2}$,
the explicit form of the transformation is
\begin{subequations}
\begin{align}
g^S_{tR} =&  g_{tx} \sin\bar{\theta} \cos\theta 
           + g_{ty} \sin\bar{\theta} \sin\theta 
           + g_{tz}\cos\bar{\theta} 
\,,\\
g^S_{t\bar{\theta}} =& z (g_{tx} \cos\theta 
                      +g_{ty}\sin\theta ) 
                      -\rho g_{tz} 
\,,\\
g^S_{t\theta} =&-y g_{tx} + x g_{ty}
\,, \\
g^S_{RR} =& g_{xx} \sin^2\bar{\theta} \cos^2\theta 
           +2g_{xy} \sin^2\bar{\theta} \cos\theta \sin\theta 
           +2g_{xz} \sin\bar{\theta} \cos\theta  \cos\bar{\theta}   
\nonumber\\ &
           +g_{yy} \sin^2\bar{\theta} \sin^2\theta 
           +2 g_{yz}  \sin\bar{\theta} \sin\theta  \cos\bar{\theta} 
           +g_{zz}\cos^2\bar{\theta}  
\,, \\
g^S_{R\bar{\theta}} =& z ( g_{xx} \sin\bar{\theta} \cos^2\theta   
                      +2g_{xy} \sin\bar{\theta} \cos\theta  \sin\theta 
                      +g_{yy} \sin\bar{\theta} \sin^2\theta 
\nonumber\\ &
                      +g_{xz} \cos\bar{\theta}  \cos\theta        
                      +g_{yz}\cos\bar{\theta}  \sin\theta   ) 
                      -(x g_{xz} + y g_{yz} + z g_{zz})\sin\bar{\theta}
\,, \\
g^S_{R\theta} =& ( -y  g_{xx} \sin\bar{\theta} \cos\theta  
                +x g_{xy}  \sin\bar{\theta} \cos\theta                
                -y g_{xy} \sin\bar{\theta} \sin\theta 
\nonumber\\ &
                +x g_{yy} \sin\bar{\theta} \sin\theta                 
                -y g_{xz} \cos\bar{\theta} 
                +x g_{yz} \cos\bar{\theta}  )
\,, \\
g^S_{\bar{\theta}\bar{\theta}} =& z^2 ( g_{xx} \cos^2\theta 
                                       +2 g_{xy} \cos\theta \sin\theta 
                                       +  g_{yy}\sin^2\theta )  
                                 -2z (x g_{xz} + y  g_{yz}) 
                                 +\rho^2  g_{zz} 
\,,\\
g^S_{\bar{\theta}\theta} =& z (-y g_{xx} \cos\theta 
                               +x g_{xy}    \cos\theta      
                               -y g_{xy} \sin\theta  
                               +x g_{yy} \sin\theta )   
                           + \rho  (y g_{xz} - x  g_{yz}) 
\,, \\
g^S_{\theta\theta} =& R^2\sin^2\bar\theta ( g_{xx} \sin^2\theta 
                                           -2g_{xy}  \cos\theta  \sin\theta  
                                           +g_{yy} \cos^2\theta  ) 
\,.
\end{align}
\end{subequations}
Henceforth, we will drop the superscript $S$ and use $g_{\mu\nu}$ for the metric 
in spherical coordinates.

The areal radius $r$ is related to $R$ by a reparametrization $R=R(r)$, given by Eq.~\eqref{reparam}, which depends on
the components $g_{\bar\theta\bar\theta}$, $g_{\theta\theta}$ only. As shown in Secs.~\ref{ssec:HDRCCWE4D} 
and~\ref{sec:D5coordinatecheck},
we find that this reparametrization is nearly constant throughout our numerical simulations.
Therefore, the quantities $g_{rr},g_{tr}$, $g_{r\bar\theta},g_{r\theta}$ can be obtained from $g_{RR},g_{tR},g_{R\bar\theta},g_{R\theta}$
by a simple rescaling: because
\begin{equation}
\frac{dR}{dr}\simeq1 \,,
\end{equation}
we have $g_{rr}\simeq g_{RR}$, and similar relations hold for the other components.

\section{Harmonic expansion of axisymmetric tensors in $D$ dimensions}\label{appintegrals}
As discussed in Section \ref{implax}, scalar spherical harmonics in $D$ dimensions 

${\cal  S}_l(\bar\theta,\theta,\phi^1,\dots,\phi^{D-4})$ are solutions of Eq.~\eqref{eqscal}
\begin{equation}
\Box{\cal S}_l=\gamma^{\bar i\bar j}{\cal S}_{l\,:\bar i\bar j}=-k^2{\cal S}_l \, , 
\label{eqscal2}
\end{equation} 
with $k^2=l(l+D-3)$. Axisymmetric scalar spherical harmonics are functions of the coordinate $\bar\theta$ only, 
${\cal S}_l={\cal S}_l(\bar\theta)$. Therefore, Eq.~\eqref{eqscal2} becomes
\begin{equation}
\Box{\cal S}_l(\bar\theta)={\cal S}_{l\,,\bar\theta\bar\theta}+(D-3)\cot\bar\theta{\cal S}_{l\,,\bar\theta}
=-k^2{\cal S}_l \, ,\label{eqscal3}
\end{equation}
since
\begin{align}
{\cal S}_{l\,:\bar\theta\bar\theta}&={\cal S}_{l\,,\bar\theta\bar\theta} \, \\
{\cal S}_{l\,:\theta\theta}&=-\Gamma_{\theta\theta}^{\bar\theta}{\cal S}_{l\,,\bar\theta}=
\sin\bar\theta\cos\bar\theta{\cal S}_{l\,,\bar\theta} \, ,\\
{\cal S}_{l\,:\phi^1\phi^1}&=-\Gamma_{\phi^1\phi^1}^{\bar\theta}{\cal S}_{l\,,\bar\theta}=
\sin^2\theta \sin\bar\theta\cos\bar\theta{\cal S}_{l\,,\bar\theta}\, , 
\end{align}
etc. The quantities ${\cal S}_{l\,\bar i\bar j}$ defined in Eq.~\eqref{defharm} are then
\begin{align}
{\cal S}_{l\,\bar i\bar j} =& \frac{1}{k^2}{\cal S}_{l\,:\bar i\bar j}+\frac{1}{D-2}\gamma_{\bar i\bar j}{\cal S}_l 
  \nonumber\\
  =& \frac{1}{k^2(D-2)}\left((D-2){\cal S}_{l\,:\bar i\bar j}+k^2 \gamma_{\bar i\bar j}{\cal S}_l\right)
  \nonumber\\
  =& \frac{1}{k^2(D-2)}{\rm diag}\left((D-3){\cal W}_l,
     -\sin^2\bar\theta {\cal W}_l,-\sin^2\bar\theta\sin^2
     \theta {\cal W}_l,\dots \right)
\label{defharm2}
\end{align}
where 
\begin{equation}
{\cal W}_l(\bar\theta) = {\cal S}_{l\,,\bar\theta\bar\theta}-\cot\bar\theta{\cal S}_{l\,,\bar\theta}
=\sin\bar\theta\left(\frac{{\cal S}_{l\,,\bar\theta}}{\sin\bar\theta}\right)_{,\bar\theta}\,.\label{defww}
\end{equation}
Indeed, using Eq.~\eqref{eqscal3} one finds
\begin{align}
k^2(D-2) {\cal S}_{l\,\bar\theta\bar\theta} =& (D-2)  {\cal S}_{l\,,\bar\theta\bar\theta}+k^2{\cal S}_l 
  = (D-3)( {\cal S}_{l\,,\bar\theta\bar\theta}-\cot\bar\theta{\cal S}_{l\,,\bar\theta})
\,, \\
k^2(D-2) {\cal S}_{l\,\theta\theta} = & (D-2) {\cal S}_{l\,,\theta\theta}+k^2\sin^2\bar\theta{\cal S}_l 
  =  \sin^2\bar\theta((D-2)\cot\bar\theta{\cal S}_{l\,,\bar\theta}+k^2{\cal S}_l)
\nonumber\\
  = &  \sin^2\bar\theta(-{\cal S}_{l\,,\bar\theta\bar\theta}+\cot\bar\theta
     {\cal S}_{l\,,\bar\theta}) 
\,,
\end{align}
and therefore
\begin{align}
{\cal S}_{l\,\bar\theta\bar\theta}=&\frac{D-3}{k^2(D-2)}{\cal W}_l 
\,,\quad
{\cal S}_{l\,\theta\theta} = -\frac{\sin^2\bar\theta}{k^2(D-2)}{\cal W}_l 
\, ,
\end{align}
and likewise for the other components.

Axisymmetric scalar spherical harmonics, as discussed in Sec. \ref{implax}, can be written in terms of Gegenbauer
polynomials (cf. \eqref{gp}):
\begin{equation}
{\cal S}_l(\bar\theta)=(K^{lD})^{-1/2}C_l^{(D-3)/2}(\cos\bar\theta) \,.\label{defSC}
\end{equation} 
If we define
\begin{equation}
W_l(\cos\bar\theta)=C_{l\,,\bar\theta\bar\theta}^{(D-3)/2}(\cos\bar\theta) -\cot\bar\theta C_{l\,,\bar\theta}^{(D-3)/2}(\cos\bar\theta) \,,
\end{equation}
we have
\begin{equation}
{\cal W}_l(\bar\theta)=(K^{lD})^{-1/2}W_l^{(D-3)/2}(\cos\bar\theta) \,.\label{defWC}
\end{equation}

We impose the normalization \eqref{normalK} 
\begin{equation}
\int d\Omega^{D-2}{\cal S}_l{\cal S}_{l'}=\delta_{ll'}\,,~~~
\int d\Omega^{D-2}{\cal S}_{l\,,\bar\theta}{\cal S}_{l'\,,\bar\theta}=\delta_{ll'}k^2\,.\label{normalK1}
\end{equation}
Using
\begin{align}
\int_0^{\pi}d\bar\theta (\sin\bar\theta)^{D-3}
C_l^{(D-3)/2}(\cos{\bar\theta}) C_{l'}^{(D-3)/2}(\cos{\bar\theta}) = & 
\delta_{ll'}\hat K^{lD}
\, ,\\
\int_0^{\pi}d\bar\theta (\sin\bar\theta)^{D-3}  
C_{l\,,\bar\theta}^{(D-3)/2}(\cos{\bar\theta}) C_{l'\,,\bar\theta}^{(D-3)/2}(\cos{\bar\theta}) = &
\delta_{ll'}k^2\hat K^{lD} 
\, , 
\end{align} 
and
\begin{equation}
\hat K^{lD}=\frac{2^{4-D}\pi\Gamma(l+D-3)}{\left(l+\frac{D-3}{2}\right)\left(\Gamma\left(\frac{D-3}{2}
\right)\right)^2\Gamma(l+1)}\,,\label{defhK}
\end{equation}
we have
\begin{equation}
K^{lD}=\hat K^{lD}{\cal A}_{D-3}\,,\label{defK}
\end{equation}
where
\begin{equation}
{\cal A}_{D-3}= \frac{2\pi^{(D-2)/2}}{\Gamma\left(\frac{D-2}{2}\right)} \, ,
\label{eq:HDWEareasphere}
\end{equation}
is the surface of the $(D-3)$-sphere $S^{D-3}$. 

Note, that 
$\int d\Omega^{D-2}(\cdots)={\cal A}_{D-3}\int d\bar\theta (\sin\bar\theta)^{D-3}(\cdots)$.
With the definitions (\ref{defharm}) ${\cal S}_{l\,\bar i}=-\frac{1}{k}{\cal S}_{l\,,\bar i}$,
\begin{align}
\label{intS0}
 \int_0^{\pi}d\tilde\theta (\sin\bar\theta)^{D-3}
 {\cal S}_l(\tilde\theta) {\cal S}_{l'}(\bar\theta) = & \delta_{ll'}{\cal A}_{D-3}^{-1}
\,,\\
\label{intS}
\int_0^{\pi}d\tilde\theta (\sin\tilde\theta)^{D-3}\gamma^{\bar i\bar j}
{\cal S}_{l\,\bar i}{\cal S}_{l'\,\bar j} 
 = & \int_0^{\pi}d\tilde\theta (\sin\tilde\theta)^{D-3}
     {\cal S}_{l\,\bar\theta}(\bar\theta){\cal S}_{l'\,\bar\theta}(\bar\theta) 
 =   \delta_{ll'}{\cal A}_{D-3}^{-1}
\,.
\end{align}
Furthermore, we note that Eqs.~\eqref{eqscal3} and \eqref{defww} imply
\begin{equation}
{\cal W}_l+(D-2)\cot\bar\theta{\cal S}_{l\,,\bar\theta}+k^2{\cal S}_l=0 \,,
\end{equation} 
so that
\begin{align}
& {\cal W}_{l\,,\bar\theta}+(D-2)\cot\bar\theta{\cal S}_{l\,,\bar\theta\bar\theta}
-\frac{D-2}{\sin^2\bar\theta}{\cal S}_{l\,,\bar\theta}+k^2{\cal S}_{l\,,\bar\theta}
\nonumber\\
& = {\cal W}_{l\,,\bar\theta}+(D-2)\cot\bar\theta{\cal W}_{l}+(k^2-D+2){\cal S}_{l\,,\bar\theta}
\,,
\end{align}
and therefore
\begin{align}
\int_0^\pi d\bar\theta(\sin\bar\theta)^{D-3}{\cal W}_l {\cal W}_{l'}
 = & \int_0^\pi d\bar\theta(\sin\bar\theta)^{D-3}{\cal W}_l\sin\bar\theta
 \left(\frac{{\cal S}_{l'\,,\bar\theta}}{\sin\bar\theta}\right)_{,\bar\theta}\nonumber\\
 =&-(D-2) \int_0^\pi d\bar\theta(\sin\bar\theta)^{D-3}{\cal W}_l\cot\bar\theta{\cal S}_{l'\,,\bar\theta}
\nonumber\\ &
   -\int_0^\pi d\bar\theta(\sin\bar\theta)^{D-3}{\cal W}_{l\,,\bar\theta}{\cal S}_{l'\,,\bar\theta}
\nonumber\\
=&(k^2-D+2)\int_0^\pi d\bar\theta(\sin\bar\theta)^{D-3}{\cal S}_{l\,,\bar\theta}{\cal S}_{l'\,,\bar\theta}\nonumber\\
=&\delta_{ll'}{\cal A}_{D-3}^{-1}\,k^2(k^2-D+2)\,.\label{intS1}
\end{align}
We thus obtain
\begin{equation}
\int d\Omega^{D-2}{\cal W}_l {\cal W}_{l'}=\delta_{ll'}k^2(k^2-D+2)\,.\label{normalK2}
\end{equation}
The perturbations $f^l_{ab}(t,r)$, $f^l_{a}(t,r)$, $H^l_L (t,r)$, $H^l_T (t,r)$ appearing in the expansion 
of the metric perturbations \eqref{eq:HDWEmpert}
\begin{align}
h_{ab}&=f^l_{ab}{\cal S}_l(\bar\theta)\, , \\
h_{a\bar\theta}&=-\frac{1}{k} rf^l_{a}{\cal S}_l(\bar\theta)_{,\bar\theta}\,  ,\\
h_{\bar\theta\bar\theta}&=2r^2\left(H^l_L {\cal S}_l(\bar\theta)
+H^l_T \frac{D-3}{k^2(D-2)}{\cal W}_l(\bar\theta)\right)\, , \\
h_{\theta\theta}&=2r^2\sin^2\bar\theta\left(H^l_L {\cal S}_l(\bar\theta)
-H^l_T \frac{1}{k^2(D-2)}{\cal W}_l(\bar\theta)\right) \, .
\end{align}\label{mpert3}
are given by the following integrals, as follows from Eqs.~\eqref{defSC}, \eqref{defWC}, \eqref{normalK1}, \eqref{normalK2}:
\begin{subequations}
\begin{align}
f^l_{ab}(t,r) =&\int d\Omega^{D-2}h_{ab}{\cal S}_l 
 = \frac{{\cal A}_{D-3}}{\sqrt{K^{lD}}}\int_0^{\pi}
   d\bar\theta (\sin\bar\theta)^{D-3}h_{ab} C_l^{(D-3)/2}
\, , \\
f_{a}(t,r) =&-\frac{1}{kr}\int d\Omega^{D-2}h_{a\bar\theta}{\cal S}_{l\,,\bar\theta} 
 = -\frac{1}{kr}\frac{{\cal A}_{D-3}}{\sqrt{K^{lD}}}\int_0^{\pi}d\bar\theta (\sin\bar\theta)^{D-3} 
    h_{a \bar \theta }C_{l\,,\bar\theta}^{(D-3)/2}
\, ,\\
H_L(t,r) =& \frac{1}{2(D-2)r^2}\int d\Omega^{D-2}\left[h_{\bar\theta\bar\theta}
           +\frac{D-3}{\sin^2\bar\theta}h_{\theta\theta}\right] {\cal S}_l 
\nonumber\\
         =& \frac{1}{2(D-2)r^2}\frac{{\cal A}_{D-3}}{\sqrt{K^{lD}}}\int_0^{\pi}d\bar\theta 
            (\sin\bar\theta)^{D-3} 
            \times \left[h_{\bar\theta\bar\theta}+\frac{D-3}{\sin^2\bar\theta}h_{\theta\theta}\right]
            C_l^{(D-3)/2}
\, ,\\
H_T(t,r) =& \frac{1}{2r^2(k^2-D+2)}\int d\Omega^{D-2} 
            \left[h_{\bar\theta\bar\theta}-\frac{1}{\sin^2\bar\theta}h_{\theta\theta}\right]{\cal W}_l 
\nonumber\\
         =& \frac{1}{2r^2(k^2-D+2)}\frac{{\cal A}_{D-3}}{\sqrt{K^{lD}}}\int_0^{\pi}d\bar\theta (\sin\bar\theta)^{D-3} 
           \left[h_{\bar\theta\bar\theta}-\frac{1}{\sin^2\bar\theta}h_{\theta\theta}\right] W_l\,,
\label{integralsfH}
\end{align}
\end{subequations}
where 
$h_{ab} = h_{ab}(t,r,\bar\theta)$, $h_{a\bar\theta} = h_{a\bar\theta}(t,r,\bar\theta)$, 
$h_{\bar\theta\bar\theta} = h_{\bar\theta\bar\theta}(t,r,\bar\theta)$, 
$h_{\theta\theta} = h_{\theta\theta} (t,r,\bar\theta)$, 
$C_l^{(D-3)/2} = C_l^{(D-3)/2}(\cos{\bar\theta})$ and $W_l = W_l(\cos\bar\theta)$.

We also note that the background Tangherlini metric depends on the $l=0$ harmonic only; 
the integral of its components over $l\ge2$ harmonics vanish. 
Therefore, if we decompose the spacetime metric (see Appendix
\ref{app:coordtrafo}) as $g_{\mu\nu}=g^{(0)}_{\mu\nu}+h_{\mu\nu}$ with $\mu,\nu=(t,r,\bar\theta,\theta)$ and
$g^{(0)}_{\mu\nu}$ is the Tangherlini background metric, we can compute the integrals (\ref{integralsfH}) in terms of the metric
$g_{\mu\nu}$
\begin{subequations}
\begin{align}
\label{eq:pertftt}
f_{tt} =& \frac{1}{\pi} \frac{{\cal A}_{D-3}}{\sqrt{K^{lD}}}
               \int d\bar{\theta}
               (\sin\bar{\theta})^{D-3} C_l^{(D-3)/2}
               \int d\theta 
               g_{tt}(\bar{\theta},\theta) 
\, ,\\
\label{eq:pertftr}
f_{tr} =& \frac{1}{\pi} \frac{{\cal A}_{D-3}}{\sqrt{K^{lD}}}
               \int d\bar{\theta}
               (\sin\bar{\theta})^{D-3} C_l^{(D-3)/2}
               \int d\theta 
               g_{tr}(\bar{\theta},\theta) 
\, , \\
\label{eq:pertfrr}
f_{rr} =& \frac{1}{\pi} \frac{{\cal A}_{D-3}}{\sqrt{K^{lD}}} 
               \int d\bar{\theta}
               (\sin\bar{\theta})^{D-3} C_l^{(D-3)/2}
               \int d\theta 
               g_{rr}(\bar{\theta},\theta) 
\, ,\\
\label{eq:pertfft}
f_{t} =& -\frac{1}{k r \pi} 
          \frac{{\cal A}_{D-3}}{\sqrt{K^{lD}}} 
          \int d\bar{\theta}
                (\sin\bar{\theta})^{D-3} \partial_{\bar{\theta}} C_l^{(D-3)/2}
                \int d\theta  
                g_{t\bar{\theta}}(\bar{\theta},\theta)
\, , \\
\label{eq:pertffr}
f_{r} =& -\frac{1}{ kr \pi} 
          \frac{{\cal A}_{D-3}}{\sqrt{K^{lD}}} 
          \int d\bar{\theta}
               (\sin\bar{\theta})^{D-3} \partial_{\bar{\theta}} C_l^{(D-3)/2}
               \int d\theta  
               g_{r\bar{\theta}}(\bar{\theta},\theta)
\, , \\
\label{eq:pertHL}
H_L =& \frac{1}{ 2 (D-2) r^2 \pi } 
       \frac{{\cal A}_{D-3}}{\sqrt{K^{lD}}} 
       \int d\bar{\theta}
               (\sin\bar{\theta})^{D-3} C_l^{(D-3)/2} 
\nonumber\\    & \times 
        \int d\theta 
              \left(   g_{\bar{\theta}\bar{\theta}}(\bar{\theta},\theta) 
                + (D-3) \frac{g_{\theta\theta}(\bar{\theta},\theta)}{\sin^2\bar\theta}  \right)
\, ,\\
\label{eq:pertHT}
H_T =& \frac{1}{ 2 (k^2 - D + 2) r^2 \pi } 
       \frac{{\cal A}_{D-3}}{\sqrt{K^{lD}}} 
\nonumber\\    & \times 
                \int d\bar{\theta}
               (\sin\bar{\theta})^{D-3} W_l
               \int d\theta 
               \left(  g_{\bar{\theta}\bar{\theta}}(\bar{\theta},\theta) 
                - \frac{g_{\theta\theta}(\bar{\theta},\theta)}{\sin^2\bar\theta} \right) 
\, .
\end{align}
\end{subequations}
Furthermore, from Eqs.~\eqref{eq:HDWEfunc1} and~\eqref{eq:pertftt}-\eqref{eq:pertHT} we deduce
\begin{align}
\label{eq:Ft}
F_{,t} =&  \partial_t H_L + \frac{1}{D-2} \partial_t H_T 
         + \frac{1}{k}f(r)\left( \partial_t f_r 
         + \frac{r}{k}\partial_t \partial_r H_T \right)\, , \\
\label{eq:Furt}
F^r_t =& f(r) \left( f_{rt} + \frac{r}{k} (\partial_t f_r + \partial_r f_t)
               +\frac{1}{k} f_t \right. \nonumber\\
       & \left. + \frac{2r}{k^2}(\partial_t H_T + r\partial_t\partial_r H_T)\right)  
        -\frac{r}{k}\partial_r f(r)\left( f_t + \frac{r}{k}\partial_t H_T \right)\,.
\end{align}
Conversely, since the perturbations do not depend on the $l=0$ harmonic, the background metric $g_{\mu\nu}$ can be
obtained as follows:
\begin{align}
 g_{tt}^{(0)} &=\frac{1}{K^{0D}\pi} \int_0^{\pi} d\bar{\theta} \sin^{D-3}\bar{\theta} 
  \int_0^{\pi} d \theta  g_{tt}(\bar{\theta},\theta) \,, \\
 g_{tr}^{(0)}&=0=\frac{1}{K^{0D}\pi} \int_0^{\pi} d\bar{\theta} \sin^{D-3}\bar{\theta} 
  \int_0^{\pi} d \theta  g_{tr}(\bar{\theta},\theta) \,, \\
g_{rr}^{(0)} &= \frac{1}{K^{0D}\pi} \int_0^{\pi} d\bar{\theta} \sin^{D-3}\bar{\theta} 
  \int_0^{\pi} d \theta  g_{rr}(\bar{\theta},\theta) \,.
\end{align}
Finally, to compute the areal radius $r$ we note that $g_{\bar\theta\bar\theta}=r^2+h_{\bar\theta\bar\theta}$ and
$g_{\theta\theta}=r^2\sin^2\bar\theta+h_{\theta\theta}$. Both the perturbations $h_{\bar\theta\bar\theta}$ and
$h_{\theta\theta}$ contain harmonics of different type (${\cal S}_l$, ${\cal S}_{l\,,\bar i\bar j}$); to extract the
background we need the combination in Eq.~\eqref{eq:pertHL}:
\begin{align}
\label{reparam}
r^2 =& \frac{1}{(D-2)K^{0D}\pi} \int_0^{\pi} d\bar{\theta} \sin^{D-3}\bar{\theta} \int_0^\pi d\theta 
        \left[g_{\bar\theta\bar\theta}+(D-3)\frac{g_{\theta\theta}}{\sin^2\bar\theta}\right]
\,.
\end{align}

\chapter{Appendix -- Black holes in a box}
\section{Evolutions in a cubic-shaped box}\label{sec:appcubebox}

We have also performed simulations using condition~\eqref{eq:Boxrefbc} on a cubic
outer boundary.  These simulations represent the inspiral of nonspinning BH
binaries with initial separation $d/M = 6.514$ and initial linear momentum
$P_{y_i}/M = \pm 0.133$.  
\begin{table}[t!bh]
\begin{center}
\begin{tabular}{|c|c|c|c|c|}
\hline
Run  &  Grid Setup &  $h_f/M$  & $R_B/M$   & b.c. \\
\hline
O1 & (192, 96, 48, 24, 12, 6)(1.5, 0.75) & $1/40$ & $192$ & O \\
\hline
C1.1 & (24,12,6)(1.5, 0.75) & $1/40$ & $24$ & R \\
C1.2 & (24,12,6)(1.5, 0.75) & $1/44$ & $24$ & R \\
C1.3 & (24,12,6)(1.5, 0.75) & $1/48$ & $24$ & R \\
\hline
\end{tabular}
\caption{\label{tab:Boxcubicruns} Parameters for a set of models evolved using a
  cubical boundary with reflective boundary condition (``R b.c.'') for models C1.1, C1.2 and
  C1.3 and an outgoing (``O b.c.'') Sommerfeld condition for model O1.  }
\end{center}
\end{table}
The grid setup for these runs is listed in Table
\ref{tab:Boxcubicruns} together with a {\em reference} model O1 which describes the
inspiral of the same binary in an asymptotically flat spacetime using outgoing
radiation boundary conditions.
Gravitational waves have been extracted at $r_{\rm ex}=20\,M$
in the form of the Newman-Penrose scalar $\Psi_4$.
In Fig.~\ref{fig:BoxComp24cart}, we compare the $l=m=2$ mode of $\Psi_4$ obtained
by the evolution of models C1.1 and O1.
 In Fig.~\ref{fig:BoxComp24cart} we present
the convergence plot of the $l=m=2$ mode of $\Psi_4$ obtained from
evolutions of models C1.1, C1.2 and C1.3. The difference between the
medium and fine resolution result has been amplified by the factor
$Q=1.58$ corresponding to fourth-order convergence. While the
overall convergence is about fourth-order as in the case of a
spherical shell,
the cubical outer boundary
introduces a substantial amount of numerical noise which is also demonstrated
in Fig.~\ref{fig:BoxComparePsi4_lm} which displays the $l=2$, $m=2$ and
the $l=4$, $m=4$ multipoles of $\Psi_4$ for models C1.3 and IN1.
We believe that this is at least partly a consequence of mode mixing
in the case of the cubical boundary shell which is not well-suited for
the geometric shape of the gravitational wave pulse. For this reason,
we have exclusively used a spherical shell in the main part of 
Chapter~\ref{chapter:BBbox}.
\begin{figure}[h!tpb]
\begin{center}
\begin{tabular}{cc}
\hspace{-0.5cm}
\includegraphics[clip=true,width=0.45\textwidth]{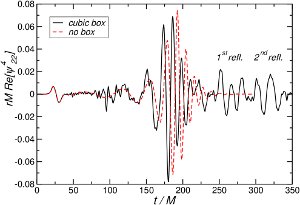} &
\includegraphics[clip=true,width=0.45\textwidth]{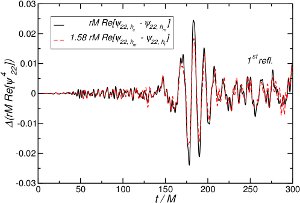}
\end{tabular}
\end{center}
\caption{\label{fig:BoxComp24cart} 
Left:
Comparison of the $l=2,m=2$ modes of
$\Psi_4$ obtained for models C1.1 (solid) and O1 (dashed curve).
The expected range in time for subsequent wave pulses
resulting from first and second reflections are indicated in the figure.
Right:
Convergence analysis of the $l=m=2$
mode of
$\Psi_4$ obtained for model {\it C1} of Table \ref{tab:Boxcubicruns}.
}
\end{figure}

\begin{figure}[h!tbp]
\begin{center}
\begin{tabular}{cc}
\hspace{-0.5cm}
\includegraphics[width=0.45\textwidth]{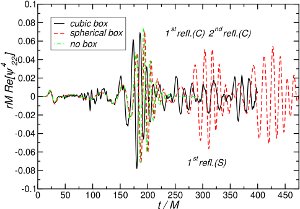} &
\includegraphics[width=0.45\textwidth]{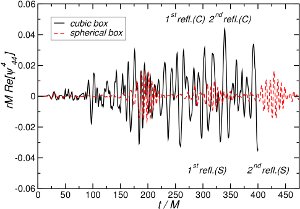}
\end{tabular}
\end{center}
\caption{\label{fig:BoxComparePsi4_lm} 
Comparison of the $l=m=2$ (left) and $l=m=4$ (right) modes of $\Psi_4$ obtained
for models C1.3 (cubic boundary), IN1 (spherical boundary) and O1 (outgoing condition).
}
\end{figure}
\clearpage
\section{Snapshots of black holes in a box}\label{ssec:appsnapshotsBox}
In this section we present snapshots of the simulation that evolves BH binaries enclosed by a mirror-like box,
presented in Chapter~\ref{chapter:BBbox}.
In Fig.~\ref{fig:Boxsnapshots} we illustrate the emission of the gravitational
wave signal during the inspiral and merger
and its evolution in the closed (confined) system containing a central,
spinning BH.
We display snapshots of the waveforms by superposing (the
real part of) $\Psi_0$ and $\Psi_4$ as obtained for
model VIS of Table \ref{tab:Boxsphereruns}.
We show a slice of the orbital plane with $x, y = -48M,...,48M$
during an interval $t/M = 150, ..., 540$.
The difference in time between the individual pictures is $\Delta t/M = 10$.
The series of snapshots starts in the late inspiral phase shortly before
the plunge
and we see a strong gravitational wave signal that is emitted throughout
the merger (first row and first two columns of the second row).
This signal reaches the spherical boundary and is reflected back as
can be seen in the final three panels of the second row and first two
panels of the third row.
Starting with the third snapshot in the third row we see a second pulse
going outwards again after it has been scattered off the BH.
This process is repeated several times and the series of snapshots ends
with the
fourth outgoing wave pulse. An animation constructed from the numerical data can be found in~\cite{webpage}.
\begin{figure}[p]
\begin{center}
\includegraphics[width=0.9\textwidth]{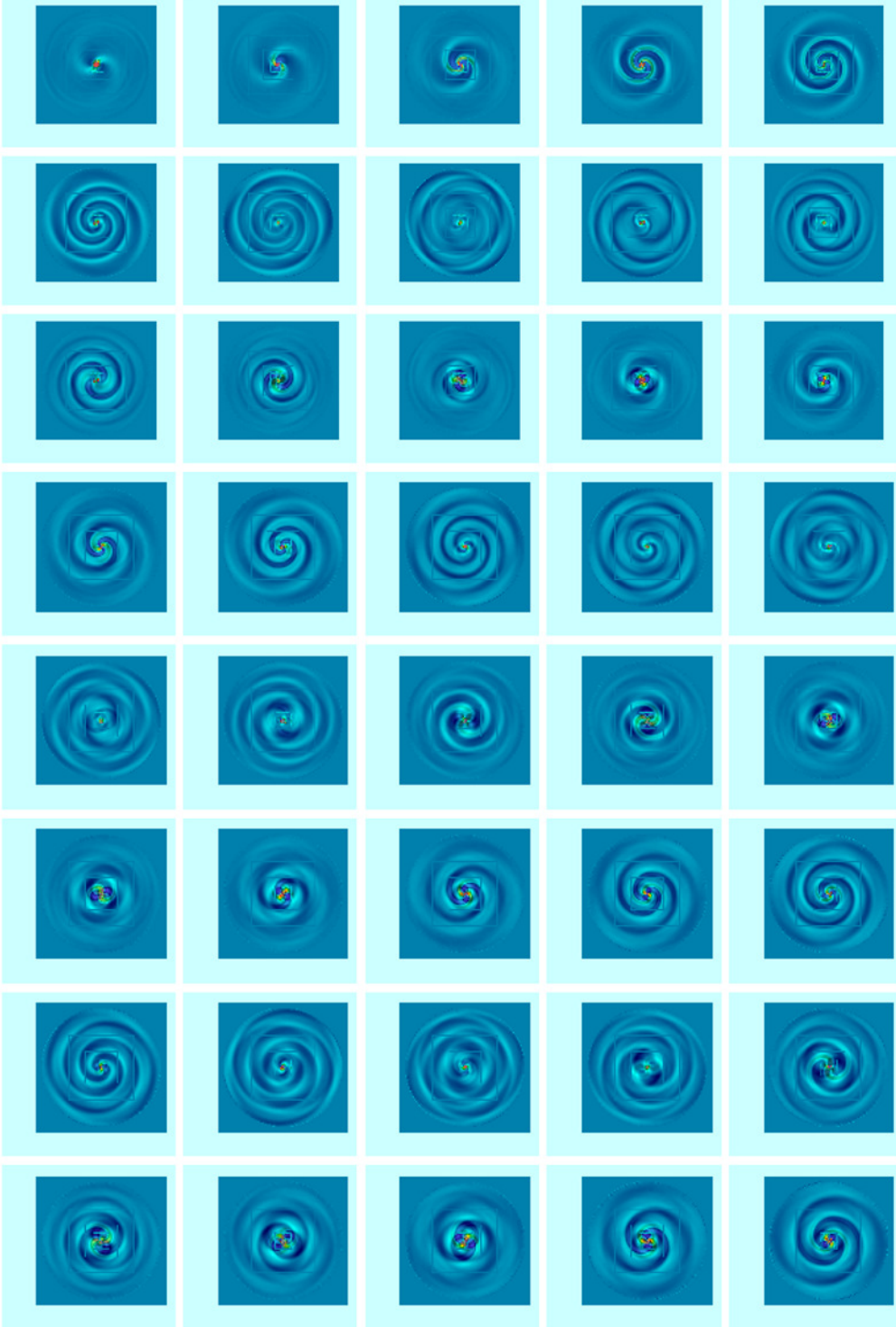} 
\end{center}
\caption{\label{fig:Boxsnapshots} 
Snapshots of $\Re(\Psi_4)$ superposed by $\Re(\Psi_0)$.
The snapshots show the evolution from  $t=150M$ until $t=540M$ and
have a time interval of $\Delta t = 10M$. We show a slice of the orbital
plane with both coordinates going from $-48M,...,48M$. 
}
\end{figure}
%



\bibliographystyle{h-physrev4}
\bibliography{Thesis_Bib}

\end{document}